\documentclass{ws-ijmpa}
\usepackage[super,compress]{cite}
\usepackage{graphicx}
\usepackage{hyperref}
\usepackage{amsmath}
\usepackage{amssymb}
\usepackage{amsfonts}
\usepackage{pdflscape}
\usepackage{subfigure}
\usepackage{xspace}
\usepackage{color}
\usepackage{mathbbol}
\usepackage{dsfont}
\usepackage{bbm}
\usepackage{bm}
\include{tetra-sym}
\usepackage{multirow}

\newcommand{\vett}[1]{\boldsymbol{#1}}
\newcommand{\qbar}{\bar q}
\newcommand{\cbar}{\bar c}

\begin{document}
\markboth{Esposito, Guerrieri, Piccinini, Pilloni, Polosa}
{Four-Quark Hadrons: an Updated Review}

\catchline{}{}{}{}{}

\title{Four-Quark Hadrons: an Updated Review}

\author{ANGELO ESPOSITO}

\address{Department of Physics, 538W 120th Street\\
Columbia University, New York, NY, 10027, USA\\
aesposito2458@columbia.edu}

\author{ANDREA L. GUERRIERI}

\address{Dipartimento di Fisica and INFN, Universit\`a di Roma `Tor Vergata'\\
 Via della Ricerca Scientifica 1, I-00133 Roma, Italy\\
andrea.guerrieri@roma2.infn.it}

\author{FULVIO PICCININI}

\address{INFN Pavia, Via A. Bassi 6, I-27100 Pavia, Italy\\
fulvio.piccinini@pv.infn.it}

\author{ALESSANDRO PILLONI and ANTONIO D. POLOSA}
\address{Dipartimento di Fisica and INFN, `Sapienza' Universit\`a di Roma\\
P.le Aldo Moro 5, I-00185 Roma, Italy\\
alessandro.pilloni@roma1.infn.it, antonio.polosa@roma1.infn.it}

\maketitle

\begin{history}
\received{Day Month Year}
\revised{Day Month Year}
\end{history}

\begin{abstract}
The past decade witnessed a remarkable proliferation of exotic charmonium-like resonances discovered at accelerators. In particular, the recently observed charged states are clearly not interpretable as $q\bar q$ mesons. Notwithstanding the considerable advances on the 
experimental side, conflicting theoretical descriptions do not seem to provide a definitive picture about the nature of the so called $XYZ$ particles.  We present here a comprehensive review about this intriguing topic, discussing both those experimental and theoretical aspects which we consider relevant to make further progress in the field.  At this state of progress, $XYZ$ phenomenology speaks in favour of the existence of compact four-quark particles (tetraquarks) and we believe that realizing this instructs us in the quest for a firm theoretical framework.

\end{abstract}

\ccode{PACS numbers: 14.40.Pq, 14.40.Rt.}
\clearpage
\tableofcontents

\section{Introduction}
\label{sec:intro}
Since the discovery of the $X(3872)$,  a decade ago, more than 20 new charmonium-like resonances have been registered. Most of them have features which do no match what expected from standard charmonium theory. A few resonances have been found in the beauty sector too. Some authors just claim that most of the so called $XYZ$ states  are not even resonances but kind of effects of kinematical or dynamical origin, due to the intricacies of strong interactions. According to them, data analyses are na\"{i}vely describing and fitting as resonances what are indeed the footprints of such  complicated effects. 

On the other hand, the $X(3872)$, for example, is an extremely narrow state, $\Gamma\lesssim 1$~MeV, and it is very difficult, in our understanding,  to imagine how this could be described with some sort of strong rescattering mechanism. We do not know of other clear examples of such phenomena in the field of  high-energy physics and  in this review we will give little space to this kind of interpretations, which we can barely follow. We shall assume instead that what experiments agree to be a resonance is indeed a resonance. 

Moreover, we find very confusing the approach of mixing the  methods proper of nuclear theory to discuss what we learned with the observations of  $XYZ$ resonances especially at Tevatron and LHC. It is true that $X$ seems to be an extreme  version of deuterium as its mass happens to be fine-tuned on the value of the $D^0D^{0*}$ threshold, but one cannot separate this observation from the fact that $X$ is observed at CMS after imposing kinematical transverse momentum cuts as  large as $p_T\simeq 15$~GeV on hadrons produced.  
Is there any evidence of a comparable prompt production of deuterium within  the same kinematical cuts, in the same experimental conditions? The ALICE experiment could provide in the near future a compelling measurement of this latter  rate (and some preliminary estimates described in the text are informative of what the result will be). 

Some of the $XYZ$, those happening to be close to some threshold, are interpreted as loosely-bound molecules, regardless of the great difficulties in explaining their production mechanisms in high energy hadron collisions. Some of them are described just as bound hadron  molecules, once they happen to be below a close-by open flavor meson threshold. Other ones, even if sensibly above the close-by thresholds, have been interpreted as molecules as well: in those cases subtle mistakes in the experimental analysis of the mass have been advocated. 

As a result the field of the theoretical description of $XYZ$ states appears as an heterogeneous mixture of ad-hoc explanations, mainly post-dictions and contradictory statements which is rather confusing to the experimental community and probably self-limiting in the direction of making any real progress. 

It is our belief instead that a more simple and {\it fundamental} dynamics is at work in the hadronization of such particles. More quark body-plans occur with respect to usual mesons and baryons: compact tetraquarks. The diquark-antidiquark model in its updated version, to be described in Section.~7, is just the most simple and economical description (in terms of new states predicted) that we could find and we think that the recent confirmation of $Z(4430)^+$ especially, and of some more related charged $J^{PG}=1^{++}$ states, is the smoking gun for the intrinsic validity of this idea.    

The charged $Z(4430)$ was the most uncomfortable state for  the molecular interpretation for at least two reasons: $i)$ it is charged and molecular models have never provided any clear and consistent prediction about charged states; $ii)$ it is far from open charm thresholds. However, if what observed (by \belle first and confirmed very recently by \lhcb) is not an ``effect'' but a real resonance,  we should find the way to explain and put it in connection to all other ones. 

The $Z(4430)$ appears extremely natural in the diquak-antidiquark model, which in general was the only approach strongly suggesting the existence of charged states years before their actual discovery. 

We think otherwise that open charm/bottom meson thresholds should likely play a role in the formation of $XYZ$ particles. We resort to the Feshbach resonance mechanism, as mediated by some classic studies in atomic physics, to get a model on the nature of this role. The core of our preliminary analysis, as discussed in \sectionname{~\ref{sec:tetraquarks}}, is the postulated  existence of a discrete spectrum of compact tetraquark levels in the fundamental strong interaction Hamiltonian. The occurrence of open charm/beauty meson thresholds in the vicinity of any of these levels might result in an enhanced probability of resonance formation.   

Tetraquarks and multiquarks in general, have been for a long time expected to be extremely broad states on the basis of large-$N$ QCD considerations -- see \sectionname{~\ref{sec:largeN}}. A recent discussion has removed this theoretical obstacle suggesting that even tetraquarks might have order $1/\sqrt{N}$ decay amplitudes for they occur as poles in the connected diagrams of the $1/N$ expansion.

Besides this, we underscore that a genuine tetraquark appears in the physical spectrum, the $Z(4430)$, as discussed at the beginning of \sectionname{~\ref{sec:experimental}}, which is devoted to a comprehensive experimental overview. 

Lattice studies have also started to play a role in the $XYZ$ field, but appear to be still in their infancy, as discussed in \sectionname{~\ref{sec:lattice}}. Lattices of $2\div3$~fm in size cannot by definition allow loosely bound molecules and it is not yet tested how those deeply bound lattice-hadron-molecules, that some studies claim to observe in lattice simulations, will tend to become loosely bound states in some large volume limit.  Moreover it is not clear how one can safely distinguish on the lattice between a tetraquark operator, a standard charmonium and a meson-meson operator, as they all happen to mix with each other.  

\sectionname{~\ref{sec:pheno}} is devoted to the description of the various phenomenological models in the literature: mainly nuclear-theory inspired molecular models, hybrids, hadro-quarkonia. 

The discriminative problem of producing loosely bound molecules at hadron colliders is discussed in \sectionname{~\ref{sec:prompt}} and this is considered as one of the most compelling motivations to go towards compact tetraquarks, to be described in \sectionname{~\ref{sec:tetraquarks}}. 

More exotic states are discussed in \sectionname{~\ref{sec:exotic}}, inspired by the problem of simulating tetraquarks on lattices. Some hints from the physics of heavy-ion collisions are also considered.  

\section{Four-Quark states in Large-$N$ QCD}
\label{sec:largeN}
\subsection{A short guide to Large-$N$ QCD}
\begin{figure}
\centering
\subfigure[]{
\includegraphics[scale=0.6]{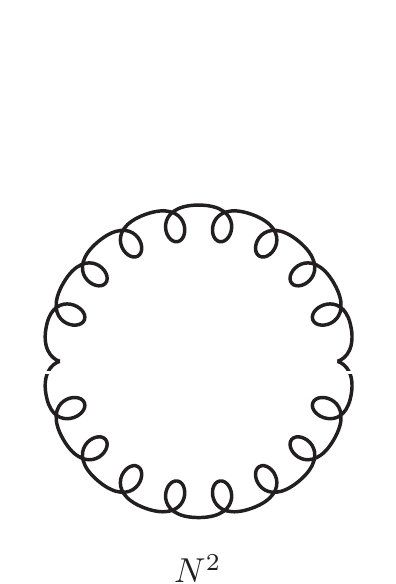}
\label{fig:Nrules(a)}
}
\subfigure[]{
\includegraphics[scale=0.6]{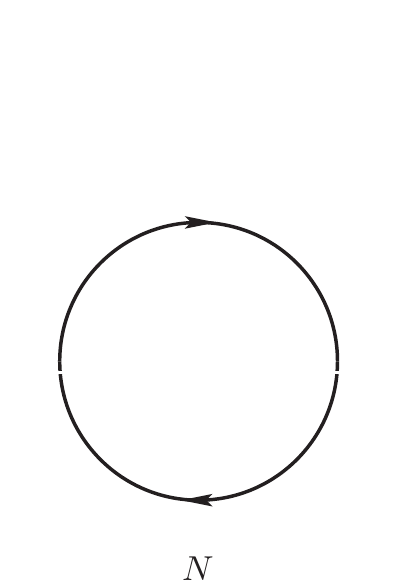}
\label{fig:Nrules(b)}
}
\subfigure[]{
\includegraphics[scale=0.6]{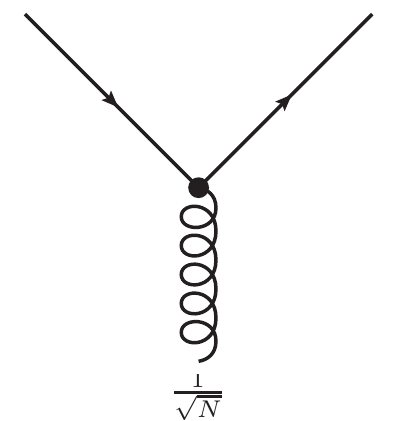}
\label{fig:Nrules(c)}
}
\subfigure[]{
\includegraphics[scale=0.6]{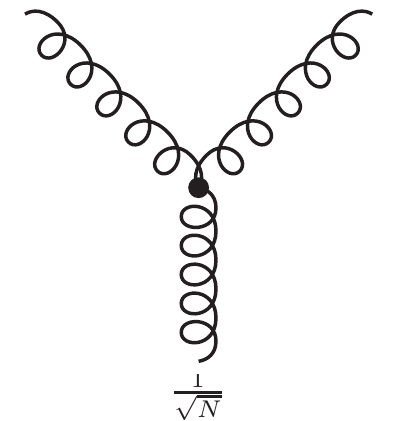}
\label{fig:Nrules(d)}
}
\subfigure[]{
\includegraphics[scale=0.6]{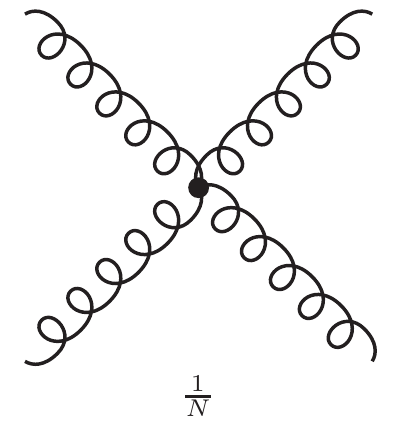}
\label{fig:Nrules(e)}
}
\label{fig:Nrules}
\caption{Basic rules for the counting of color factors in Feynman diagrams.}
\end{figure}
Quantum Chromodynamics (QCD) in the limit of a large number, $N$, of colors\cite{'tHooft:1974hx} has been used in the past 
$40$ years as a simplified though reliable model of the strong interaction phenomena\cite{Witten:1979kh}.
The perturbative expansion in Feynman diagrams is simplified by a number of selection rules holding when $N \to \infty$.
Nevertheless, the theory thus obtained is non-trivial and shows asymptotic freedom, being non-perturbative in the 
infrared region.
Assuming that confinement persists also in the $N \to \infty$ limit, it can be shown 
that the following peculiar properties hold:
\begin{itemize}
\item Mesons and glueballs (bound states of just gluons as explained in Sec.~\ref{sec:hybrids}) are stable and non-interacting at leading order in the $1/N$ expansion.
\item Meson decay amplitudes are of order $1/\sqrt{N}$ and meson-meson elastic scattering amplitudes are of order $1/N$.
\item OZI rule is exact and the mixing of mesons with glue states is suppressed.
\item Baryons are heavier than mesons: they decouple from the spectrum having a mass growing as $N$.
\end{itemize}
All of these statements can be proven without computing explicitly Feynman diagrams, but simply counting their color factors. 
To do this and to follow the theoretical arguments reported in this section, it is first necessary to analyze in greater detail the content of QCD with $SU(N)$ gauge group.

\begin{figure}[t]
\centering
\subfigure[]{
\includegraphics[scale=0.6]{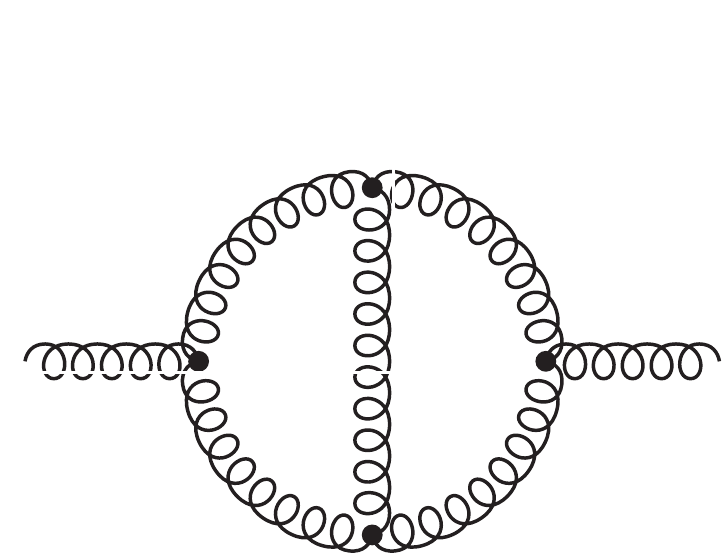}
}
\subfigure[]{
\includegraphics[scale=0.6]{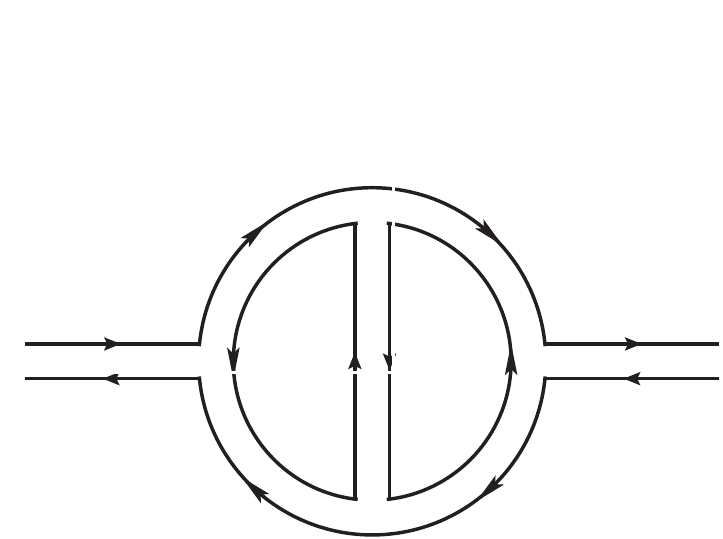}
}
\caption{Planar diagram contributing to the gluon self energy at leading order in $1/N$ expansion. The counting of color factors and couplings gives 
$N^2\times g^4/N^2=g^4$.}
\label{fig:combo1}
\subfigure[]{
\includegraphics[scale=0.6]{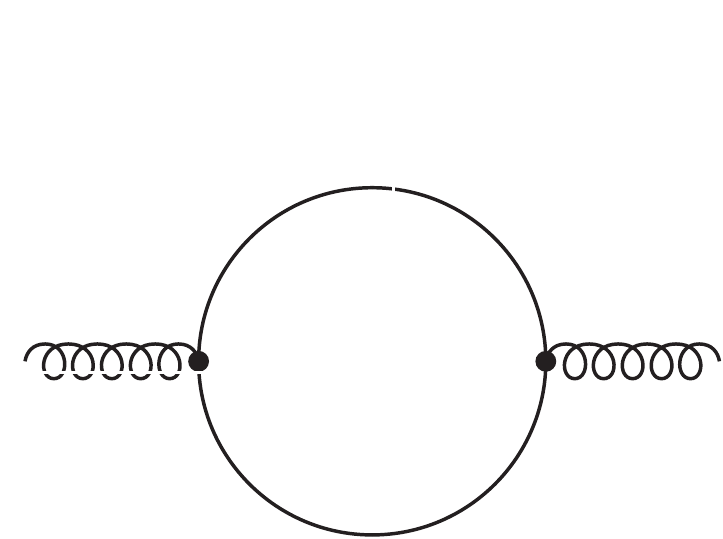}
}
\subfigure[]{
\includegraphics[scale=0.6]{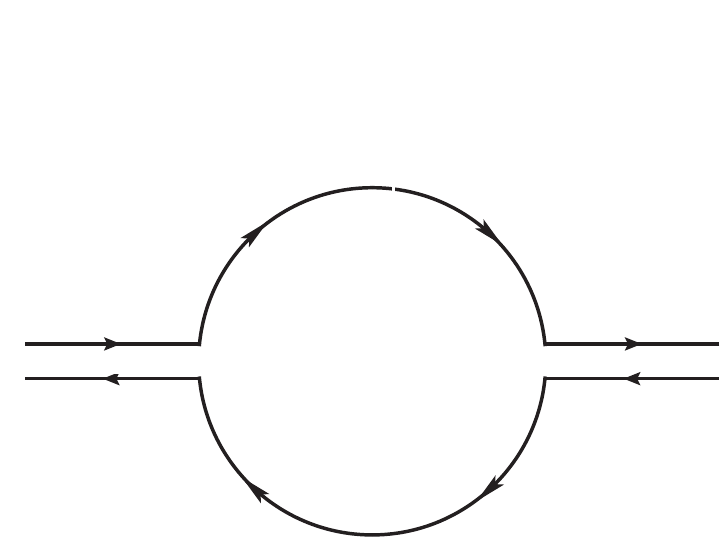}
}
\caption{Example of a diagram with a quark loop, suppressed woth respect to a planar diagram with gluon internal lines. The counting of color factors and couplings 
gives $g^2/N$, showing that although this diagram is leading in the coupling $g$ expansion compared to the diagram in \figurename{~\ref{fig:combo1}}, it is 
sub-leading in the $1/N$ counting.}
\label{fig:combo2}
\subfigure[]{
\includegraphics[scale=0.6]{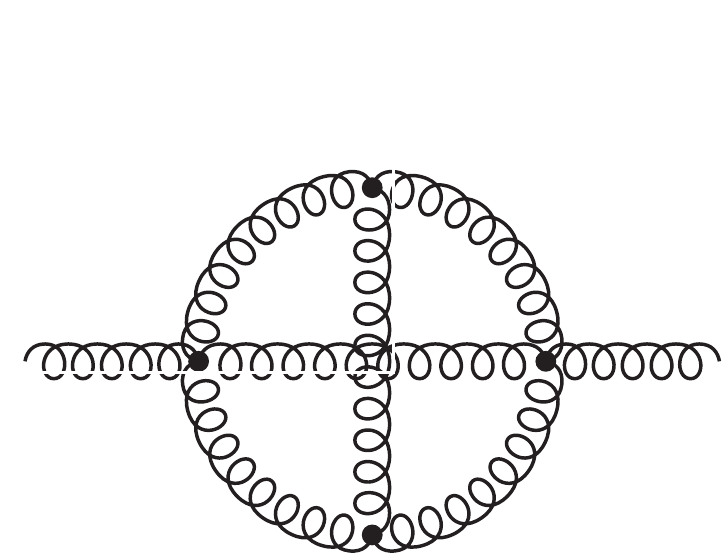}
}
\subfigure[]{
\includegraphics[scale=0.6]{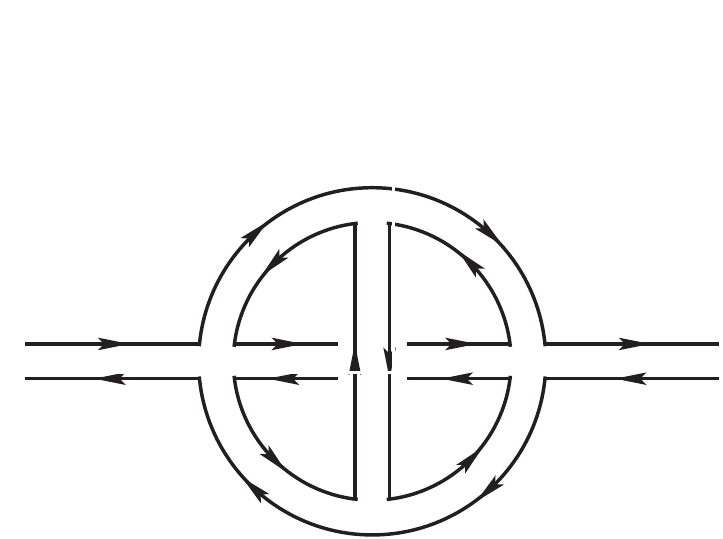}
}
\caption{Non-planar contribution to the gluon self energy. The counting of color factors and couplings gives $N\times g^2/N \times g^2/N^2=g^4/N^2$.}
\label{fig:combo3}
\end{figure}

Quark and antiquark fields have $N$ color components, while gluon fields are $N \times N$ matrix-valued fields with $(N^2-1) \sim N^2$ independent components\footnote{This approximation is justified because, as shown by `t~Hooft\cite{'tHooft:1974hx}, 
the traceless condition plays no role in the limit $N \to \infty$.}.
As a result, the gluon bubble diagram, \figurename{~\ref{fig:Nrules(a)}}, brings a color factor $N^2$ since that is the number of possible 
intermediate gluon states. In contrast, a quark bubble diagram, \figurename{~\ref{fig:Nrules(b)}}, brings a color factor $N$ being that the 
number of possible intermediate quarks.
The interaction vertices $g \bar q q$ and $ggg$ scale as $1/\sqrt{N}$ -- Figs.~\ref{fig:Nrules(c)} and \ref{fig:Nrules(d)} -- and the four-gluon vertex as $1/N$ -- \figurename{~\ref{fig:Nrules(e)}}.
These factors appearing in the interaction vertices are a consequence of the rescaling of the coupling constant, $\lambda = g/\sqrt{N}$, necessary to avoid further positive powers of $N$ in the perturbative expansion in the rescaled Yang-Mills coupling $\lambda$. 
For instance, the perturbative expansion in $\lambda$ of the gluon propagator is at most of order $N^0$ although, at this order in $1/N$, infinite diagrams with different orders in $g$ contribute.

The simplest way to take properly into account all the combinatoric color factors is to introduce the `t~Hooft double line representation\cite{'tHooft:1974hx}.
As already observed, gluons are $N \times N$ matrices, thus, as far as the color factors are concerned, they are indistinguishable from $q \bar q$ pairs when $N$ is large. 
For this reason one can substitute each gluon line with a couple of lines oriented in opposite directions.
Examples of this representation are shown in Figs.~\ref{fig:combo1}, \ref{fig:combo2} and \ref{fig:combo3}.
The gluon self energy diagram in \figurename{~\ref{fig:combo1}} is of order $N^2 \times g^4/N^2 = g^4$: the factor $N^2$ is consequence of the 
two color loops and $g^4/N^2$ of the four interaction vertices. 

The diagram in \figurename{~\ref{fig:combo2}} is of order $g^2/N$, arising from the vertices only. In fact, in the double line representation, this diagram has no closed fermion lines and hence no powers $N$ coming from loops.
The comparison of this diagram with that in \figurename{~\ref{fig:combo1}} shows the difference between the 
weak-coupling and the Large-$N$ expansion: a sub-leading term in the former, \figurename{~\ref{fig:combo1}}, is not so in the latter, \figurename{~\ref{fig:combo2}}, and viceversa.
Moreover, $1/N$ is a good expansion parameter regardless of the running of $g$.

In \figurename{~\ref{fig:combo3}} it is shown an example of non-planar diagram. Non-planar means that it is impossible to draw it  
without line crossings. In this case the counting of color factors gives $N \times g^2/N \times g^2/N^2=g^4/N^2$. Again, beside the vertices, the single factor of $N$ comes from the only fermionic closed line present in the diagram.
It can be shown that this relative suppression of the non-planar diagrams compared to the planar ones is true in general. This is one of the most important simplifications induced by the Large-$N$ expansion.

The above discussion can be summarized in few important rules:
\begin{itemize}
\item Planar diagrams with only gluon internal lines are all of the same order in the $1/N$ expansion.
\item Diagrams containing quark loops are subleading: the theory is quenched in the limit $N\to \infty$.
\item Non-planar diagrams are also subleading.
\end{itemize}

\subsection{Weinberg's observation}
\label{weinberg}

In his classic Erice lectures\cite{Coleman}, Coleman justifies the non-existence of exotic mesons noticing that the application of local gauge-invariant quark quadri-linear operators to the vacuum state
creates meson pairs and nothing else. The argument was as follows.     

By Fierz rearrangement of fermion fields, any color-neutral operator formed from two quark and two antiquark fields
\begin{equation}
Q(x)=\epsilon^{abc}\epsilon^{ade} q^b q^c \,\, \bar q^{\,d} \bar q^{\,e}
\end{equation}
can be rewritten in the form
\begin{equation}
Q(x)=\sum_{ij} C_{ij} B_i(x)B_j(x),
\end{equation}
where $C_{ij}$ are numerical coefficients and
\begin{equation}
B_i(x)=\bar q(x) \Gamma_i q(x)
\end{equation}
is some generic color-neutral quark bilinear with spin-flavor structure determined by the matrix $\Gamma_i$.

Let us look at the two-point correlation function of the $Q$ operators, $\left\langle T\left(Q(x) Q^\dag(y)\right) \right\rangle$, and perform the fermonic Wick contractions first\footnote{This is always possible because the fermionic action is quadratic.}.
For simplicity, also suppose that the expectation value of single fermion bilinears vanishes, \ie $\left\langle B_i(0) \right\rangle=0$.
The two-point function $\left\langle T\left(Q(x) Q^\dag(y)\right) \right\rangle$ is a sum of terms that can be grouped in two different classes:
double trace terms of the form 
\begin{subequations} \label{eq:doubletrace}
\begin{align}
&\left\langle \left\langle T\left(B_i(x) B_k(y)\right) \right\rangle_{\psi} \left\langle T\left(B_j(x) B_l(y)\right) \right\rangle_{\psi} \right\rangle_{A}   \\
&\qquad= \left\langle \tr\left[S(x-y)\Gamma_i S(y-x)\Gamma_k\right] \right\rangle_A \left\langle \tr\left[S(x-y)\Gamma_j S(y-x)\Gamma_l\right] \right\rangle_A
\end{align}
\end{subequations}
and single trace terms
\begin{subequations} \label{eq:singletrace}
\begin{align}
&\left\langle T\left(B_i(x) B_j(x) B_k(y) B_l(y)\right) \right\rangle_{\psi,A} \\
&\qquad=\left\langle \tr \left[S(x-y) \Gamma_i S(y-x) \Gamma_j S(x-y) \Gamma_k S(y-x) \Gamma_l \right]\right\rangle_{A},
\end{align}
\end{subequations}
where the flavor of the quark propagator $S(x-y)$ is implied to simplify the notation. 
The subscripts $A$ and $\psi$ indicate a functional integration over the corresponding fields (gauge fields and fermions respectively).
It is worth noticing that in the Large-$N$ limit the contribution in \eqref{eq:doubletrace} has a perturbative expansion in $1/N$ of the form
\begin{subequations} \label{eq:BB_BB}
\begin{align}
&\left\langle \left\langle T\left(B_i(x) B_k(y)\right) \right\rangle_{\psi} \left\langle T\left(B_j(x) B_l(y)\right) \right\rangle_{\psi} \right\rangle_A \\
&\qquad = \left\langle T\left(B_i(x) B_k(y)\right) \right\rangle_{\psi,A}  \left\langle T\left(B_j(x) B_l(y)\right) \right\rangle_{\psi,A} + \mathcal{O}(N^0).
\end{align}
\end{subequations}

\begin{figure}
\centering
\includegraphics[scale=0.8]{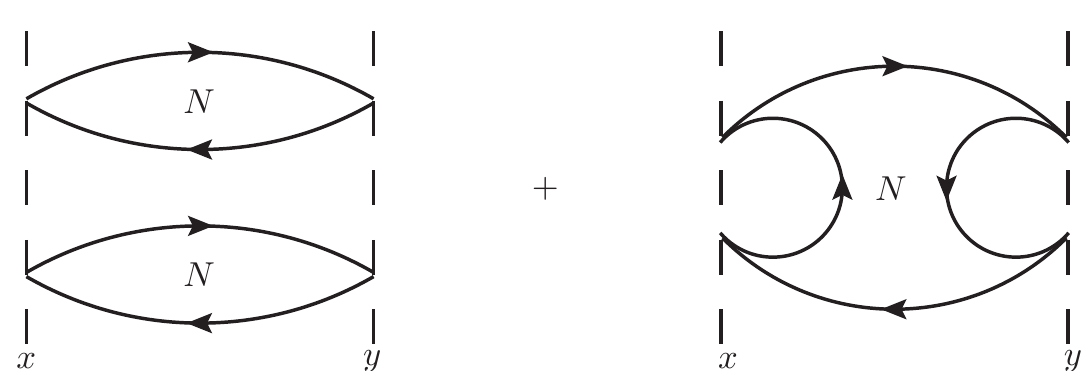}
\caption{Diagrammatic representation of the two-point correlation function of tetraquark operators at the leading and next-to-leading order in the Large-$N$ expansion.
Everywhere it is understood that the insertion of any number of planar gluon \emph{internal} lines does not change the order of the diagrams.} 
\label{fig:QxQy}
\end{figure}
\begin{figure}
\centering
\subfigure[]{
\includegraphics[scale=0.8]{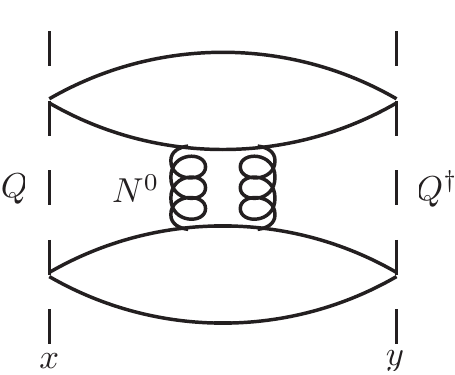}
}
\subfigure[]{
\includegraphics[scale=0.8]{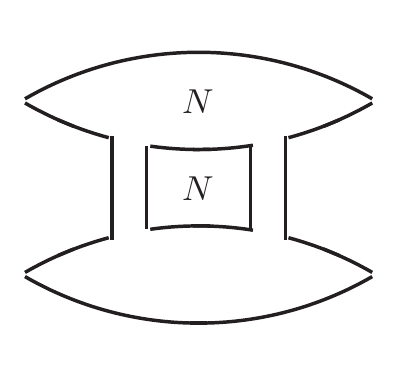}
}
\caption{Example of subleading contribution to the double-trace term in the tetrquark two-point correlation function. The color factor is $N^2 \times g^4/N^2=g^4$.} 
\label{fig:subbolle}
\end{figure}

The first term of the right hand side in Eq.~\eqref{eq:BB_BB} is the product of two non-interacting meson bubbles (\figurename{~\ref{fig:QxQy}}, left panel) and is 
of order $N^2$. The next-to-leading order term for this double trace contribution is the sum of planar diagrams like that in \figurename{~\ref{fig:subbolle}}
that are at most of order $N^2\times 1/N^2=1$. For this reason we will refer to the leading contribution of this correlation function as ``disconnected", implying that 
there are no gluon lines connecting the two meson bubbles.
The contribution in Eq.~\eqref{eq:singletrace} is, instead, of order $N$: in \figurename{~\ref{fig:QxQy}}, right panel, it is shown one of these possible single trace subleading terms 
(the shape of the diagram depends on the flavor structure of the quark bilinears). 
It is important to stress that the addition of any number of gluon lines \emph{internal} to the fermion loops does not spoil the order of the diagram in the $1/N$ expansion because of the cancellation between the positive powers coming from the additional loops and the negative powers coming from the additional quark-gluon vertices. Therefore, as long as the $1/N$ expansion is considered, it will always be understood that any possible number of internal gluon lines, not changing the order of the expansion itself, will be included. Hence, since the leading order contribution to the single trace term in Eq.~\eqref{eq:singletrace} is made of all the possible planar diagrams with internal gluon lines we will refer to it as the ``connected" contribution.
The complete two-point tetraquark correlation function can then be written as
\begin{subequations} \label{QxQy}
\begin{align} 
\left\langle T\left(Q(x) Q^\dag(y) \right) \right\rangle =
 \sum_{ijkl} & C_{ij} C_{kl} \Big[ \left\langle T\left(B_i(x) B_k(y) \right)\right\rangle_{\psi,A} \left\langle T\left(B_j(x) B_l(y) \right)\right\rangle_{\psi,A}  \\
 &+\left\langle T\left(B_i(x) B_j(x) B_k(y) B_l(y)\right) \right\rangle_{\text{conn},\psi,A} \Big] + \mathcal{O}(N^0),
\end{align}
\end{subequations}
where the subscript ``conn'' stands for connected.
The disconnected term contains only information about the propagation of two non-interacting mesons. 
This can be seen cutting in half the first diagram in \figurename{~\ref{fig:QxQy}}: assuming confinement, the only way to put 
on-shell the two bubbles is to form two non-interacting mesons.
In contrast, there are different ways of cutting in half the second diagram of \figurename{~\ref{fig:QxQy}} and it is possible to 
put on-shell simultaneously four quark propagators.
For this reason, if a one-tetraquark pole exists, it contributes only to the connected term of the correlation function, which is of order $N$, relatively vanishing if compared to the disconnected one, which is of order $N^2$.
Consequently, Coleman~\cite{Coleman} concludes that such exotic mesons do not exist when $N\to \infty$.

\vspace{1em}

However, Weinberg recently pointed out\cite{Weinberg:2013cfa} that Coleman's argument seems not to be conclusive.
To make an analogy, consider the meson-meson scattering in the Large-$N$ limit. The scattering amplitude is 
dominated by the analogous of the disconnected term in Eq.~\eqref{QxQy}, with the difference that all quark bilinears are now located in different points.
Terms of interaction between meson bubbles are subleading in the $1/N$ expansion.
This essentially means that, when $N \to \infty$, the mesons are non-interacting, but surely we do not infer that mesons do not scatter at all in the physical world, when $N=3$.
In other words, one should compute a scattering amplitude first and then take the Large-$N$ limit, otherwise the result would vanish right from the beginning, being the mesons non-interacting in this limit.

To better understand this point and what follows we will give some further details.
Consider the scattering amplitude with $B_1$, $B_2$ ingoing and $B_3$, $B_4$ outgoing mesons:
\begin{equation} \label{eq:ampl}
\widetilde{\mathcal{A}}_{B_1B_2 \to B_3 B_4}=\lim_{N\to\infty} \prod_{i=1}^4\lim_{q^{2}_{i}  \to  m^{2}_i} (q^2_i - m^2_i) \frac{1}{\sqrt{Z_{B_{i}}}}\widetilde{G}_4(q^2_i;s,t),
\end{equation}
where $\widetilde{G}_4$ is the Fourier transform of the four-point correlation function
\begin{equation}
G_4(x,y,z,w)=\left\langle T\left(B_1(x) B_2(y) B_3(z) B_4(w)\right) \right\rangle,
\end{equation}
$s$, $t$ are the two independent Mandelstam variables characterizing the scattering process.
Moreover, $G_4$ can be expanded in the parameter $1/N$, as the two-point function in Eq.~\eqref{QxQy}.
The renormalization constants $Z_{B_i}$ bring a color factor of $N$.
This follows from the definition of $Z_{B_i}$ in terms of the two-point function, $G_{2}(x)=\left\langle T\left(B_i(x)B_i(0)\right) \right\rangle$, whose Fourier tansform reads: 
\begin{equation}
\widetilde{G}_2(p^2)=\frac{iZ_{B_i}}{p^2 - m_i^2 + i\epsilon} + \dots,
\end{equation}
where the dots stand for additional poles or cut contributions.
Since the leading order contribution to $G_2(x)$ is the bubble in \figurename{~\ref{fig:Nrules(b)}}, with the insertion of any number of internal gluon lines not changing the $N$-counting, it follows that
\begin{equation}
\langle T\left(B_i (x) B_i (y)\right) \rangle \propto Z_{B_i} \propto N.
\label{BB}
\end{equation}
Therefore, the $Z_{B_i}$ in Eq.~\eqref{eq:ampl} bring a factor of $1/N^2$. As in Eq.~\eqref{QxQy}, the leading disconnected contribution to $G_4$ is proportional to $N^2$ and hence it produces a term of order one in the amplitude. However, this term corresponds to 
a couple of freely-propagating mesons. 
For this reason, it doesn't contribute to the cross section since it corresponds to the identity part of the $S$-matrix, $S=\mathbb{1} + iT$, that is subtracted in the LSZ formalism.
The connected subleading term is, instead, of order $N$ and thus contributes as $1/N$ to the amplitude. This is the real leading term in the scattering matrix.
On the other hand, if we had taken the limit $N\to \infty$ first and applied the LSZ formalism after, we would have got $\mathcal{A}=0$ because, as we mentioned, QCD in the Large-$N$ limit is a theory of non-interacting mesons and glueballs. In other words, taking $N\to\infty$ beforehand kills all the contribution but the one describing the two mesons propagating without interacting.

In this spirit, Weinberg shows that, admitting the existence of a one-tetraquark pole in some connected correlation function of the kind mentioned above, the Large-$N$ expansion can actually be used 
to learn more about the phenomenology of tetraquark in the physical situation of finite $N$.
Consider the decay amplitude of a tetraquark into two ordinary mesons. 
As just observed, the quark bilinear entering in the LSZ formulation 
has to be normalized as $N^{-1/2} B(x)$. The same happens for tetraquark interpolating operators, where, for the connected term, 
holds
\begin{equation}
\langle T\left(Q(x) Q^\dag(y)\right) \rangle \propto Z_Q \propto N,
\label{eq:ZQ}
\end{equation}
$Z_Q$ being the residue at the tetraquark pole.
The properly normalized operator for the creations or annihilation of a 
tetraquark is $N^{-1/2} Q(x)$, as for an ordinary meson.

The amplitude for the decay of a tetraquark is then proportional to a suitable Fourier transform, $\widetilde{G}_3$, of the three-point function
\begin{align}
\frac{1}{N^{3/2}}\left\langle T(Q(x) B_n(y) B_m(z)) \right\rangle &=\frac{1}{N^{3/2}}\sum_{ij} C_{ij} \left\langle T(B_i(x) B_n(y)) \right\rangle_{\psi,A} \left\langle T(B_j(x) B_m(z))  \right\rangle_{\psi,A} \nonumber\\
&\qquad +\frac{1}{N^{3/2}} \left\langle T(Q(x) B_n(y) B_m(z)) \right\rangle_{\text{conn},\psi,A} + \mathcal{O}\left(\frac{1}{N}\right).
\label{Qdecay}
\end{align}
The disconnected term is leading and hence the decay width has a color factor proportional to ${(N^2 \times1/N^{3/2})}^2=N$. Therefore, it seems that tetraquarks would be very broad states, \ie they would be unobservable in the Large-$N$ limit. However, when we amputate the tetraquark external leg,  
\begin{equation}
\mathcal{A}(Q\to B_1B_2)
\propto \lim_{q^2 \to m^2_Q} \frac{1}{\sqrt{Z_Q}} \left( q^2-m^2_Q\right) \widetilde{G}_3,
\label{eq:QBB}
\end{equation}
 this term vanishes: $\widetilde{G}_3$ at leading order is just the convolution of two meson propagators, and thus contains meson poles only. Therefore the factor $q^2-m_Q^2$ makes the amplitude vanish in the on-shell limit.
On the other hand, if a tetraquark pole actually exists in the connected subleading term (the second term in~\eqref{Qdecay}), it would have a decay rate 
proportional to ${(N \times1/N^{3/2})}^2 = 1/N$ and it would be stable in the Large-$N$ limit, just like an ordinary meson.

\subsection{Flavor structure of narrow tetraquarks} \label{sec:narrow}

In the previous section we showed that if a tetraquark exists, then it has a decay width proportional, at most, to $1/N$. In this kind of analysis the flavor quantum numbers play a crucial role in predicting their decay widths.

Here we summarize some recent results about the classification of all possible 
flavor structures of tetraquarks\cite{Knecht:2013yqa}.
In order to simplify the discussion, we define a quark bilinear with flavor quantum numbers $A$ and $B$ as
\begin{equation}
B_{AB}(x) = \bar q_A(x) \Gamma q_B(x).
\end{equation}
We will also assume that the flavor indices $A,B$ are different so that the vacuum expectation value $\langle B_{AB}(x) \rangle$ identically vanishes. 
The tetraquark interpolating field will be denoted as
\begin{align}
Q_{AB;CD}(x)=B_{AB}(x)B_{CD}(x).
\end{align}
Since $A \neq B$ and $C\neq D$, we have only three non-trivial possibilities for the flavor structure of the couple $CD$:
\begin{equation}
C=B,\hspace{10mm}C=A,\hspace{10mm}A\neq B \neq C \neq D.
\end{equation}

\begin{figure}
\centering
\subfigure[]{
\includegraphics[scale=0.8]{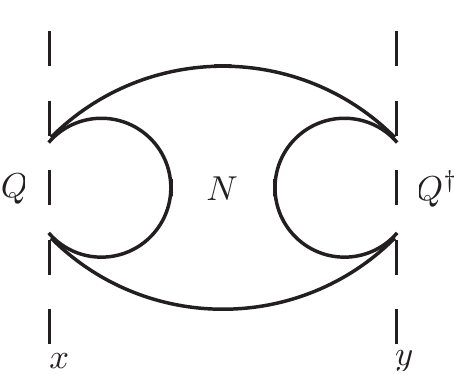}
}
\caption{Tetraquark correlation function with flavor $B=C$, for instance $(\bar u c)(\bar c d)$. Cutting vertically this diagram 
reveals the contribution from both tetraquark and meson intermediate states.}
\label{fig:diagramma(a)}
\subfigure[]{
\includegraphics[scale=0.8]{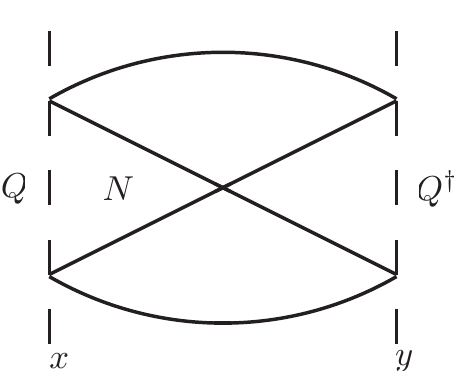}
}
\caption{Tetraquark correlation function with flavor $A=C$, for instance $(\bar c u)(\bar c d)$\,\cite{Esposito:2013fma}. }
\label{fig:diagrammi(b)}
\subfigure[]{
\includegraphics[scale=0.8]{Figures/TypeB-crop.pdf}
}
\caption{Tetraquark correlation function with all different flavors, $(\bar c u)(\bar s d)$. The leading order connected diagrams, in this case, need the exchange of at least two gluons between the two meson bubbles.}
\label{fig:diagrammi(c)}
\end{figure}

These three possibilities imply different quark contractions in the correlation functions involving tetraquark operators since contractions between different flavors are forbidden, and hence determine different Large-$N$ behaviors.
The resulting predictions are summarized in \tablename{~\ref{tab:flavor}} and can be derived looking at
 Figs.~\ref{fig:diagramma(a)}, \ref{fig:diagrammi(b)} and \ref{fig:diagrammi(c)}. For a detailed derivation one should refer to the original work\cite{Knecht:2013yqa}.

The remarkable aspect of this analysis is that a careful treatment of the flavor quantum numbers reveals the presence of even narrower tetraquarks than those decaying as $1/N$. This happens in those situations in which the tetraquark is made out of quarks with all different flavors, for instance $\left[cs\right]\left[\bar u \bar d\right]$\footnote{The notation $[q_1 q_2][\bar q_1 \bar q_2]$ is introduced to distinguish between a tetraquark written in the diquark-antidiquark basis -- see Sec.~\ref{sec:tetraquarks} -- against the notation $(\bar q_1 q_2)(\bar q_3 q_4)$ -- see Table~\ref{tab:flavor}.}. 
Let us perform a detailed analysis for this case, the extension to the other flavor structures follows straightforwardly. 

In order to determine the decay width of such a tetraquark it is sufficient to take only the leading connected contribution to the amplitude, in the sense specified in \sectionname{~\ref{weinberg}}, with properly normalized operators.
In this case, $Z_Q \sim N^0$ because (\figurename{~\ref{fig:subbolle}}) the color factor of $\left \langle T\left(Q^\dag(x) Q(y)\right) \right \rangle$ is $N^2\times g^4/N^2\sim N^0$ -- also recall that the Large-$N$ order of $Z_Q$ must be evaluated from the two-point function as in \sectionname{~\ref{weinberg}}.
The color factor of the decay amplitude, Eq.~\eqref{eq:QBB}, is:
\begin{equation} \label{eq:peris}
\mathcal{A}_{Q\to BB}\sim \frac{1}{\sqrt{Z_Q}}\frac{1}{Z_B} \widetilde{G}_3 \sim \frac{1}{N^0} \times \frac{1}{N} \times N^0 \sim \frac{1}{N},
\end{equation}
where the color factor of $\widetilde{G}_3$ is $N^0$ since the decay diagram of a tetraquark of this kind is analogous to \figurename{~\ref{fig:diagrammi(c)}}
-- see again the original reference \cite{Knecht:2013yqa} for details.
Finally, in this case a tetraquark-meson mixing is absent: cutting vertically the diagram in \figurename{~\ref{fig:diagrammi(c)}} 
it is impossible to have a cut involving only two color lines, thus there is no meson intermediate state contribution. 
The decay width of this kind of tetraquarks goes as $1/N^2$, being the decay amplitude of order $1/N$, Eq.~\eqref{eq:peris}.

As a final note, we comment the Large-$N$ behavior of the diagram in \figurename{~\ref{fig:diagrammi(b)}}. At a first look, it seems a 
subleading non-planar contribution, since there are line crossings. However, the planarity is a topological property 
of the diagram, being its $1/N$ order associated to its Euler characteristic $\chi$, as shown by t'Hooft\cite{'tHooft:1974hx, Coleman} ,
through the power law 
\begin{equation}
N^{\chi} \hspace{2em} \text{ with } \hspace{2em} \chi=L-I+V,
\end{equation}
where $L$ is the number of loops, $I$ the internal lines and $V$ the vertices of the diagrams.
It hence follows that the one showed in \figurename{~\ref{fig:diagrammi(b)}} is clearly a connected planar diagram (there are many examples of 
non-trivial planar diagrams, for instance in Maiani~\etal\cite{Maiani:scalars}).
\begin{table}[t]
\tbl{Flavor structure and associated decay width for tetraquarks as reported by Knecht and Peris~\cite{Knecht:2013yqa}. The notation $(\bar q_1 q_2)(\bar q_3 q_4)$ is used when the tetraquark is written in the meson-meson basis.}
{
\begin{tabular}{ccccc}
\hline\hline
Type & Decay width  & $\sqrt{Z_Q}$ & Tetraquark-Meson mixing & Example\\ \hline 
$C=B$ & $1/N$  & $\sqrt{N}$ & $N^0$ & $(\bar u c) (\bar c d)$\\
$C=A$ & $1/N$  & $\sqrt{N}$ & absent & $(\bar c u) (\bar c d) $\\
$A\neq B \neq C \neq D$ & $1/N^2$  & $N^0$ & absent & $(\bar c u) (\bar s d)$ \\ \hline\hline
\end{tabular}
\label{tab:flavor}
}
\end{table}
\subsection{Hypothetical non-perturbative contributions to tetraquark operators}

In a recent paper by Lebed\cite{Lebed:2013aka} a potential incongruence is found in considering the normalization of the tetraquark wave functions created by LSZ normalized operators. 

The fact that $Z_B \sim N$ is equivalent to the conclusion that LSZ reduction identifies the operator 
$N^{-1/2}B(x)$ as the one creating or destroying properly normalized asymptotic states. 
This prefactor also produces correctly normalized meson states 
\begin{equation} 
\frac{1}{\sqrt{N}}B_i \left | 0 \right\rangle=\frac{1}{\sqrt{N}}\sum_{a=1}^{N} \bar q^a \Gamma_i q^a \left | 0 \right\rangle.
\end{equation}
Nevertheless, also $Z_Q \sim N$ -- see Eq.~\eqref{eq:ZQ}, leading to a properly normalized tetraquark operator $N^{-1/2}Q(x)$. However, its application to the vacuum creates states with norm squared $N$
\begin{equation} 
\frac{1}{\sqrt{N}}Q\left | 0 \right\rangle=\frac{1}{\sqrt{N}}\sum_{ij}\sum_{a=1}^{N}\sum_{b=1}^N C_{ij} \,\, \bar q^{\,a}\Gamma_i q^a \,\, \bar q^{\,b}  \Gamma_j q^{b} \left | 0 \right\rangle.
\label{eq:Qnorma}
\end{equation}
Lebed\cite{Lebed:2013aka} pointed out that, in order to obtain the additional $1/\sqrt{N}$ suppression needed for the correct normalization of the state in Eq.~\eqref{eq:Qnorma}, 
the definition of the tetraquark operator as a local product of fermion bilinears must be revisited.
We have already shown that, when the Large-$N$ limit is involved together with another different limit procedure, they must be treated carefully.
In particular, in the LSZ formalism one has to take the infrared limit first, otherwise all the scattering amplitudes would identically vanish.
In this spirit, one could ask if the Large-$N$ commutes with the definition of composite operators: an operation that involves a limit 
procedure. For example, in the $\phi^4$ scalar theory one may define the composite operator $:\phi^2(0):$ as
\begin{equation}
: \phi^2(0): = \lim_{x\to 0} \left( \phi(x)\phi(0) - \frac{C}{x^2} \mathbf{1} \right),
\label{eq:phi2}
\end{equation}
in order to obtain finite renormalized correlation functions with its insertions~\footnote{The operator product expansion is blind to contact terms of the form $\Box^n \delta(x)$, for some power $n$: an additional reason 
to define composite operators through a limit procedure.}.

Lebed\cite{Lebed:2013aka} suggests that the non-commutativity of the limit $N\to\infty$ and the local limit in the definition of the composite tetraquark 
operator $Q(x)$ is crucial in resolving the lack of the additional $1/\sqrt{N}$ suppression factor in the 
tetraquark wave function in Eq.~\eqref{eq:Qnorma}.
Consider the product of operators
\begin{equation}
B_i(x) B_j(0) \sim \dots + C_{ij}(x) B_i(0)B_j(0) + \dots,
\end{equation}
when $x \to 0$~\footnote{We are ignoring the mixing with operators of dimension less than 6 because we are dealing with properly defined composite 
operators, Eq.~\eqref{eq:phi2}.} and 
allow the coefficients $C_{ij}(x)$ to have a contribution 
of the form 
\begin{equation}
\delta C_{ij} \sim e^{-N^m \Lambda_{QCD}^2 x^2}.
\end{equation}
For any finite separation $x^2 \geq \frac{1}{\Lambda_{QCD}^2}$ this contribution is vanishing when $N \to \infty$, in order to preserve the usual 
$N$-counting for the correlation function of two mesons.

If one defines the tetraquark operator smearing the product $B(x)B(0)$ over a small spatial region 
of size $\mathcal{O}(1/\Lambda_{QCD})$
\begin{equation}
Q(x)=\int_{\Lambda_{QCD}^3} d^3y \sum_{ij} C_{ij}(y-x)B_i(x)B_j(x),
\end{equation}
the above mentioned four-point correlation function, in the limit $x \to 0$, gets a contribution of $N^{-m/2}$ for each spatial integral
of the gaussian factor $\delta C_{ij}$.
If $m=1/3$ one obtains precisely the desired additional suppression $1/\sqrt{N}$ 
in order to obtain properly normalized tetraquark wave functions.

We remark that the coefficients $\delta C_{ij}$ are non-perturbative in the $N$-counting and cannot be inferred from perturbation theory. 
The mechanism proposed is only a possibility and there are no reasons to believe that it happens in this precise way.
Nevertheless, it suggests that in order to allow the existence of one-tetraquark poles in the connected piece of 
the correlation functions considered so far, some non-perturbative mechanism in the $1/N$ expansion must occur.

\subsection{Flavored tetraquarks in Corrigan-Ramond Large-$N$ limit}

So far we discussed in detail the Large-$N$ physical behavior of meson and tetraquark states 
in the so-called `t~Hooft limit\cite{'tHooft:1974hx}. Another well studied limit is that of Veneziano\cite{Veneziano:1976wm}, with $N_f \to \infty$ flavors, $N_c\to\infty$ colors,
provided that $N_f/N_c$ is fixed. 
In both these formulations, a simple definition of baryons does not exist, in contrast with what happens for mesons and tetraquarks. 
In fact, in generic $SU(N)$ gauge theories color-neutral states composed of only quarks -- \ie baryons -- are made of $N$ quarks in a totally antisymmetric combination
\begin{equation}
\epsilon_{i_1 i_2 \dots i_N} q^{i_1}q^{i_2} \dots q^{i_N}.
\end{equation}
As shown by Witten\cite{Witten:1979kh}, they have very distinctive properties in the Large-$N$ limit. As already mentioned, their mass goes as $N$,
thus they disappear from the hadronic spectrum when $N\to\infty$. 

As firstly proposed by Corrigan and Ramond\cite{Corrigan:1979xf}, it could be important to have, for every value of $N$, color-neutral bound states 
composed of only three quarks. 
A simple way to do it is to introduce new fermions, originally called ``larks"\cite{Corrigan:1979xf}, transforming as the $N(N-1)/2$ (antisymmetric) representation of $SU(N)$.
This choice is motivated by the observation that, when $N=3$, the dimension of this representation is $3$ and coincides with the ${\bf \bar 3}_c$ conjugate representation
\begin{equation} \label{eq:larks}
q_{ij}=\epsilon_{ijk}q^k\,\,\,\,\,\,i,j,k=1,2,3.
\end{equation} 
In this formulation, the baryons for Large-$N$ are constructed out of 
\begin{equation}
\bar q_{ij}q^iq^j
\end{equation}
color-neutral states, which look more like physical baryonic states. Moreover, just like quarks, larks only couple to gluons with a minimal coupling.
The introduction of a lark sector in the Large-$N$ extrapolation of QCD, not only allows to define three-quark states, but also modifies the $N$-counting. 
The reason is simple. If we introduce the `t~Hooft double line representation\cite{'tHooft:1974hx} to understand the color flow of this theory, 
we notice immediately that lark lines split in 
two with arrows pointing in the same direction since both color indices in Eq.~\eqref{eq:larks} belong to the same representation, in contrast with gluons, represented as two oriented lines pointing in opposite directions since their color indices always appear as a color-anticolor combination.
Apart from the different orientation of the color flows, lark loops have a color factor $N^2$ like gluon loops. This implies that leading order planar diagrams contain any possible internal gluons as well as 
lark loop corrections to these gluon lines. In other words, we can also introduce an arbitrary number of lark ``bubbles'' in the middle of a gluon propagator. The reason is again simple: each insertion of a lark loop in a gluon line counts as $N$ but each lark-gluon vertex counts as $1/\sqrt{N}$ and hence a lark loop in the middle of a gluon line does not change the $N$-counting of the considered diagram -- see Sec.~\ref{weinberg}. This is in contrast with quark loops, whose insertion inside a gluon propagator suppresses the order of the diagram in the $1/N$ expansion. 
For this reason we also see how including other antisymmetric tensors is dangerous for the $1/N$ expansion since they contribute in loops as $N^m$, where $m$ is the number of antisymmetric color indices, thus spoiling the perturbative expansion of the correlators of the theory
\footnote{Larks can be considered as additional quark species: they couple to gluons with the 
same coupling constant of the quarks, but with the $SU(N)$ generators in the covariant derivative belonging to the antisymmetric representation.
The same can be done for other species belonging to different representations of $SU(N)$.}. In fact, the fine cancellation between positive powers coming from loops and negative powers coming from couplings does not work for $m\neq2$.

\vspace{1em}

It was recently shown\cite{Cohen:2014via} that in the Corrigan-Ramond Large-$N$ formulation, {\it i.e.} the t'Hooft limit with quarks 
in the antisymmetric $SU(N)$ representation, (sometimes called QCD(AS)) it is possible to unambiguously define 
narrow tetraquark states. Consider a source operator of the form
\begin{equation}
Q(x)=C_{AB} \, \bar q^{ij} \Gamma_A q_{jk} \,\bar q^{kl} \Gamma_B q_{li},
\label{QCR}
\end{equation}
where lowercase letters indicate color indices and $\Gamma_{A,B}$ are matrices in the Dirac and flavor space. Spin and flavor quantum numbers of the operator are fixed by a suitable choice of $C_{AB}$.
This combination is gauge invariant, \ie is a color singlet, if we notice that
\begin{subequations}
\begin{align}
q_{ij} \to& \,\Omega_i^k \Omega_j^l q_{kl}, \\
\bar q^{ij} \to& \,\bar q^{kl} (\Omega^\dag)_k^i (\Omega^\dag)_l^j,
\end{align}
\end{subequations}
under a generic $SU(N)$ gauge transformation $\Omega$.
Moreover, lark color indices are saturated in such a way that the operator given by Eq.~\eqref{QCR} 
can never be splitted into two independent color singlets for $N>3$\footnote{As already mentioned, the case $N=3$ is equivalent to the diquark-antidiquark formulation.}.
 In other words, the two-point correlation function with sources as in Eq.~\eqref{QCR} cannot be separated in disconnected pieces.
\begin{figure}[t]
\centering
\subfigure[]{
\includegraphics[scale=0.7]{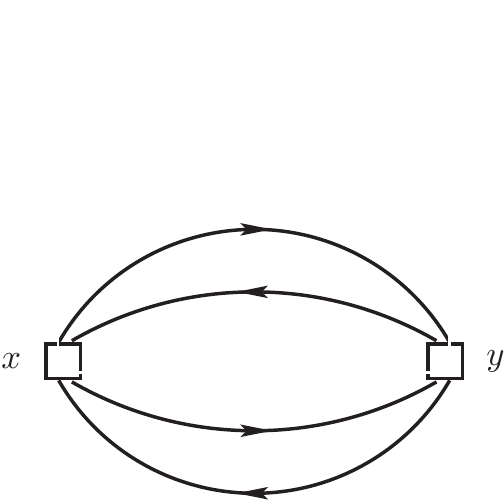}
\label{fig:4larks(a)}
}
\subfigure[]{
\includegraphics[scale=0.7]{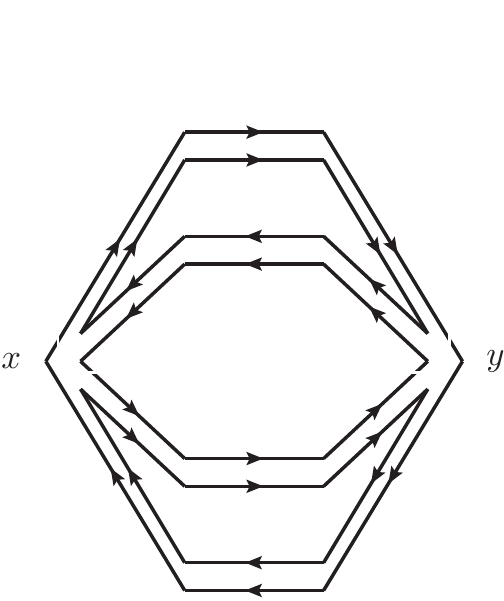}
\label{fig:4larks(b)}
}
\caption{``Tetralark" operator two-point correlation function. It is clear from the diagram that it is not possible to separate 
the lark lines in gauge invariant subdiagrams.}
\end{figure}
In \figurename{~\ref{fig:4larks(a)}} we show the leading order contribution to the correlation function $\langle Q(x) Q^\dag(0) \rangle$.
Using the `t~Hooft double line representation, \figurename{~\ref{fig:4larks(b)}}, we see that it is impossible to rearrange the colors in order to obtain two disconnected color singlet 
diagrams. This essentially means that operators like the one in Eq.~\eqref{QCR} unambigously interpolate tetraquarks (or tetralarks in a generic $SU(N)$) states, without mixing with ordinary mesons.

Counting the lark loops we find that the color factor of this diagram is $N^4$.
As usual in the Large-$N$ expansion, the $N$-counting does not change if we consider all the planar diagrams with the insertion of an arbitrary number of gluon internal lines.
It is remarkable to notice again that this counting does not change even if we add any number of planar lark loops in the middle of these gluon lines for the reasons explained before.

At this point, we can easily show that an operator source like that in Eq.~\eqref{QCR} can create out of the vacuum
single tetralarks at leading order in the Large-$N$ expansion. We must be careful that, if the flavor of the sources allows lark-antilark annihilations, there are other diagrams besides that in \figurename{~\ref{fig:4larks(a)}}. In that case, however, it would not be possible to unambiguously disentangle the contribution of tetralark intermediate states from those of mesons.
Imagine cutting vertically the diagram in \figurename{~\ref{fig:4larks(b)}}. 
Assuming confinement, the only way to form a gauge invariant combination of the lark lines involved in the cut 
is to group the four larks in a single hadron.
This statement holds even if we insert an arbitrary number of gluon internal lines in the diagram in figure name{~\ref{fig:4larks(b)}}. In that case 
the generic gauge-invariant contribution to the cut would have the form
\begin{equation}
\text{Tr} \left[ q \, A\dots A \, \bar q \,  A \dots A \,  q\, A \dots A \, \bar q \right],
\end{equation}
with $A$ the gluon field. This is still a color-singlet combination with four larks.
Since the two-point function $\langle T\left(Q(x) Q^\dag(0)\right)\rangle$ brings a color factor $N^4\sim Z_Q$ (as shown by the power counting of Fig.~\ref{fig:4larks(b)}) we also find that the LSZ properly normalized
tetralark operators are $N^{-2}Q(x)$.
Since lark loops are not suppressed, there are also flavor-singlet contribution of states with an arbitrary number of lark-antilark couples.
This means that the tetralark states found are not pure, but rather they are a superposition of infinite states, with arbitrary even number of larks, \ie the analogous of ``sea quarks'' in the lark sector.
However, they are unambiguously exotic states, since mesons made of larks cannot contribute to their two-point correlation function.

\vspace{1em}

\begin{figure}[hbtp!]
\centering
\subfigure[]{
\includegraphics[scale=0.7]{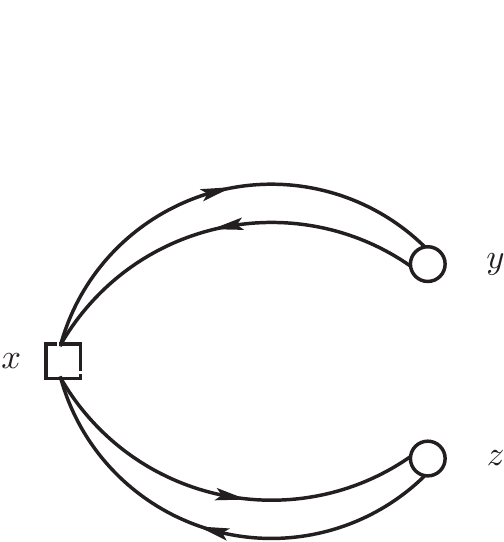}
}	
\subfigure[]{
\includegraphics[scale=0.7]{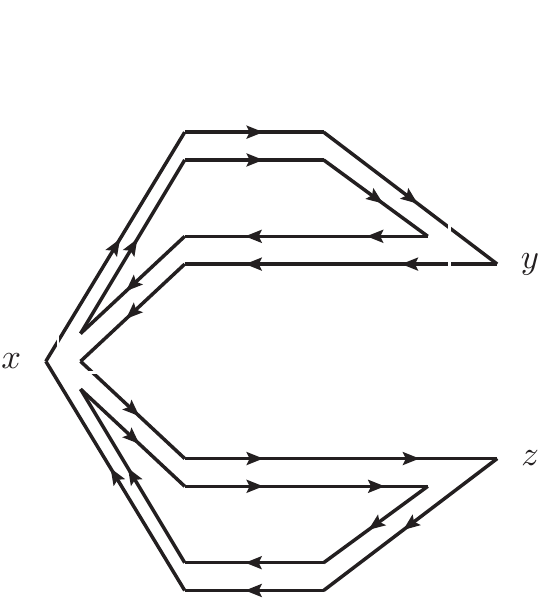}
}
\caption{Three point correlation function $\langle Q(x) B(y) B(z) \rangle$ for the decay of  a ``tetralark" in two mesons.} \label{fig:4larksdecay}
\end{figure}

Finally, we show that tetralarks in the Corrigan-Ramond Large-$N$ limit are actually narrow states.
To see this, consider the three point correlation function
\begin{equation}
\langle T\left(Q(x) B(y) B(z)\right)\rangle,
\label{QBB}
\end{equation}
where the operator $B$ interpolates a lark-antilark couple (we use the same notation of ordinary meson operators to stress the analogy among them). 
One of the leading order diagrams contributing to Eq.~\eqref{QBB} is shown in \figurename{~\ref{fig:4larksdecay}}. 
The counting of the color loops gives a factor $N^3$, instead of $N^4$ obtained for the two-point correlation function of ``tetralark" operators.
Normalizing properly the amplitude, we obtain a total color factor $1/N^2 \times \left(1/N\right)^2 \times N^3=1/N$, where we have used the result that 
$N^{-1}B(x)$ is the LSZ normalized lark-antilark meson operator (we have not shown this, but it can be easily proven).
The resulting decay width is then proportional to $1/N^2$, showing that the in the limit $N \to \infty$ Corrigan-Ramond tetralarks are narrow states.

\subsection{Flavored tetraquarks in `t~Hooft Large-$N$ limit}
\label{sec:cohen}
From a field theory point of view, it is a challenging task to identify 
operators interpolating only tetraquarks with flavor content $c \bar c u \bar d$.
This is because such an operator would interpolate also mesonic states as $u \bar d$ having the same quantum numbers.

This is one of the major difficulties in treating these states using Lattice QCD, although some recent works on the subject has been presented
\cite{Prelovsek:2013cra, Prelovsek:2013xba, Prelovsek:2014swa,Wagner:2013vaa,Alexandrou:2012rm}. 
Nevertheless, a class of operators that do not overlap with ordinary mesons and that unambiguously contain four valence quarks can be found\cite{Esposito:2013fma}. 

It is the same class of operators 
already introduced in Corrigan-Ramond QCD(AS) formulation and 
considered in a recent work by Cohen and Lebed\cite{Cohen:2014tga}. The authors show that the leading order connected diagrams contributing to the scattering amplitude of mesons with the appropriate exotic quantum numbers do not contain 
any tetraquark $s$-channel cut. A careful and detailed analysis of the argument can be found in the original paper\cite{Cohen:2014tga}. Here we will sketch the same argument referring to a specific case, in order to make the discussion more concrete.

\begin{figure}[t]
\centering
\subfigure[]
{
\includegraphics[scale=0.7]{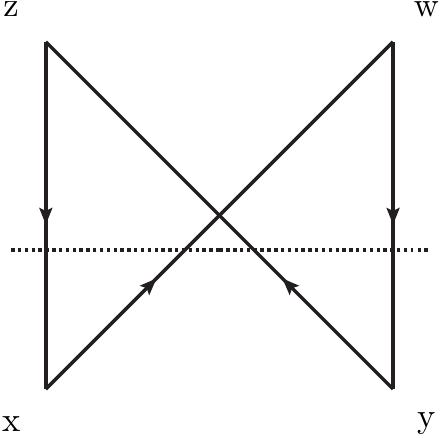}
\label{fig:cohen(a)}
}
\subfigure[]
{
\includegraphics[scale=0.7]{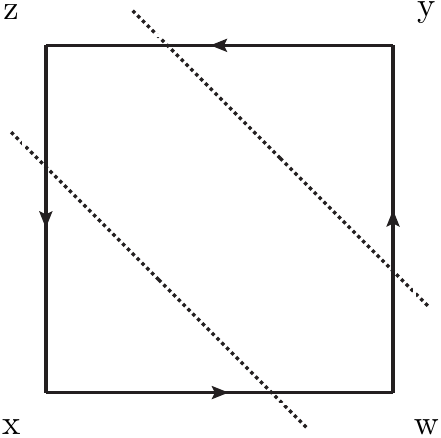}
\label{fig:cohen(b)}
}
\caption{Leading order connected contribution to the meson meson scattering amplitude with exotic quantum numbers. The dashed lines represent the cut in the $s$-channel.} \label{fig:cohen}
\end{figure}

Consider the following four-point correlation function -- see Fig.~\ref{fig:cohen} -- that can be used, for instance, to compute the elastic scattering amplitude of $D_s^+\, D^0$ mesons
\begin{equation}
G_4(x,y,z,w)=\left\langle T\left(\bar c \Gamma_1 s (z) \,\,\, \bar c \Gamma_2 u (w) \,\,\, \bar s \Gamma_3 c(x) \,\,\, \bar u \Gamma_4 c(y) \right)\right\rangle,
\label{G4}
\end{equation}
with $\Gamma_1$, $\Gamma_2$, $\Gamma_3$, $\Gamma_4$ appropriate Dirac matrices.
We are looking for a possible $s$-channel cut contributing to the four-point amplitude in which an on-shell tetraquark with flavor content $\left[ cc \right] \left[ \bar s \bar u \right]$ propagates. We also notice that a $t$-channel cut cannot reveal the presence 
of a tetraquark with these exotic flavor quantum numbers.

\begin{table}[b]
\tbl{The current status of Large-$N$ tetraquarks.}
{
\begin{tabular}{ll}
\hline\hline
\multirow{2}{*}{Coleman\cite{Coleman}/Witten\cite{Witten:1979kh} (1979)}& Tetraquarks do not exist,\\
&they are subleading in the large-$N$ QCD expansion.\\
\dots & \dots \\
\multirow{2}{*}{Weinberg\cite{Weinberg:2013cfa} (2013) }& Even if subleading, tetraquarks can exist. In that case \\
& they are narrow as $1/N$ (like mesons).\\
\hline
Knecht-Peris\cite{Knecht:2013yqa} & Tetraquarks with 4 different flavors are as narrow as $1/N^2$.\\
Lebed\cite{Lebed:2013aka} & Non-perturbative effects in $1/N$ could affect tetraquark wave function.\\
\multirow{2}{*}{Cohen-Lebed 1\cite{Cohen:2014tga}} & Tetraquarks naturally exist in Corrigan-Ramond limit \\
& (quarks in the antisymmetric representation).\\
Cohen-Lebed 2\cite{Cohen:2014via} & Production of tetraquarks in scattering amplitudes is only sub-subleading.\\
\hline\hline
\end{tabular}
\label{tab:largeN_summary}
}
\end{table}
Imagine to cut the diagram in \figurename{~\ref{fig:diagrammi(b)}} separating the incoming mesons in $x,y$ from the outgoing mesons in $z,w$ .
The resulting cut is shown in \figurename{~\ref{fig:cohen(a)}}. Apparently, we are led to say that in the considered scattering amplitude there is a contribution from a tetraquark cut.
However, drawing the diagram in a different, topologically equivalent, way (\figurename{~\ref{fig:cohen(b)}}), we see that the effect 
of the cut is to put on shell the corners of the diagram, thus separating all the meson sources from each other.

Recalling that the scattering amplitude is obtained, in momentum space, from Eq.~\eqref{G4} multiplying it for the inverse of the propagators of the mesons in the external legs,
we have
\begin{equation}
\mathcal{A}(s,t)=\prod_i \lim_{q_i^2\to m_i^2} (q_i^2 - m_i^2) \frac{1}{\sqrt{Z_i}} \widetilde{G}_4(\{q_i^2\};s,t),
\end{equation}
with $\widetilde{G}_4(\{q_i^2\};s,t)$ the Fourier transform of $G_4$ in Eq.~\eqref{G4}.
The factors $q_i^2 - m_i^2$ cancel exactly the contribution of the sources to the $s$-channel and the on-shell contributions come only from 
meson intermediate states. 

Drastically different is the situation for a tetraquark with flavor $\left(\bar c u\right) \left( \bar d  c \right)$, as the recently discovered charged resonance $Z(4430)$. 
The resulting connected leading order diagram is similar to the diagram in \figurename{~\ref{fig:diagramma(a)}}. In that case a cut in the $s$-channel reveal either a meson or 
a tetraquark intermediate state. 

\vspace{1em}

To summarize, a pure tetraquark intermediate state, \ie with flavor quantum numbers that can only be interpreted as exotic, cannot contribute to the leading order connected contribution to meson-meson elastic scattering amplitude.
It is remarkable to notice that the experimental situation is drastically different.
Until now, there is no evidence for such exotic resonances with all four different flavors 
and the considerations illustrated here are not applicable to the current experimental situation.

The key points of this section are schematically summarized in Table~\ref{tab:largeN_summary}.

\section{Experimental overview}
\label{sec:experimental}
\begin{figure}[b]
\centering
\includegraphics[width=.75\textwidth]{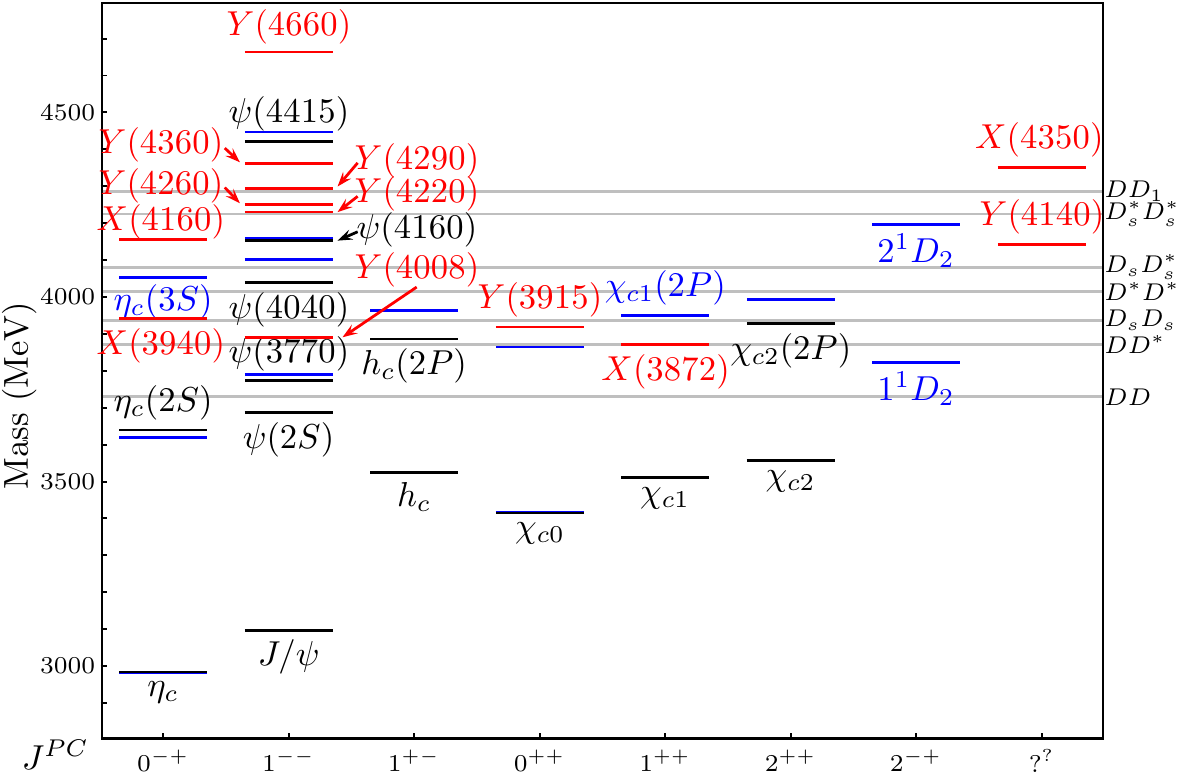}
\includegraphics[width=.75\textwidth]{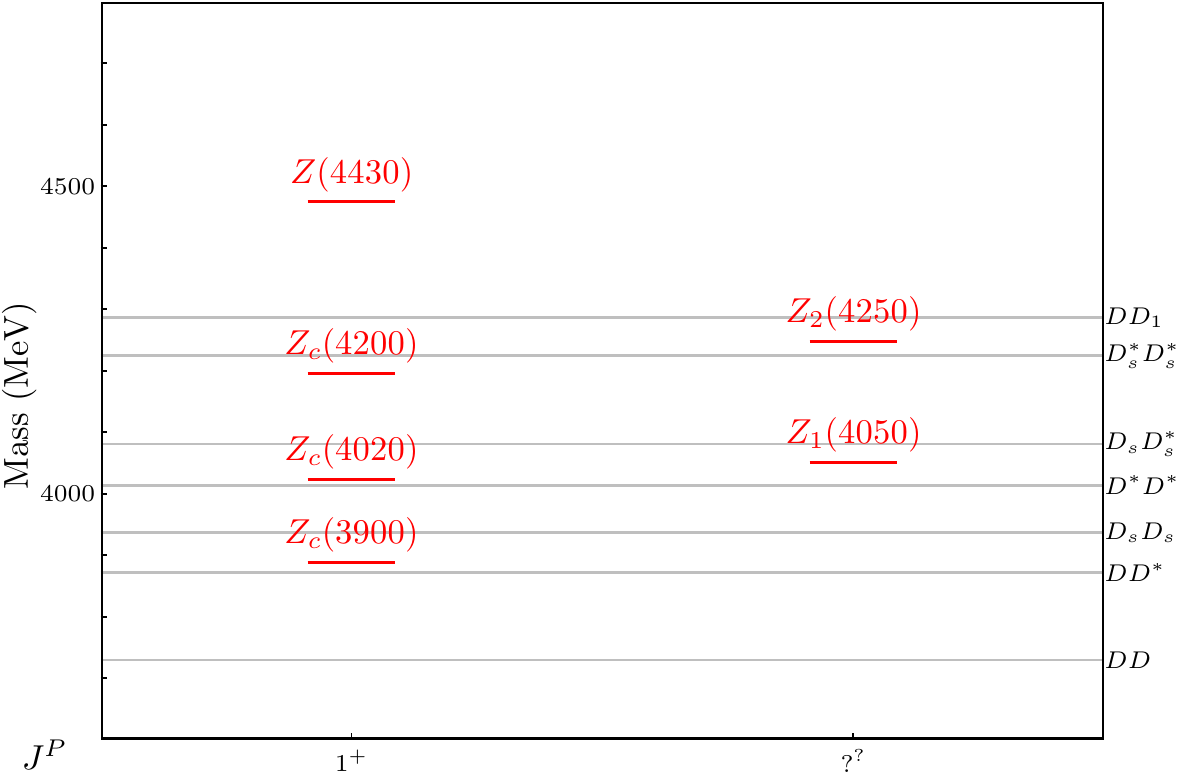}
\caption{Charmonium sector. In the upper panel, we show ordinary charmonia 
and neutral exotic states, in the lower panel charged exotic states. 
Black lines represent observed charmonium levels, blue lines represent 
predicted levels according to Radford and Repko\cite{Radford:2007vd}, 
red line are exotic states. The open charm thresholds are reported on the right.}.
 \label{fig:charmonium_levels}
\end{figure}
As shown in the previous section, recently a good deal of work has 
been done to understand the phenomenology of tetrquark states in Large-$N$ 
QCD. In particular, some doubts were raised about a possible broadness of 
these particles, that would make them experimentally undetectable. 
However, as we will show in the following sections, in the past eleven 
years many different experiments, both at lepton and hadron colliders, 
reported evidences for a large number of particles having properties which 
can hardly be embedded in the known charmonia frameworks. A pictorial 
representation of this is visible 
in \figurename{~\ref{fig:charmonium_levels}}. In particular, the charged 
states reported in the second panel are manifestly exotic. Some states, 
like the $X(3872)$ or the $X(3915)$, have more or less the correct mass 
and quantum numbers to be identified with (otherwise missing) ordinary 
charmonia; on the other hand, in the vector sector we have much more levels 
than expected. In any case, the decay pattern of these states is not 
compatible with charmonia predictions, and so it needs some exotic assignment.

Besides finding the states, the measurement of the quantum numbers is 
needed to establish their exotic nature.
While prompt production at hadron colliders can produce particles with any 
quantum numbers, exclusive production modes, in particular at $e^+e^-$ 
colliders, can constrain the assignment. For example, a generic state $X$ 
could be produced:
\begin{itemize}
 \item Directly with $e^+e^- \to X$, if the center-of-mass energy 
coincides with the mass of the state (typically at $\tau$-$c$ factories), 
or in association with Initial State Radiation (ISR) which lowers the center 
of mass energy, $e^+e^- \to e^+e^- \gamma_\text{ISR} \to X \gamma_\text{ISR}$, 
typically at $B$-factories. In the first case an invariant mass distribution can 
be studied by varying the energy of the beam, which does not allow to collect 
many data points with high statistics, while in the second the same distribution 
is studied as a function of the $\gamma_\text{ISR}$ energy. In both cases, the 
quantum numbers must be the same as the photon, \ie $J^{PC} = 1^{--}$.
  \item In the fusion of two quasi-real photons, 
$e^+e^- \to e^+ e^- \gamma\gamma \to e^+ e^- X$, where $e^+$ and $e^-$ are 
scattered at a small angle and are not detected; the signal events have no 
tracks and neutral particles but the daughters of $X$. If the photons are 
quasi-real, Landau-Yang theorem holds\cite{Yang:1950rg}, and $J \neq 1$; 
moreover $C= +$ is costrained.
  \item In double charmonium production, for example $e^+e^- \to \jpsi X$, 
which constrains $X$ to have $C$ opposite to the one of the associated charmonium.
\end{itemize}

The production in $B$ decays allows $X$ to have any $J^{PC}$, 
albeit low values of the spin are preferred.

Hadron colliders, instead, produce charmonia states both directly and in $B$
decays, and the search is typically carried out inclusively. 

A summary of the resonances we will talk about is reported in \tablename{~\ref{tab:allexp}}. We start our review from the charged ones, first the recently confirmed 
$Z(4430)$ in \sectionname{~\ref{sec:Z4430}}, then we move to the charged 
states in the $3900$-$4200$\mev region (\sectionname{~\ref{sec:Z3900}}) 
and the corresponding ones in the bottomonium sector 
(\sectionname{~\ref{sec:Zb}}). The $X(3872)$ is extensively described 
in \sectionname{~\ref{sec:X3872}}, as well as the vector states in  
\sectionname{~\ref{sec:vectors}}. Finally, the other resonances around $3940$\mev 
are described in \sectionname{~\ref{sec:X3940}}, and the remaining ones 
in \sectionname{~\ref{sec:otherplus}}. Summary and future perspectives are 
in \sectionname{~\ref{sec:expsummary}}. Other 
information can be found in 
older reviews by Faccini~\etal\cite{Drenska:2010kg,Faccini:2012pj}; 
a complete treatise about the physics of \babar and \belle can be found 
in the recent review book\cite{pbf}.

\begin{landscape}
 \begin{table}[t]
\footnotesize
\tbl{Summary of quarkonium-like states. 
For charged states, the $C$-parity is given for the neutral members of the corresponding 
isotriplets.}{
\begin{tabular}{lrcllc}\hline\hline
State & $M$ (\mev) & $\Gamma$ (\mev) & $J^{PC}$ & Process (mode) & Experiment (\#$\sigma$) \\
\hline
      $X(3823)$ & $3823.1\pm1.9$ & $<24$ & $?^{?-}$ &
      $B\to K(\chi_{c1}\gamma)$ & \belle\cite{Bhardwaj:2013rmw} (4.0) \\
      $X(3872)$ & $3871.68\pm0.17$ & $<1.2$ & $1^{++}$ &
      $B\to K(\pi^+\pi^-\jpsi)$ & \belle\cite{Choi:2003ue,Choi:2011fc} ($>$10), \babar\cite{Aubert:2008gu} (8.6) \\
      & & & &
      $p\bar{p}\to(\pi^+\pi^-\jpsi)\,...$ & \cdf\cite{Abulencia:2006ma,Aaltonen:2009vj} (11.6), D0\cite{Abazov:2004kp} (5.2)  \\
      & & & &
      $pp\to(\pi^+\pi^-\jpsi)\,...$ & \lhcb\cite{Aaij:2011sn,Aaij:2013zoa} (np)  \\
      & & & &
      $B\to K(\pi^+\pi^-\pi^0\jpsi)$ & \belle\cite{Abe:2005ix} (4.3), \babar\cite{delAmoSanchez:2010jr} (4.0) \\
      & & & &
      $B\to K(\gamma\, \jpsi)$ & \belle\cite{Bhardwaj:2011dj} (5.5), \babar\cite{Aubert:2008rn} (3.5) \\
      & & & & & \lhcb\cite{Aaij:2014ala}~($>10$) \\
      & & & &
      $B\to K(\gamma\, \psiprime)$ & \babar\cite{Aubert:2008rn} (3.6), \belle\cite{Bhardwaj:2011dj} (0.2) \\
      & & & & & \lhcb\cite{Aaij:2014ala}~(4.4) \\
      & & & &
      $B\to K(D\bar{D}^*)$ & \belle\cite{Adachi:2008su} (6.4), \babar\cite{Aubert:2007rva} (4.9) \\
      $Z_c(3900)^+$ & $3888.7\pm3.4$ & $35\pm7$ & $1^{+-}$ &
      $Y(4260)\to\pi^-(D\bar{D}^*)^+$ & \bes\cite{Ablikim:2013xfr} (np) \\
      & & & &
      $Y(4260)\to\pi^-(\pi^+\jpsi)$ & \bes\cite{Ablikim:2013mio} (8), \belle\cite{Liu:2013dau} (5.2) \\
      & & & & & \cleo data\cite{Xiao:2013iha} ($>$5) \\
      $Z_c(4020)^+$ & $4023.9\pm2.4$ & $10\pm6$ & $1^{+-}$ &
      $Y(4260)\to\pi^-(\pi^+h_c)$ & \bes\cite{Ablikim:2013wzq} (8.9) \\
      & & & &
      $Y(4260)\to\pi^-(D^*\bar{D}^*)^+$ & \bes\cite{Ablikim:2013emm} (10) \\
      $Y(3915)$ & $3918.4\pm1.9$ & $20\pm5$ & $0^{++}$ &
      $B\to K(\omega \jpsi)$ & \belle\cite{Abe:2004zs} (8), \babar\cite{Aubert:2007vj,delAmoSanchez:2010jr} (19) \\
      & & & &
      $e^+e^-\to e^+e^-(\omega \jpsi)$ & \belle\cite{Uehara:2009tx} (7.7), \babar\cite{Lees:2012xs} (7.6) \\
      $Z(3930)$ & $3927.2\pm2.6$ & $24\pm6$ & $2^{++}$ &
      $e^+e^-\to e^+e^-(D\bar{D})$ & \belle\cite{Uehara:2005qd} (5.3), \babar\cite{Aubert:2010ab} (5.8) \\
      $X(3940)$ & $3942^{+9}_{-8}$ & $37^{+27}_{-17}$ & $?^{?+}$ &
      $e^+e^-\to \jpsi\,(D\bar{D}^*)$ & \belle\cite{Abe:2007jn,Abe:2007sya} (6) \\
      $Y(4008)$ & $3891\pm42$ & $255\pm42$ & $1^{--}$ &
      $e^+e^-\to (\pi^+\pi^-\jpsi)$ & \belle\cite{Yuan:2007sj,Liu:2013dau} (7.4) \\
      $Z(4050)^+$ & $4051^{+24}_{-43}$ & $82^{+51}_{-55}$ & $?^{?+}$ &
      $\bar{B}^0\to K^-(\pi^+\chi_{c1})$ & \belle\cite{Mizuk:2008me} (5.0), \babar\cite{Lees:2011ik} (1.1) \\
      $Y(4140)$ & $4145.6 \pm 3.6$ & $14.3\pm5.9$ & $?^{?+}$ &
      $B^+\to K^+(\phi \jpsi)$ & \cdf\cite{Aaltonen:2009tz,Aaltonen:2011at} (5.0), \belle\cite{Shen:2009vs} (1.9), \\
      & & & & & \lhcb\cite{Aaij:2012pz} (1.4), \cms\cite{Chatrchyan:2013dma} ($>$5) \\
      & & & & & \Dzero\cite{Abazov:2013xda} (3.1) \\
      $X(4160)$ & $4156^{+29}_{-25}$ & $139^{+113}_{-65}$ & $?^{?+}$ &
      $e^+e^-\to \jpsi\,(D^*\bar{D}^*)$ & \belle\cite{Abe:2007sya} (5.5) \\
     $Z(4200)^+$ & $4196^{+35}_{-30}$ & $370^{+99}_{-110}$ & $1^{+-}$ &
     $\bar{B}^0\to K^-(\pi^+\jpsi)$ & \belle\cite{Chilikin:2014bkk} (7.2) \\
      \hline\hline
\end{tabular}\label{tab:allexp} }
\end{table}
\end{landscape}
\begin{landscape}
 \begin{table}[t]
\footnotesize
\tbl{({\it Continued}).}{
\begin{tabular}{lrcllc}\hline\hline
State & $M$ (\mev) & $\Gamma$ (\mev) & $J^{PC}$ & Process (mode) & Experiment (\#$\sigma$) \\
\hline
     $Y(4220)$ & $4196^{+35}_{-30}$ & $39\pm32$ & $1^{--}$ &
     $e^+e^-\to (\pi^+\pi^- h_c)$ & \bes data\cite{yuan1,yuan2} (4.5) \\
     $Y(4230)$ & $4230 \pm 8$ & $38\pm12$ & $1^{--}$ &
     $e^+e^-\to (\chi_{c0}\omega)$ & \bes\cite{Ablikim:2014qwy} ($>$9) \\
      $Z(4250)^+$ & $4248^{+185}_{-45}$ & $177^{+321}_{-72}$ & $?^{?+}$ &
      $\bar{B}^0\to K^-(\pi^+\chi_{c1})$ & \belle\cite{Mizuk:2008me} (5.0), \babar\cite{Lees:2011ik} (2.0) \\
      $Y(4260)$ & $4250\pm9$ & $108\pm12$ & $1^{--}$ &
      $e^+e^-\to (\pi\pi \jpsi)$ & \babar\cite{Aubert:2005rm,Lees:2012cn} (8), CLEO\cite{Coan:2006rv,He:2006kg} (11) \\
      & & & & & \belle\cite{Yuan:2007sj,Liu:2013dau} (15), \bes\cite{Ablikim:2013mio} (np) \\
      & & & &
      $e^+e^-\to(f_0(980)\jpsi)$ & \babar\cite{Lees:2012cn} (np), \belle\cite{Liu:2013dau} (np) \\
      & & & &
      $e^+e^-\to(\pi^-Z_c(3900)^+)$ & \bes\cite{Ablikim:2013mio} (8), \belle\cite{Liu:2013dau} (5.2) \\
      & & & &
      $e^+e^-\to(\gamma\,X(3872))$ & \bes\cite{Ablikim:2013dyn} (5.3) \\
     $Y(4290)$ & $4293\pm9$ & $222\pm67$ & $1^{--}$ &
     $e^+e^-\to (\pi^+\pi^- h_c)$ & \bes data\cite{yuan1,yuan2} (np) \\
      $X(4350)$ & $4350.6^{+4.6}_{-5.1}$ & $13^{+18}_{-10}$ & $0/2^{?+}$ &
      $e^+e^-\to e^+e^-(\phi \jpsi)$ & \belle\cite{Shen:2009vs} (3.2) \\
      $Y(4360)$ & $4354\pm11$ & $78\pm16$ & $1^{--}$ &
      $e^+e^-\to(\pi^+\pi^-\psiprime)$ & \belle\cite{Wang:2007ea} (8), \babar\cite{Lees:2012pv} (np) \\
      $Z(4430)^+$ & $4478\pm17$ & $180\pm 31$ & $1^{+-}$ &
      $\bar{B}^0\to K^-(\pi^+\psiprime)$ & \belle\cite{Mizuk:2009da,Chilikin:2013tch} (6.4), \babar\cite{Aubert:2008aa} (2.4) \\
      & & & & & \lhcb\cite{Aaij:2014jqa} (13.9) \\
      & & & &
      $\bar{B}^0\to K^-(\pi^+\jpsi)$ & \belle\cite{Chilikin:2014bkk} (4.0) \\
      $Y(4630)$ & $4634^{+9}_{-11}$ & $92^{+41}_{-32}$ & $1^{--}$ &
      $e^+e^-\to(\Lambda_c^+\bar{\Lambda}_c^-)$ & \belle\cite{Pakhlova:2008vn} (8.2) \\
      $Y(4660)$ & $4665\pm10$ & $53\pm14$ & $1^{--}$ &
      $e^+e^-\to(\pi^+\pi^-\psiprime)$ & \belle\cite{Wang:2007ea} (5.8), \babar\cite{Lees:2012pv} (5) \\ \hline
      $Z_b(10610)^+$ & $10607.2\pm2.0$ & $18.4\pm2.4$ & $1^{+-}$ &
      $\Upsilon(5S)\to\pi(\pi\Upsilon(nS))$ & \belle\cite{Belle:2011aa,Krokovny:2013mgx} ($>$10) \\
      & & & & 
      $\Upsilon(5S)\to\pi^-(\pi^+h_b(nP))$ & \belle\cite{Belle:2011aa} (16) \\
      & & & & 
      $\Upsilon(5S)\to\pi^-(B\bar{B}^*)^+$ & \belle\cite{Adachi:2012cx} (8) \\
      $Z_b(10650)^+$ & $10652.2\pm1.5$ & $11.5\pm2.2$ & $1^{+-}$ &
      $\Upsilon(5S)\to\pi^-(\pi^+\Upsilon(nS))$ & \belle\cite{Belle:2011aa} ($>$10) \\
      & & & & 
      $\Upsilon(5S)\to\pi^-(\pi^+h_b(nP))$ & \belle\cite{Belle:2011aa} (16) \\
      & & & & 
      $\Upsilon(5S)\to\pi^-(B^*\bar{B}^*)^+$ & \belle\cite{Adachi:2012cx} (6.8) \\
      \hline\hline
\end{tabular}\label{tab:allexp2} }
\end{table}
\end{landscape}

\subsection{$Z(4430)$}
\label{sec:Z4430}
In April 2014, LHCb confirmed the existence of a charged resonance in 
the $\psiprime \pi^-$ channel\cite{Aaij:2014jqa}.~\footnote{Unless specified, 
the charged conjugated modes are understood.} This announcement solved a 
controversy between \belle, which discovered\cite{Choi:2007wga} and 
confirmed\cite{Mizuk:2009da,Chilikin:2013tch} the existence of this state, 
and \babar, which did not observe any new structure and criticized some aspects 
of \belle's analysis\cite{Aubert:2008aa}.
A state decaying into a charmonium and a charged light meson has undoubtly a 
four-quark content, being the production of a heavy quark pair from vacuum 
OZI suppressed. As we will discuss later, the very existence of such an exotic 
state far from usual open-charm thresholds is extremely interesting for 
phenomenological interpretations. We now briefly review the experimental history 
of this and other charged states.
\begin{figure}[t]
  \begin{center}
    \includegraphics[width=6cm]{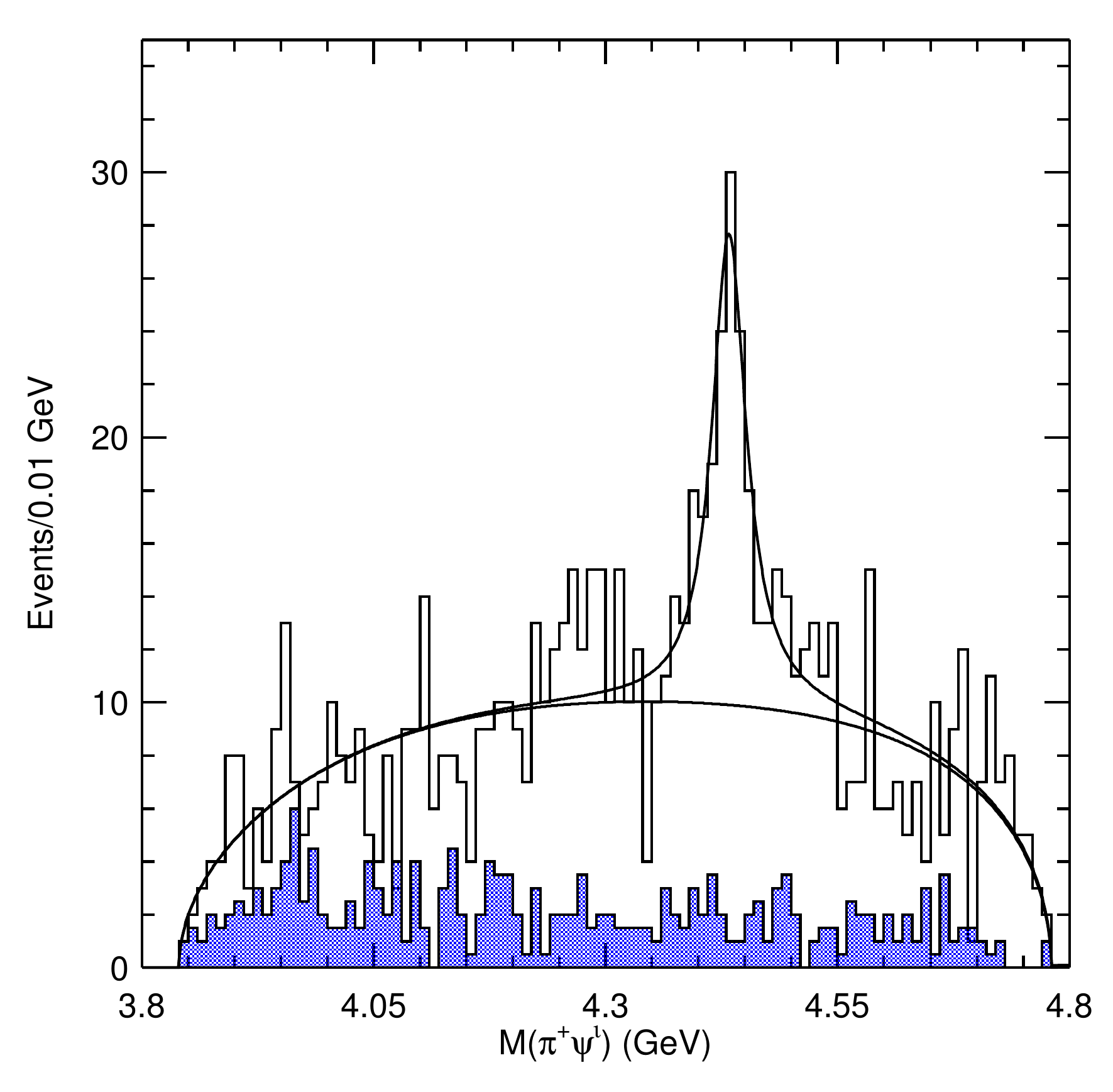} 
    \includegraphics[width=6.5cm]{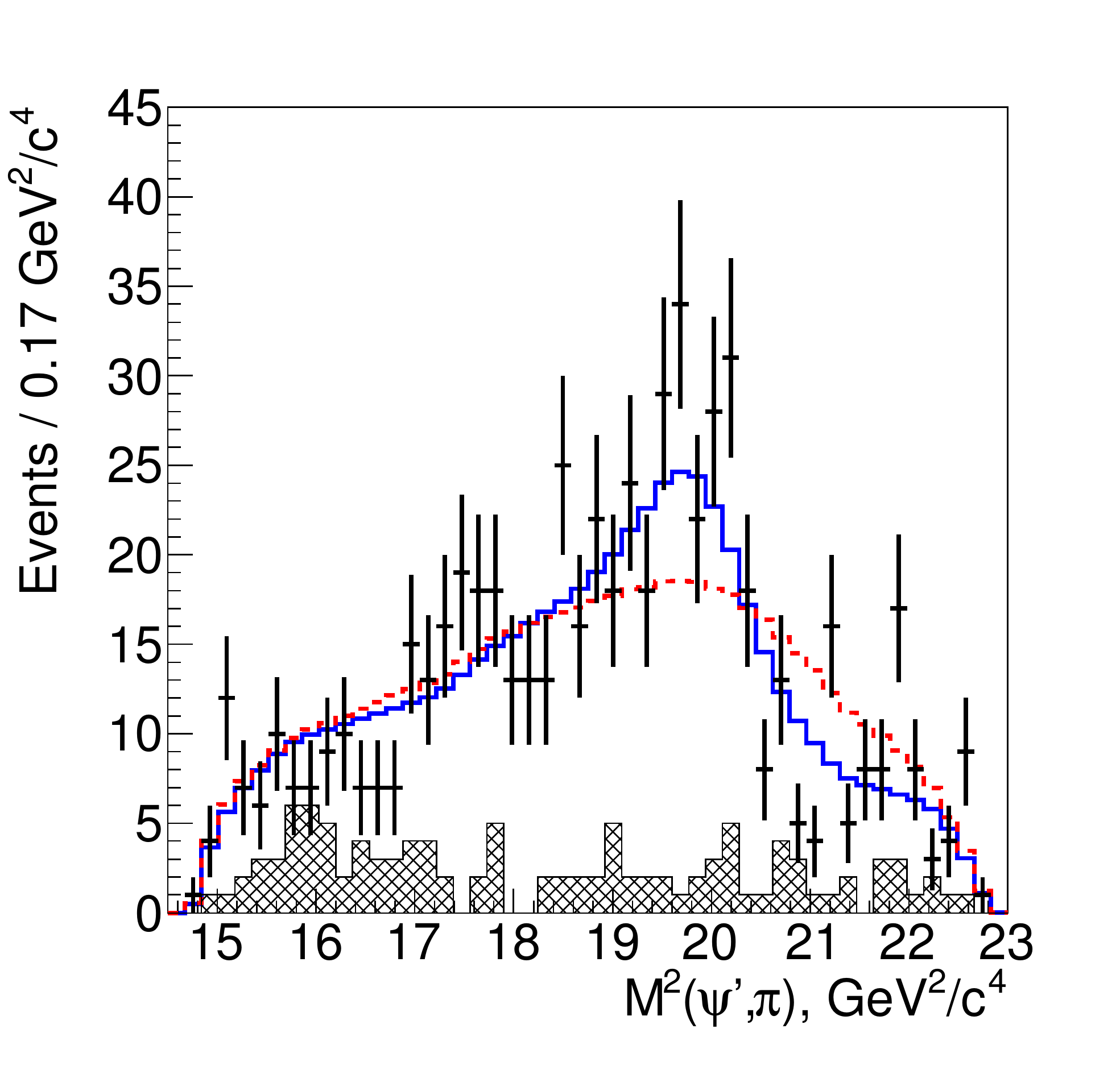} 
  \end{center}
   \caption{Invariant mass distributions in $\psiprime \pi^-$ channel according to first\cite{Choi:2007wga} (left), and
   last\cite{Chilikin:2013tch} (right) \belle analyses. The fit shows that 
an additional resonances is needed to describe the data. In the right panel, 
the blue solid (red dashed) curve shows the fit with (without) the additional 
$Z(4430)$ resonance. In both figures, a $K^*$ veto has been applied.}
  \label{fig:z4430belle}
\end{figure}

The original \belle paper\cite{Choi:2007wga} studies the $B\to\psiprime \pi K$ 
decays, and reports a peak in the $\psiprime \pi$ invariant mass distribution, 
with $M=(4430 \pm 4 \pm 2) \mev$ and $\Gamma = (45^{+18}_{-13}{}^{+30}_{-13})\mev$ \figref{fig:z4430belle}. This kind of analysis is particularly difficult, because the rich structure of $K\pi$ resonances could reflect into the $\psiprime \pi$ channel and create many fake peaks. However, \belle considered that the events with $M(\psiprime\pi^-)\sim 4430\mev$ correspond to events with $\cos\theta_{K\pi}\simeq 0.25$, \ie an angular region where interfering  $L=0,1,2$ partial waves cannot produce a single peak without creating other larger structures elsewhere. \belle named this state $Z(4430)$, and reported the product branching fractions
\begin{equation}
\BR\left(\Bz \to K^+ Z(4430)^-\right) \times \BR\left(Z(4430)^- \to \psiprime \pi^-\right) = \left(4.1 \pm 1.0 \pm 1.4\right) \times 10^{-5}.
\end{equation}

\babar reviewed this analysis\cite{Aubert:2008aa}, by studying in detail the efficiency corrections and the shape of the background, relying for the latter on data as much as possible. Hints of a structure near $4430\mev$ appeared, even though not statistically significant, thus leading to a 95\% C.L. upper limit on the production branching fraction
\begin{equation}
\BR(B^0 \to K^+ Z(4430)^-)\times\BR(Z(4430)^-\to\psiprime\pi^-) <3.1\times 10^{-5}.
\end{equation}

\begin{figure}[b]
  \begin{center}
    \includegraphics[width=7cm]{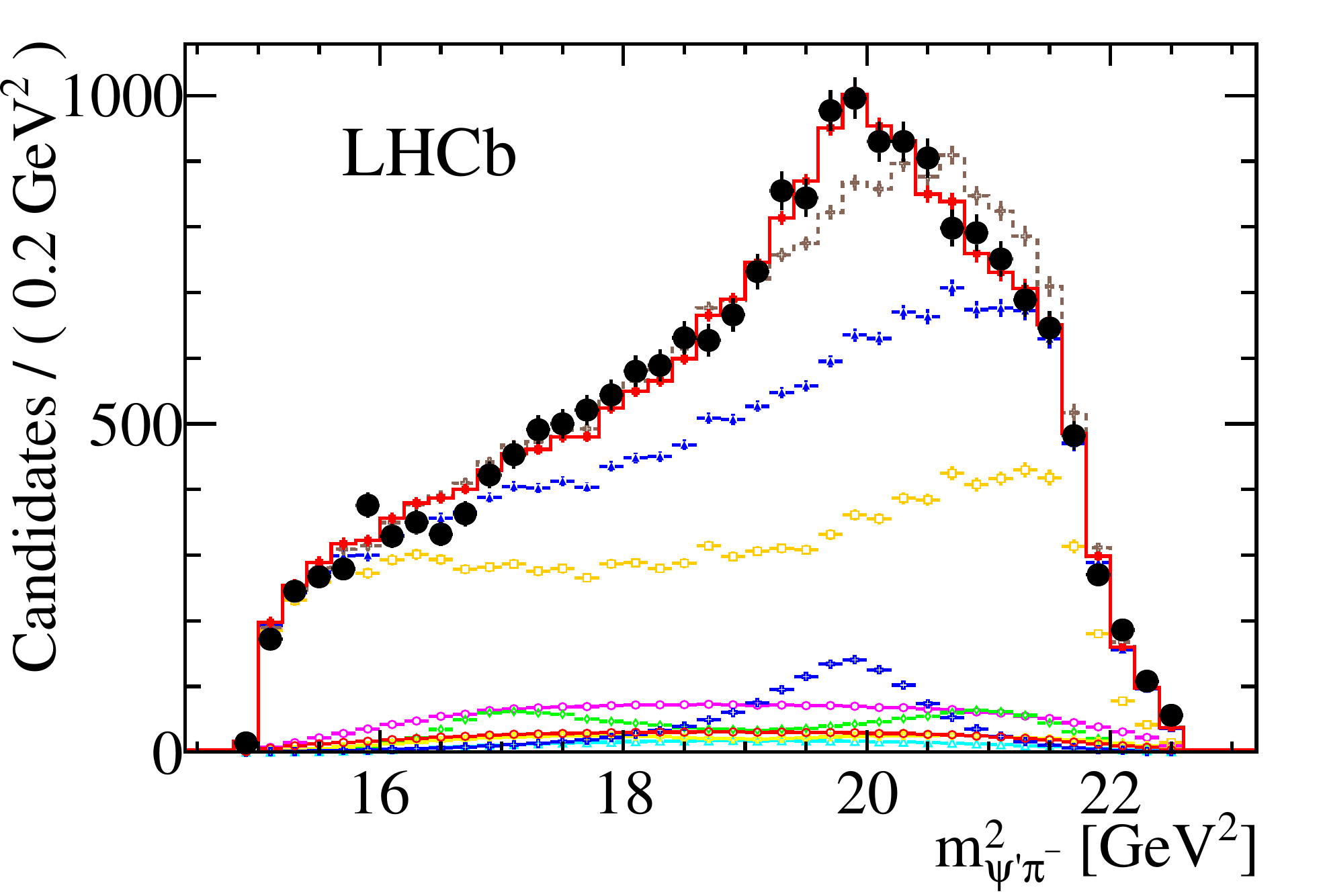} 
    \includegraphics[width=5cm]{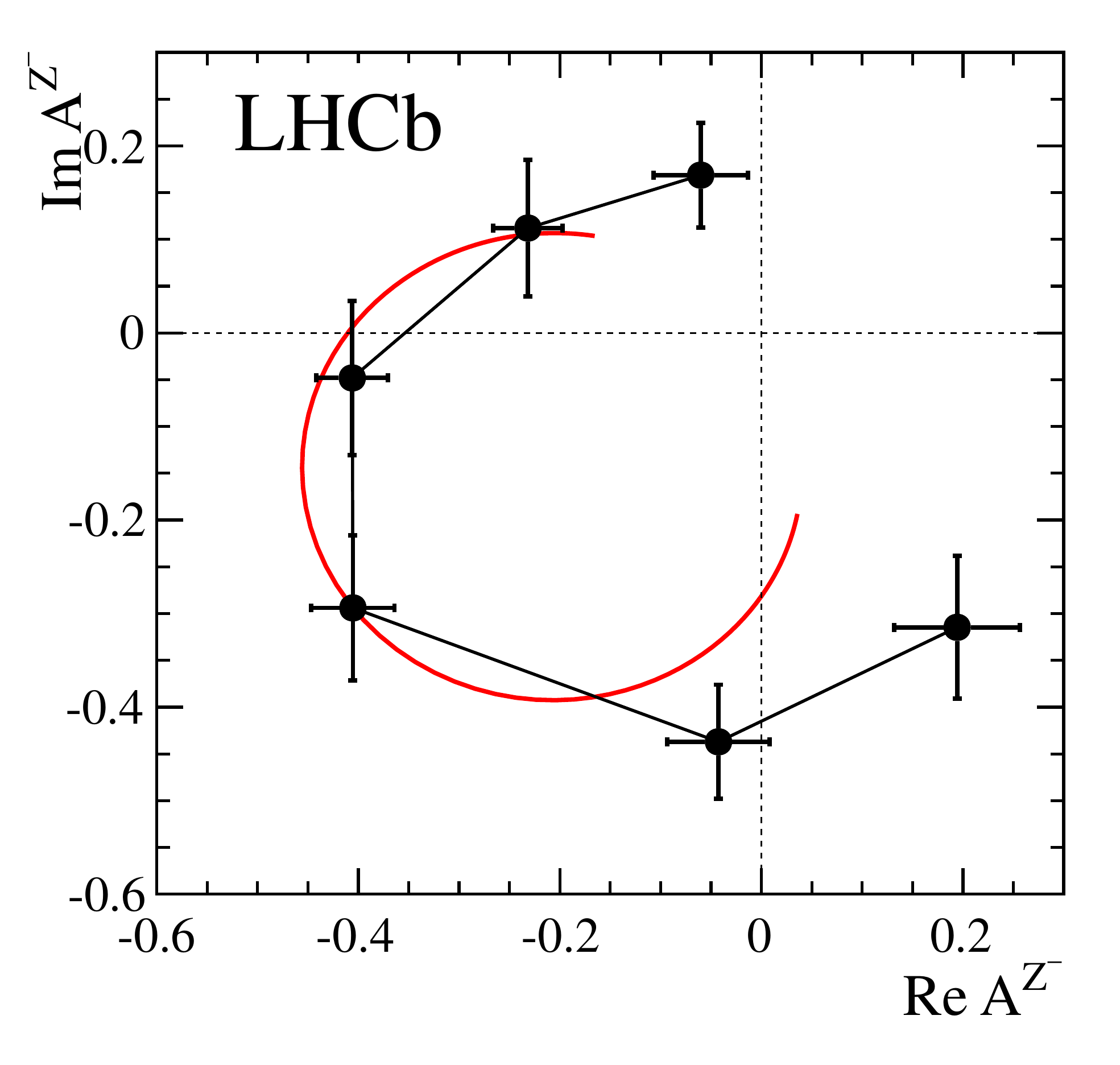} 
  \end{center}
   \caption{Invariant mass distributions in $\psiprime \pi^-$ channel (left) and resonant behaviour (right) according to \lhcb\cite{Aaij:2014jqa}. In the left panel, the red solid (brown dashed) curve shows the fit with (without) the additional $Z(4430)$ resonance. In the right panel, the complex value of the $Z(4430)$ fitted amplitude for six bins of $M(\psiprime \pi)$ is shown. The red curve is the
prediction from the Breit-Wigner formula with a resonance mass (width) of $4475$ ($172$)\mev.}
  \label{fig:z4430lhcb}
\end{figure}
After that, \belle revised the analysis\cite{Mizuk:2009da} studying in detail the 3-body Dalitz plot, and adding all known $K\pi$ resonances, both with and without a coherent amplitude for the $Z(4430)$ in the $\psiprime \pi^-$ channel. \belle confirmed the presence of a peak with a statistical significance of $6.4\sigma$. The Breit-Wigner parameter from the Dalitz analysis are $M=(4443^{+15}_{-12}{}^{+19}_{-13})\mev$ and  $\Gamma = (109^{+86}_{-43}{}^{+74}_{-56})\mev$. A more recent 4D re-analysis by \belle\cite{Chilikin:2013tch} shows that the $J^P=1^+$ hypothesis is favored, modifying mass and width values 
to $M = 4485^{+22 +28}_{-22 -11}\mev$ and 
$\Gamma = 200^{+41 +26}_{-46 -35}\mev$ \figref{fig:z4430belle}. 
The production branching fraction is instead
\begin{equation}
\BR\left(B^0\to K^+ Z(4430)^-\right)\times\BR\left(Z(4430)^- \to\psiprime\pi^-\right) = \left(6.0^{+1.7}_{-2.0}{}^{+2.5}_{-1.4}\right)\times10^{-5}.
\end{equation}

\lhcb confirmed this last result with a similar 4D analysis of the same decay channel. The $Z(4430)^+$ is confirmed with a significance of $13.9\sigma$ at least, and the fitted mass and width are  $M=(4475\pm 7^{+15}_{-25})\mev$
and $\Gamma = (172\pm 13^{+37}_{-34})\mev$. Also the $J^P=1^+$ signature is confirmed with high significance. The average {\it \`a la} PDG of \belle's and \lhcb's mass and width are:
\begin{equation}
M=(4478 \pm 17)\mev,\quad\Gamma =(180\pm31) \mev. 
\end{equation}

Since some theoretical papers\cite{Bugg:2008wu} cast doubts on the resonant nature of the peak, in this analysis the complex value of the $Z(4430)$ amplitude has been plotted as a function of $M(\psiprime \pi)$ \figref{fig:z4430lhcb}. The behaviour is compatible with the Breit-Wigner prediction with the fitted values of mass and width. The same analysis also shows hints for a $Z(4200)$ peak with quantum numbers likely $J^P=0^-$, mass and width $M= (4239 \pm 18^{+45}_{-10}) \mev$, $\Gamma=(220 \pm 47^{+108}_{-74}) \mev$; however, since the Argand diagram is not conclusive about its resonant nature, \lhcb has decided not to claim the discovery of another state.

Recently, \belle published a similar analysis of the $B\to\jpsi \pi K$ decays\cite{Chilikin:2014bkk}. Hints of a $Z(4430)$ have been reported in $M(\jpsi \pi)$ invariant mass, with branching fraction
\begin{equation}
\BR\left(B^0\to K^+ Z(4430)^-\right)\times\BR\left(Z(4430)^- \to\jpsi\pi^-\right) = \left(5.4^{+4.0}_{-1.0}{}^{+1.1}_{-0.6}\right)\times10^{-6}.
\end{equation}

The fact that the $Z(4430)$ is found in different decay channels gives solidity to its existence.
In the same analysis, \belle claimed the discovery of a broad $Z(4200)$ state with quantum numbers likely $J^P=1^+$, mass and width $M= (4196^{+31}_{-29}{}^{+17}_{-13}) \mev$, $\Gamma=(370^{+70}_{-70}{}^{+70}_{-132}) \mev$, with a significance of $6.2\sigma$, possibly related to the \lhcb hint. The reported branching fraction is
\begin{equation}
\BR\left(B^0\to K^+ Z(4200)^-\right)\times\BR\left(Z(4200)^- \to\jpsi\pi^-\right) = \left(2.2^{+0.7}_{-0.5}{}^{+1.1}_{-0.6}\right)\times10^{-5}.
\end{equation}

\subsection{Charged states in the $3900$-$4300$\mev region}
\label{sec:Z3900}
\begin{figure}[t]
  \begin{center}
    \includegraphics[width=6cm]{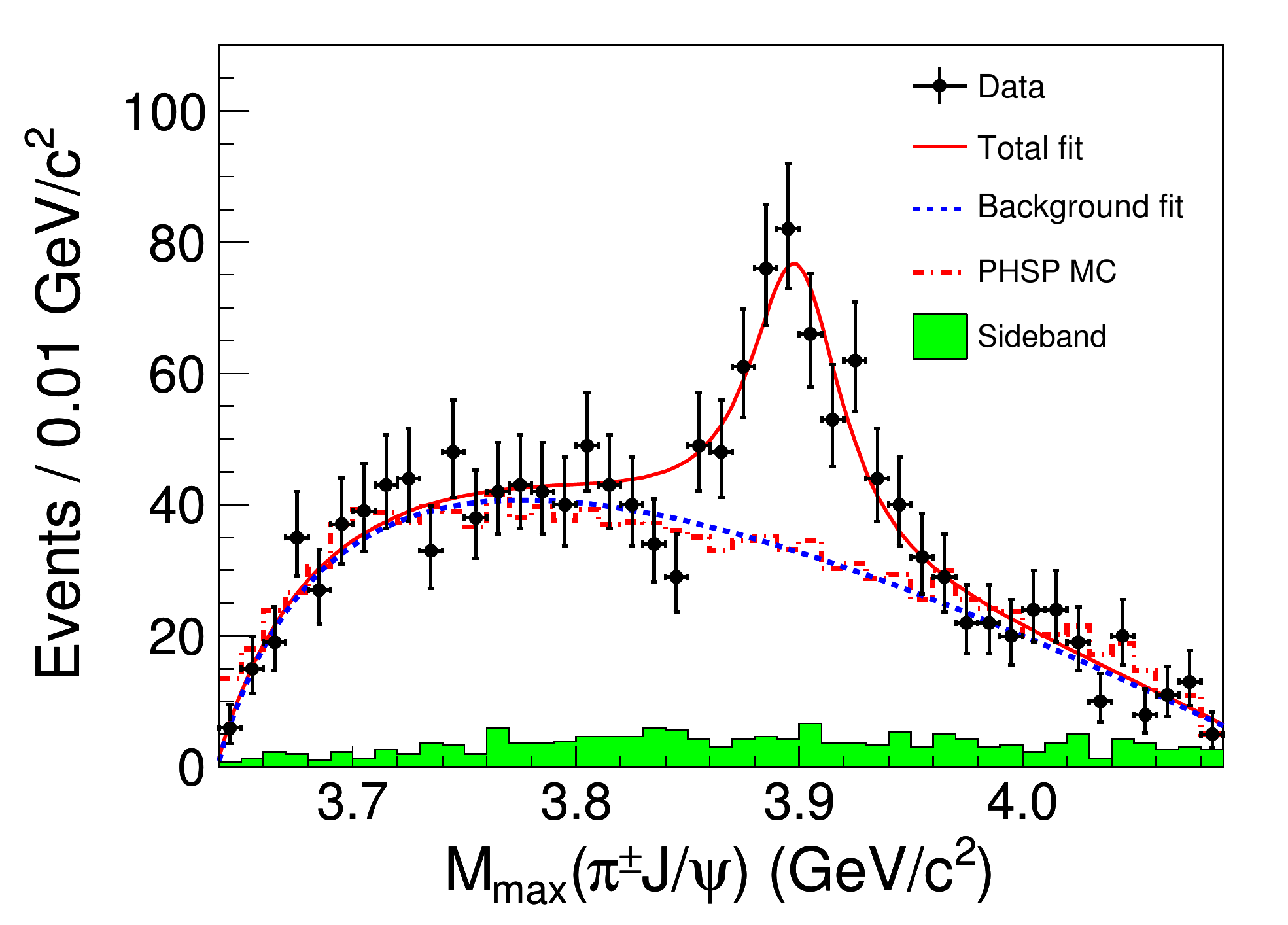} 
    \includegraphics[width=6cm]{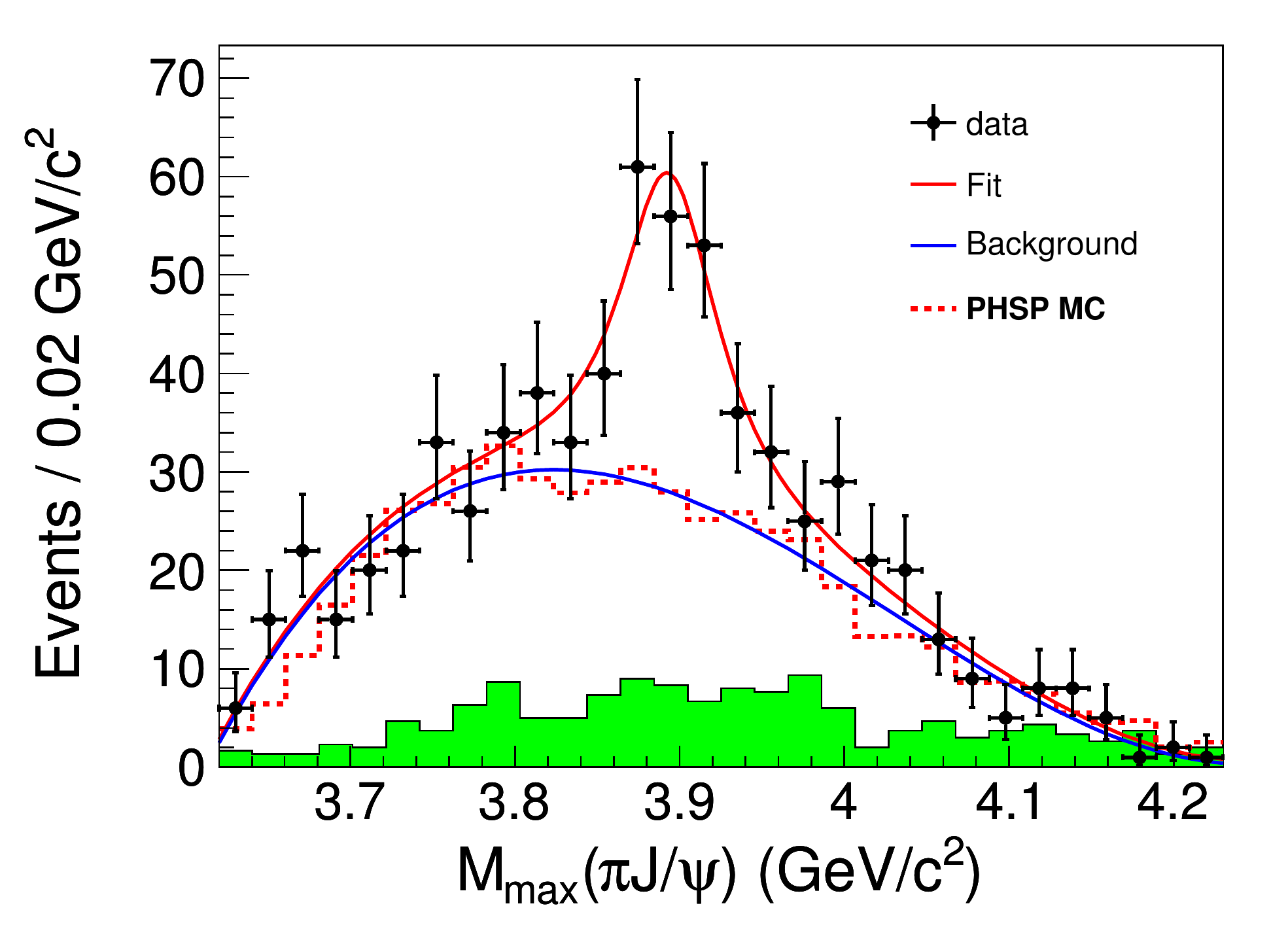} 
  \end{center}
   \caption{Distributions of $M_\text{max}(\jpsi \pi^\pm)$, \ie the larger one of the two $M(\jpsi \pi^\pm)$ in each event, according to \bes\cite{Ablikim:2013mio} (left) and \belle\cite{Liu:2013dau} (right) in the $Y(4260)\to\jpsi\pi^+\pi^-$ decay. The red solid curve is the result of the fit, the blue dotted curve is the background component, the green histogram shows the normalized \jpsi sideband events.}
  \label{fig:z3900jpsipi}
\end{figure}

In March 2013, \bes\cite{Ablikim:2013mio} and \belle\cite{Liu:2013dau} claimed the discovery of a charged resonance in the channel $\jpsi \pi^+$ at a mass of about $3900\mev$, \ie slightly above the $DD^*$ threshold \figref{fig:z3900jpsipi}. \bes takes data at the $Y(4260)$ pole, and analyzes the process $e^+e^- \to Y(4260) \to \jpsi \pi^+\pi^-$; \belle instead produces $Y(4260)$ in addition with initial state radiation (ISR), and analyzes the process $e^+e^- \to Y(4260) \gamma_\text{ISR} \to \jpsi \pi^+\pi^- \gamma_\text{ISR}$. The measured mass and width of the resonance are
\begin{subequations}
\begin{align}
 M &= (3899.0 \pm 3.6 \pm 4.9) \mev, & \Gamma &= (46 \pm 10 \pm 20) \mev\quad\text{(\bes),}\\
M &= (3894.5 \pm 6.6 \pm 4.5) \mev, &  \Gamma &= (63 \pm 24 \pm 26) \mev\quad\text{(\belle),}
\end{align}
\end{subequations}
and production branching fractions
\begin{multline}
 \frac{\BR\left(Y(4260)\to Z_c(3900)^+ \pi^-\right)\times\BR\left(Z_c(3900)^+ \to\jpsi\pi^+\right)}{\BR\left(Y(4260) \to\jpsi\pi^+\pi^-\right)}\\ = (21.5\pm3.3)\% \text{~(\bes)}= (29.0\pm8.9)\% \text{~(\belle)}.
\end{multline}

\begin{figure}[t]
  \begin{center}
    \includegraphics[width=6cm]{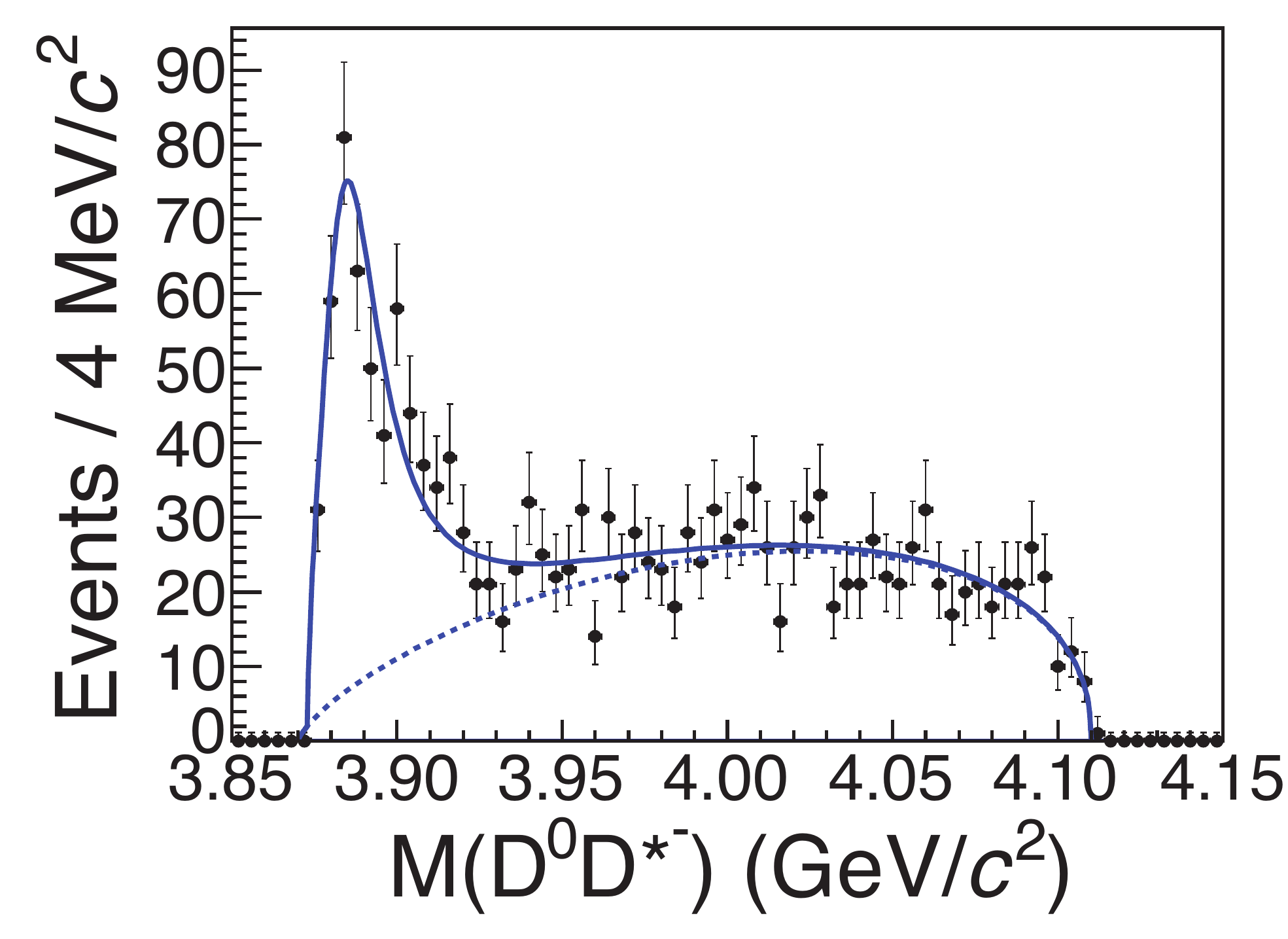} 
    \includegraphics[width=6cm]{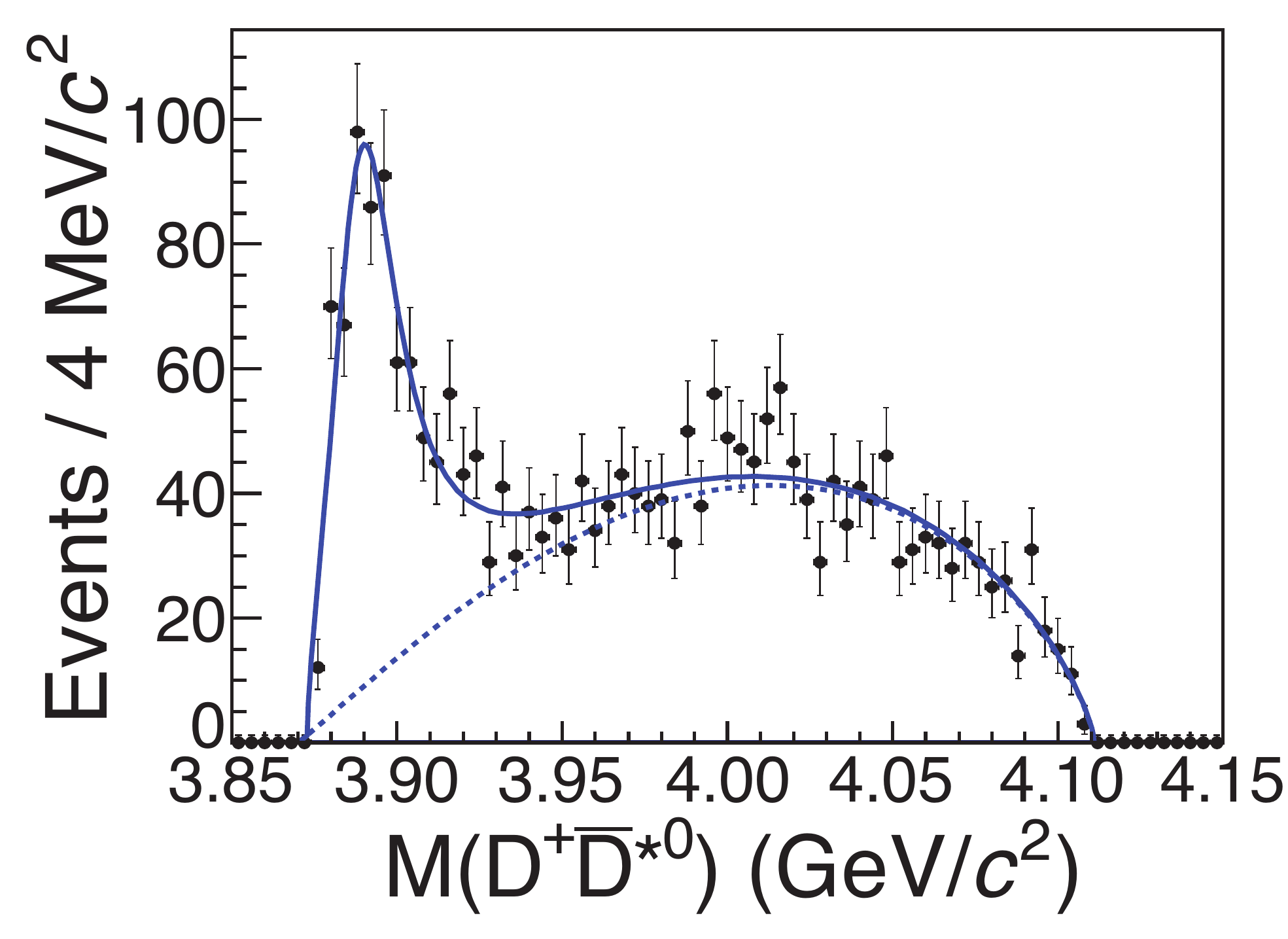} 
  \end{center}
   \caption{Invariant mass distributions of $D^0 D^{*-}$ (left) and $D^+ \Dstarzb$ (right), according to \bes\cite{Ablikim:2013xfr}. The solid curve is the result of the fit, the blue dotted curve is the background component.}
  \label{fig:z3900DDstar}
\end{figure}
This is the first time that a charged manifestly exotic state has been confirmed by two independent experiments, which has given some excitement to the 
charmonium community. The resonance was called $Z_c(3900)$. 
No measurement of quantum numbers has been performed, but $J^P=1^+$ is most likely if the decay $Z_c(3900)\to\jpsi\pi^+$ is assumed to be in $S$-wave.
Soon after, an analysis of \cleoc data confirms\cite{Xiao:2013iha} the presence of the charged $Z_c(3900)^+$ in the $\psi(4160) \to \jpsi\pi^+\pi^-$ decay and provides evidence for a neutral partner in the $\psi(4160) \to \jpsi\pi^0\pi^0$ decay, with fitted parameters
\begin{subequations}
\begin{align}
M(Z_c^+) &= (3886 \pm 4 \pm 2) \mev, & \Gamma &= (37 \pm 4 \pm 8)\mev,\\
M(Z_c^0) &= (3904 \pm 9 \pm 5) \mev, &  \Gamma &= 37 \mev \text{ (fixed).}
\end{align}
\end{subequations}
A preliminary result by \bes confirm the existence of the neutral partner in $Y(4260)\to Z_c(3900)\pi^0\to \jpsi \pi^0\pi^0$\cite{bes3900preliminary}. 
A similar signal has been observed by \bes in $e^+e^-\to (D\Dstarb)^+ \pi^-$, as a resonance in the $(D\Dstarb)^+$ invariant mass\cite{Ablikim:2013xfr}, with mass and width $M = (3883.9\pm 1.5 \pm4.2) \mev$ and $\Gamma = (24.8 \pm 3.3 \pm 11.0)\mev$ \figref{fig:z3900DDstar}. The signature $J^P=1^+$ is favored, and if this state is assumed to be the same as in the $\jpsi \pi^+$ channel, we have
\begin{equation}
 \frac{\BR\left(Z_c(3900) \to D\Dstarb\right)}{\BR\left(Z_c(3900) \to\jpsi\pi\right)} = 6.2 \pm 1.1 \pm 2.7.
\end{equation}

The resulting PDG averaged mass and width are\cite{pdg}:
\begin{align}
 M &= (3888.7 \pm 3.4) \mev, & \Gamma &= (35\pm7) \mev\quad\text{(PDG).}
\end{align}

\begin{figure}[t]
  \begin{center}
    \includegraphics[width=6cm]{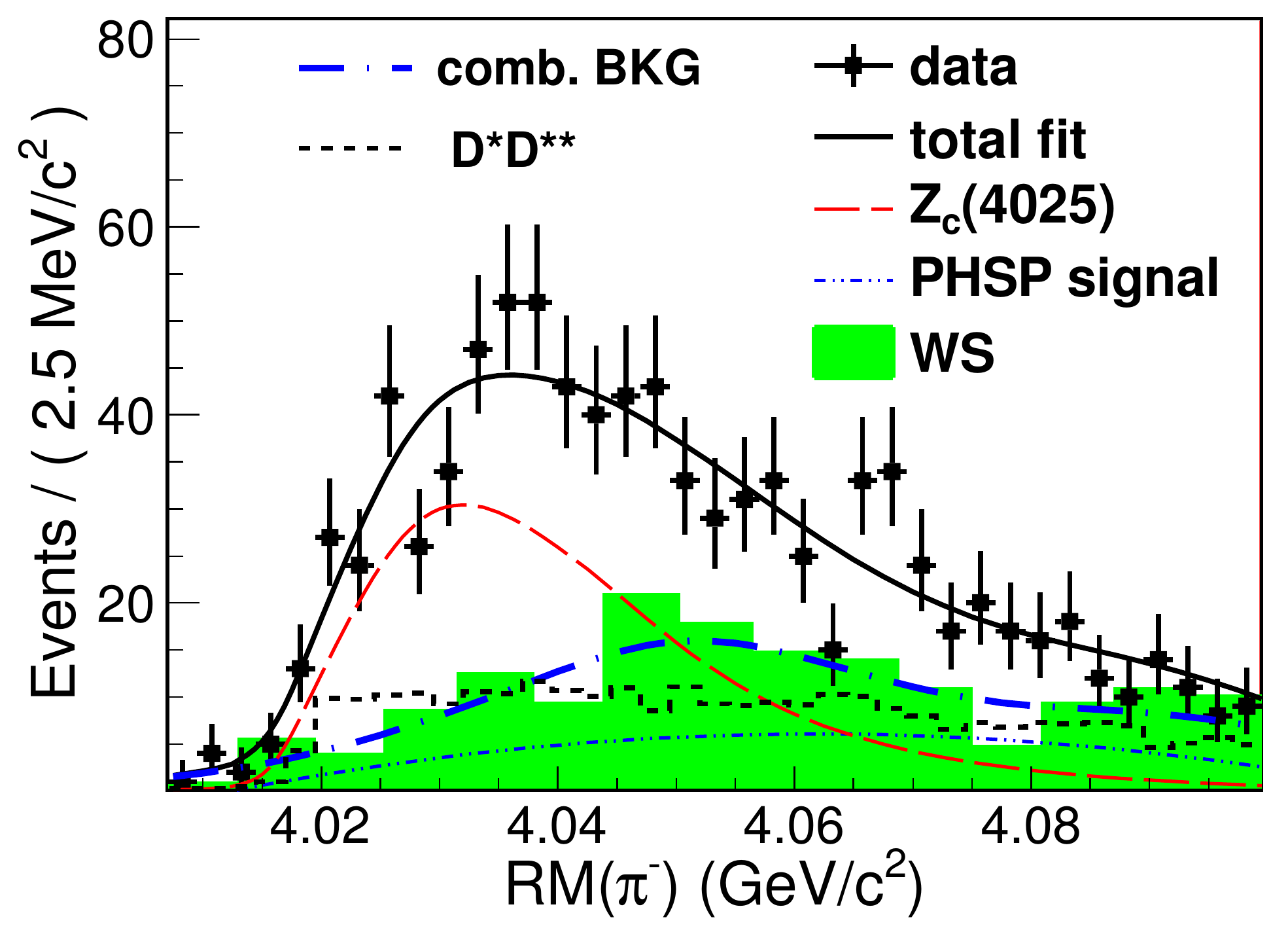} 
    \includegraphics[width=6cm]{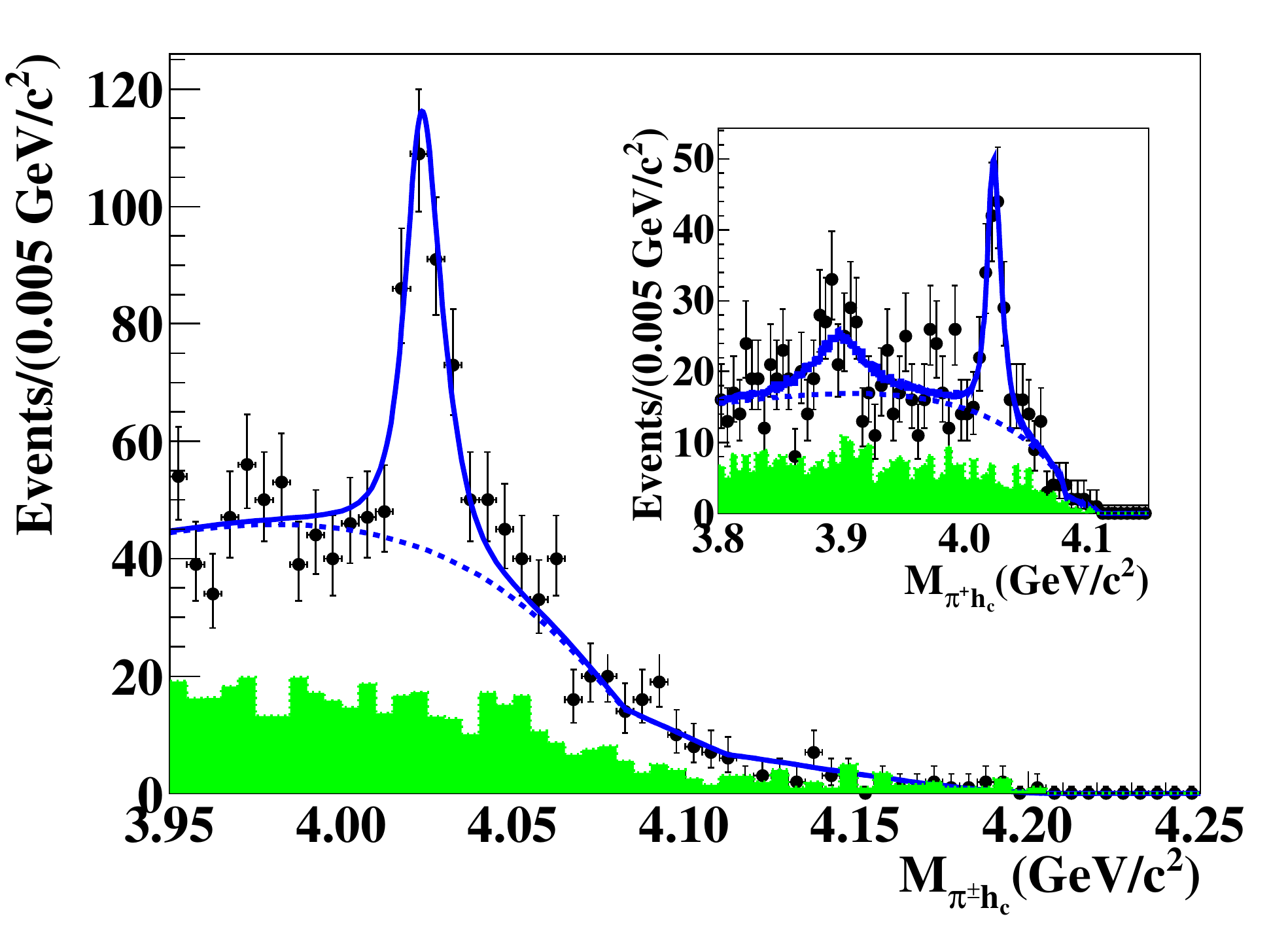} 
  \end{center}
   \caption{Left panel: unbinned maximum likelihood fit to the $\pi$ recoil mass spectrum, in the $e^+e^- \to (\Dstar\Dstarb)^+\pi^-$ analysis by \bes\cite{Ablikim:2013emm}. Right panel: fits to the $M(h_c \pi)$ distributions by \bes\cite{Ablikim:2013wzq}; the inset shows the sum fits if allowing for an additional $Z_c(3900)$ resonance.}
  \label{fig:z4020}
\end{figure}

In the same period, \bes studied the $e^+e^- \to (\Dstar\Dstarb)^+ \pi^-$ process, and observed another charged resonance in the $\Dstar\Dstarb$ channel\cite{Ablikim:2013emm}, at a mass slightly above the $D^*D^*$ threshold, with quantum numbers likely $J^P=1^+$. Soon after, \bes reported a similar peak in the $e^+e^- \to h_c \pi^+\pi^-$ reaction as a resonance in $h_c\pi^+$ invariant mass\cite{Ablikim:2013wzq}. This state is dubbed $Z_c^\prime(4020)$ \figref{fig:z4020}, and the measured masses and widths are:
\begin{subequations}
\begin{align}
M &= (4026.3 \pm 2.6 \pm 3.7) \mev, & \Gamma &= (24.8 \pm 5.6 \pm 7.7)\mev \hspace{0.3em} (Z_c^\prime\to \Dstar\Dstarb),\\
M &= (4022.9 \pm 0.8 \pm 2.7) \mev, & \Gamma &= (7.9 \pm 2.7 \pm 2.6)\mev \hspace{0.8em} (Z_c^\prime\to h_c \pi),\\
M &= (4023.9 \pm 2.4) \mev, & \Gamma &= (10\pm 6)\mev \hspace{4.4em} (\text{PDG}).
\end{align}
\end{subequations}

Moreover, \bes has recently reported some evidence for the neutral isospin partner $Z_c^\prime(4020)^0$, with $M=(4023.9 \pm 2.2 \pm 3.89)\mev$ and the width fixed to $\Gamma(Z_c^\prime(4020)^+)$\cite{Ablikim:2014dxl}.
The $Z_c(3900)$ is also searched\cite{Ablikim:2013wzq} in the $h_c \pi$ 
final state. A peak occurs at $2.1\sigma$ level, thus not statistically significant. A 90\% C.L. upper bound on the production cross section is established:
\begin{align}
\sigma\left(e^+e^- \to Z_c(3900)^+ \pi^- \to h_c \pi^+ \pi^-\right) &< 11\pb,
\intertext{to be compared with}
\sigma\left(e^+e^- \to Z_c(3900)^+ \pi^- \to \jpsi \pi^+ \pi^-\right) &= (13.5\pm2.1)\pb.\text{\cite{Ablikim:2013mio}}
\end{align}

Similarly, no $Z_c^\prime(4020)$ has been seen by \bes and \belle decaying into $\jpsi \pi$, as it is shown in \figurename{~\ref{fig:z3900jpsipi}}.

It is worth noticing that no $Z_c(3900)$ has been seen by \belle in the $B \to K \jpsi \pi$ channel\cite{Chilikin:2014bkk}, and the 90\% C.L. upper bound on the branching fraction is:
\begin{equation}
\BR\left(B^0\to K^+ Z(3900)^-\right)\times\BR\left(Z(3900)^- \to\jpsi\pi^-\right) < 9\times10^{-7}.
\end{equation}

Moreover, the COMPASS collaboration reported a search for $\gamma N \to Z_c^+(3900) N$, where the photon is obtained with scattering of positive muons at 160 and 200\gev on a target of LiD or NH$_3$\cite{Adolph:2014hba}. No signal is observed, and a 90\% C.L. upper bound is put:
\begin{equation}
\frac{\BR\left(Z_c(3900)  \to \jpsi \pi^+\right) \times \sigma\left(\gamma N \to Z_c^+(3900) N \right) }{\sigma\left(\gamma N \to \jpsi  N \right)} < 3.7 \times 10^{-3}
\end{equation}
at $\sqrt{s_{\gamma N}}\simeq 13.8 \gev$.

In a Dalitz-plot analysis of $B\to \chi_{c1}\pi^+ K$ decays, \belle could get an acceptable fit only by adding two resonances
in the  $\chi_{c1}\pi^+$ channel, which were named $Z_1(4050)$ and $Z_2(4250)$\cite{Mizuk:2008me}. The fitted masses and widths are
\begin{subequations}
\begin{align}
M & =(4051 \pm 14^{+20}_{-41})\mev & \Gamma & =(82^{+21+47}_{-17-22})\mev &&(Z_1^+),\\
M &=(4248^{+44+180}_{-29-35})\mev & \Gamma &=(177^{+54+316}_{-39-61})\mev &&(Z_2^+),
\end{align}
\end{subequations}
and reported the production branching fractions
\begin{subequations}
\begin{align}
  \BR(B\to Z_1^-K^+)\times \BR(Z_1^-\to\chi_{c1}\pi^-)	& =(3.0^{+1.2+3.7}_{-0.8-1.6})\times 10^{-5},  \\
  \BR(B\to Z_2^-K^+)\times \BR(Z_2^-\to\chi_{c1}\pi^-)	& =(4.0^{+2.3+19.7}_{-0.9-0.5})\times 10^{-5}.
\end{align}
\end{subequations}

\begin{figure}[t]
  \begin{center}
    \includegraphics[width=5.5cm]{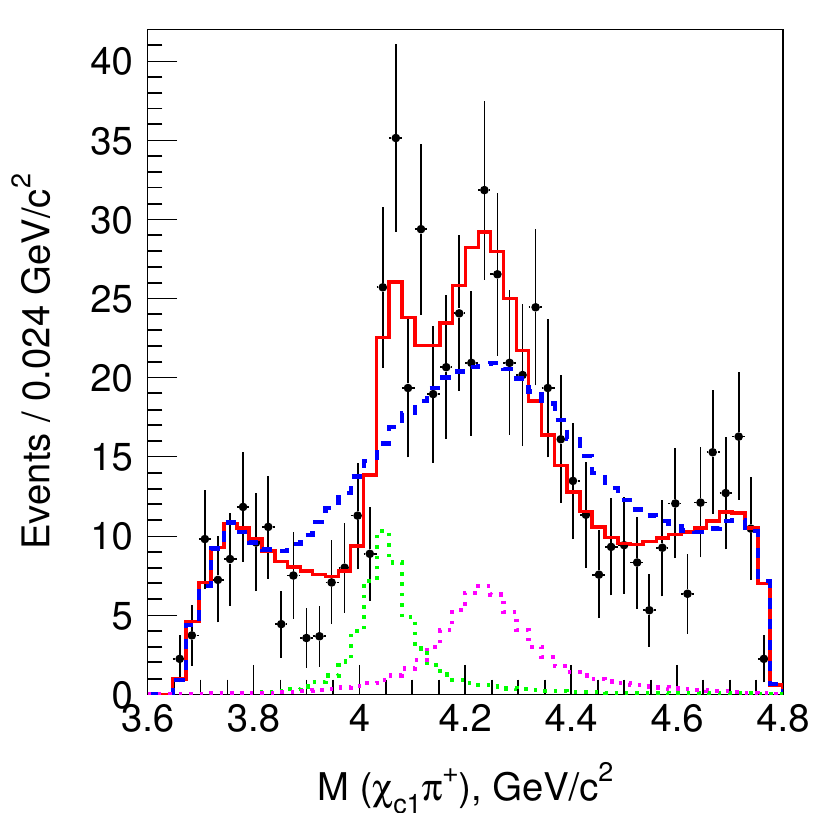} 
    \includegraphics[width=7cm]{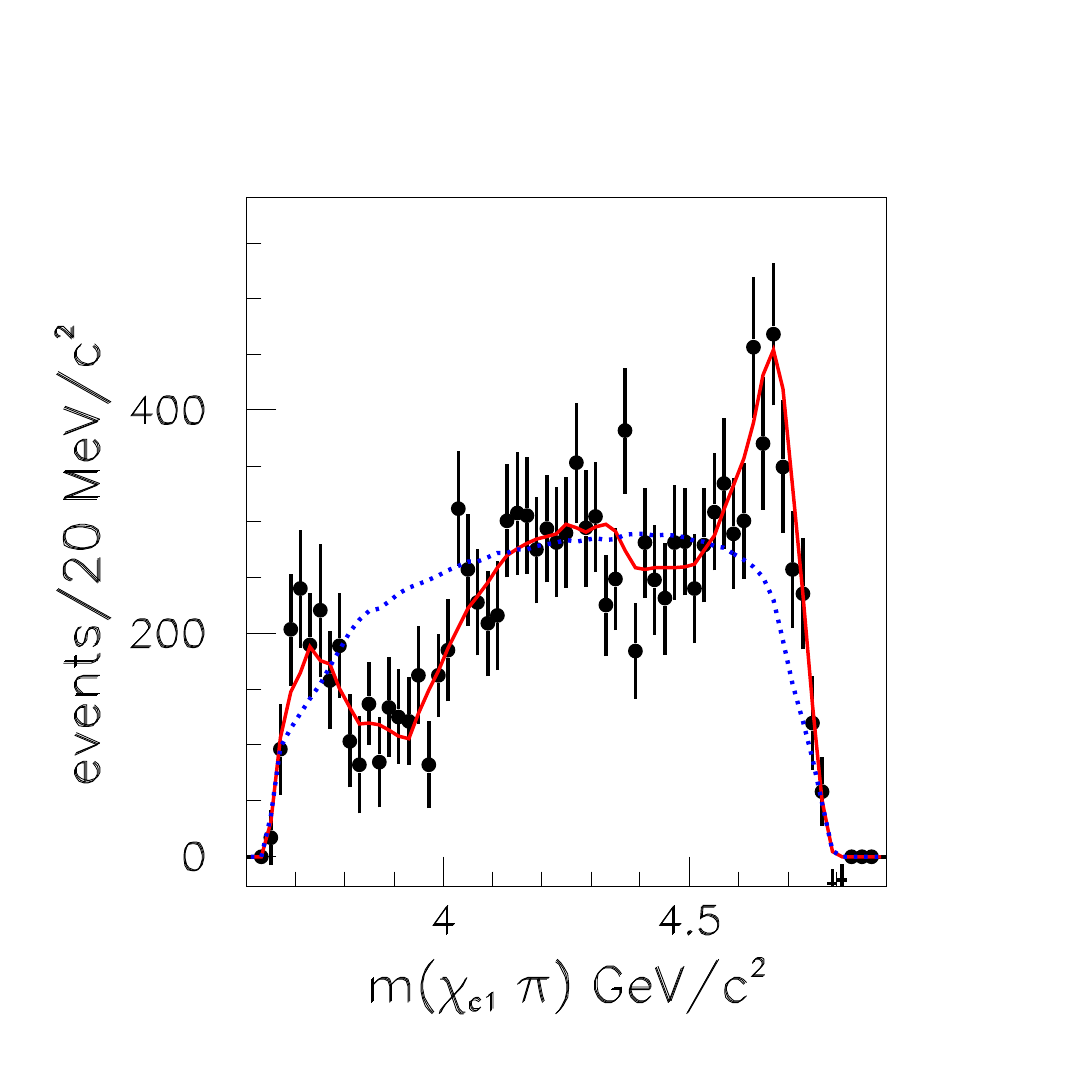} 
  \end{center}
   \caption{Invariant mass distributions of 
	$\chi_{c1}\pi^\pm$,
	with fit results showing the charged resonances in the \belle (left)\cite{Mizuk:2008me} and \babar (right)\cite{Lees:2011ik} analyses.
	The region of the
	$K^{\ast}(890)$ and $K^{\ast}(1410)$ peaks are removed.
	In left panel, the solid red histogram shows the results
	of the fit that includes coherent $Z_1$ and $Z_2$ amplitudes;
	the dashed blue curve is the result of the fit using $K\pi$
	amplitudes only. In right panel, the solid curve fits data using $K\pi$ amplitudes only.}
  \label{fig:z12belle}
\end{figure}
The same decay was investigated by \babar, which carefully
studied the effects of interference between resonances in the $K\pi$
system\cite{Lees:2011ik}. Considering interfering resonances in the $K\pi$ channel only, \babar obtained good fits to data without adding any $\chi_c \pi$ resonance. Upper limits  at 95\% C.L. on the product branching fractions of $Z_{1}$ and $Z_{2}$ can  be evaluated if incoherent resonant amplitudes for these two
states are added to the fit:
\begin{subequations}
\begin{align}
  \BR(B\to Z_1^-K^+)\times \BR(Z_1^-\to\chi_{c1}\pi^-)	& <1.8\times 10^{-5}, 	\\
  \BR(B\to Z_2^-K^+)\times \BR(Z_2^-\to\chi_{c1}\pi^-)	& <4.0\times 10^{-5}.
\end{align}
\end{subequations}
Part of the discrepancy between the two experiments may be due to the fact that
in the \babar analysis the $Z_1$ and $Z_2$ terms are added
incoherently and do not interfere with the $K\pi$ amplitudes, while in the \belle analysis,  significant constructive
and destructive interference between the $Z_{1,2}$ amplitudes
and the $K\pi$ resonances is more relevant (see the dips and peaks of the solid red curve in \figurename{\ref{fig:z12belle}).

Finally, we report a $3.5\sigma$ peak in the $\psiprime \pi^+$ invariant mass, in the $e^+e^- \to \psiprime \pi^+\pi^-$ full statistics analysis by \belle\cite{Wang:2014hta}, with best fit parameters $M=(4054 \pm 3 \pm 1)\mev$ and $\Gamma =(45 \pm 11\pm 6)\mev$.

\subsection{Charged bottomonium states: $Z_b(10610)/Z_b^\prime(10650)$}
\label{sec:Zb}
\begin{figure}[b]
\centering
  \includegraphics[width=0.6\textwidth]{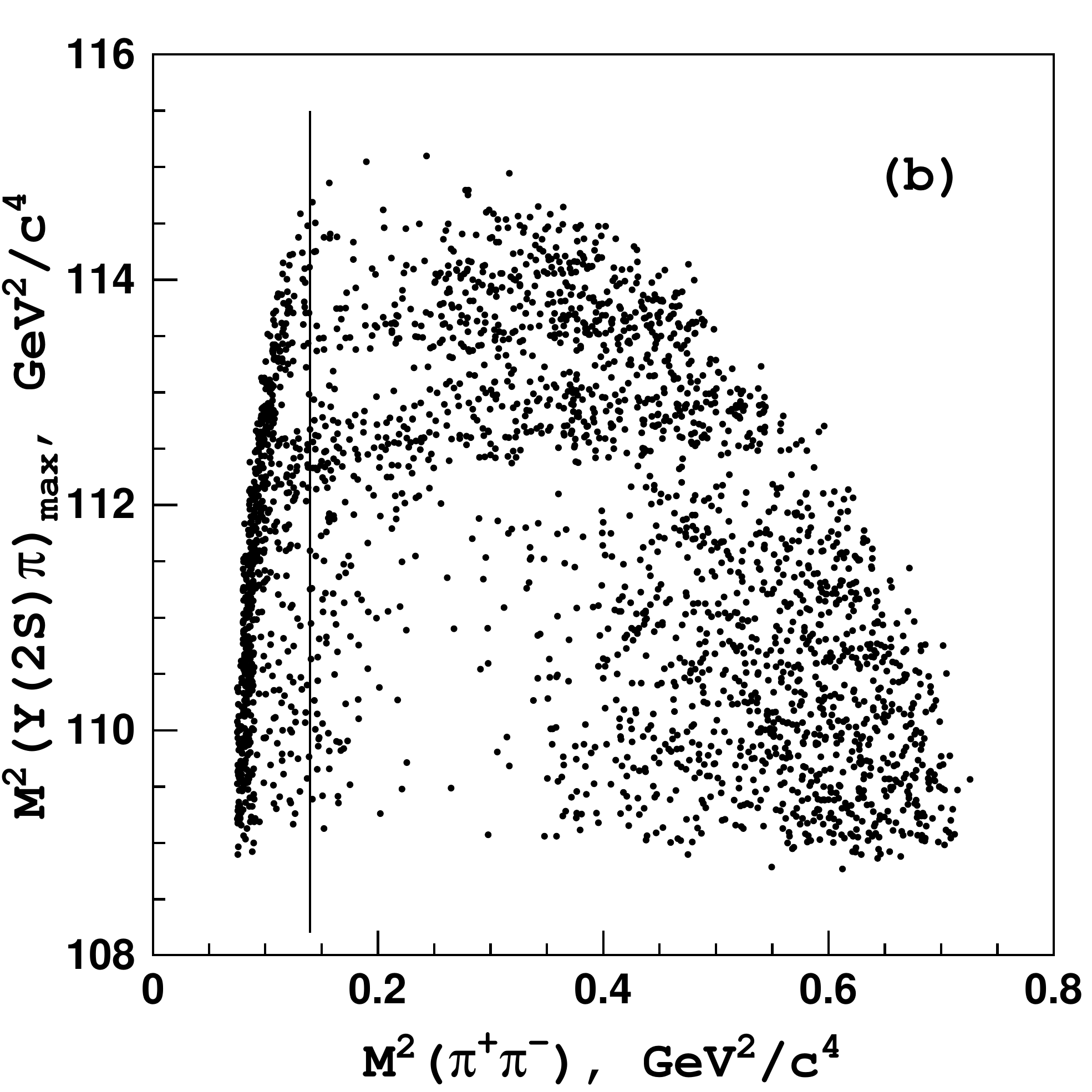}
  \caption{Dalitz plot for $\Upsilon(2S)$ events in the signal region. Events to the left of the vertical line are excluded. From \belle\cite{Belle:2011aa}.
}
\label{fig:ynspp-b-dp}
\end{figure}
The $Z_c(3900)$ and the $Z_c^\prime(4020)$ could have their counterparts in the bottomonium sector. 
\belle reported the observation of
anomalously high rates for the $\Upsilon(5S) \to \pi^+\pi^- \Upsilon(nS)$ 
$(n = 1,2,3)$\cite{Abe:2007tk} and  $\Upsilon(5S) \to \pi^+\pi^- h_b(nP)$ 
$(n = 1,2)$\cite{Adachi:2011ji}
transitions. The measured partial decay widths
$\Gamma\left(\Upsilon(5S) \to \Upsilon (nS)\pi^+\pi^-\right)\simeq 0.5 \mev$ are about two orders of magnitude larger than typical widths for dipion
transitions among the four lower $(nS)$ states. Furthermore, the observation of $\pi^+\pi^- h_b(nP)$ final states with rates comparable to $\pi^+\pi^- \Upsilon(nS)$ violates heavy-quark spin conservation. \belle searched for exotic resonant substructures in these decays\cite{Belle:2011aa}. In order to have a relatively background-free sample, the $\Upsilon(nS)$ states are observed in their
$\mu^+\mu^-$ decay only, whereas the $h_b(nP)$ are reconstructed inclusively.

The Dalitz plots in the signal region (see for example \figurename{~\ref{fig:ynspp-b-dp}}) is fitted with a sum of interfering resonances: the $f_0(980)$, the $f_2(1270)$ in $\pi\pi$ channel, two new charged resonances in the $\Upsilon(nS)\left[h_b(nP)\right] \pi^\pm$ channel, and a nonresonant background. The result of each fit is reported in \tablename{~\ref{tab:zb}}; all the studied channels show the highly significant presence of both charged resonances, dubbed $Z_b(10610)$ and $Z_b^\prime(10650)$, with compatible masses and widths. The one-dimensional invariant mass projections for events in each $\Upsilon(nS)$ and $h_b(nP)$ signal region are shown in \figurename{~\ref{fig:mhbpi}}. 
The average of all channels gives for $Z_b(10610)$ 
a mass and width of $M=(10607.2 \pm 2.0) \mev$,  $\Gamma = (18.4 \pm 2.4) \mev$, 
and for $Z_b^\prime(10650)$ a mass and width 
of $M=(10652.2 \pm 1.5) \mev$, $\Gamma = (11.5 \pm 2.2) \mev$.

\begin{table}[b]
\tbl{List of branching fractions for the $Z^+_b(10610)$ and 
         $Z^+_b(10650)$ decays. From \belle\cite{Adachi:2012cx}.}
{
\begin{tabular}{lcc}  \hline \hline
   ~Channel~\hspace*{80mm}  & \multicolumn{2}{c}{Fraction, \%}   \\
               & $Z_b(10610)$  & $Z_b^\prime(10650)$      \\
\hline 
 $\Upsilon(1S)\pi^+$      & $0.32\pm0.09$ & $0.24\pm0.07$      \\
 $\Upsilon(2S)\pi^+$      & $4.38\pm1.21$ & $2.40\pm0.63$      \\
 $\Upsilon(3S)\pi^+$      & $2.15\pm0.56$ & $1.64\pm0.40$      \\
 $h_b(1P)\pi^+$           & $2.81\pm1.10$ & $7.43\pm2.70$      \\
 $h_b(2P)\pi^+$           & $4.34\pm2.07$ & $14.8\pm6.22$      \\
 $B^+\bar{B}^{*0}+\bar{B}^0B^{*+}$
                          & $86.0\pm3.6$  &    $-$             \\
 $B^{*+}\bar{B}^{*0}$     &    $-$        & $73.4\pm7.0$       \\
\hline \hline
  \end{tabular}
  \label{tab:zbratios}
}
\end{table}
The $Z_b(10610)$
production rate is similar to that of the $Z_b^\prime(10650)$ for each of the five decay
channels. Their relative phase is consistent with zero for the final states
with the $\Upsilon(nS)$ and consistent with $180$ degrees for the final states with
$h_b(nP)$. Production of the $Z_b$'s saturates the $\Upsilon(5S) \to h_b(nP) \pi^+\pi^-$ transitions
and accounts for the high inclusive $h_b(nP)$ production rate reported by \belle
\cite{Adachi:2011ji}. Analyses of charged pion angular 
distributions\cite{Belle:2011aa,Garmash:2014dhx} favor the $J^P=1^+$ spin-parity assignment
for both the states. 
\begin{landscape}
\begin{figure}[t]
  \centering
  \includegraphics[width=0.30\textwidth]{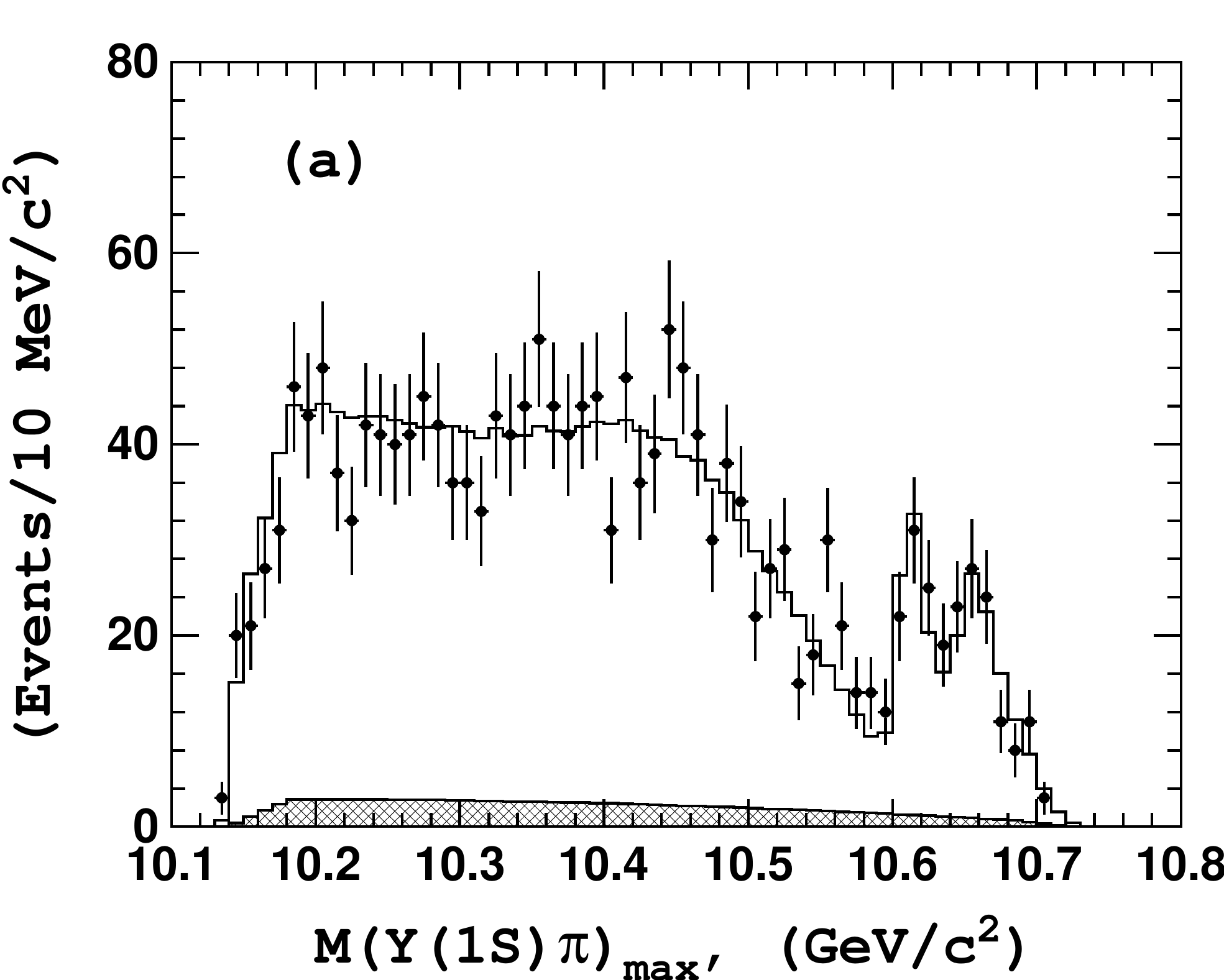} 
  \includegraphics[width=0.30\textwidth]{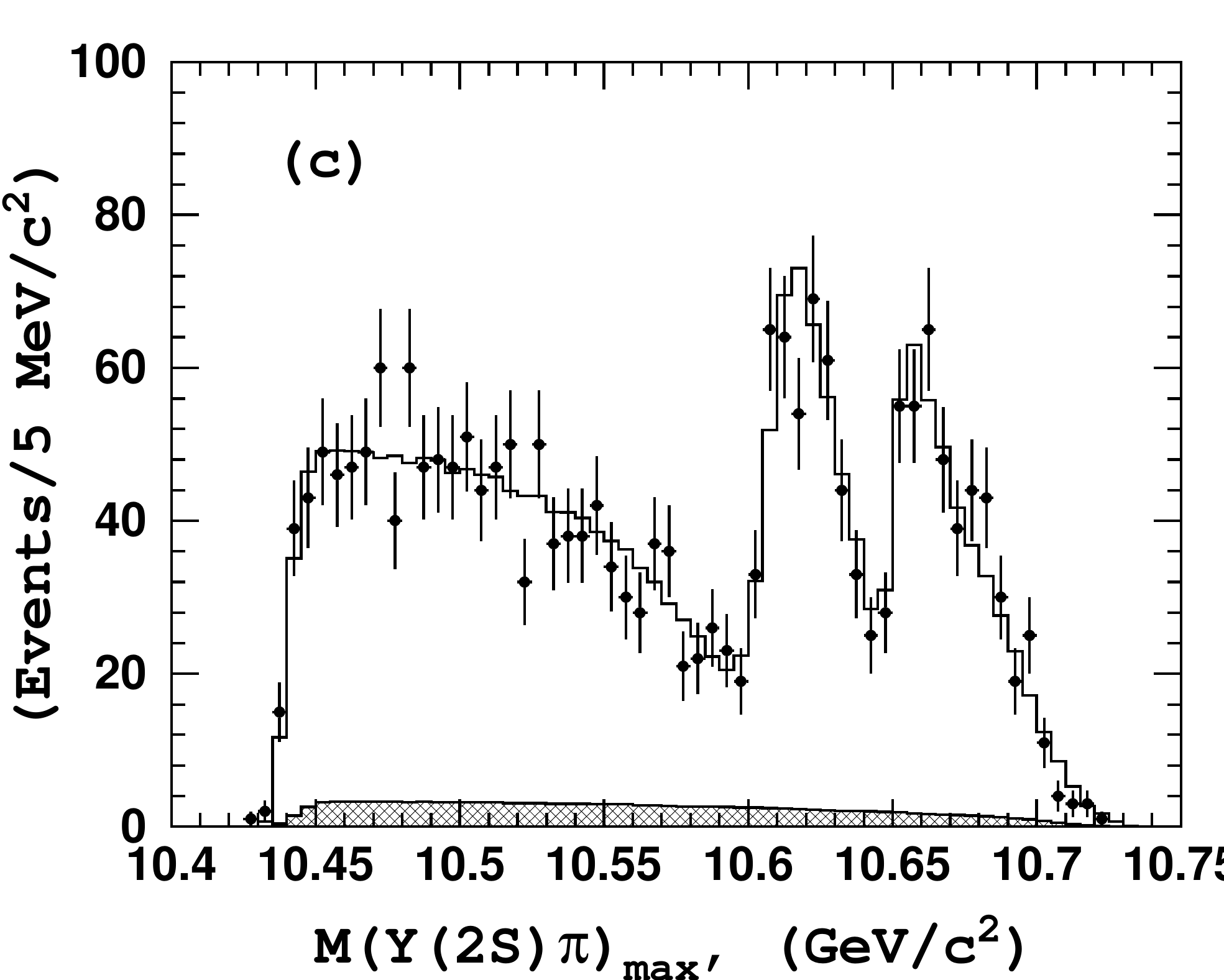} 
  \includegraphics[width=0.30\textwidth]{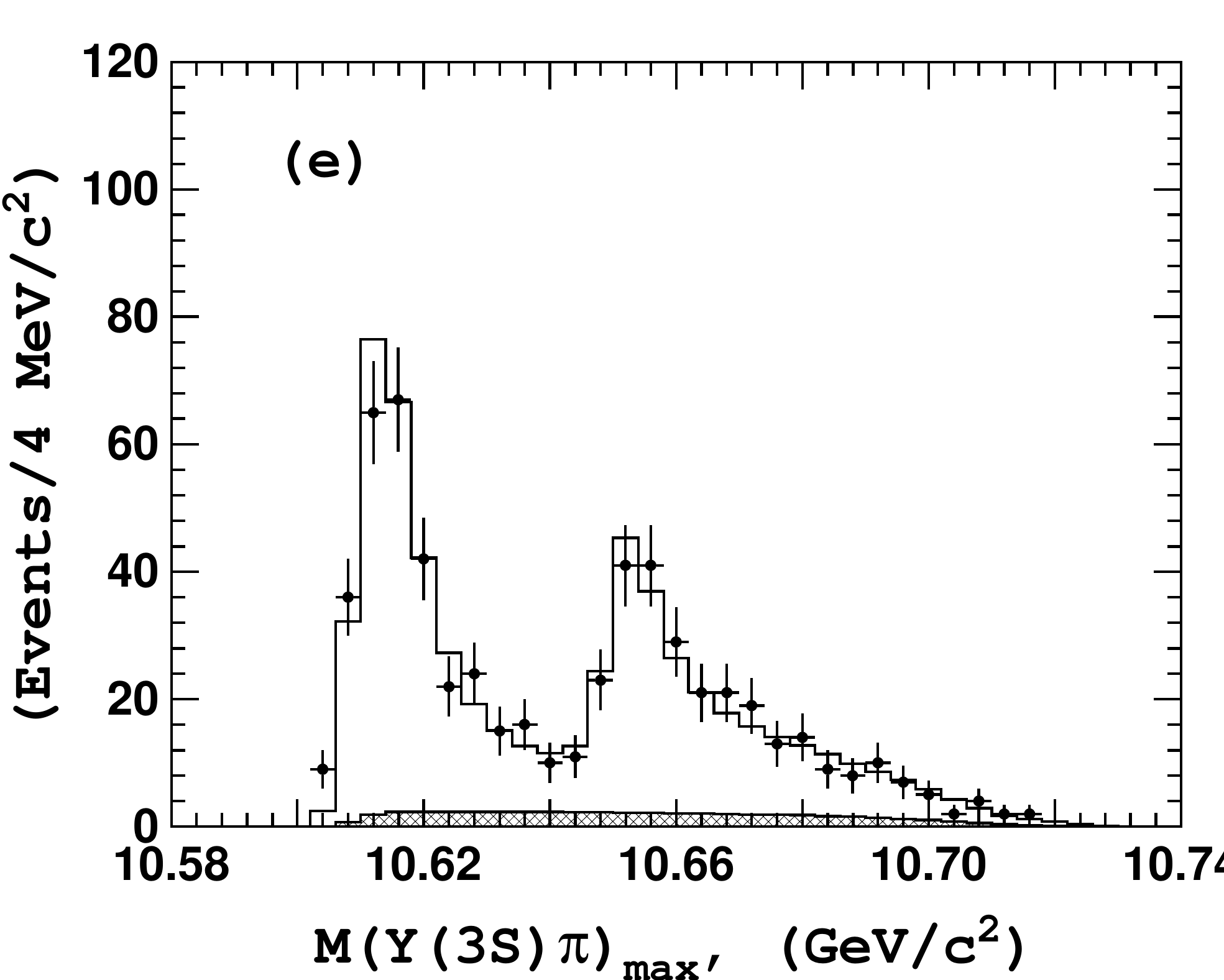} \\ \vspace{.5cm}
\includegraphics[width=0.30\textwidth]{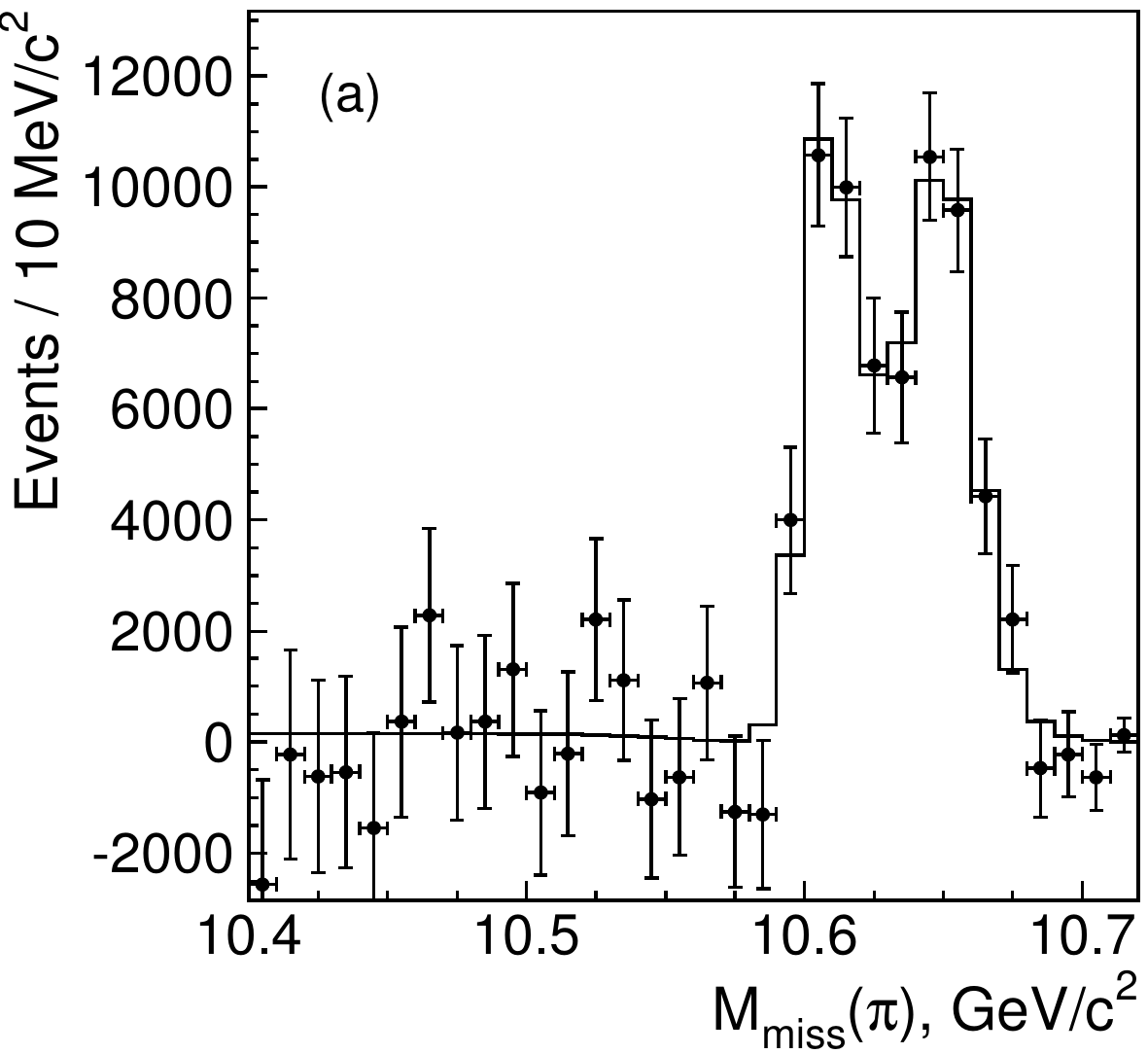} 
\includegraphics[width=0.30\textwidth]{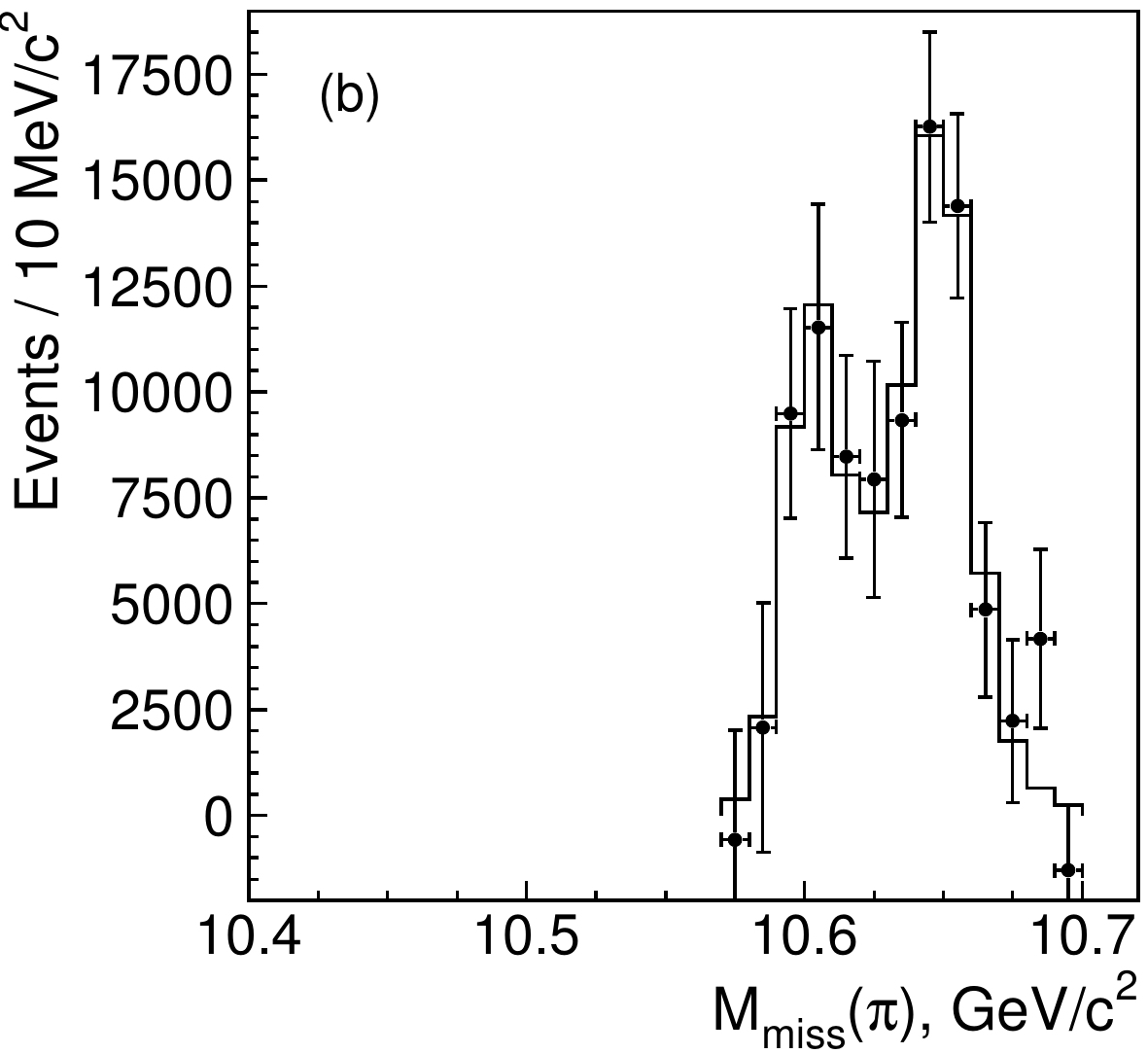}
  \caption{Comparison of fit results (open histogram) with
    experimental data (points with error bars) for events in the $\Upsilon(nS)$ (upper plots) and $h_b(nP)$ (lower plots) regions. From \belle\cite{Belle:2011aa}.
}
\label{fig:mhbpi}
\end{figure}

\begin{table}[b]
  \tbl{Comparison of results on $Z_b(10610)$ and $Z_b^\prime(10650)$ parameters
           obtained from $\Upsilon(5S)\to\Upsilon(nS)\pi^+\pi^-$ and $\Upsilon(5S) \to h_b(nP)\pi^+\pi^-$ 
analyses\cite{Belle:2011aa}. }
{
\begin{tabular}{lccccc} \hline \hline
Final state & $\Upsilon(1S)\pi^+\pi^-$      &
              $\Upsilon(2S)\pi^+\pi^-$      &
              $\Upsilon(3S)\pi^+\pi^-$      &
              $h_b(1P)\pi^+\pi^-$           &
              $h_b(2P)\pi^+\pi^-$
\\ \hline
           $M[Z_b(10610)]$, \!\mev        &
           $10611\pm4\pm3$                  &
           $10609\pm2\pm3$                  &
           $10608\pm2\pm3$                  &
           $10605\pm2^{+3}_{-1}$            &
           $10599{^{+6+5}_{-3-4}}$
 \\
           $\Gamma[Z_b(10610)]$, \!\mev  &
           $22.3\pm7.7^{+3.0}_{-4.0}$       &
           $24.2\pm3.1^{+2.0}_{-3.0}$       &
           $17.6\pm3.0\pm3.0$               &
           $11.4\,^{+4.5+2.1}_{-3.9-1.2}$   &
           $13\,^{+10+9}_{-8-7}$
 \\
           $M[Z_b(10650)]$, \!\mev        &
           $10657\pm6\pm3$                  &
           $10651\pm2\pm3$                  &
           $10652\pm1\pm2$                  &
           $10654\pm3\,{^{+1}_{-2}}$        &
           $10651{^{+2+3}_{-3-2}}$
 \\
           $\Gamma[Z_b(10650)]$, \!\mev     &
           $16.3\pm9.8^{+6.0}_{-2.0}$~      &
           $13.3\pm3.3^{+4.0}_{-3.0}$       &
           $8.4\pm2.0\pm2.0$                &  
           $20.9\,^{+5.4+2.1}_{-4.7-5.7}$   & 
           $19\pm7\,^{+11}_{-7}$ 
 \\
           Rel. normalization               &
           $0.57\pm0.21^{+0.19}_{-0.04}$    &
           $0.86\pm0.11^{+0.04}_{-0.10}$    &
           $0.96\pm0.14^{+0.08}_{-0.05}$    &
           $1.39\pm0.37^{+0.05}_{-0.15}$    &
           $1.6^{+0.6+0.4}_{-0.4-0.6}$
 \\
           Rel. phase, degrees              &
           $58\pm43^{+4}_{-9}$              &
           $-13\pm13^{+17}_{-8}$            &
           $-9\pm19^{+11}_{-26}$            &
           $187^{+44+3}_{-57-12}$           &
           $181^{+65+74}_{-105-109}$   
\\
\hline \hline
\end{tabular}
\label{tab:zb} }
\end{table}
\end{landscape}
\belle searched these states also in pairs of open bottom mesons\cite{Adachi:2012cx}. The Dalitz plots of $\Upsilon(5S)\to (\B\Bstar)^- \pi^+$ and  $\Upsilon(5S)\to (\Bstar\Bstar)^- \pi^+$ report a $8\sigma$ signal of $Z_b^-(10610) \to (\B \Bstar)^-$ and a $6.5\sigma$ signal of $Z_b^{\prime-}(10650) \to (\Bstar \Bstar)^-$, 
respectively, whereas $Z_b^{\prime-}(10650) \to (\B \Bstar)^-$ is compatible with zero\footnote{$Z_b^{-}(10610) \to (\Bstar \Bstar)^-$ is phase-space forbidden.}. The best estimate for the branching ratios are reported in \tablename{~\ref{tab:zbratios}}.

Recently, \belle has been able to find the neutral isospin partner $Z_b^0(10610)$\cite{Krokovny:2013mgx} in $\Upsilon(5S)\to\Upsilon(2,3S)\pi^0\pi^0$ decays, at a significance of $6.5\sigma$ if mass and width are fixed to the averaged values 
of the $Z_b^+(10610)$. If the mass is let free, the fitted value is $M=(10609\pm 4 \pm 4) \mev$, consistent with the charged partner mass. On the other hand, no significant signal of $Z_b^{\prime\,0}(10650)$ is seen.

\subsection{The $X(3872)$ saga}
\label{sec:X3872}
\begin{figure}[b]
\begin{center}
\includegraphics[width=8cm]{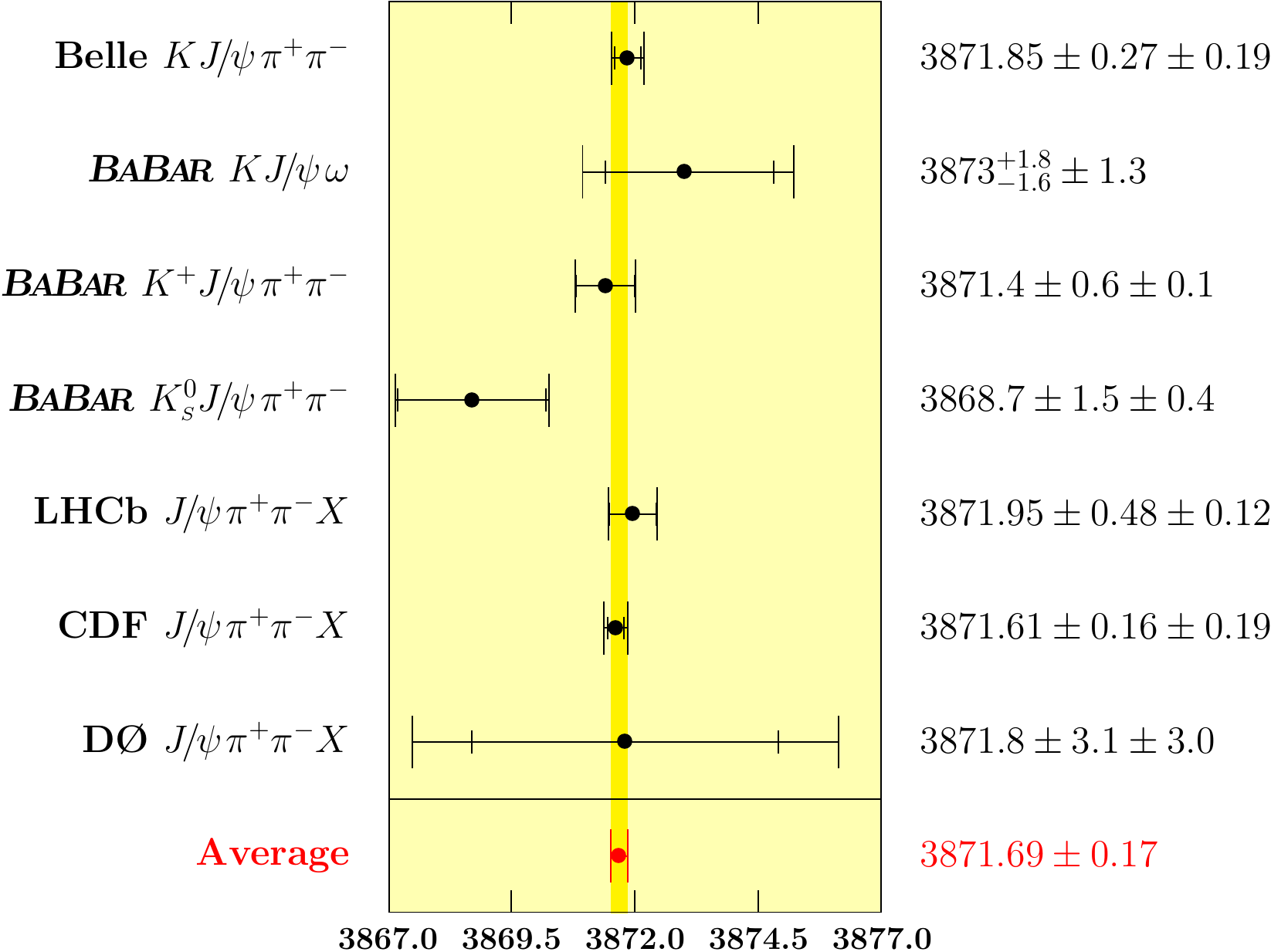} 
 \caption{Measured mass of the $X(3872)$. We show the measurements which contribute to the average in PDG\cite{pdg}.  
}
    \label{FIG:XYZ:X-MASS}
    \end{center}
\end{figure}
The queen of exotic states is the $X(3872)$. It was discovered by \belle while studying the $B \to K \jpsi \pi^+ \pi^-$ decays\cite{Choi:2003ue}, as an unexpected resonance in the $\jpsi\pi^+\pi^-$ invariant mass distribution (see \figurename{~\ref{FIG:x-invariant}}, left panel). It was then confirmed both in $B$ decays\cite{Aubert:2004ns} and in inclusive prompt $p\bar{p}$\cite{Acosta:2003zx,Abazov:2004kp} and $pp$ production\cite{Aaij:2011sn,Chatrchyan:2013cld} -- see \sectionname{~\ref{sec:prompt}} for a long-standing controversy about the theoretical interpretation of that.
First of all, an exotic nature was suggested by its narrow width, $\Gamma < 2.3\mev$ at 90\% C.L.\cite{Choi:2003ue}, despite being above threshold for the decay into a charmed meson pair. Furthermore, both $\pi^+\pi^-$ invariant mass distribution\cite{Choi:2003ue,Abulencia:2005zc} and angular analyses\cite{Abulencia:2006ma} show that the $\pi^+\pi^-$ amplitude is dominated by the $\rho$ meson, \ie a $I=1$ resonance. If the $X(3872)$ were an ordinary charmonium with $I=0$, such a decay would badly violate isospin symmetry. The size of isospin breaking was 
quantified by the measurement of the $X(3872) \to \jpsi\omega$ branching fraction by \belle\cite{Abe:2005ix} and \babar\cite{delAmoSanchez:2010jr}:
\begin{equation}
 \frac{\Gamma\left(X(3872)\to\jpsi\omega\right)}{\Gamma\left(X(3872)\to\jpsi\pi^+\pi^-\right)} = 0.8 \pm 0.3.
\end{equation}

The $C=+$ assignment was confirmed by the observation of the $X(3872)\to\jpsi\gamma$ decay\cite{Abe:2005ix,Aubert:2006aj}, and by the non-observation of $X(3872)\to \chi_{c1}\gamma$\cite{Choi:2003ue}. 
As for the spin, a preliminary angular analysis of the $X(3872)\to\jpsi\pi^+\pi^-$ by \belle\cite{Abe:2005iya} favored $1^{++}$ assignment. Soon after, a more detailed analysis by \cdf\cite{Abulencia:2006ma} was able to rule out all but the $1^{++}$ and $2^{-+}$ assignments. The latter could not be excluded because of the additional complex parameter given by the ratio between the two independent amplitudes for $X(2^{++})\to\jpsi\pi^+\pi^-$, which could not be constrained in inclusive $X(3872)$ production; on the other hand, the former was preferred by theoretical models. Instead, the analysis of the $\jpsi \omega$ invariant mass distribution by \babar\cite{delAmoSanchez:2010jr} favored the $2^{-+}$ hypothesis, and stimulated a discussion on its theoretical feasibility\cite{Burns:2010qq,Hanhart:2011tn,Faccini:2012zv,Faccini:2012pj,Braaten:2013poa}. However, a $J=2$ assignment would allow $X(3872)$ to be produced in $\gamma\gamma$ fusion, but \cleo has found no significant signal in $\gamma\gamma\to X(3872)\to\jpsi \pi^+\pi^-$\cite{Dobbs:2004di}. A statistically improved analysis of angular distributions in $X(3872)\to\jpsi \pi^+\pi^-$ has been made by \belle\cite{Choi:2011fc}, again favoring $1^{++}$. The limited statistics forced \belle to consider three different one-dimensional projections of the full angular distribution, which were not able to rule out $2^{-+}$. 

Finally, \lhcb has recently published an analysis of a large $B^+ \to K^+ X(3872)$ sample\cite{Aaij:2013zoa}. This study is based on an event-by-event likelihood ratio test of $1^{++}$ and $2^{-+}$
hypotheses on the full 5D angular distribution, and favors the $1^{++}$ over $2^{-+}$ at $8\sigma$ level. The additional complex parameter in the $2^{-+}$ distributions is treated as  a nuisance parameter; its best value extracted from the fit is found to be consistent with
the value obtained if the events are MC generated with a $1^{++}$ assumption; this is consistent with the \belle's result too\cite{Choi:2011fc}. It is worth noticing that the only analysis which favored the $2^{-+}$ assignment was the $\jpsi \omega$ \babar analysis and an independent analysis of the same channel by other experiments would be very interesting.

\begin{figure}[b]
\begin{center}
\includegraphics[width=.25\columnwidth]{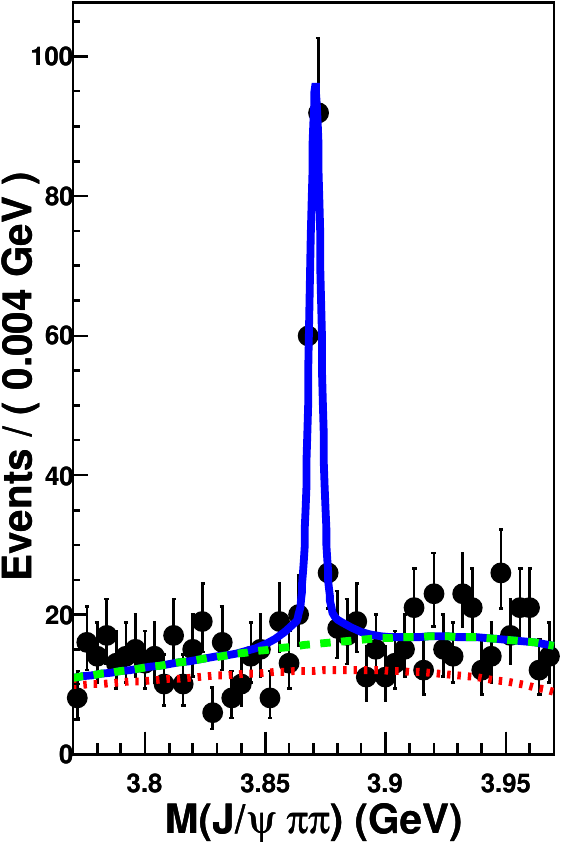} \hspace{1cm}
\includegraphics[width=.55\columnwidth]{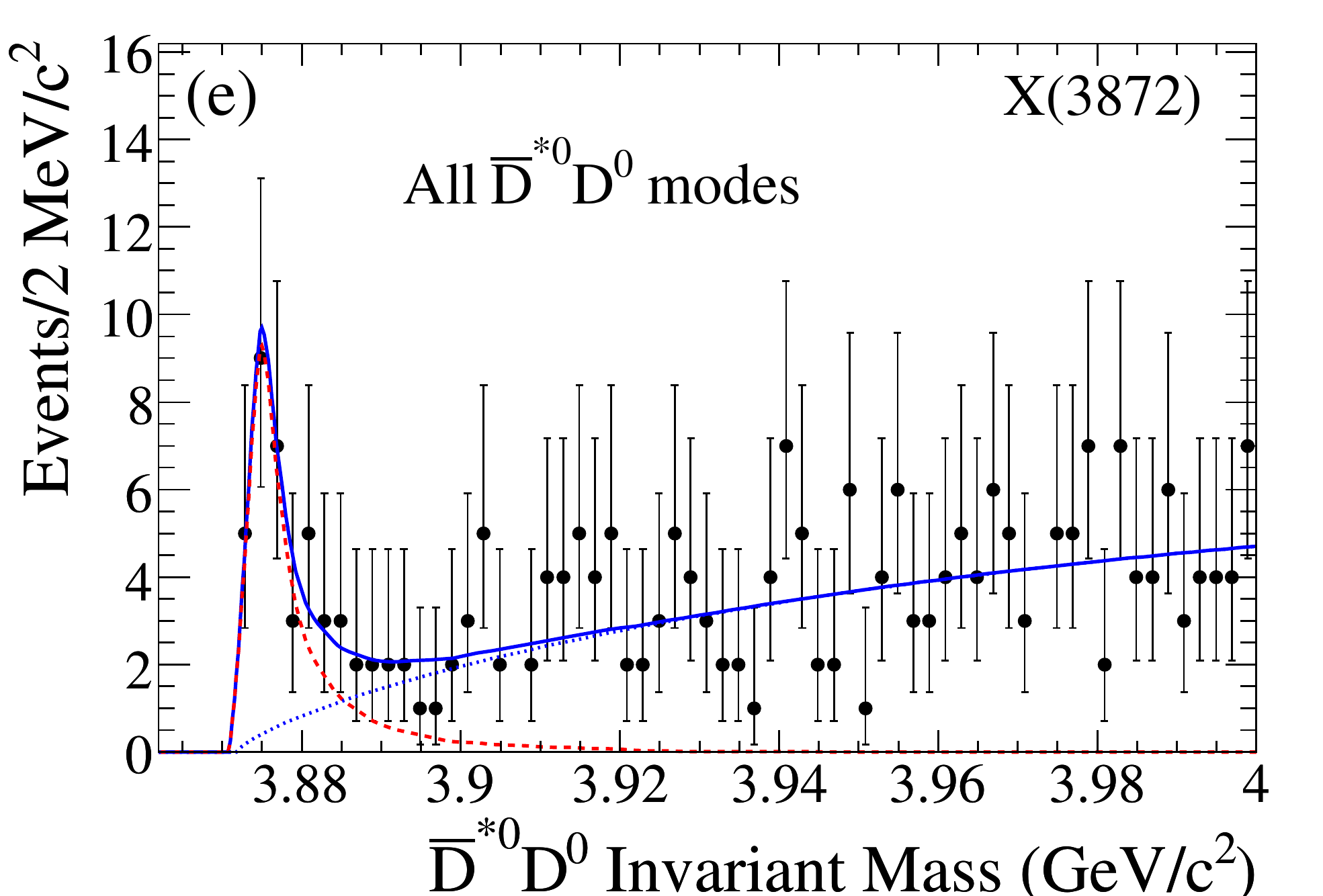} 
 \caption{Invariant mass spectra of  $\jpsi \pi^+\pi^-$ 
in $B^+ \to \jpsi\pi^+\pi^- K^+$ decay by \belle\cite{Choi:2011fc} (left panel) and of the $\Dstarzb \Dz$ system in $B \to \Dstarzb \Dz K$ decays
by \babar\cite{Aubert:2007rva} (right panel).
}
    \label{FIG:x-invariant}
    \end{center}
\end{figure}
In~\figurename{~\ref{FIG:XYZ:X-MASS} we report a list of the most recent mass measurements.
The current world average, considering only $X(3872)$ decays into final states
including the \jpsi, is $M = (3871.69 \pm 0.17)\mev$\cite{pdg}.
The most precise measurements are those of
\cdf\cite{Aaltonen:2009vj},
\belle\cite{Choi:2011fc},
the new measurement from \lhcb\cite{Aaij:2011sn}, and
\babar\cite{Aubert:2008gu},
all in the channel $\jpsi\pi^+\pi^-$; 
the hadronic machines measure inclusive production in $pp (\bar p)$,
while the $B$-factories measurements are dominated by $B^+\to K^+ \jpsi\pi^+\pi^-$.  

\belle observed the decay $X(3872)\to\Dstarz\Dzb$
in the $\pi^0\Dz\Dzb$ final state
at the higher mass $M = (3875.2 \pm 0.7 ^{+0.3}_{-1.6} \pm 0.8)\mev$
\cite{Gokhroo:2006bt}. This was confirmed by \babar\cite{Aubert:2007rva} (see \figurename{~\ref{FIG:x-invariant}}, right panel)
and again by \belle\cite{Adachi:2008su},
leading to an average mass of $M = (3873.8 \pm 0.5)\mev$.
As this is significantly larger than the value observed in the discovery
mode $\jpsi \pi^+\pi^-$, there has been some discussion about the possibility 
that $X(3875)\to\Dstarz\Dzb$ and $X(3872)\to\jpsi\pi^+\pi^-$ are distinct particles. However, some papers\cite{Artoisenet:2010va,Hanhart:2010wh,Hanhart:2011jz} argued that, since the $\Dstarz$ will in general be off-shell, a detailed study of the $\pi^0\Dz\Dzb$ and $\gamma\Dz\Dzb$ lineshapes is needed to distinguish between a below- and above-threshold $X(3872)$ (see Sec.~\ref{sec:lowenergy}). Moreover, in order to improve the 
resolution, the experimental analyses constrain the $\Dstar$ mass, 
and this yields to a reconstructed $X(3872)$ mass which is above threshold by construction. Because of these biases, this channel has been dropped from mass averages in PDG\cite{pdg}. 

As far as the width is concerned, the $X(3872)$ was known to be narrow since the very first analysis,
with a limit $\Gamma < 2.3\mev$ at 90\% C.L.\cite{Choi:2003ue}.
The best current upper limit for the width is given by \belle\cite{Choi:2011fc},
which finds $\Gamma < 1.2\mev$ at 90\% C.L.
based on a 3D fit to \mes, \DeltaE, and $M(\pi^+\pi^-\jpsi)$, which allows the limit to be constrained below the experimental resolution on invariant mass: 
the distributions in \mes and \DeltaE provide constraints
on the area of the $M(\pi^+\pi^-\jpsi)$ peak, which make the peak height
sensitive to the natural width. 

In addition to $\jpsi \pi^+\pi^- (\pi^0)$ and $\Dstarz\Dzb$ final states, the $X(3872)$ has been sought in many other different channels, which we list in \tablename{~\ref{tab:xdecays}}.

We just discuss the case of $X(3872) \to \psiprime \gamma$, which is of interest for theoretical interpretations.
\babar\cite{Aubert:2008rn} and \lhcb\cite{Aaij:2014ala} find a signal with a relative branching fraction of:
\begin{subequations}
\begin{align}
  \frac{\BR\left(X(3872)\to\psiprime\gamma\right)}{\BR\left(X(3872)\to\jpsi\gamma\right)}
		   & = 3.4 \pm 1.4 \hspace{4.9em} \text{(\babar)}, \\
		   & = 2.46 \pm 0.64 \pm 0.29 \hspace{1em} \text{(\lhcb)}, \\
		   & < 2.1 \hspace{7.5em} \text{(\belle)}.
\end{align}
\end{subequations}
In particular, for the decay $X(3872)\to\psiprime\gamma$, 
\belle\cite{Bhardwaj:2011dj} sees no significant signal and puts a 90\% C.L. upper limit (see \figurename{~\ref{FIG:x-psigamma}}).

\begin{figure}[t]
\begin{center}
\includegraphics[width=.45\columnwidth]{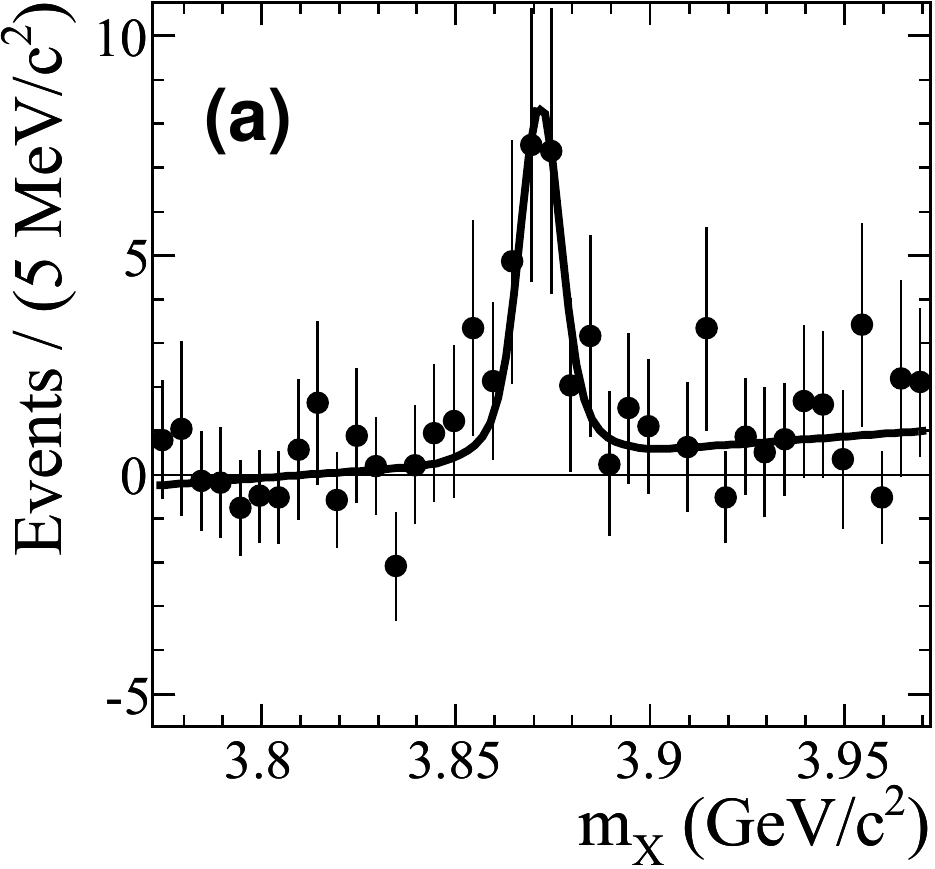} \includegraphics[width=.45\columnwidth]{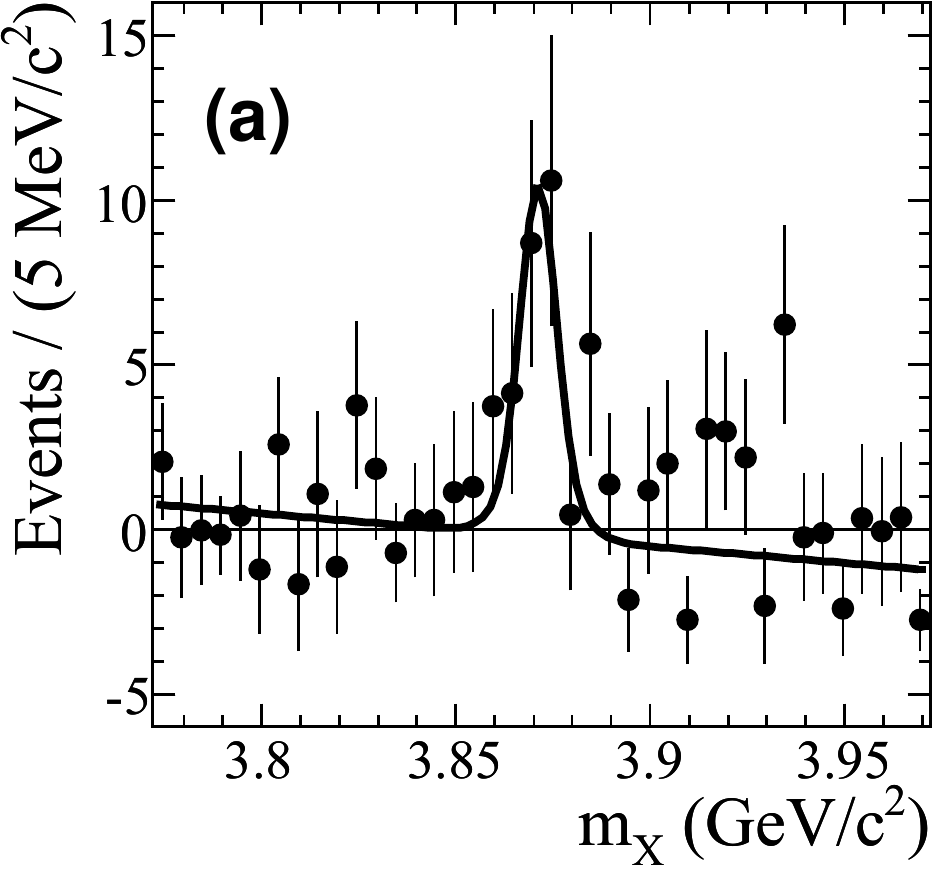}\\
\includegraphics[width=.45\columnwidth]{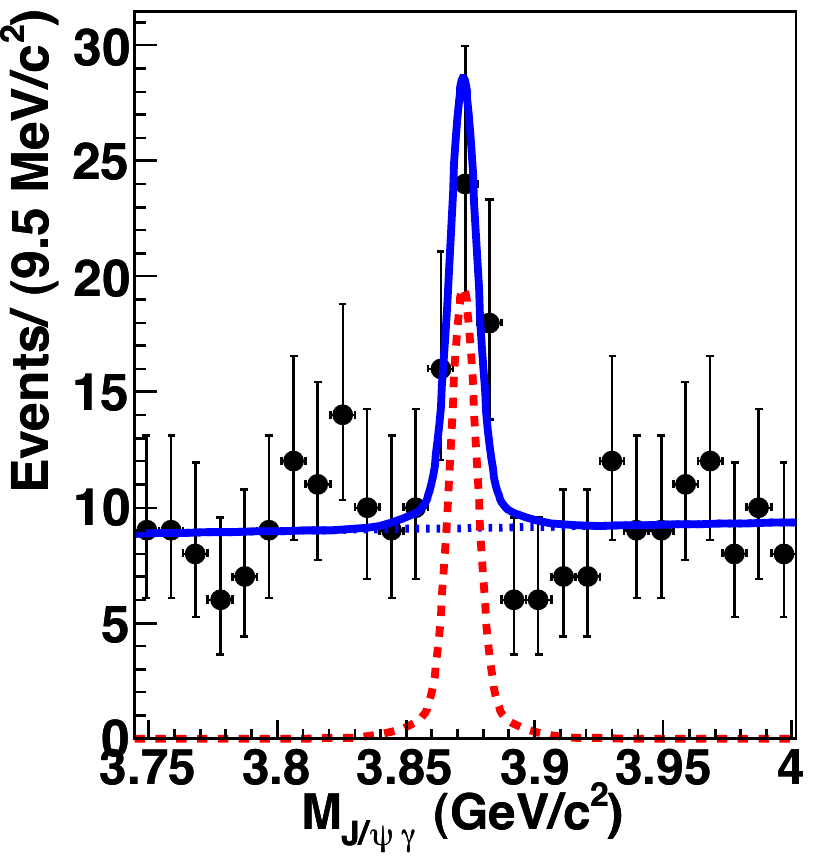} \includegraphics[width=.45\columnwidth]{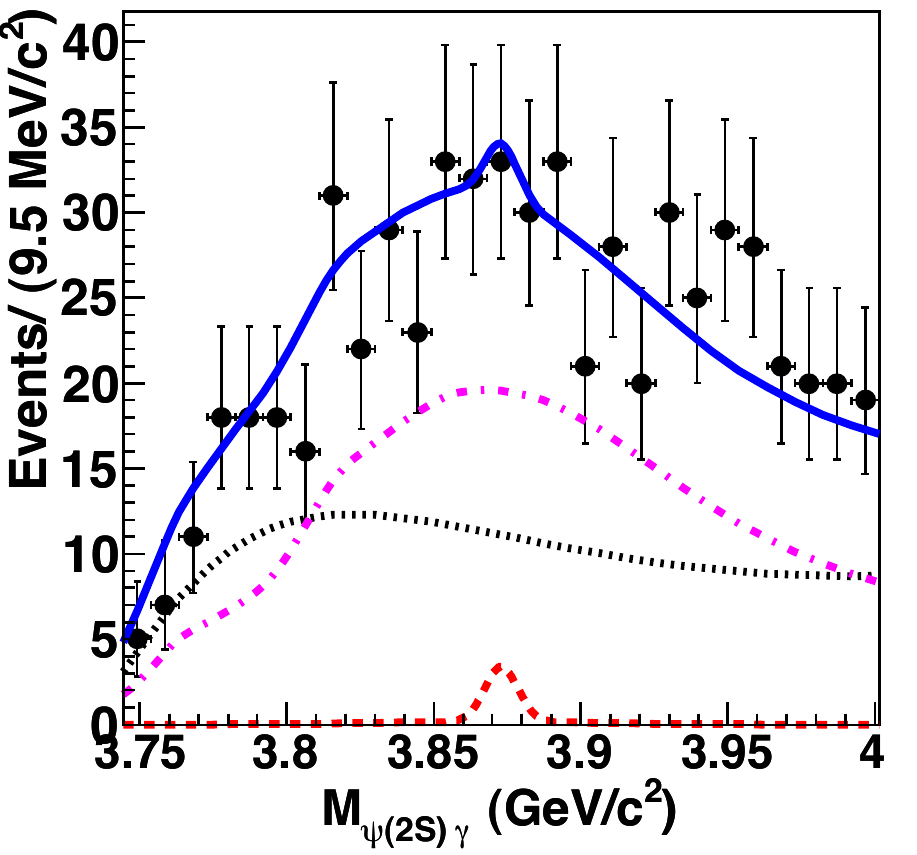}\\
\includegraphics[width=.45\columnwidth]{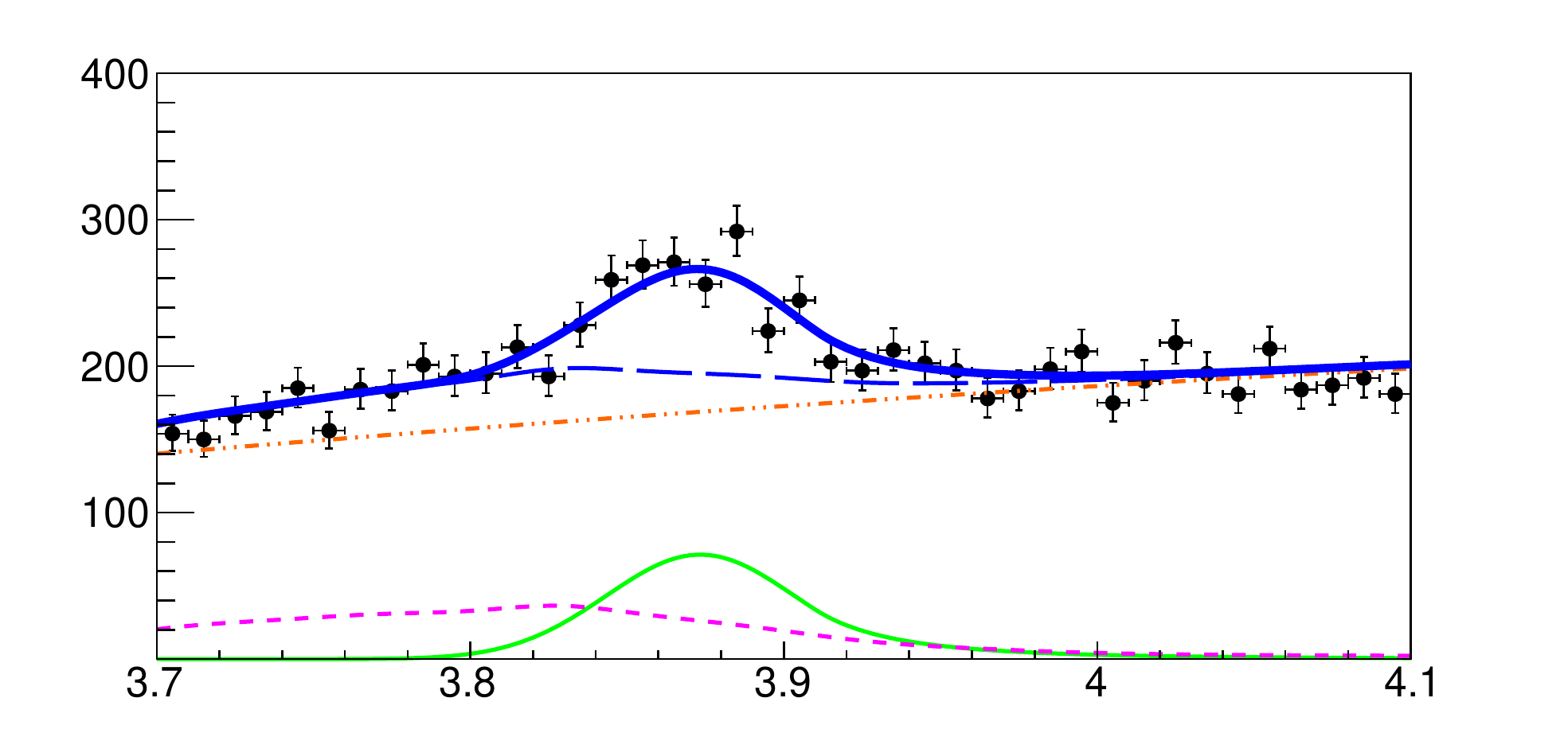} \includegraphics[width=.45\columnwidth]{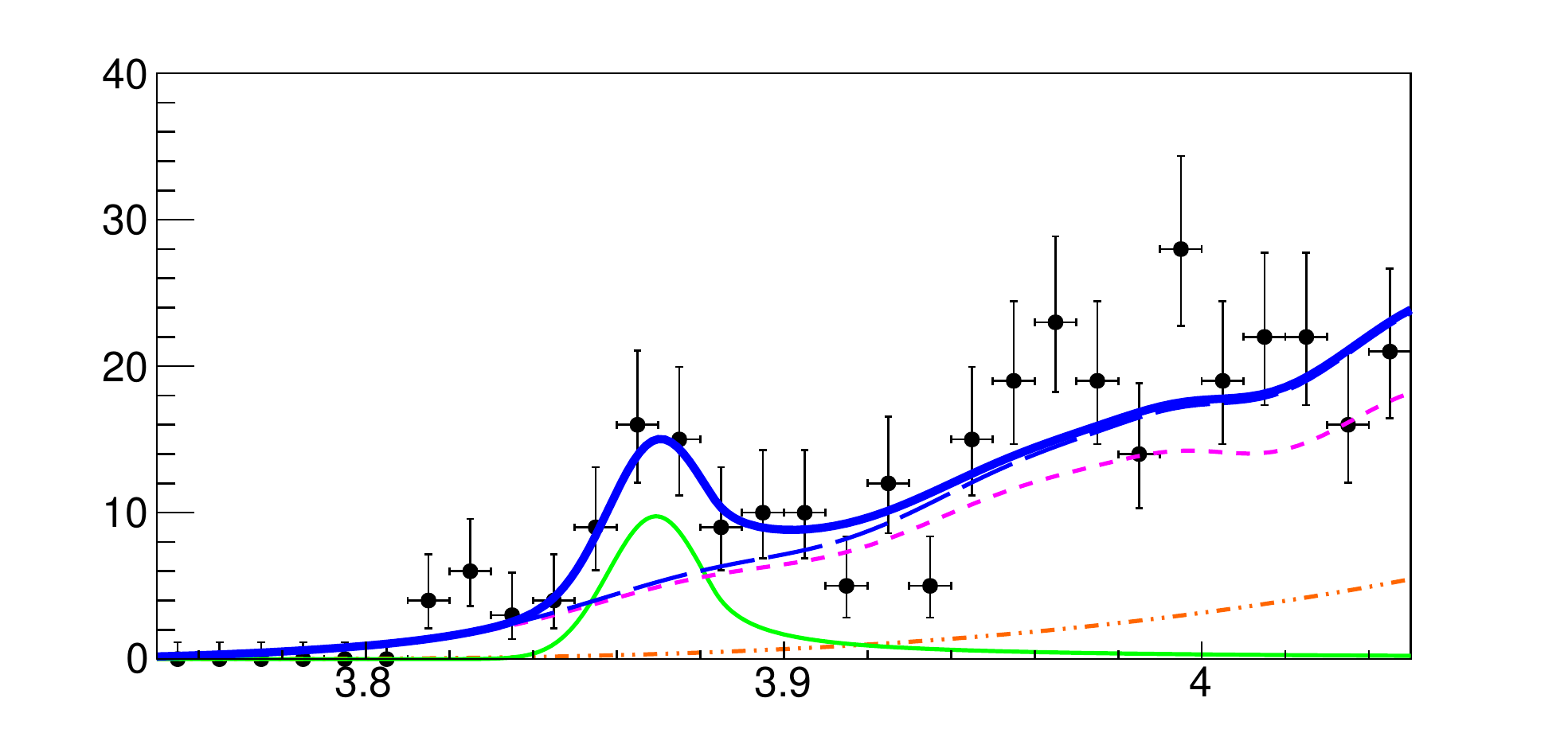}\\

 \caption{Invariant mass spectra of $\jpsi \gamma$ (left) and $\psiprime \gamma$ (right) in $B^+ \to \psi^{(\prime)} \gamma K^+$, according to \babar\cite{Aubert:2008rn} (upper), \belle\cite{Bhardwaj:2011dj} (middle) and \lhcb\cite{Aaij:2014ala} (lower).}
    \label{FIG:x-psigamma}
    \end{center}
\end{figure}

Other production mechanisms like $B^0\to K^+ \pi^- X(3872)$
have also been studied. Such decays are seen, but with a smooth distribution
in $K^+\pi^-$ invariant mass; an upper limit is set on $\BR\left(B^0 \to \Kstar(892)^0 X(3872)\right)$\cite{Adachi:2008te}.
This is in contrast to ordinary charmonium states,
where $B\to\Kstar\ccbar$ and $K\ccbar$ branching fractions are comparable,
and \Kstar\ dominates over nonresonant $K\pi$. We also mention the decay $Y(4260) \to \gamma X(3872)$ seen by \bes\cite{Ablikim:2013dyn}, with a production cross section of $\sigma\left(e^+ e^- \to Y(4260) \to \gamma X(3872)\right) \times \BR\left(X(3872) \to \jpsi \pi^+ \pi^-\right) = 
(0.33 \pm 0.12 \pm 0.02) \pb$.

In \tablename{~\ref{tab:xdecays}} we update the results of Drenska \etal\cite{Drenska:2010kg} on the absolute branching fractions of the $X(3872)$. These can be obtained from measured product branching fractions of $X(3872)$ by exploiting the upper limit on $B \to X(3872) K$ measured by \babar
from the spectrum of the kaons recoiling against fully reconstructed $B$ mesons\cite{Aubert:2005vi},
$\BR(B^\pm \to K^\pm X(3872)) < 3.2 \times 10^{-4}$ at 90\% C.L.. Combining the likelihood from the
measurements of the product branching fractions in the observed channels, the $B \to X(3872) K$
upper limit and the $X(3872)$ width distribution\cite{Liu:2013dau}, with a bayesian procedure we extracted
the likelihood for the absolute $X(3872)$ branching fractions and the widths in each of the decay modes. Then,
we used the probability distributions obtained with this procedure to set limits on the
not observed channels. The full shape of the experimental likelihoods was used whenever
available, while gaussian errors and poissonian counting distributions have been assumed
elsewhere. The 68\% confidence intervals (defined in such a way that the absolute value of
the PDF is the same at the upper and lower bound, unless one of them is at the boundary
of the physical range) are summarized in \tablename{~\ref{tab:xdecays}} for each of the decay modes. Some of the likelihoods are shown in \figurename{~\ref{fig:likelihoods}}.
\begin{figure}[t]
\begin{center}
\includegraphics[width=.45\columnwidth]{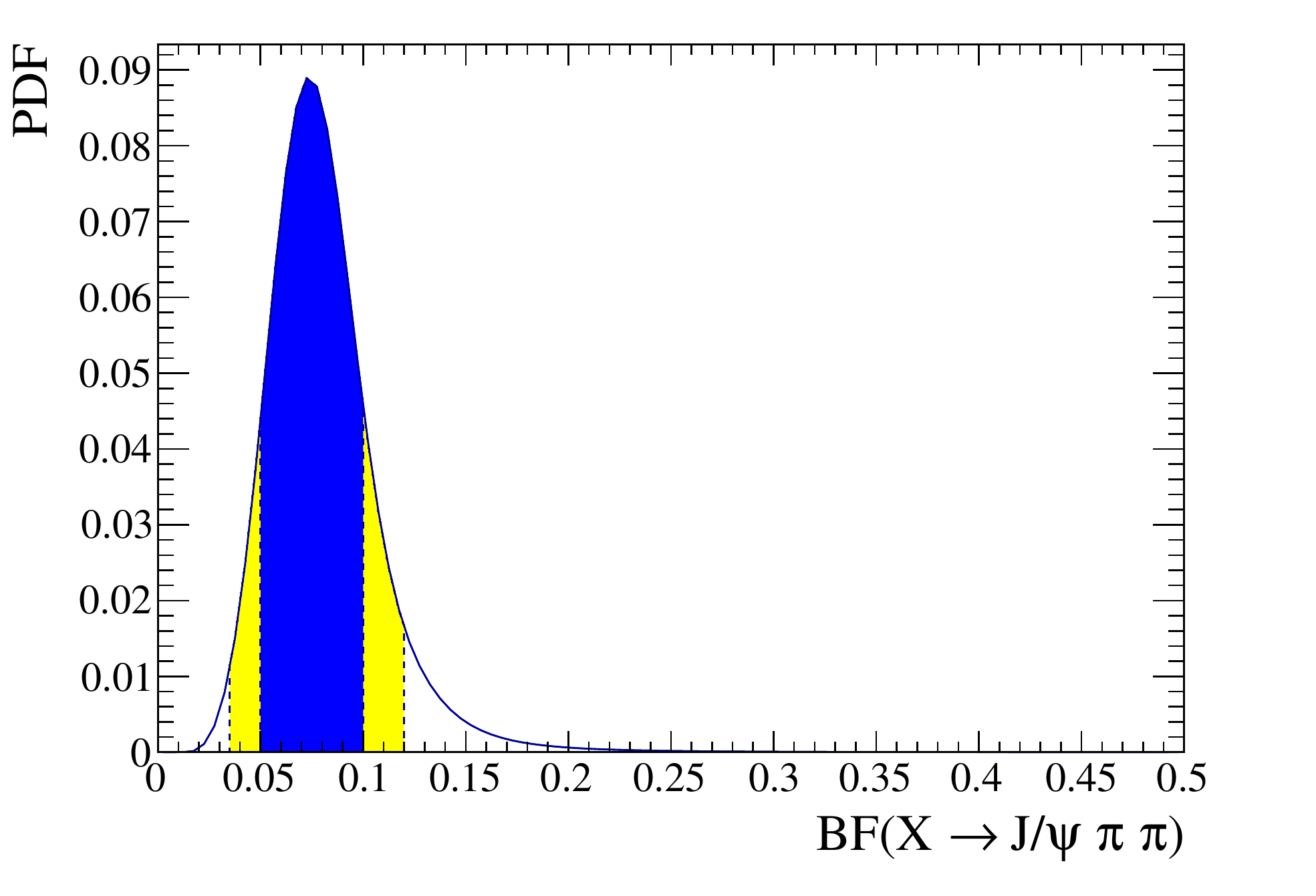} \includegraphics[width=.45\columnwidth]{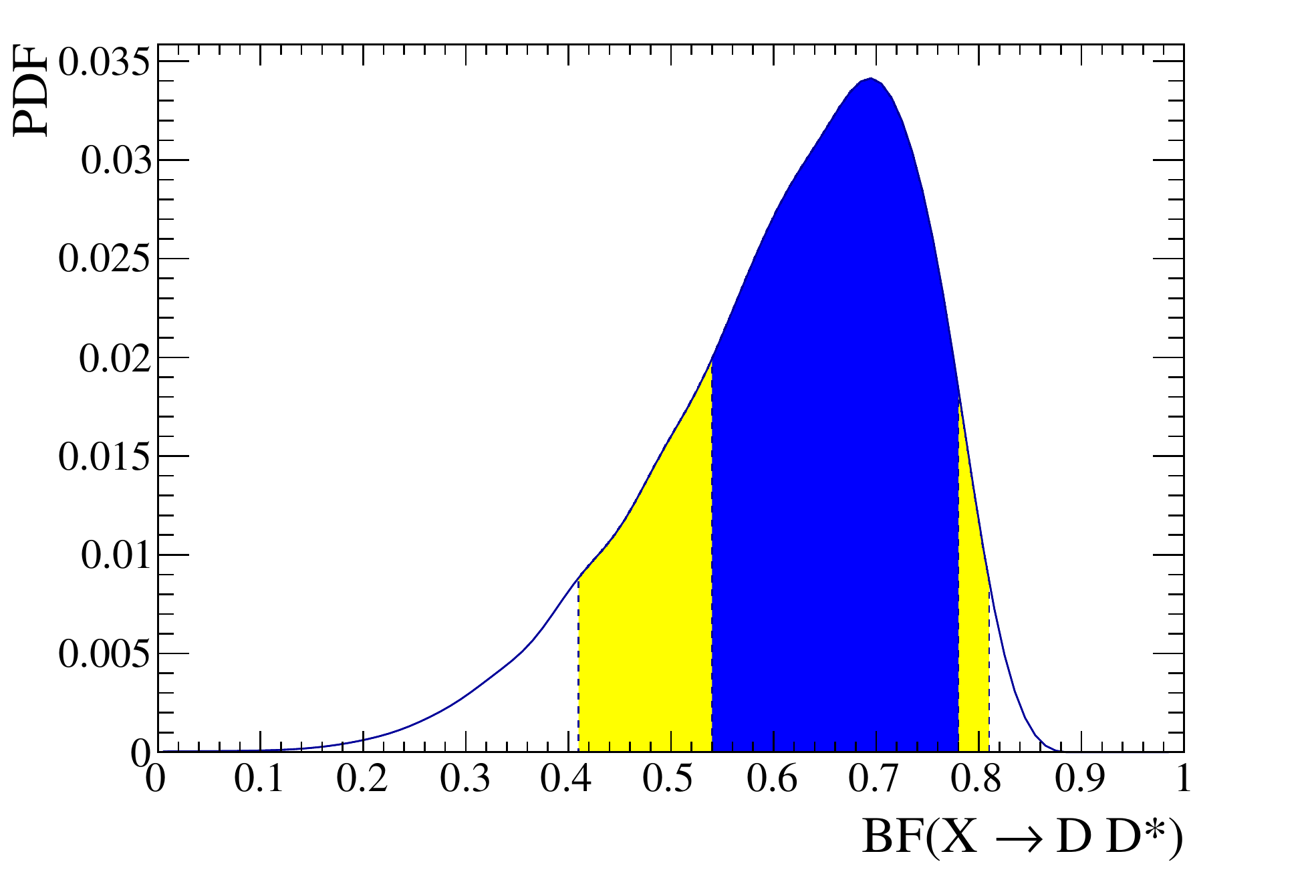}\\
\includegraphics[width=.45\columnwidth]{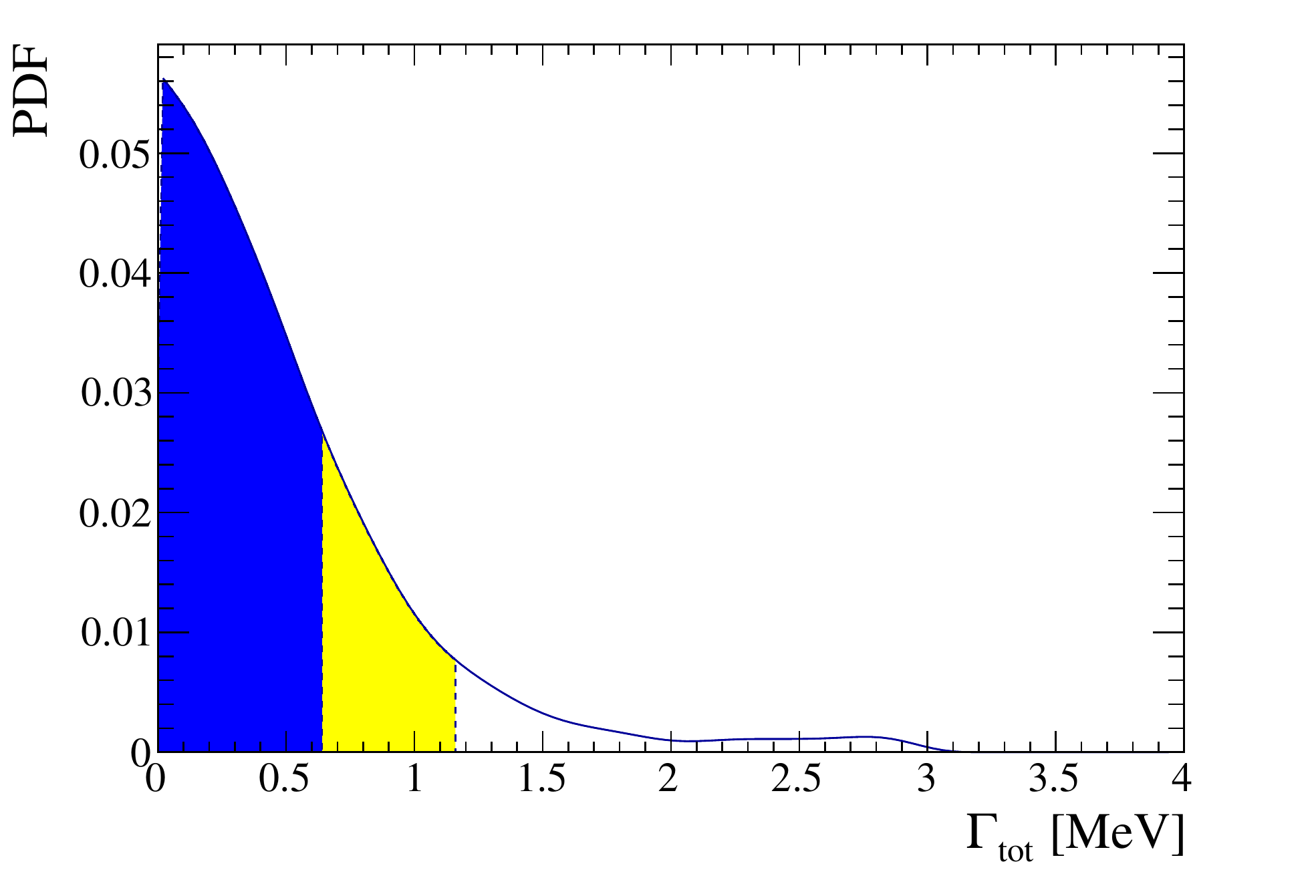} \includegraphics[width=.45\columnwidth]{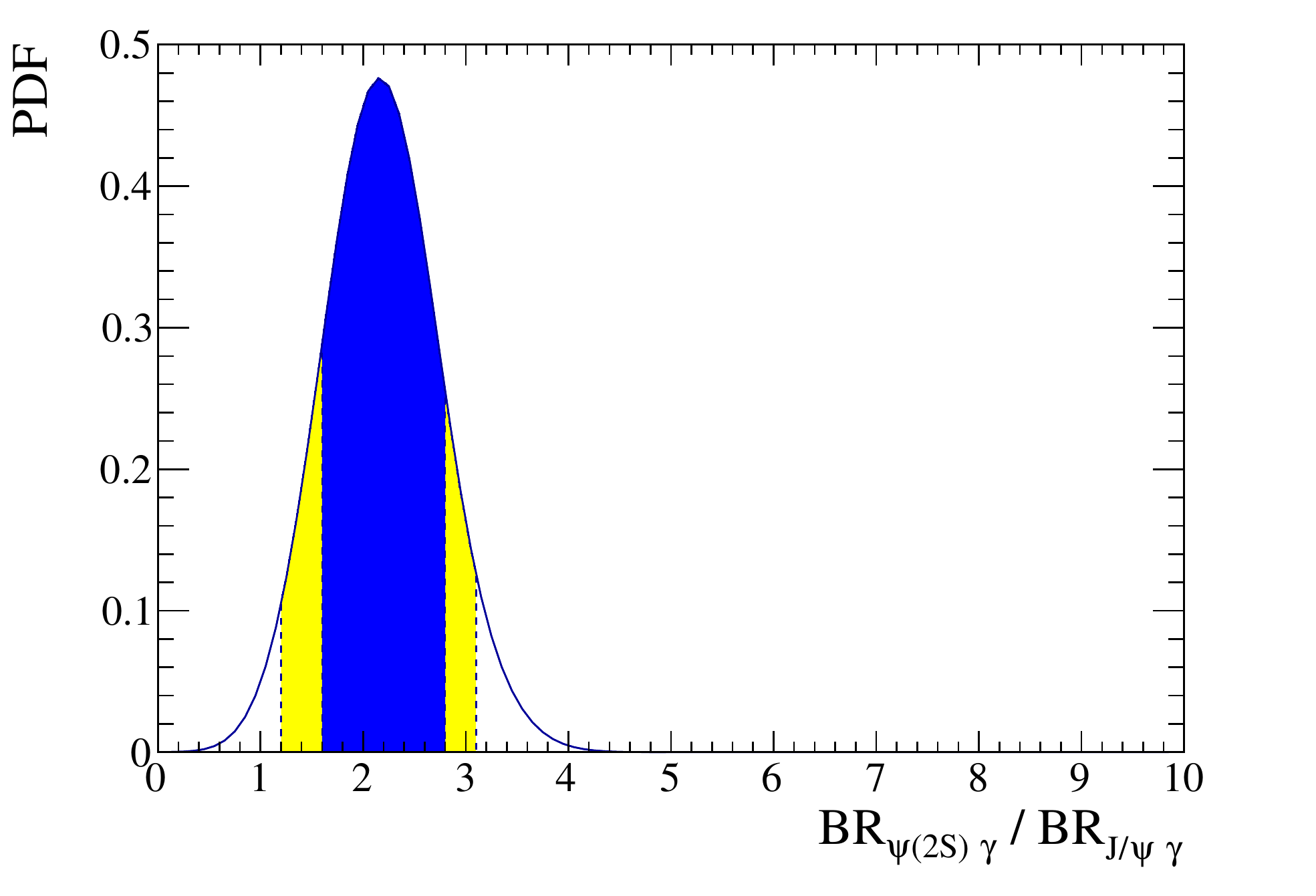}\\

 \caption{Updated likelihood (w.r.t. Drenska \etal\cite{Drenska:2010kg}) 
function of the $X(3872)$ branching fraction in $\jpsi\pi\pi$ and $\Dstarz\Dzb$, the
total width and the ratio $\BR(X(3872)\to \psiprime \gamma) / \BR(X(3872)\to \jpsi \gamma)$. See text for a description of the
combination method. The dark (light) filled area corresponds to the 68\% (90\%) C.L.
region.}
    \label{fig:likelihoods}
    \end{center}
\end{figure}

The searches for partner states of the $X(3872)$ have been motivated by the predictions of the tetraquark model (see \sectionname{~\ref{sec:tetraquarks}}). 
For example, it has been hypothesized that the $X$ state produced in $B^+$ decays was different from the $X$ state produced in $B^0$ decays. If so, the two $X$ should have different masses. 
Both \babar\cite{Aubert:2005zh,Aubert:2008gu}
and \belle\cite{Adachi:2008te,Choi:2011fc} have performed
analyses distinguishing the two samples. The most recent results
set the mass difference of the two $X$ at
\begin{subequations}
\begin{align}
  \delta M
	&\equiv	M(X\,|\,\Bp\to\Kp X) - M(X\,|\,\Bz\to\Kz X)	\nonumber \\
	& = 	(+2.7 \pm 1.6 \pm 0.4)\mev \hspace{2em} \text{(\babar)}, 	 \\
	& =	(-0.7 \pm 1.0 \pm 0.2)\mev \hspace{2em} \text{(\belle)},	 \\
	& = 	(+0.2 \pm 0.8)\mev\phantom{\pm 0.22} \hspace{2em} \text{(mean)}.
\end{align}
\end{subequations}

Moreover, an inclusive analysis by \cdf\cite{Aaltonen:2009vj},
of the $\jpsi \pi^+\pi^-$ spectrum, gives no evidence for
any other neutral state, setting an upper limit on the mass difference of
$3.6\mev$ at the 95\% C.L.. 

The same analyses provide measurements of
the ratio of product branching fractions
\begin{subequations}
\begin{align}
  \frac	{\BR(\Bz\to\Kz X) \times \BR(X\to\pi^+\pi^-\jpsi)}
			{\BR(\Bp\to\Kp X) \times \BR(X\to\pi^+\pi^-\jpsi)}
	& =	0.41 \pm 0.24 \pm 0.05\quad\text{(\babar)},		 \\
	& =	0.50 \pm 0.14 \pm 0.04\quad\text{(\belle)}.
\end{align}
\end{subequations}
Searches for charged partners have also been performed by both 
\babar\cite{Aubert:2004zr} and
\belle\cite{Choi:2011fc}.
No evidence for such a state is seen, with limits on the product branching
fractions of
\begin{subequations}
\begin{align}
 \BR(\Bzb\to\Km X^+) \times \BR(X^+ \to \rho^+\jpsi)	& < 5.4 \times 10^{-6}\quad\text{(\babar)},  \\
 & < 4.2 \times 10^{-6}\quad\text{(\belle)},  \\
 \BR(\Bp\to\Kz X^+) \times \BR(X^+ \to \rho^+\jpsi)	& < 22 \times 10^{-6}\quad\text{(\babar)},  \\
  & < 6.1 \times 10^{-6}\quad\text{(\belle)},
\end{align}
\end{subequations}
to be compared with
\begin{subequations}
\begin{align}
  \BR(\Bp\to\Kp X)
	& \times \BR(X \to \rho^0\jpsi)				  \nonumber \\
	& = (8.4 \pm 1.5 \pm 0.7) \times 10^{-6}\quad\text{(\babar)},  \\
	& = (8.6 \pm 0.8 \pm 0.5) \times 10^{-6}\quad\text{(\belle)}
\end{align}
\end{subequations}
for the discovery mode,
measured by \babar\cite{Aubert:2008gu} and \belle\cite{Choi:2011fc}.
\begin{landscape}
\begin{table}[h]
  \tbl{Measured $X(3872)$ product branching fractions,
	separated by production and decay channel. Our averages are in boldface.
	The last two columns report the results in terms of
	absolute $X(3872)$ branching fraction ($B_{fit}$) and in terms of 
	the branching fraction normalized to $\jpsi\pi\pi$ ($R_{fit}$)
	as obtained from the global likelihood fit described in the text.
	For non-zero measurements we report the mean value, and the 68\% C.L. range in form of asymmetric errors. The limits are provided at 90\% C.L. 
	The $X(3872)\to\pi\pi\pi^0\jpsi$ is dominated by $\omega\jpsi$, but no limits on the non-resonant $\pi\pi\pi^0\jpsi$ component have been set.
	The ratio $R^\prime$ given by \lhcb\cite{Aaij:2013rha} is the ratio $\BR\left(X(3872)\to \psiprime\gamma\right)/ \BR\left(X(3872)\to \jpsi\gamma\right)$.
  }
 {
\small
  \centering 
  \begin{tabular}{llllcc} \hline\hline 
$B$ decay mode & $X$ decay mode & \multicolumn{2}{c}{product branching fraction ($\times 10^5$)} & $B_{fit}$ & $R_{fit}$ \\ \hline
$\Kp X$ & $X\to \pi\pi\jpsi$ & $\mathbf{0.86\pm0.08}$ & (\babar\cite{Aubert:2008gu}, \belle\cite{Choi:2011fc}) & $0.081^{+0.019}_{-0.031}$ & 1\\
 &  & $0.84\pm 0.15\pm 0.07$ & \babar\cite{Aubert:2008gu} &  & \\
 &  & $0.86\pm 0.08\pm 0.05$ & \belle\cite{Choi:2011fc} &  & \\
$K^0 X$ & $X\to \pi\pi\jpsi$ & $\mathbf{0.41\pm0.11}$ & (\babar\cite{Aubert:2008gu}; \belle\cite{Choi:2011fc}) &  & \\
 &  & $0.35\pm 0.19\pm 0.04$ & \babar\cite{Aubert:2008gu} &  & \\
 &  & $0.43\pm 0.12\pm 0.04$ & \belle\cite{Choi:2011fc} &  & \\
$(K^+\pi^-)_{NR}X$ & $X\to \pi\pi\jpsi$ & $0.81\pm 0.20^{+0.11}_{-0.14}{}$ & \belle\cite{Adachi:2008te} &  & \\
$\Kstarz X$ & $X\to \pi\pi\jpsi$ & $<0.34$, 90\% C.L. & \belle\cite{Adachi:2008te} &  & \\ \hline
$K X$ & $X\to\omega\jpsi$ & $R=0.8\pm0.3$ & \babar\cite{delAmoSanchez:2010jr} & $0.061^{+0.024}_{-0.036}$ & $0.77^{+0.28}_{-0.32}$\\
$\Kp X$ &  & $0.6\pm 0.2\pm 0.1$ & \babar\cite{delAmoSanchez:2010jr} &  & \\
$\Kz X$ &  & $0.6\pm 0.3\pm 0.1$ & \babar\cite{delAmoSanchez:2010jr} &  & \\ 
$K X$ & $X\to\pi\pi\pi^0\jpsi$ & $R= 1.0 \pm 0.4 \pm 0.3$ & \belle\cite{Abe:2005ix} &  & \\ 
\hline
$\Kp X$ & $X\to \Dstarz\Dzb$ & $\mathbf{8.5\pm2.6}$ & (\babar\cite{Aubert:2007rva}; \belle\cite{Adachi:2008su}) & $0.614^{+0.166}_{-0.074}$ & $8.2^{+2.3}_{-2.8}$\\
 &  & $16.7\pm 3.6\pm 4.7$ & \babar\cite{Aubert:2007rva} &  & \\
 &  & $7.7\pm 1.6 \pm 1.0$ & \belle\cite{Adachi:2008su} &  & \\
$\Kz X$ & $X\to \Dstarz\Dzb$ & $\mathbf{12\pm4}$ & (\babar\cite{Aubert:2007rva}; \belle\cite{Adachi:2008su}) &  & \\
 &  & $22\pm 10 \pm 4$ & \babar\cite{Aubert:2007rva} &  & \\
 &  & $9.7\pm 4.6\pm1.3$ & \belle\cite{Adachi:2008su} &  & \\ \hline\hline

					\end{tabular} \label{tab:xdecays} }
\end{table}
\end{landscape}
\begin{landscape}
\begin{table}[h]
  \tbl{({\em Continued}).
  }
 {
\small
  \centering 
  \begin{tabular}{llllcc} \hline\hline 
$B$ decay mode & $X$ decay mode & \multicolumn{2}{c}{product branching fraction ($\times 10^5$)} & $B_{fit}$ & $R_{fit}$ \\ \hline
$\Kp X$ & $X\to \gamma\jpsi$ & $\mathbf{0.202\pm0.038}$ & (\babar\cite{Aubert:2008rn}; \belle\cite{Bhardwaj:2011dj}) & $0.019^{+0.005}_{-0.009}$ & $0.24^{+0.05}_{-0.06}$\\
$\Kp X$ &  & $0.28\pm 0.08\pm 0.01$ & \babar\cite{Aubert:2008rn} &  & \\
 &  & $0.178^{+0.048}_{-0.044}\pm 0.012$ & \belle\cite{Bhardwaj:2011dj} &  & \\
$\Kz X$ &  & $0.26\pm 0.18\pm 0.02$ & \babar\cite{Aubert:2008rn} &  & \\
 &  & $0.124^{+0.076}_{-0.061}\pm 0.011$ & \belle\cite{Bhardwaj:2011dj} &  & \\ \hline
$\Kp X$ & $X\to \gamma\psiprime$ & $\mathbf{0.44\pm0.12}$ & \babar\cite{Aubert:2008rn} & $0.04^{+0.015}_{-0.020}$ & $0.51^{+0.13}_{-0.17}$\\
$\Kp X$ &  & $0.95\pm 0.27\pm 0.06$ & \babar\cite{Aubert:2008rn} &  & \\
 &  & $0.083^{+0.198}_{-0.183}\pm 0.044$ & \belle\cite{Bhardwaj:2011dj} &  & \\
 &  & $R^\prime=2.46 \pm 0.64 \pm 0.29$ & \lhcb\cite{Aaij:2014ala} &  & \\
$\Kz X$ &  & $1.14\pm 0.55\pm 0.10$ & \babar\cite{Aubert:2008rn} &  & \\
 &  & $0.112^{+0.357}_{-0.290}\pm 0.057$ & \belle\cite{Bhardwaj:2011dj} &  & \\ \hline
$\Kp X$ & $X\to\gamma\chi_{c1}$ & $< 9.6 \times 10^{-3}$ & \belle\cite{Bhardwaj:2013rmw} & $< 1.0 \times 10^{-3}$ & $<0.014$\\ \hline
$\Kp X$ & $X\to\gamma\chi_{c2}$ & $< 0.016$ & \belle\cite{Bhardwaj:2013rmw} & $< 1.7 \times 10^{-3}$ & $<0.024$\\ \hline
$\kaon X$ & $X\to\gamma\gamma$ & $<4.5\times 10^{-3}$ & \belle\cite{Abe:2006gn} & $< 4.7 \times 10^{-4}$ & $< 6.6 \times 10^{-3}$\\ \hline
$\kaon X$ & $X\to \eta\jpsi$ & $<1.05$ & \babar\cite{Aubert:2004fc} & $< 0.11$ & $<1.55$\\ \hline
$\Kp X$ & $X\to p\bar p$ & $<9.6 \times 10^{-4}$ & \lhcb\cite{Aaij:2013rha} & $< 1.6\times10^{-4}$ & $<2.2 \times 10^{-3}$\\ \hline\hline

					\end{tabular} \label{tab:xdecays2} }
\end{table}
\end{landscape}

We conclude this section on the $X(3872)$ with the inclusive production at hadron colliders: the prompt production has been studied at \cdf\cite{cdfnote} and \cms\cite{Chatrchyan:2013cld}, giving
\begin{subequations}
\begin{align}
 \frac{\sigma^\text{prompt}\left(p\bar p \to X(3872) + \text{all}\right)}{\sigma\left(p\bar p \to X(3872) + \text{all}\right)} &= (83.9 \pm 4.9 \pm 2.0)\%\; \text{ at }\; \sqrt{s}=1.96\tev,\\
 \frac{\sigma^\text{prompt}\left(p p \to X(3872) + \text{all}\right)}{\sigma\left(p p \to X(3872) + \text{all}\right)} &= (73.7 \pm 2.3 \pm 1.6)\%\; \text{ at }\; \sqrt{s}=7\tev.
\end{align}
\end{subequations}
\begin{figure}[t]
\centering
\includegraphics[width=.5\textwidth]{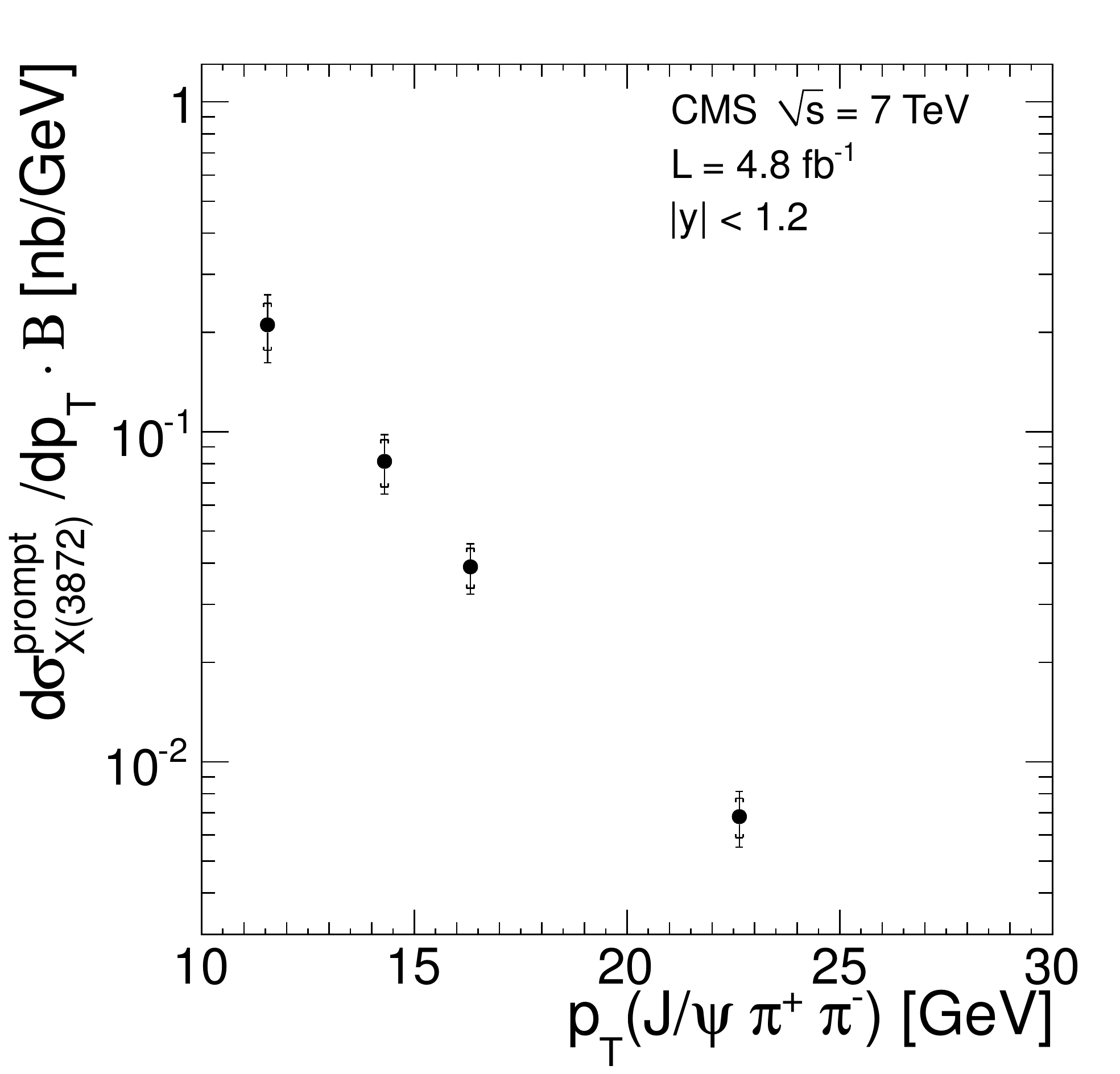}
\caption{Measured differential cross section for prompt $X(3872)$ production times branching
fraction $\BR( X(3872) \to \jpsi \pi^+\pi^-)$ as a function of $p_T$, from \cms\cite{Chatrchyan:2013cld}.}
 \label{fig:xcmscrosssection}
\end{figure}

\cms published also the value for the prompt production cross section, $\sigma^\text{prompt}\left( pp \to X(3872) + \text{all}\right) \times \BR\left( X(3872) \to \jpsi\pi^+\pi^- \right) = (1.06 \pm 0.11\pm 0.15)\nb$ at $\sqrt{s}=7\tev$ (see \figurename{~\ref{fig:xcmscrosssection}}). 

The same measurement is not present in the \cdf note, 
but it has been estimated by Bignamini \etal\cite{Bignamini:2009sk}: 
$\sigma^\text{prompt}\left( pp \to X(3872) + \text{all}\right) \times \BR\left( X(3872) \to \jpsi\pi^+\pi^- \right) = (3.1 \pm 0.7)\nb$ at $\sqrt{s}=1.96\tev$.

\subsection{Vector resonances}
\label{sec:vectors}
Many states with unambiguous $J^{PC}=1^{--}$ have been discovered via direct production in $e^+ e^-$ collisions. The $B$-factories can investigate a large mass range, by searching events with an additional energetic photon $\gamma_\text{ISR}$ emitted by the initial state, which lowers the center-of-mass energy down to the mass of the particle. The $\tau-c$ factories can instead scan the mass range by varying their center-of-mass energy. A graphic summary of all this states is in \figurename{~\ref{fig:ipsilon}}.

\begin{figure}[b]
\centering
\includegraphics[width=.75\textwidth]{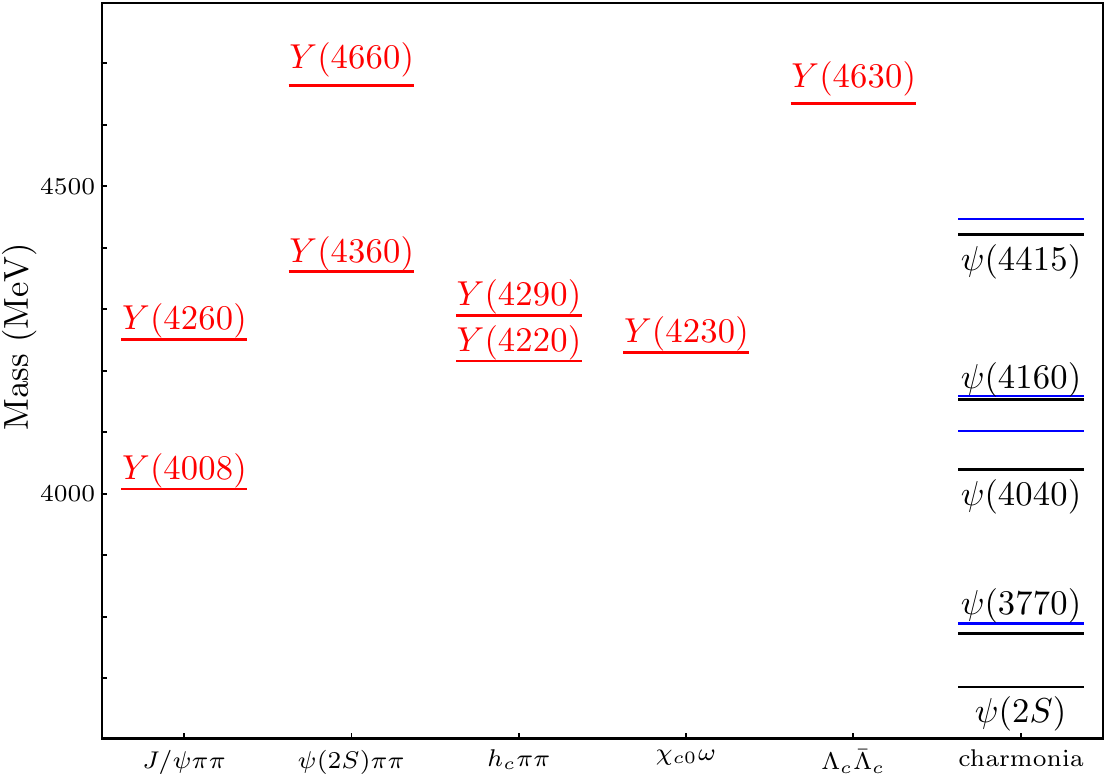}
\caption{Exotic vector states divided by decay channel. In the right column, we report observed (black) and predicted (blue) charmonium levels. 
Red line are exotic states.}
 \label{fig:ipsilon}
\end{figure}
In 2005 \babar observed an unexpected vector charmonium state decaying into $\jpsi \pi^+\pi^-$ named $Y(4260)$\cite{Aubert:2005rm}, with a mass of $M= (4259 \pm 8^{+2}_{-6})\mev$ and a
width of $\Gamma = (88 \pm 23_{-4}^{+6})\mev$. Soon after it was confirmed by \cleo\cite{He:2006kg,Coan:2006rv}, which reported evidence also for $Y(4260)\to \jpsi\pi^0\pi^0$. \babar performed a similar analysis in the $e^+e^-\to \psiprime\pi^+\pi^-$ channel\cite{Aubert:2006ge}, finding no evidence of $Y(4260)$; instead, a heavier state was observed at a mass $M= (4324 \pm 24)\mev$ and a width $\Gamma =(172 \pm 33)\mev$, dubbed $Y(4360)$. The absence of $Y(4360)\to  \jpsi\pi^+\pi^-$ is significant: $\BR(Y(4360)\to \jpsi\pi^+\pi^-)/\BR(Y(4360)\to \psiprime\pi^+\pi^-) < 3.4\times 10^{-3}$ at the 90\% C.L.\cite{Cotugno:2009ys}, and is hard to understand in an ordinary charmonium framework. This pattern has been confirmed in an update of \babar's analysis\cite{Lees:2012pv}.

\belle confirmed both these vector states\cite{Yuan:2007sj,Wang:2007ea}, and observed another resonance, called $Y(4660)$,
in the $\psiprime\pi^+\pi^-$ channel, which \babar was not
able to see because of limited statistics, with mass $M=(4664 \pm 11 \pm 5)\mev$ and  width $\Gamma = ( 48 \pm 15 \pm 3)\mev$. It also found a broad structure in $\jpsi\pi^+\pi^-$ named $Y(4008)$, at mass $M=(4008 \pm 40_{-28} ^{+114})\mev$ and width $\Gamma = ( 226 \pm 44 \pm 87)\mev$. This last state has not been seen by \babar\cite{Lees:2012cn}, but it has been confirmed in the full statistics analysis by \belle\cite{Liu:2013dau}, with $M=(3890.8 \pm 40.5 \pm 11.5)\mev$ and $\Gamma = (254.5 \pm 39.5 \pm 13.6)\mev$. The PDG\cite{pdg} averaged mass and width for the $Y(4260)$ are based on the most recent analyses by \belle\cite{Liu:2013dau}, \babar\cite{Lees:2012cn} and \cleo\cite{He:2006kg} and are $M= (4251\pm 9) \mev$ and $\Gamma = (120 \pm 12)\mev$. The full statistics analysis in $\psiprime \pi^+\pi^-$ by \belle\cite{Wang:2014hta} gives for the $Y(4360)$ a mass and width of $M=(4347 \pm 6 \pm 3)\mev$ and $\Gamma = (103 \pm 9 \pm 5)\mev$, and for the $Y(4660)$ a mass and width of $M=(4652 \pm 10 \pm 8)\mev$ and $\Gamma = (68 \pm 11 \pm 1)\mev$. In \figurename{~\ref{fig:ipsilonspectra}} we report some distributions of $\jpsi\pi^+\pi^-$ and $\psiprime \pi^+\pi^-$ by \belle.

\begin{figure}[b]
\centering
\includegraphics[width=.45\textwidth]{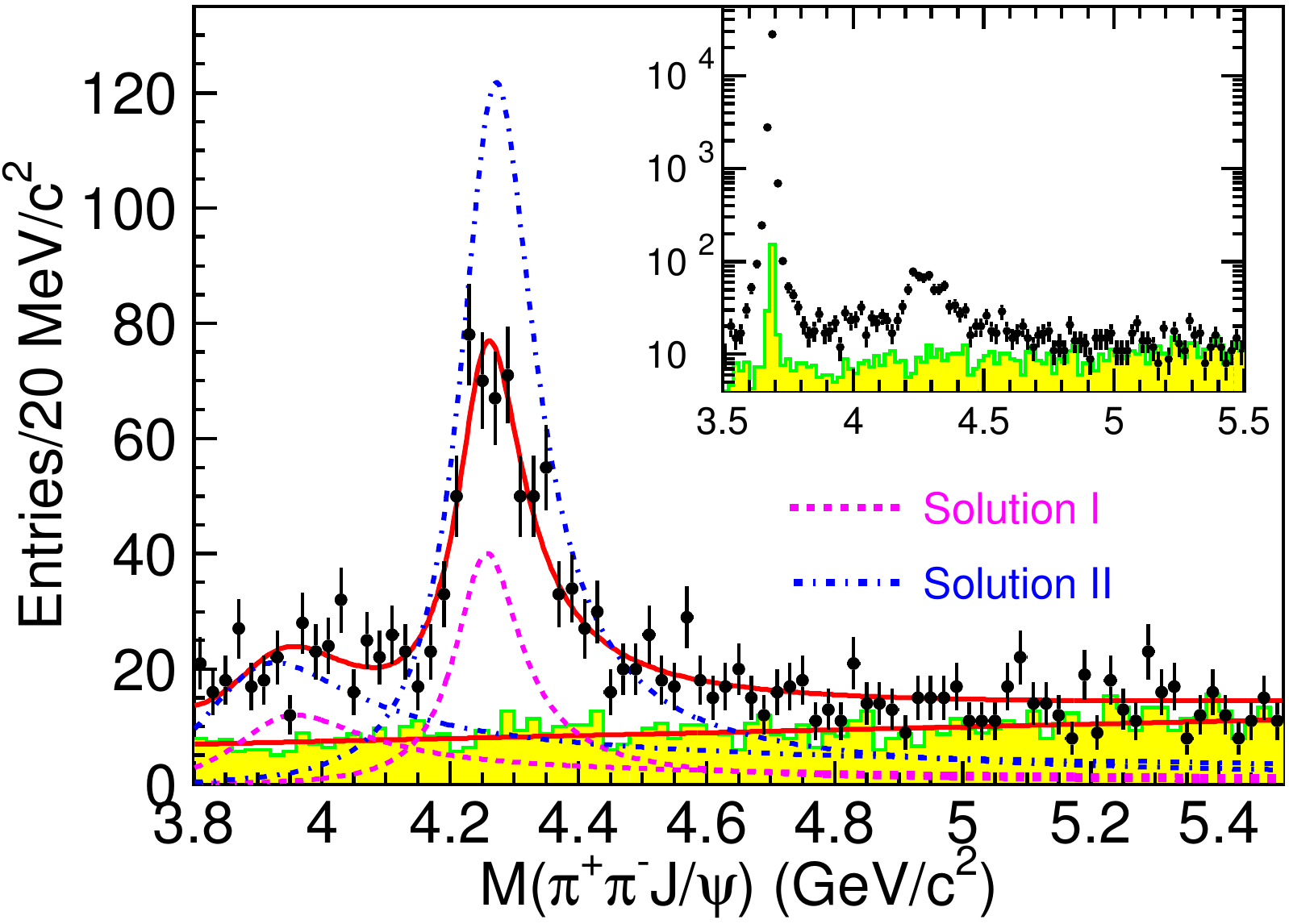}
\includegraphics[width=.45\textwidth]{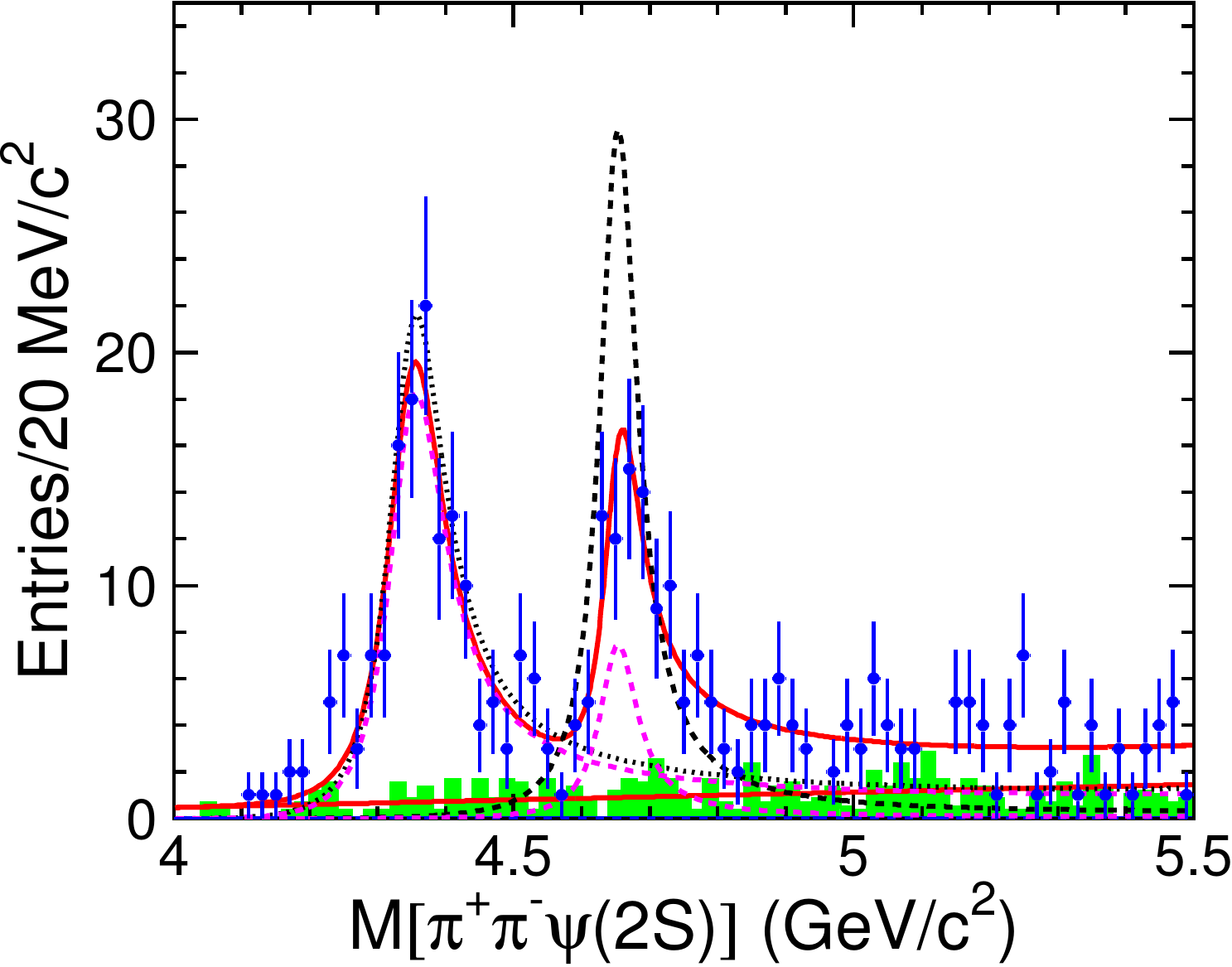}
\caption{\belle analyses of $e^+e^- \to \jpsi \pi^+\pi^-$ (left)\cite{Liu:2013dau} and $\to \psiprime \pi^+\pi^-$ (right)\cite{Wang:2014hta}.}
 \label{fig:ipsilonspectra}
\end{figure}
Motivated by the tetraquark predictions, \belle searched for vector resonances decaying into $\Lambda_c\Lbar_c$\cite{Pakhlova:2008vn}.
A structure (the $Y(4630)$) has actually been found near the baryon threshold, with Breit-Wigner parameters $M=(4634^{+8}_{-7}{}_{-8}^{+5})\mev$ and $\Gamma = (92^{40}_{-24}{}^{+10}_{-21})\mev$.
A combined fit of the $\psiprime \pi^+\pi^-$ and $\Lambda_c\Lbar_c$ spectra concluded that the two structures $Y(4630)$ and $Y(4660)$
can be the same state, with a strong preference for the baryonic decay mode:
$\BR(Y(4660)\to\Lambda_c\Lbar_c)/ \BR(Y(4660)\to\psiprime\pi^+\pi^-)= 25\pm7$\cite{Cotugno:2009ys}.

The vector states mentioned before are considered to be exotic. In fact, there are no unassigned $1^{--}$ charmonia below $4500\mev$, and the branching ratios into open charm mesons are too small for above-threshold charmonia: \babar sees no evidence for a signal\cite{Aubert:2007pa,delAmoSanchez:2010aa}, and set 90\% C.L. upper limits:
\begin{subequations}
\begin{align} 
\BR(Y(4260)\to \D\Dbar)/\BR(Y(4260)\to \jpsi\pi^+\pi^-)	& < 1.0, \\
\BR(Y(4260)\to \Dstar\Dbar)/\BR(Y(4260)\to \jpsi\pi^+\pi^-)	& < 34,	 \\
\BR(Y(4260)\to \Dstar\Dstarb)/\BR(Y(4260)\to \jpsi\pi^+\pi^-)	& < 40,  \\
\BR(Y(4260)\to \Dsp\Dsm)/\BR(Y(4260)\to \jpsi\pi^+\pi^-)	& < 0.7, \\
\BR(Y(4260)\to \Dsm\Dssm)/\BR(Y(4260)\to \jpsi\pi^+\pi^-)	& < 44,  \\
\BR(Y(4260)\to \Dssp\Dssm)/\BR(Y(4260)\to \jpsi\pi^+\pi^-)	& < 30,
\end{align}
\end{subequations}
whereas the limits set by \belle\cite{Pakhlova:2009jv} are:
\begin{subequations}
\begin{align} 
\BR(Y(4260) \to \Dz\Dstarm\pi^+)/\BR(Y(4260) \to \jpsi\pi^+ \pi^-) &< 9, \\
\BR(Y(4360) \to \Dz\Dstarm\pi^+)/\BR(Y(4360) \to \psiprime\pi^+ \pi^-) &< 8, \\
\BR(Y(4660) \to \Dz\Dstarm\pi^+)/\BR(Y(4660) \to \psiprime\pi^+ \pi^-) &< 10,
\end{align}
\end{subequations}
to be compared with $\BR(\psi(3770) \to \D\Dbar)/\BR(\psi(3770) \to \jpsi\pi^+ \pi^-) \gtrsim 480$ for an ordinary above-threshold vector charmonium. As for radiative decays, $Y(4260) \to \gamma X(3872)$ has been observed by \bes. Some clean events of $e^+e^- \to \gamma X(3872)$ have been measured. Moreover, the production cross section $\sigma\left(e^+ e^-\to \gamma X(3872)\right) \times \BR\left(X(3872) \to \jpsi \pi^+ \pi^-\right)$ scales as a function of the center-of-mass energy consistently with a Breit-Wigner with $Y(4260)$ mass and width as parameters, consequently the observed events come from the intermediate resonant state and not from the continuum.  
The $Y(4260)$ has been searched without success in many other final states, which we report in \tablename{~\ref{tab:y4260}}. 

Another important question to understand the nature of these vector states is whether or not the 
pion pair comes from any resonance. The updated \babar analysis in $\jpsi \pi^+\pi^-$\cite{Lees:2012cn} finds some evidence of a $\jpsi f_0(980)$ component. Since the decay $Y(4260)\to \psiprime f_0(980)$ is phase-space forbidden, this could partially explain why the $Y(4260)$ does not decay into $\psiprime\pi^+\pi^-$ (although the relevant non-resonant component could allow this decay). Some indications of an $f_0(980)$ component in the $Y(4660)$ appear in \belle's $\psiprime\pi^+\pi^-$ analysis\cite{Wang:2007ea}, while no definite structure is recognizable for the other resonances.

\begin{table}[t]
\tbl{Upper limits for $Y(4260)$ into different final states. The decays into open charm mesons are discussed in the text.}
{ \begin{tabular}{lll}
  Final state & Upper limit (90\% C.L.)& Experiment \\ \hline\hline
 \multicolumn{3}{l}{ $\Gamma_{ee} \times \BR\left(Y(4260)\to f\right)$ (\ev) } \\ \hline
  $\jpsi K^+K^-$ & $1.2$ & \belle\cite{Yuan:2007bt} \\
  $\jpsi\eta$ & $14.2$ & \belle\cite{Wang:2012bgc} \\
  $\phi\pi^+\pi^-$ & $0.4$ & \babar\cite{Aubert:2007ur} \\
  $\KS K^+\pi^-$ & $0.5$ & \babar\cite{Aubert:2007ym} \\
  $K^+ K^-\pi^0$ & $0.6$ & \babar\cite{Aubert:2007ym} \\ \hline
  \multicolumn{3}{l}{  $\BR\left(Y(4260)\to f\right) / \BR\left(Y(4260)\to\jpsi\pi^+\pi^-\right)$ } \\ \hline
  $h_c \pi^+\pi^-$ & $1.0$ & \cleo\cite{CLEO:2011aa} \\
$p\bar p$ & $0.13$ & \babar\cite{Aubert:2005cb} \\ \hline
  \multicolumn{3}{l}{  $\sigma\left(e^+e^-\to f\right)$ (\pb) } \\ \hline
  $\chi_{c1}\omega$ & $18$ $(\sqrt{s}=4.31\gev)$ & \bes\cite{Ablikim:2014qwy} \\
$\chi_{c2}\omega$ & $11$ $(\sqrt{s}=4.36\gev)$ & \bes\cite{Ablikim:2014qwy} \\ \hline\hline
 \end{tabular}
\label{tab:y4260}
}
\end{table}
\begin{figure}[b]
\centering
\includegraphics[width=.75\textwidth]{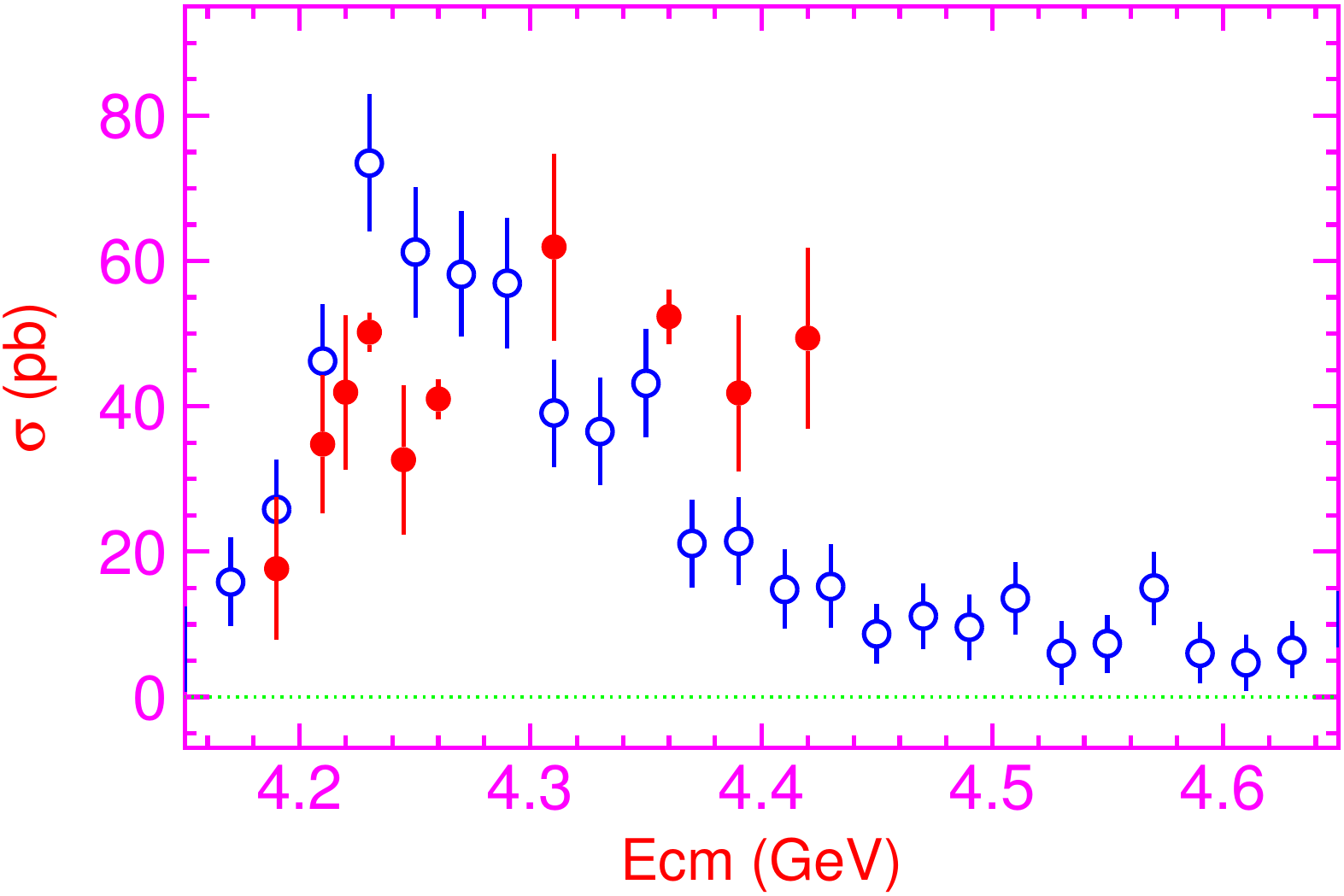}
\caption{\bes data of $e^+e^-\to h_c \pi^+\pi^-$ (red dots)\cite{Ablikim:2013wzq} compared to \belle data of $e^+e^-\to \jpsi \pi^+\pi^-$ (blue circles)\cite{Liu:2013dau}. From Yuan\cite{yuan1,yuan2}.}
 \label{fig:yuan}
\end{figure}

\bes also measured the $e^+e^- \to h_c \pi^+\pi^-$ cross sections at center-of-mass energies varying from $3.9$ to $4.42\gev$\cite{Ablikim:2013wzq} (see \figurename{~\ref{fig:yuan}}). The values of the cross sections are similar to the $e^+ e^- \to \jpsi \pi^+\pi^-$, but the line shape is completely different and does not show any signal for the $Y(4260)$. The $h_c \pi^+\pi^-$ has been fitted by Yuan\cite{yuan1,yuan2}, which found a significant signal for a new $Y(4220)$ state. The fit improves if a second $Y(4290)$ resonance is added, however the lack of experimental data above $4.4\gev$ makes hard to distinguish this second peak from a non-resonant background. The values of the mass and width according to the one peak hypothesis are $M=(4216\pm7)\mev$ and $\Gamma=(39\pm17)\mev$. If there are two peaks, the best fitted values are $M_1=(4216\pm18)\mev$, $\Gamma_1=(39\pm32)\mev$ and $M_2=(4293\pm9)\mev$, $\Gamma_2=(222\pm67)\mev$.

A somewhat similar signal has been seen by \bes in $e^+ e^-\to \chi_{c0}\omega$\cite{Ablikim:2014qwy} at a mass of $M = (4230 \pm 8)\mev$ and a width of $\Gamma = (38\pm12)\mev$, again not compatible  with $Y(4260)$ parameters.

\subsection{The $3940$ family}
\label{sec:X3940}
Some resonances with $C=+$ have been observed around $3940\mev$. Even if they could be likely interpred as ordinary charmonium states, some peculiarities in 
their decay patterns favor a more exotic assignment.

The $X(3940)$ was observed by \belle in double-charmonium production events as a peak in the $M_{\jpsi}$ recoiling mass\cite{Abe:2007jn,Abe:2007sya}, with $M=(3942^{+7}_{-6} \pm 6)\mev$ and $\Gamma = (37^{+26}_{-15} \pm 8)\mev$. A partial reconstruction technique in this production channel
showed that $X(3940)\to \Dstar\Dbar$ is a prominent decay mode (see \figurename{~\ref{fig:3940}}, right panel), whereas $X(3940)\to \D\Dbar,\, \jpsi\omega$ show no signal.
The production mechanism $e^+e^- \to \gamma^* \to \jpsi X(3940)$ costrains the state to have $C=+$. All known states observed via this production mechanism have $J=0$, so a tentative $J^{PC}$ assignment for this state is $0^{-+}$, where the parity is suggested by the absence of $\D\Dbar$ decays.   

\belle observed another state at a similar mass in $B\to\jpsi\omega K$ decays as a resonance in the $\jpsi\omega$ invariant mass, with $M=(3943 \pm 11 \pm13)\mev$ and $\Gamma = (87 \pm 22 \pm 26)\mev$\cite{Abe:2004zs}. The fact that such a state 
is not seen in $B\to \Dstar\Dbar K$ strongly suggests that it is not the $X(3940)$, whence it was dubbed $Y(3940)$. The decay into two vectors costrains a $C=+$ assignment, whereas $J=0,1,2$ and $P=\pm$ are equally allowed. \babar confirmed the state in  $B\to \jpsi\omega K$\cite{Aubert:2007vj,delAmoSanchez:2010jr}, even if at a lower mass and with narrower width, $M=(3919.4^{+3.8}_{-3.4}\pm2.0)\mev$ and $\Gamma = (31^{+10}_{-8} \pm5)\mev$, compatible at $2\sigma$ level with \belle measurement  (see \figurename{~\ref{fig:3940}}, left panel). This discrepancy could be due to different assumptions about the shape of the background. Another state called $Y(3915)$ was observed in $\gamma\gamma$ fusion by both \belle\cite{Uehara:2009tx} and \babar\cite{Lees:2012xs}, with mass and width compatible with the \babar $Y(3940)$ result. The PDG, which assumes the resonances seen in $\gamma\gamma$ fusion and in $B$ decays to be the same state (called $Y(3915)$), gives an averaged mass and width of $M=(3918.4\pm1.9)\mev$ and $\Gamma = (20 \pm5)\mev$\cite{pdg}. The study of angular correlations by \babar favors a $J^{PC}=0^{++}$ assignment\cite{Lees:2012xs}, which would make this state a candidate for $\chi_{c0}(2P)$. 
However, the $\chi_{c0}(2P)$ is expected to have $\Gamma(\chi_{c0}(2P) \to \D\Dbar)\sim 30\mev$, \ie wider than the total width measured of the $Y(3915)$. Even if no upper bound on $\BR(Y(3915)\to \D\Dbar)$ has been reported, 
no signs of a signal for such a decay appear in the
measured $\D\Dbar$ invariant mass distributions for $B\to \D\Dbar K$ decays
published by \babar\cite{Aubert:2007rva} and \belle\cite{Brodzicka:2007aa}. Moreover, if the $Z(3930)$ (see below) is identified as the $\chi_{c2}(2P)$ state, the hyperfine splitting $\chi_{c2}(2P) - \chi_{c0}(2P)$ would be only $6\%$ with respect to the $\chi_{c2}(1P) - \chi_{c0}(1P)$ splitting. This is much smaller than the similar ratio in the bottomonium system ($r \sim 0.7$), and than the potential model 
predictions\cite{Barnes:2005pb} ($0.6 < r <0.9$). These facts challenge the ordinary charmonium interpretation\cite{Guo:2012tv,Olsen:2014maa}. 

Another state, at the time called $Z(3930)$, was seen by \belle in
$\gamma\gamma \to \D\Dbar$\cite{Uehara:2005qd}, and confirmed by \babar\cite{Aubert:2010ab}, at an averaged mass and width of $M=(3927.2\pm 2.6) \mev$ and $\Gamma =  (24\pm 6) \mev$. The angular analysis by \babar favors a $2^{++}$ assignment. This state is compatible with the $\chi_{c2}(2P)$ assignment.

\begin{figure}[t]
\centering
\includegraphics[width=.40\textwidth]{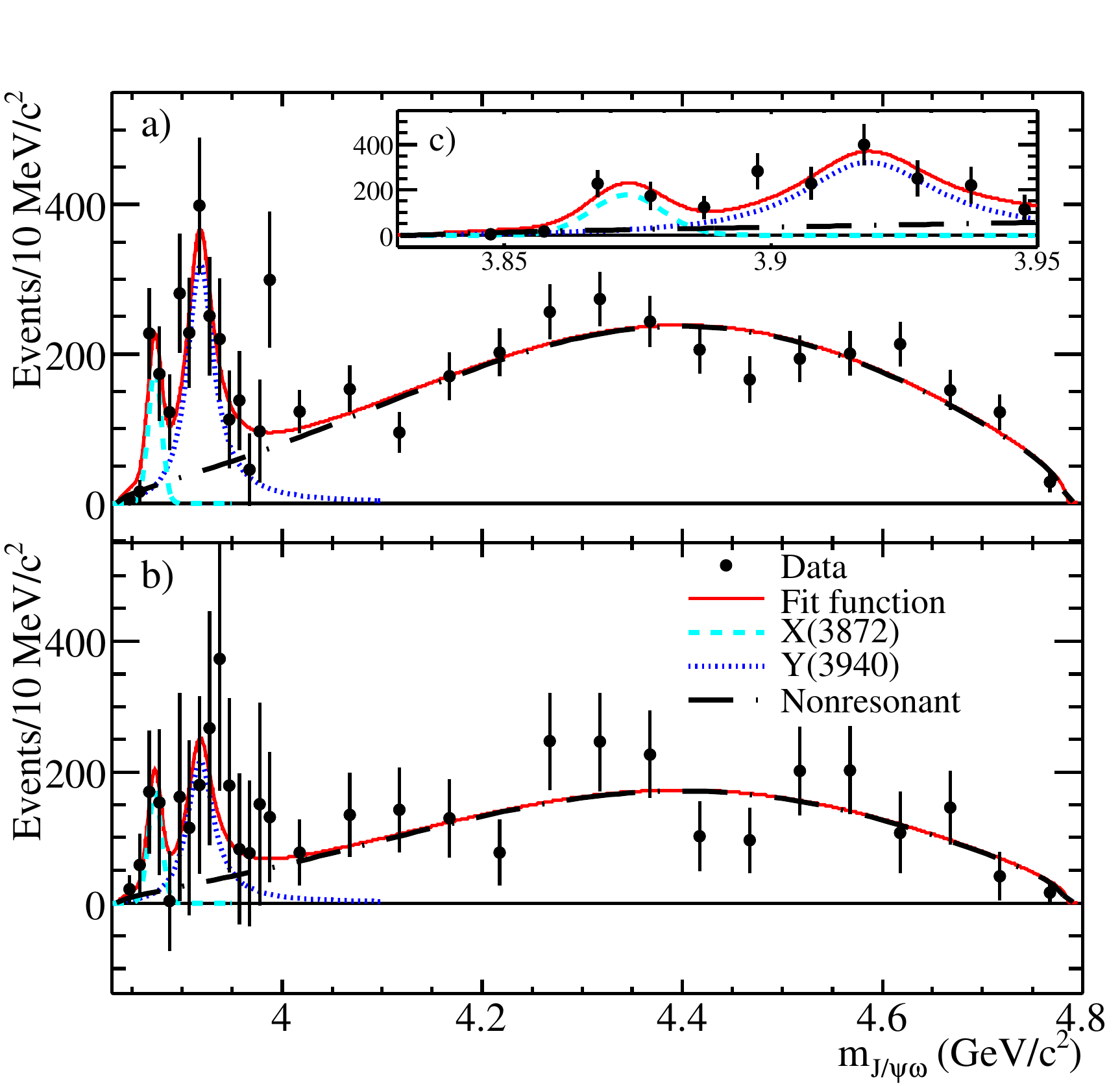}
\includegraphics[width=.50\textwidth]{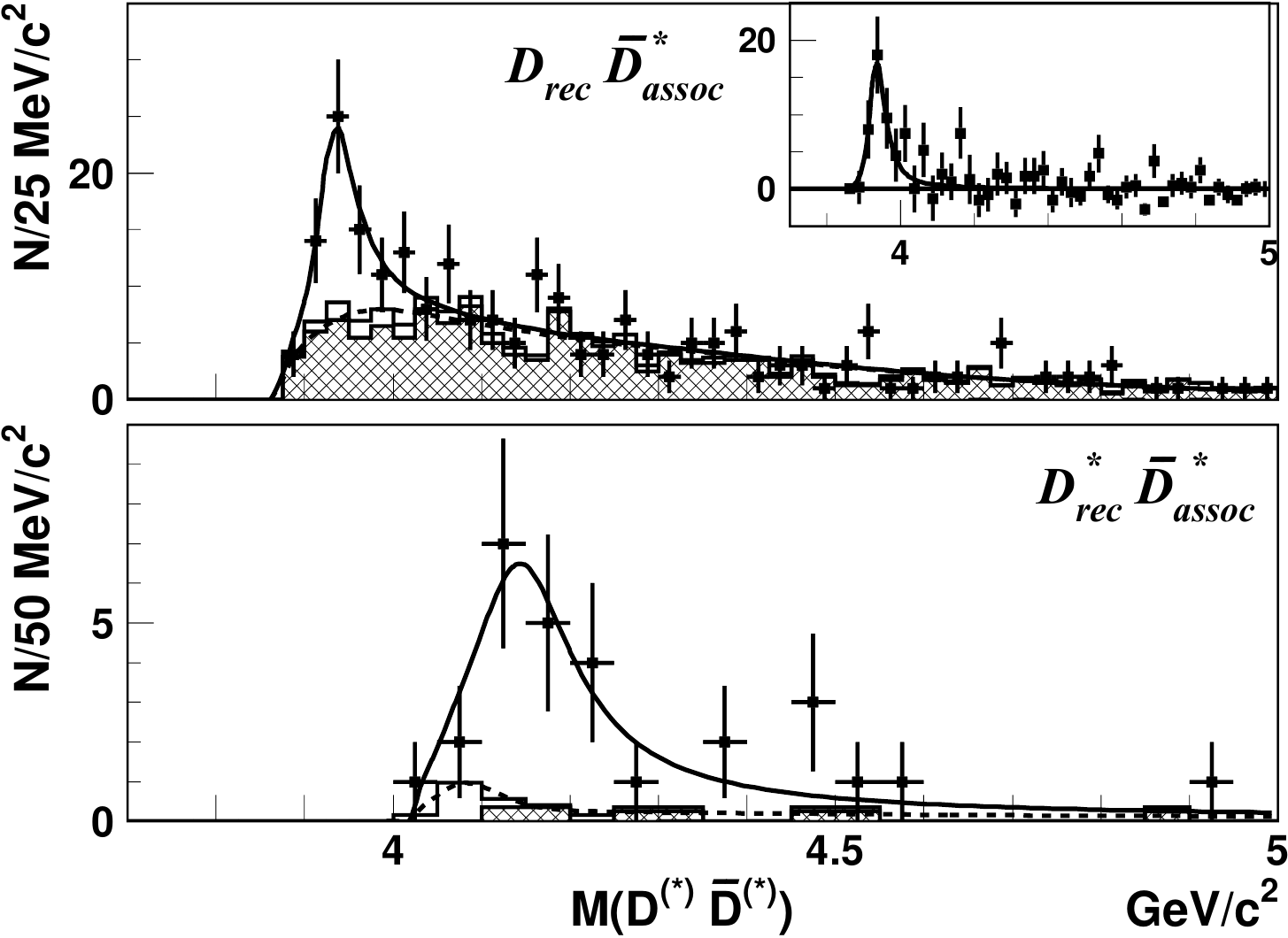}
\caption{Left panel: observation of $Y(3915)$ (at the time called $Y(3940)$) in the invariant mass distribution of $\jpsi\omega$ in $B\to \jpsi \omega K$ decay, by \babar\cite{delAmoSanchez:2010jr}. Right Panel: observation of $X(3940)$ and $X(4160)$ in the invariant mass distribution of $\Dstar\Dbar$ (upper) and $\Dstar\Dstarb$ (lower) in $e^+e^-\to \jpsi \Dstar\Dbar^{(*)}$ events, by \belle\cite{Abe:2007sya}.}
 \label{fig:3940}
\end{figure}
\subsection{Other states}
\label{sec:otherplus}
The analysis by \belle of double charmonium events which discovered the $X(3940)$ observed also a state called $X(4160)$ in the $\Dstar\Dstarb$ invariant mass\cite{Abe:2007sya}  (see \figurename{~\ref{fig:3940}}, right panel). The fitted mass and width are $M=(4156^{+25}_{-20}\pm15)\mev$ and $\Gamma=(139^{+111}_{-61}\pm21)\mev$. 
The production mechanism costrins $C=+$ and favors $J=0$, thus making this state a good candidate for a a $\eta_c(nS)$ state.

The \cdf experiment announced a resonance close to threshold in $\jpsi \phi$ invariant mass, in the channel $B\to \jpsi \phi K$\cite{Aaltonen:2009tz,Aaltonen:2011at}. Since the creation of a $s\bar s$ pair is OZI suppressed, the very existence of such states likely requires exotic interpretations.
This state is called $Y(4140)$, and has mass and width $M=(4143.0\pm2.9\pm 1.2)\mev$ and $\Gamma=(11.7^{+8.3}_{-5.0}\pm3.7)\mev$. The natural quantum number would be $J^{PC}=0^{++}$,
but the exotic assignment $J^{PC}=1^{-+}$ is not excluded. 
\belle searched this state in $\gamma\gamma$ fusion, driven by a molcular prediction\cite{Branz:2009yt}, but found no $Y(4140)$ signal and put a 90\% C.L. upper bound for $\Gamma_{\gamma\gamma}\times \BR(\phi \jpsi)<41\,(6)\ev$ for
$J^P=0^+\, (2^+)$\cite{Shen:2009vs}. Instead, a peak with a $3.2\sigma$ significance was seen at $M=(4350.6^{+4.6}_{-5.1}\pm 0.7)\mev$ and
$\Gamma=(13^{+18}_{-9}\pm 4)\mev$ (see \figurename{~\ref{fig:spectra4350}), and dubbed $X(4350)$.

\begin{figure}[t]
\centering
  	\includegraphics[width=.6\textwidth]{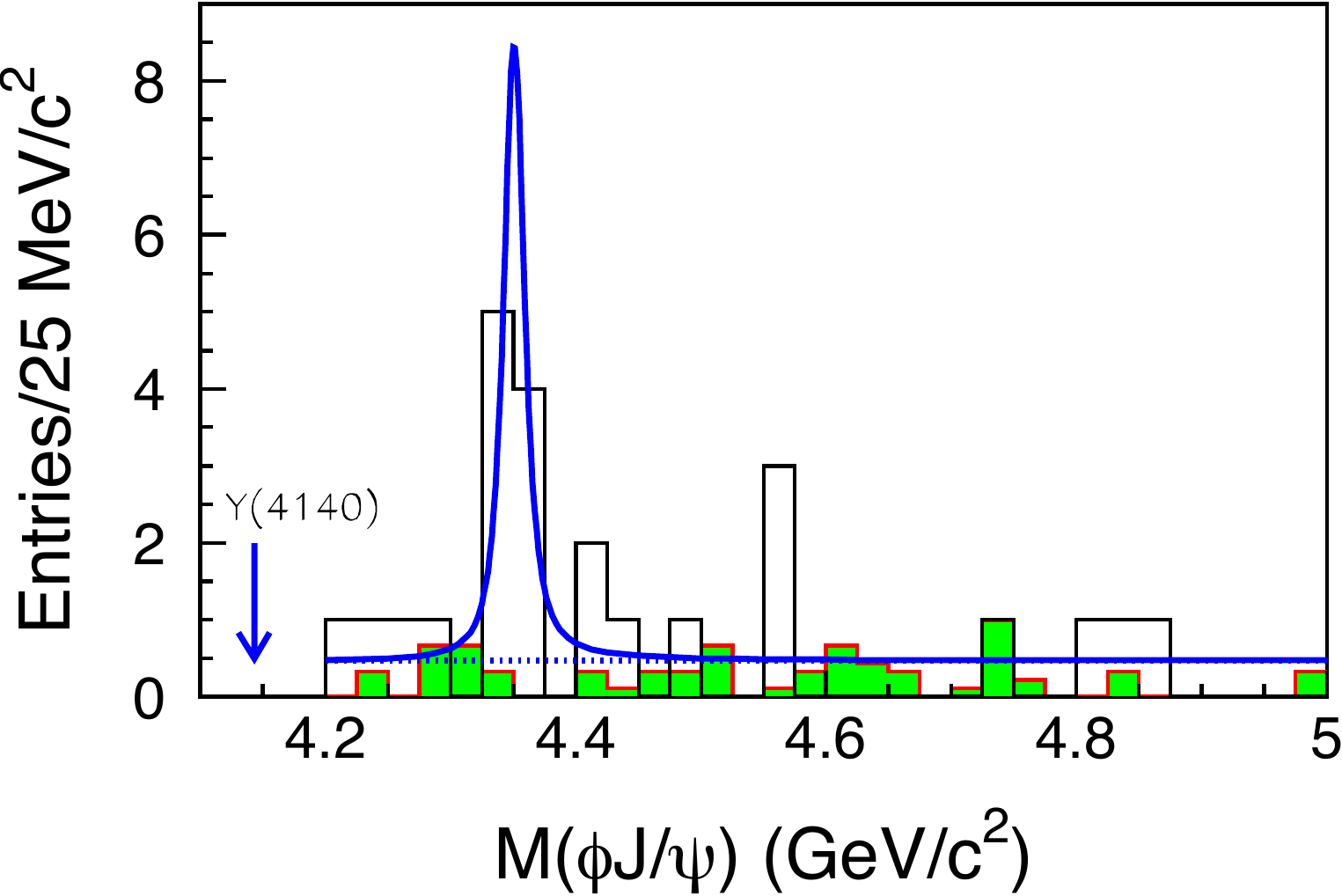}
  \caption{Invariant mass distributions of $\jpsi\phi$, from \belle\cite{Shen:2009vs}. No evidence for $Y(4140)$ is seen, whereas the peak of $X(4350)$ is fitted.}
  \label{fig:spectra4350}
\end{figure}
Several experiments have searched for the $Y(4140)$: \Dzero\cite{Abazov:2013xda} and \cms\cite{Chatrchyan:2013dma} have recently confirmed the observation, and reported mass and width $M=(4159.0\pm 4.3\pm6.6)\mev,\,\Gamma=(19.9\pm 12.6_{-8.0}^{+3.0})\mev$, and $M=(4148.0 \pm 2.4 \pm 6.3 \mev,\,\Gamma = (28^{+15}_{-11}\pm 19) \mev$, with significances of $\sim 3\sigma$ and $>5\sigma$, respectively. On the other hand, neither \lhcb\cite{Aaij:2012pz} nor \babar\cite{Lees:2014lra} are able to see any significant signal, and put 90\% C.L. upper limits on the relative branching fractions of
\begin{subequations}
\begin{align}
 \frac{\BR\left(B^+ \to Y(4140) K^+\right) \times \BR\left(Y(4140) \to \jpsi \phi\right)}{\BR\left(B^+ \to \jpsi \phi K^+\right)} &< 0.07\,\text{(\lhcb)},\\
 &< 0.135\,\text{(\babar)},
\end{align}
\end{subequations}
to be compared with a $\sim 0.1$ measured by \cms. We mention also a preliminary null result of \bes in the $\jpsi\phi$ invariant mass in $e^+e^-\to \gamma Y(4140)$ process\cite{bes4140preliminary}.

The averaged values of mass and width {\it \`a la} PDG\cite{pdg} from the experiments that have claimed the observation are $M = (4145.6 \pm 3.6)\mev$ and $\Gamma = (14.3\pm5.9)\mev$.

Last state we review is the $X(3823)$ seen by \belle in $B\to (\chi_{c1}\gamma) K$ radiative decays, with mass and width of $M=(3823.1\pm1.8\pm 0.7)\mev$ and $\Gamma< 24\mev$ at 90\% C.L., with a significance of $4\sigma$\cite{Bhardwaj:2013rmw}. Nothing prevents the identification of this state as the $2^{--}$ ordinary charmonium.

\subsection{Summary}
\label{sec:expsummary}
We show in tabular form the state-of-the-art of experimental searches of exotic states, organized by production mechanism: 
\B decay (\tablename{~\ref{tab:obsmatrB}}),
$e^+e^-$ direct production (with or without ISR, \tablename{~\ref{tab:obsmatrISR}}),
double charmonium production (\tablename{~\ref{tab:obsmatrRec}}),
and two-photon production (\tablename{~\ref{tab:obsmatrGG}});
studies of charged states are summarized in \tablename{~\ref{tab:obsmatrCh}}.
It appears that our knowledge is quite fragmentary: all exotic resonances but the $X(3872)$ and the $Y(3915)$ have been observed in one production mechanism only, and many in one final state only.
Analyses of particular combinations of production mechanism and final state
are missing, or not performed in the relevant range of invariant mass, 
or there is no fit to the data to test for the presence of 
already discovered exotic states.
To conclude, a systematic study of the exotic spectrum is needed
to form a definite experimental picture of these states and their structure.
New results from the LHC and from high-luminosity $B$- and $\tau$-$c$ factories will be important to settle our understanding of this sector.

\renewcommand{\psiprime}{\ensuremath{\psi^\prime}}

\begin{table}[b]
\tbl{ Status of searches for the new charged states, for several final states $f$,
updated with respect to Drenska \etal\cite{Drenska:2010kg}. The meaning of the symbols is explained in the caption of \tablename{~\ref{tab:obsmatrB}}. For these states we expect $G$ parity to be conserved in the decay.}
{
\begin{tabular}{ l l *{9}{c}}  \hline\hline
\multicolumn{11}{l}{Charged states}\\ \hline
State & $J^{PG}$ & $\psi\pi$ & $\psi\pi\pi^0$ & $\psiprime\pi$ & $\psiprime\pi\pi^0$ & $\chi_{c1}\pi$ & $h_c\pi$ & $\D\Dbar$ & $\D\Dstarb$ & $\Dstar\Dstarb$\\ \hline
$X(3872)^+$ & $1^{+-}$ & \Xforbid & \Xnotseen & \Xforbid & \Xforbid & \Xnofit & \Xforbid & \Xnotdone & \Xnofit & \Xforbid\\
$Z(3900)^+$ & $1^{++}$ & \Xseen & \Xforbid & \Xnofit & \Xforbid & \Xforbid & \Xnotseen & \Xnotdone & \Xseen & \Xforbid\\
$Z(3930)^+$ & $2^{+-}$ & \Xnofit & \Xnotdone & \Xnofit & \Xnotdone & \Xnofit & \Xnofit & \Xnotdone & \Xnotdone & \Xforbid\\
$Z(4020)^+$ & $1^{++}$ & \Xnotseen & \Xnotdone & \Xnofit & \Xforbid & \Xforbid & \Xseen & \Xnotdone & \Xnotdone & \Xseen\\
$Z(4050)^+$ & $J^{PG}$ & \Xnofit & \Xnotdone & \Xnofit & \Xnotdone & \Xseen & \Xnofit & \Xnotdone & \Xnotdone & \Xnofit\\
$Y(4140)^+$ & $J^{PG}$ & \Xnofit & \Xnotdone & \Xnofit & \Xnotdone & \Xnofit & \Xnotdone & \Xnotdone & \Xnotdone & \Xnofit\\
$Z(4200)^+$ & $1^{++}$ & \Xseen & \Xforbid & \Xnofit & \Xforbid & \Xforbid & \Xnotdone & \Xnotdone & \Xseen & \Xforbid\\
$Z(4250)^+$ & $J^{PG}$ & \Xnofit & \Xnotdone & \Xnofit & \Xnotdone & \Xseen & \Xnotdone & \Xnotdone & \Xnotdone & \Xnofit\\
$X(4350)^+$ & $J^{P-}$ & \Xnofit & \Xnotdone & \Xnofit & \Xnotdone & \Xnofit & \Xforbid & \Xnotdone & \Xnotdone & \Xnofit\\
$Z(4430)^+$ & $1^{++}$ & \Xseen & \Xforbid & \Xseen & \Xforbid & \Xforbid & \Xnotdone & \Xnotdone & \Xnotdone & \Xnofit\\
 \hline\hline 
\end{tabular}
\label{tab:obsmatrCh}
}
\end{table}

\begin{landscape}
\begin{table}[!htb]
\tbl{ Status of searches for the new states in $B$ decays, for several final states $f$,
updated with respect to Drenska \etal\cite{Drenska:2010kg}. 
Final states where each exotic state was observed (\Xseen: ``seen'') or excluded (\Xnotseen: ``not seen'') are indicated; \Xnotseennow is reserved to final states which have been searched and not seen, but are  forbidden by quantum numbers not known at the time of the analysis. A final state is marked as \Xnotdone (``not performed'') if the analysis has not been performed in a given mass range and with \Xnofit (``missing fit'') if the spectra are published but a fit to a given state has not been performed. Finally ``\Xforbid'' indicates that the known quantum numbers or available energy forbid the decay; and ``\Xtoohard'' that an analysis is experimentally too challenging. As explained in \sectionname{~\ref{sec:X3940}}, we consider a state $Y(3915)$ decaying into $\jpsi\omega$, seen both in $B$ decays and in $\gamma\gamma$ fusion, and a state $X(3940)$ seen in double charmonium production and decaying into $D\Dstarb$. ``Vectors'' indicates the $1^{--}$ states discovered via ISR not explicitly mentioned in the table.
}{
\begin{footnotesize}

\begin{tabular}{ l l*{16}{c} }  \hline\hline
\multicolumn{18}{l}{$B \to \mathcal{X} K,\,\mathcal{X}\to f$}\\ \hline
State & $J^{PC}$ & $\psi\pi\pi$ & $\psi\omega$ & $\psi\gamma$ & $\psi\phi$ & $\psi\eta$ & $\psiprime\pi\pi$ & $\psiprime\omega$ & $\psiprime\gamma$ & $\chi_c\gamma$ & $p\bar p$ & $\Lambda_c\Lbar_c$ & $\D\Dbar$ & $\D\Dstarb$ & $\Dstar\Dstarb$ & ${\D}_s^{(*)}{\Db}_s^{(*)}$ & $\gamma\gamma$\\ \hline
$X(3872)$ & $1^{++}$ & \Xseen & \Xseen & \Xseen & \Xforbid & \Xnotseennow & \Xforbid & \Xforbid & \Xseen & \Xnotseennow & \Xnotseen & \Xforbid & \Xforbid & \Xseen & \Xforbid & \Xforbid & \Xnotseennow\\
$Y(3915)$ & $0^{++}$ & \Xnofit & \Xseen & \Xnotseen & \Xforbid & \Xforbid & \Xforbid & \Xforbid & \Xnofit & \Xforbid & \Xnofit & \Xforbid & \Xnofit & \Xnotseen & \Xforbid & \Xnotdone & \Xnotdone\\
$Z(3930)$ & $2^{++}$ & \Xnofit & \Xnofit & \Xnotseen & \Xforbid & \Xforbid & \Xforbid & \Xforbid & \Xnofit & \Xforbid & \Xnofit & \Xforbid & \Xnofit & \Xnofit & \Xforbid & \Xnotdone & \Xnotdone\\
$Y(4140)$ & $J^{P+}$ & \Xnofit & \Xnofit & \Xnotdone & \Xseen & \Xforbid & \Xnotdone & \Xforbid & \Xnotdone & \Xforbid & \Xnofit & \Xforbid & \Xnofit & \Xnotdone & \Xnotdone & \Xnotdone & \Xnotdone\\
$X(4160)$ & $0^{P+}$ & \Xnofit & \Xnofit & \Xnotdone & \Xnofit & \Xforbid & \Xnotdone & \Xforbid & \Xnotdone & \Xforbid & \Xnofit & \Xforbid & \Xnofit & \Xnotdone & \Xnotdone & \Xnotdone & \Xnotdone\\
$X(4350)$ & $J^{P+}$ & \Xnofit & \Xnofit & \Xnotdone & \Xnofit & \Xforbid & \Xnotdone & \Xnotdone & \Xnotdone & \Xforbid & \Xnofit & \Xforbid & \Xnotdone & \Xnotdone & \Xnotdone & \Xnotdone & \Xnotdone\\
$Y(4260)$ & $1^{--}$ & \Xnotseen & \Xforbid & \Xforbid & \Xforbid & \Xnofit & \Xnotdone & \Xforbid & \Xforbid & \Xnotdone & \Xnofit & \Xforbid & \Xnotdone & \Xnotdone & \Xnotdone & \Xnotdone & \Xforbid\\
vectors & $1^{--}$ & \Xnofit & \Xforbid & \Xforbid & \Xforbid & \Xnofit & \Xnotdone & \Xforbid & \Xforbid & \Xnotdone & \Xnofit & \Xforbid & \Xnotdone & \Xnotdone & \Xnotdone & \Xnotdone & \Xforbid\\
$Y(4660)$ & $1^{--}$ & \Xnotdone & \Xforbid & \Xforbid & \Xforbid & \Xnofit & \Xnotdone & \Xforbid & \Xforbid & \Xnotdone & \Xnofit & \Xnofit & \Xnotdone & \Xnotdone & \Xnotdone & \Xnotdone & \Xforbid\\
\hline\hline
\end{tabular}
\end{footnotesize}
\label{tab:obsmatrB}
}\vspace{1cm}
\tbl{Status of searches for the new states  in ISR produtcion for several final states $f$, updated with respect to Drenska \etal\cite{Drenska:2010kg}. In this table we consider the $Y(4630)$ decaying into $\Lambda_c\Lbar_c$ and the $Y(4660)$ decaying into $\psiprime\pi\pi$ to be the same state. 
The meaning of the symbols is explained in the caption of \tablename{~\ref{tab:obsmatrB}}.}{
\footnotesize
\begin{tabular}{ l l *{13}{c} }  \hline\hline
\multicolumn{15}{l}{$e^+e^-\to \gamma_{ISR} \mathcal{X}$, $\mathcal{X}\to f$}\\ \hline
State & $J^{PC}$ & $\psi\pi\pi$ & $\psiprime\pi\pi$ & $h_c\pi\pi$ & $\psi\eta$ & $\chi_c\gamma$ & $\chi_c\omega$ & $p \bar p$ & $\Lambda\Lbar$ & $\Lambda_c\Lbar_c$ & $\D\Dbar$ & $\D\Dstarb$ & $\Dstar\Dstarb$ & ${\D}_s^{(*)}{\Db}_s^{(*)}$\\ \hline
$Y(4008)$ & $1^{--}$ & \Xseen & \Xnofit & \Xnofit & \Xnofit & \Xnofit & \Xforbid & \Xnofit & \Xnofit & \Xforbid & \Xnofit & \Xnofit & \Xnofit & \Xnofit\\
$Y(4220)$ & $1^{--}$ & \Xnofit & \Xnofit & \Xseen & \Xnofit & \Xnofit & \Xseen & \Xnofit & \Xnofit & \Xforbid & \Xnofit & \Xnofit & \Xnofit & \Xnofit\\
$Y(4260)$ & $1^{--}$ & \Xseen & \Xnotseen & \Xnotseen & \Xnotseen & \Xnotseen & \Xnotseen & \Xnotseen & \Xnofit & \Xforbid & \Xnotseen & \Xnotseen & \Xnotseen & \Xnotseen\\
$Y(4290)$ & $1^{--}$ & \Xnofit & \Xnotseen & \Xseen & \Xnotseen & \Xnotseen & \Xnotseen & \Xnofit & \Xnofit & \Xforbid & \Xnofit & \Xnofit & \Xnofit & \Xnofit\\
$Y(4360)$ & $1^{--}$ & \Xnotseen & \Xseen & \Xnofit & \Xnofit & \Xnofit & \Xnofit & \Xnofit & \Xnofit & \Xforbid & \Xnofit & \Xnofit & \Xnofit & \Xnofit\\
$Y(4660)$ & $1^{--}$ & \Xnotseen & \Xseen & \Xnotdone & \Xnofit & \Xnofit & \Xnofit & \Xnofit & \Xnofit & \Xseen & \Xnofit & \Xnofit & \Xnofit & \Xnofit\\
\hline\hline
\end{tabular}
\label{tab:obsmatrISR}}
\end{table}

\begin{table}[!htb]
\tbl{Status of searches for the new states in double charmonium production events, for several final states $f$,
updated with respect to Drenska \etal\cite{Drenska:2010kg}. We tentatively assign $P=-$ to $X(3940)$ because of the lack of $X(3940)\to D\Dbar$ decay mode. The meaning of the symbols is explained in the caption of \tablename{~\ref{tab:obsmatrB}}.
}{
\footnotesize
\begin{tabular}{ l l *{14}{c} }  \hline\hline
\multicolumn{16}{l}{ $e^+e^-\to \mathcal{X} \jpsi$, $\mathcal{X}\to f$}\\ \hline
State & $J^{PC}$ & $\psi\pi\pi$ & $\psi\omega$ & $\psi\gamma$ & $\psi\phi$ & $\psiprime\pi\pi$ & $\psiprime\omega$ & $\psiprime\gamma$ & $\chi_c\gamma$ & $p \bar p$ & $\Lambda\Lbar$ & $\Lambda_c\Lbar_c$ & $\D\Dbar$ & $\D\Dstarb$ & $\Dstar\Dstarb$\\ \hline
$X(3872)$ & $1^{++}$ & \Xtoohard & \Xnotdone & \Xtoohard & \Xforbid & \Xtoohard & \Xforbid & \Xtoohard & \Xtoohard & \Xtoohard & \Xtoohard & \Xforbid & \Xnofit & \Xnofit & \Xforbid\\
$X(3940)$ & $0^{-+}$ & \Xtoohard & \Xnotdone & \Xtoohard & \Xforbid & \Xtoohard & \Xforbid & \Xtoohard & \Xtoohard & \Xtoohard & \Xtoohard & \Xforbid & \Xnotseennow & \Xseen & \Xforbid\\
$Z(3930)$ & $2^{++}$ & \Xtoohard & \Xnotdone & \Xtoohard & \Xforbid & \Xtoohard & \Xforbid & \Xtoohard & \Xtoohard & \Xtoohard & \Xtoohard & \Xforbid & \Xnofit & \Xnofit & \Xforbid\\
$Y(4140)$ & $J^{P+}$ & \Xtoohard & \Xnotdone & \Xtoohard & \Xnotdone & \Xtoohard & \Xforbid & \Xtoohard & \Xtoohard & \Xtoohard & \Xtoohard & \Xforbid & \Xnofit & \Xnofit & \Xnofit\\
$X(4160)$ & $0^{P+}$ & \Xtoohard & \Xnotdone & \Xtoohard & \Xnotdone & \Xtoohard & \Xforbid & \Xtoohard & \Xtoohard & \Xtoohard & \Xtoohard & \Xforbid & \Xnofit & \Xseen & \Xnofit\\
$X(4350)$ & $J^{P+}$ & \Xtoohard & \Xnotdone & \Xtoohard & \Xnotdone & \Xtoohard & \Xnotdone & \Xtoohard & \Xtoohard & \Xtoohard & \Xtoohard & \Xtoohard & \Xnofit & \Xnofit & \Xnofit\\ \hline\hline
\end{tabular}
\label{tab:obsmatrRec}}
\vspace{1cm}
\tbl{Status of searches for the new states in $\gamma\gamma$ fusion, for several final states $f$,
updated with respect to Drenska \etal\cite{Drenska:2010kg}. The meaning of the symbols is explained in the caption of \tablename{~\ref{tab:obsmatrB}}.
}
{
\footnotesize
\begin{tabular}{ l l *{14}{c} }  \hline\hline
\multicolumn{16}{l}{ $e^+e^-\to e^+ e^- \gamma\gamma$, $\gamma\gamma\to\mathcal{X}$, $\mathcal{X}\to f$}\\ \hline
State & $J^{PC}$ & $\psi\pi\pi$ & $\psi\omega$ & $\psi\gamma$ & $\psi\phi$ & $\psiprime\pi\pi$ & $\psiprime\omega$ & $\psiprime\gamma$ & $p \bar p$ & $\Lambda\Lbar$ & $\Lambda_c\Lbar_c$ & $\D\Dbar$ & $\D\Dstarb$ & $\Dstar\Dstarb$ & ${\D}_s^{(*)}{\Db}_s^{(*)}$\\ \hline
$X(3872)$ & $1^{++}$ & \Xnotseennow & \Xforbid & \Xforbid & \Xforbid & \Xforbid & \Xforbid & \Xforbid & \Xforbid & \Xforbid & \Xforbid & \Xforbid & \Xforbid & \Xforbid & \Xforbid\\
$Y(3915)$ & $0^{++}$ & \Xnotdone & \Xseen & \Xtoohard & \Xforbid & \Xforbid & \Xforbid & \Xtoohard & \Xnofit & \Xnofit & \Xforbid & \Xnofit & \Xnotdone & \Xforbid & \Xnotdone\\
$Z(3930)$ & $2^{++}$ & \Xnotdone & \Xnofit & \Xtoohard & \Xforbid & \Xforbid & \Xforbid & \Xtoohard & \Xnofit & \Xnofit & \Xforbid & \Xseen & \Xnotdone & \Xforbid & \Xnotdone\\
$Y(4140)$ & $J^{P+}$ & \Xnotdone & \Xnofit & \Xtoohard & \Xnotseen & \Xnotdone & \Xforbid & \Xtoohard & \Xnotdone & \Xnotdone & \Xforbid & \Xnofit & \Xnotdone & \Xnotdone & \Xnotdone\\
$X(4160)$ & $0^{P+}$ & \Xnotdone & \Xnofit & \Xtoohard & \Xnotseen & \Xnotdone & \Xforbid & \Xtoohard & \Xnotdone & \Xnotdone & \Xforbid & \Xnofit & \Xnotdone & \Xnotdone & \Xnotdone\\
$X(4350)$ & $J^{P+}$ & \Xnotdone & \Xnotdone & \Xtoohard & \Xseen & \Xnotdone & \Xnotdone & \Xtoohard & \Xnotdone & \Xnotdone & \Xnotdone & \Xnotdone & \Xnotdone & \Xnotdone & \Xnotdone\\
	\hline\hline
\end{tabular}
\label{tab:obsmatrGG}
}

\end{table}
\end{landscape}

\renewcommand{\psiprime}{\ensuremath{\psiprime}}

\section{Lattice QCD status of exotics}
\label{sec:lattice}
\begin{figure}[b]
\centering
\includegraphics[width=.75\textwidth]{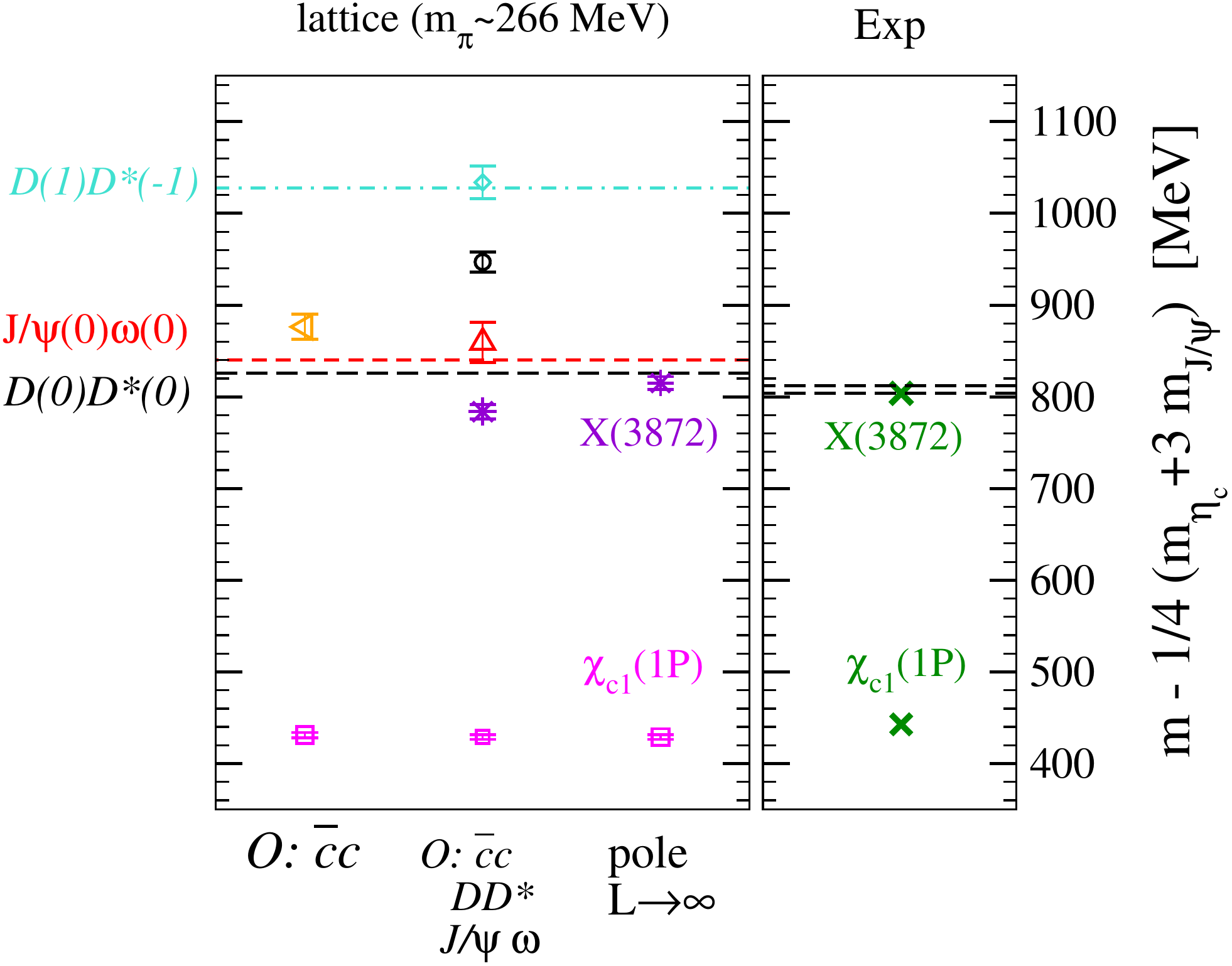}
\caption{Energy levels computed in lattice simulations in the $J^{PC}=1^{++},\,I=0$ channel (left panel), to compare with the experimental mass of the $\chi_{c1}(1P)$ and $X(3872)$ (right panel). On the $x$ axis  the operator basis used 
in simulations is sketched. From Prelovsek~\etal\cite{Prelovsek:2013xba}} \label{Xlattice}
\end{figure}
Lattice QCD has recently reached some preliminary results about exotics, albeit the non-trivial numerical and theoretical difficulties.
In fact, from a field theoretical point of view, there is no way to distinguish between a meson and a tetraquark 
with the same quantum numbers, as we discussed for Large-$N$ QCD (see \sectionname{~\ref{sec:largeN}}). For instance, the charged resonance $Z_c(3900)^+$, with quark content $c\bar c u \bar d$, has the same quantum numbers as the $a_1^+(980)$ (the lightest $I=1$ axial vector), so that any operator able to resolve the $Z_c$ interpolates also the excitations of $a_1$. In principle, the existence of the $Z_c$ can be revealed by extracting all the excited $a_1$ levels up to the mass of the $Z$, but this is not numerically feasible.
A numerically reliable approximation, widely used in heavy quarkonium spectroscopy, is to neglect charm annihilation diagrams\cite{Liu:2012ze}, which are expected to be small because of OZI suppression.
Under this approximation, it is possible to deal with these states using a field theory approach.
In current lattice simulations one considers the vacuum expectation value of two-point functions for
a set of interpolating operators with given quantum numbers. For each of them, the spectral representation gives
\begin{equation} 
C_{ij}(t)=\left\langle O_i^\dag (x ,t) O_j(0) \right\rangle=\sum_n \sqrt{Z_i^{n*} Z_j^n}e^{-E_n t}.
\label{2punti}
\end{equation}
From a single correlation function it is possible to extract only the lowest lying state using the effective mass method:
when the time $t$ is large, the function
\begin{equation}
m_\text{eff}=-\ln \frac{C_{ij}(t)}{C_{ij}(t-1)}
\end{equation}
has a plateau at the energy of the ground state.
The excited energy levels are extracted using the generalized eigenvalue problem\cite{Luscher:1990ck}.
If we have $N_{op}$ different operators with the same quantum numbers, we can compute the correlation function matrix $C_{ij}$, ($i ,j=1,\dots,N_{op}$). 
The solution of the eigenvalue problem
\begin{figure}[t]
\centering
\includegraphics[width=.65\textwidth]{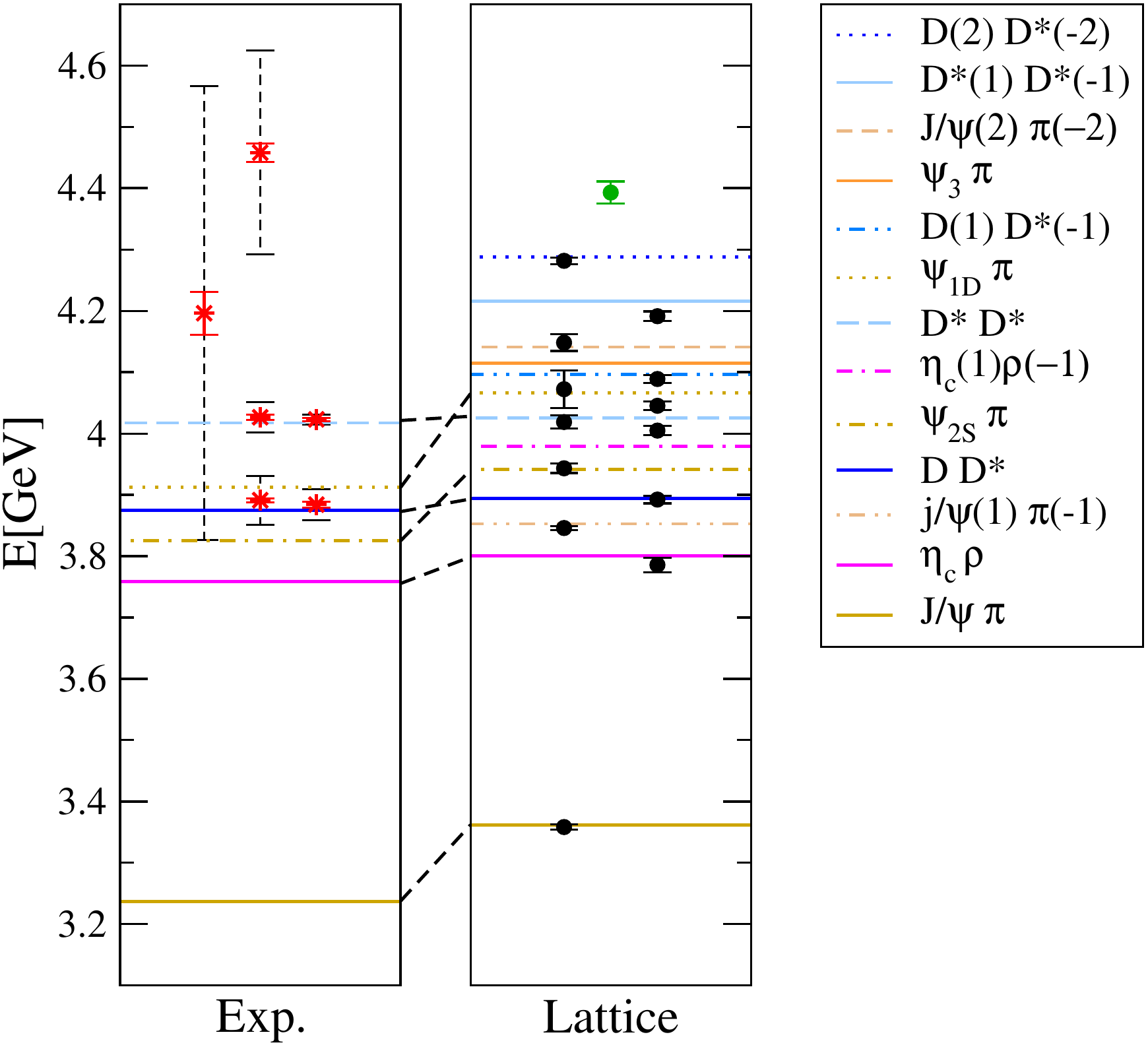}
\caption{Experimental spectrum of the charged exotic resonances (left panel), and energy levels of charged states computed in lattice simulations, in the $J^{PC}=1^{+-},\,I=1$ channel (right panel). From Prelovsek \etal\cite{Prelovsek:2014zga}} \label{Zclattice}
\end{figure}
\begin{equation}
C(t)\psi = \lambda(t,t_0)C(t_0) \psi,
\end{equation}
gives $N_{op}$ levels of the energy spectrum:
in fact, the resulting eigenvalues $\lambda_n$ decay exponentially with the $n^\text{th}$ energy level, up to exponentially suppressed deviations:
\begin{equation}
\lambda_{n}(t,t_0) \sim e^{-E_{n}(t-t_0)}.
\end{equation} 
The larger is the basis of operators, the larger is the number of computable excited levels.
For numerical reasons, the operators have to be also as different as possible.
If we were interested in below-threshold states, this is enough. 
If we instead are interested in above-threshold resonances, we have to look at all 2-particle levels with the same quantum numbers as the resonance.
While at infinite volume these levels form a continuum\footnote{In fact, no information about resonances can be deduced from Euclidean correlators in the thermodynamic limit\cite{Maiani:1990ca}.}, on the lattice these levels have a rather peculiar behavior as a function of the size of the volume. In particular, their energy is related to the infinite volume scattering phase\cite{Luscher:1990ux, Luscher:1991cf}.
Roughly speaking in fact, by varying the size of the lattice, we vary the relative momentum of the 2-particle states ($\propto \tfrac{2\pi}{L}$), hence we simulate a ``scattering'' experiment at different momenta.

Currently, the only positive result in charmonium lattice spectroscopy is the confirmation of an energy level compatible with the $X(3872)$
in the $J^{PC}=1^{++}$ channel with isospin $I=0$\cite{Prelovsek:2013xba} -- see \figurename{~\ref{Xlattice}}.
It is argued that the energy level found on the lattice is a real shallow bound state because of the large positive shift in 
energy of the state $D(0) D^*(0)$\cite{Sasaki:2006jn}. 
The signal of a level below the $D\Dstarb$ threshold seems to indicate the 
presence of the $X(3872)$ in QCD spectrum. It is worth noticing that this result is very sensitive to lattice artifacts, in particular the charm mass (and consequently the threshold) is affected by large discretization effects: for example this level could go away from threshold when approaching the physical point. Moreever, there is no way to distinguish such state 
from the ordinary $\chi_{c1}(2P)$: even if the level were confirmed, Lattice QCD cannot 
say whether it has the exotic features of the $X(3872)$.
\begin{figure}
\centering
\includegraphics[width=.90\textwidth]{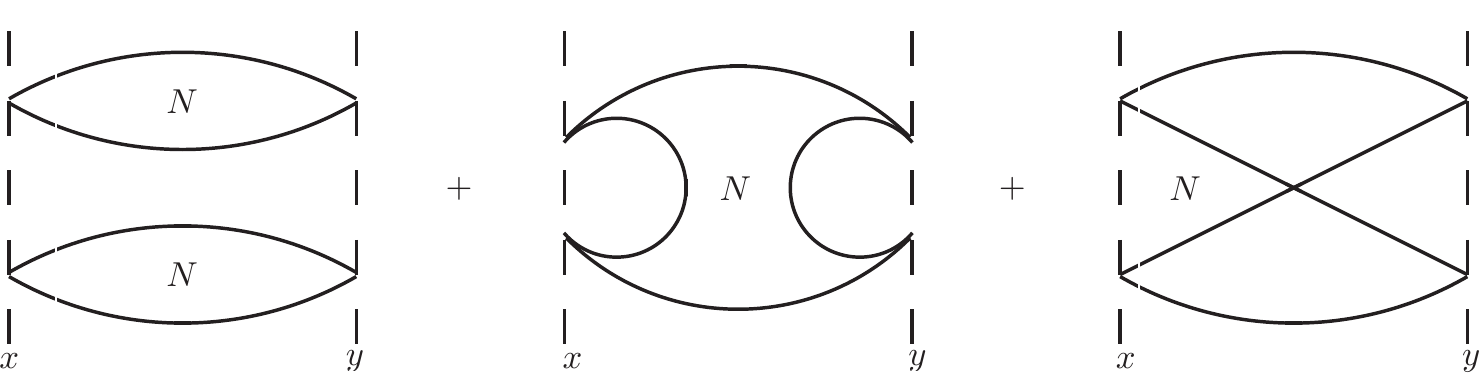}
\caption{Diagrams contributing to the 2-point correlation function $\left\langle cc\bar u\bar d(x) \, \bar c\bar c  u d(y)\right\rangle$. In the large-$N$ limit, only the second diagram could contain any tetraquark pole. However, this diagram cannot be included in lattice simulations, because the light meson content would obscure the information about heavier states. For details, see \sectionname{~\ref{sec:cohen}}. }\label{fig:tetracorr}
\end{figure}

For the $J^{PC}=1^{+-}$ sector with $I=1$ the situation is still unclear:
the analysis of the energy levels does not reveal any additional state, expected in presence of a resonance\cite{Prelovsek:2014zga,Chen:2014afa} (\figurename{~\ref{Zclattice}}).
However, this level could be obscured by the presence 
of many different two-particle mesonic channels.

Furthermore, the above-mentioned approximation of neglecting charm annihilation contributions, unavoidable in practice, 
could make tetraquark states hard to be found.
This statement is motivated by a Large-$N$ analysis. Consider the 2-point correlation function 
$\left\langle Z_c^\dag(x) Z_c(0) \right\rangle$.
The leading and subleading contributions to this correlation function are shown in \figurename{~\ref{fig:tetracorr}}.
In \sectionname{~\ref{sec:cohen}}, we showed that both the disconnected and the crossed diagrams (first and third in \figurename{~\ref{fig:tetracorr}}) receive contributions from two meson states only.
However, a tetraquark pole could appear in (subleading) diagrams which present any quark annihilations (second in \figurename{~\ref{fig:tetracorr}}). Such kind of diagrams are neglected in $I=1$ simulations (it is not numerically feasible to take into account charm annihilation), whereas in the $I=0$ channel the light quarks are able to annihilate. This suggests that the possible tetraquark pole could be out of reach of current $I=1$ simulations.

\section{Phenomenology}
\label{sec:pheno}
\subsection{Molecule}
Soon after the first observation of the $X(3872)$ in 2003, its closeness to the $D^0\bar D^{*0}$ threshold suggested to many authors that it might be the very first example of a loosely bound meson molecule. The possible existence of such states has been proposed many years ago by Tornqvist\cite{Tornqvist:1991ks, Ericson:1993wy} looking for hypotetical $K\bar K^*$, $\rho\rho$, $\rho\omega$, etc. bound states. It has been argued that the one-pion-exchange potential is likely to bind some states composed of ground state mesons, the idea being driven by the analogy with the time-honored case of deuterium where one knows that this potential is the dominant one. 
The same idea can also be extended to the heavy sector\cite{DeRujula:1976qd,Voloshin:1976ap,Tornqvist:1993ng,Manohar:1992nd}. In particular, it has been found\cite{Tornqvist:1993ng} that one-pion-exchange alone is strong enough to form at least \emph{deuteron-like} $B\bar B^*$ and $B^*\bar B^*$ states with binding energy of abount $50$ MeV. Composites made of $D\bar D^*$ and $D^*\bar D^*$ and bound by pion exchange alone -- \ie neglecting the contribution from other kinds of potential -- are expected near threshold, while molecular states composed of light mesons would require a stronger additional short range attraction and hence are likely not to be formed if only pions are taken into account. 

Using an effective Lagrangian for pions one can find the following potentials in momemtum space for the interaction between pseudo-scalar ($P$) and vector ($V$) mesons\cite{Tornqvist:1993ng}:
\begin{subequations}
\begin{align}
U_\pi^{(VV)}(\vett q)=&-U_\pi^{(V\bar V)}(\vett q)=\frac{g^2}{f_\pi^2}\left(\vett{\tau}_1\cdot\vett{\tau}_2\right)\left(\vett{\Sigma}_1\cdot\vett{q}\right)\left(\vett{\Sigma}_2\cdot\vett{q}\right)\frac{1}{\vett{q}^2+m_\pi^2}; \\
U_\pi^{(PV\to VP)}(\vett q)=&\frac{g^2}{f_\pi^2}\left(\vett{\tau}_1\cdot\vett{\tau}_2\right)\left(\vett{\epsilon}_1\cdot\vett{q}\right)\left(\vett{\epsilon}_2^*\cdot\vett{q}\right)\frac{1}{\vett{q}^2+m_\pi^2-{(m_V-m_P)}^2},
\end{align}
\end{subequations}
where $f_\pi\simeq132$ MeV is the pion decay constant and $g\simeq0.5\div0.7$ is some axial effective strong coupling. $\vett{\Sigma}$ are the spin-1 matrices, $\vett{\tau}$ are the Pauli isospin matrices and $\vett{\epsilon}$ is the polarization vector for the vector meson. It should be stressed that the $PP$ potential is forbidden by parity conservation. It is worth noticing that, in coordinate space, the potential is singular, and needs an ultraviolet cutoff $\Lambda =0.8 \div 1.2\gev$. The existence of loosely bound molecules can crucially depend on the choice of the cutoff\cite{Tornqvist:1991ks,Swanson:1992ec,Kalashnikova:2012qf}.

In \tablename{~\ref{tab:molecules}} we report the expected bound states according to this one-pion-exchange framework\cite{Tornqvist:2004qy}. As one can see the $X(3872)$ would perfectly fit into this picture. This motivated a great ammount of work done on the topic. In the following sections we will present some phenomenological models and their consequences, assuming these exotic states to be mesonic molecules.
\begin{table}[th]
\centering
\tbl{Bound states expected by the one-pion-exchange model\cite{Tornqvist:2004qy}. The masses are predicted to be near threshold for the case of $D$ mesons and about 50~\mev below threshold in the case of $B$ mesons. All states have isospin $I=0$.}
{\begin{tabular}{ccc|ccc}
\hline\hline
Bound state & $J^{PC}$ & Mass [MeV] & Bound state & $J^{PC}$ & Mass [MeV] \\
\hline
$D\bar D^*$ & $0^{-+}$ & $\simeq 3870$ & $B\bar B^*$ & $0^{-+}$ & $\simeq10545$ \\
$D\bar D^*$ & $1^{++}$ & $\simeq 3870$ & $B\bar B^*$ & $1^{++}$ & $\simeq10562$ \\
\hline
$D^*\bar D^*$ & $0^{++}$ & $\simeq 4015$ & $B^*\bar B^*$ & $0^{++}$ & $\simeq 10582$ \\
$D^*\bar D^*$ & $0^{-+}$ & $\simeq 4015$ & $B^*\bar B^*$ & $0^{-+}$ & $\simeq 10590$ \\
$D^*\bar D^*$ & $1^{+-}$ & $\simeq 4015$ & $B^*\bar B^*$ & $1^{+-}$ & $\simeq 10608$ \\
$D^*\bar D^*$ & $2^{++}$ & $\simeq 4015$ & $B^*\bar B^*$ & $2^{++}$ & $\simeq 10602$ \\
\hline \hline
\end{tabular}} \label{tab:molecules}
\end{table}

A somehow complementary approach was established by Barnes and Swanson\cite{Barnes:1991em,Swanson:2006st}: meson-meson interactions
can be obtained as the sum of effective potentials between the constituent quarks of the mesons. The hamiltonian is given by
\begin{subequations}
\begin{align}
H &= \frac{1}{2} \sum_{i\neq j} \left(U_{1g} + U_\text{conf} + U_\text{hyp}\right)_{ij} \\
&= \frac{1}{2} \sum_{i\neq j} \frac{\lambda_i}2 \frac{\lambda_j}2 \left(  \frac{\alpha_s}{r_{ij}} - \frac{3b}{4} r_{ij} - \frac{8\pi \alpha_s}{3 m_i m_j} {\bm S}_i \cdot {\bm S}_j \frac{\sigma^3}{\pi^{3/2}} e^{-\sigma^2 r_{ij}^2} \right)
\end{align}
\end{subequations}
where $U_{1g}$ is the one-gluon exchange potential at Born level, $U_\text{conf}$ is the (non-perturbative) linear potential which takes into account confinement, and $U_\text{hyp}$ parametrizes the hyperfine splitting of the charmonium levels.
Even if constituent quark models are commonly used in quarkonium physics, it is unclear whether they can describe strong interactions on the scale of loosely bound molecules ($\sim 10\fm$); it is more likely that quark can interact with each other on the typical scale of strong interactions, \ie $\sim 1\fm$, but if so, the distinction between 
hadronic molecules and tetraquarks would become just a matter of language, the only difference between the two being the way in which color is saturated. It is worth noticing that this interaction is not strong enough to bind the $X(3872)$, and a contribution from one-pion-exchange has to be added\cite{Swanson:2003tb}.

Finally, in the heavy sector one can use heavy quark spin symmetry to obtain predictions for molecular spectrum and decay patterns, regardless of the details of the binding potential~\cite{Bondar:2011ev,Guo:2009id,Guo:2013sya}.

\subsubsection{Low-energy universality and line shapes of the $X(3872)$} \label{sec:lowenergy}
As we mentioned before different potential models predict the presence of bound molecular states. Among these possible molecules the $X(3872)$, interpreted as an $S$-wave $D^0\bar D^{*0}$ state, would have a whole set of striking features due to the closeness to its constituents threshold. Its binding energy (simply given by the difference between its measured mass and the mass of its constituents) would be\cite{Tomaradze:2012iz} $E_X=(-0.142\pm0.220)$ MeV. The natural energy scale for a pionic interaction is given by $m_\pi^2/m_D\simeq10$ MeV and hence is much larger than $E_X$. 

Bound states with such a feature share some common properties -- the so-called \emph{low-energy universality}\footnote{ It should be mentioned that low-energy universality has been exploited for the first time by Voloshin\cite{Voloshin:2003nt} to compute the momentum distribution for the $X\to D^0\bar D^0\pi^0$ and $X\to D^0\bar D^0\gamma$ decays.}  -- coming from non-relativistic Quantum Mechanics and, in particular, many of their characteristic can be described via a single parameter: the scattering length, $a$. When the scattering length gets bigger and bigger (or analogously when the binding energy, $E$, gets smaller and smaller) we have that
\begin{align}
E\longrightarrow\frac{1}{2\mu a^2}.
\end{align}
For the case of the $X(3872)$ we have $\mu=966.6$ MeV and this leads to an unusually large scattering length, $a\simeq12$ fm $\gg 1/m_\pi\simeq1.5$ fm, the last one being the typical range of the interaction between the two $D$ mesons\footnote{ It has been shown\cite{Esposito:2013ada} that the scattering lenght obtained with this formalism can hardly be reconciled with the one obtained by the experimental data on the $X(3872)$ width, which appears to be smaller by (at least) a factor of $3\div4$.}. Such a striking feature necessarily requires some kind of fine tuning. Moreover, the wave function for the constituents assumes the universal form
\begin{align}
\psi_{DD^*}(r)\longrightarrow\frac{1}{\sqrt{2\pi a}}\frac{e^{-r/a}}{r}.
\end{align}
Note that this also implies that a loosely bound molecule is an extremely extended object, having a typical radius $r_0 \simeq a$.

It has been pointed out\cite{Braaten:2003he} that the most generic quantum mechanical state for the $X(3872)$ can be written as
\begin{subequations} \label{eq:Xket}
\begin{align}
\left|X\right\rangle=&\sqrt{Z_{DD^*}}\int \frac{d^3p}{{(2\pi)}^3}\tilde\psi(p)\frac{1}{\sqrt{2}}\left(\left|D^0(\vett p)\bar D^{*0}(-\vett p)\right\rangle+\left|\bar D^0(\vett p)D^{*0}(-\vett p)\right\rangle\right) \\
&+\sum_H\sqrt{Z_H}\left|H\right\rangle, 
\end{align}
\end{subequations}
where $\tilde\psi(p)$ is the wave function of the $D$ mesons in momentum space and $\left|X\right\rangle$ are other possible states (discrete or continuous) having the same quantum numbers $J^{PC}=1^{++}$, \eg} $\left|D^+(\vett p)D^{*-}(-\vett p)\right\rangle$ or $\left|\chi_{c1}(2P)\right\rangle$. The constants $Z_i$ are the probabilities for a certain configuration. 
Using an effective field theory approach, it can be shown\cite{Braaten:2003he} that such suppression factors go as $Z_H\sim 1/\nu_H a$, where $\nu_H$ is the energy gap between the state $H$ and the $D^0\bar D^{*0}$ threshold.

Two mechanisms to explain the large scattering length of the $X(3872)$ has been proposed\cite{Braaten:2003he}:
\begin{enumerate}
\item If all the other states $H$ have an energy gap $\nu_H>m_\pi^2/m_D$ then, for a fairly large $a$, $Z_H\simeq0$ and $Z_{DD^*}\simeq1$, \ie the $X(3872)$ would be purely a molecule. In this case the fine tuning necessary to explain the value of $a$ would be something related to the interaction between the two components only, \eg the depth or the width of the potential or the mass of the $D$ mesons. In particular, one can consider $m_u$ as a tuning parameter since it influences both the one-pion potential and the mass of the two mesons.
\item If one of the states $H$ has mass very close to the $D^0\bar D^{*0}$ threshold, the $\nu_H$ factor would compensate the suppression due to the scattering length and lead to an almost equal mixture of this state and of the molecule, $Z_{DD^*}\simeq 1-Z_H$. This mechanism is the analogous of the well-known Feshbach resonances which are used in atomic physics to control the scattering length\cite{book:17652}. It has been hypotesized that this state might be the (still undiscovered) charmonium $\chi_{c1}(2P)$. However, potential models predict the mass of this particle to be $\sim90$ MeV above the threshold and hence we would need a fortuitous shift of this by at least $\sim80$ MeV, in order to achieve $\nu_\chi<m_\pi^2/m_D$.
\end{enumerate}
Since the second mechanism requires a large amount of luck (the discovery of the $\chi_{c1}(2P)$ with a mass value quite smaller than the expected one) we would only consider the first one, hence assuming that all the states appearing in Eq.~\eqref{eq:Xket} can be neglected except for the molecular one.

This model is also able to explain the narrowness of the $X(3872)$. In fact, one finds that the following partial  widths are given by\cite{Braaten:2003he}
\begin{subequations}
\begin{align}
\Gamma\left(X\to D^0\bar D^{0}\pi^0\right)&=Z_{DD^*}C_\pi\Gamma\left(D^{*0}\to D^0\pi^0\right); \\
\Gamma\left(X\to D^0\bar D^0\gamma\right)&=Z_{DD^*}C_\gamma\Gamma\left(D^{*0}\to D^0\gamma\right),
\end{align}
\end{subequations}
where $C_\pi$ and $C_\gamma$ are coefficients taking into account the interference from the charge conjugate components. In particular, they both depend on the value of the binding energy of $X(3872)$ but they are of order one. While these final states receive a non-zero contribution from the decay of the $D^{*0}$ component, other channels like $\psiprime\gamma$, $\eta_c(2S)\gamma$, $\jpsi\rho$ and $\jpsi\omega$ must occur either thanks to a short distance interaction between the two components, which is suppressed by the large separation of the two $D$ mesons, or thanks to one of the charmonium states $\left|H\right\rangle$ appearing in Eq.~\eqref{eq:Xket}, which are suppressed by $1/a$. Therefore, this could explain the small width of the $X(3872)$, which would then be of order $\Gamma_X\sim\Gamma_{D^*}\simeq65$ keV.

In later works\cite{Hanhart:2007yq,Braaten:2007dw,Stapleton:2009ey,Artoisenet:2010va} the previous analysis has been extended considering the possibility for the $X(3872)$ to be an above threshold resonant state -- \ie allowing for negative scattering lenghts. It has been proposed\cite{Braaten:2007dw,Stapleton:2009ey,Artoisenet:2010va} that the discrimination between these two cases can be done using the \emph{line shapes} for different decay channel, meaning the shape of the invariant mass distributions of the final products.

It is known from non-relativistic Quantum Mechanics that the shape of a resonance near threshold is proportional to $\left|f(E)\right|^2$, $f(E)$ being the analytic continuation of the scattering amplitude as a function of the total energy of the particles in their center-of-mass system. The previously mentioned low-energy universality for $S$-wave states implies
\begin{align}
f(E)=\frac{1}{-\gamma+\sqrt{-2\mu(E+i\epsilon)}},
\end{align}
with $\gamma=1/a$ and $E$ the energy with respect to the threshold. If $\gamma>0$ the resonance shape, $\left|f(E)\right|^2$, has a peak below the $D^0\bar D^{*0}$ threshold, corresponding to a real bound state, while if $\gamma<0$ it has a pole right above it, corresponding to a virtual resonance. A more accurate analysis of the problem showed\cite{Braaten:2007dw} that, in order to include the effects of the non-zero width of the $D^{*0}$ and possible inelastic scatterings for the charmed mesons, the previous expression must be modified to
\begin{align}
f(E)=\frac{1}{-(\gamma_\text{re}+i\gamma_\text{im})+\sqrt{-2\mu(E+i\Gamma_{D^*}/2)}},
\end{align}
where we introduced the width of the $D^{*0}$ and an imaginary part for $\gamma$.

Using this approach one can study the invariant mass distribution for different decay channels and compare the experimental results with the theoretical ones under the hypotesis of a real bound state or a virtual resonance. This analysis has been performed\cite{Braaten:2007dw,Stapleton:2009ey} for the $\jpsi\pi^+\pi^-$ and $D^0\bar D^0\pi^0$ final states as reported by the \belle collaboration\cite{Choi:2003ue} and the resulting fit has favored a peak of the line shapes below the threshold, thus pointing to a possible real bound state.

It should be mentioned that a similar approach was also used to study the line shapes of the exotic $Z(4430)$ under the hypotesis of a $D_1\bar D^*$ bound state\cite{Braaten:2007xw}. However, the most recent measures of the mass of such particle\cite{Aaij:2014jqa} has casted some serious doubts on the validity of this analysis since the mass gap for the $Z(4430)$ is now shifted to a much higher value, $\nu_Z\simeq47$ MeV, which prevents from using the low-energy universality and put in jeopardy its interpretation in terms of a molecular state.

\subsubsection{Non-Relativistic Effective Field Theory} \label{sec:NREFT}
During the past years a fairly large amount of work\cite{Fleming:2007rp,Cleven:2011gp,Cleven:2013sq,Cleven:2014qka,Wang:2013cya,Guo:2013zbw,Esposito:2014hsa} has been done to develop and apply a Non-Relativistic Effective Field Theory (NREFT) for the study of exotic mesons in the molecular framework. The goal is to build a set of tools to describe the interaction between exotic, heavy and light mesons. The resulting theory combines the time-honored Heavy Meson Chiral Theory\cite{Casalbuoni:1996pg} adding terms describing the interaction of the exotic states with their constituents.

In the following we summarize the main aspects of such a formalism:
\begin{itemize}
\item The first key ingredient is that all the considered exotic mesons are intended as near-threshold molecular states and therefore the problem can be treated in a non-relativistic fashion. Since the velocities involved are small (see below) one can replace the HQET fields in the Lagrangian with their non-relativistic counterparts. Such limit is obtained by letting $v\to(1,\vett{0})$ in the usual HQET bi-spinors\cite{manohar2007heavy}. In particular, the non-relativistic Lagrangians involving the exotic mesons $X$, $Y$, $Z$ and $Z^\prime$ are
\begin{subequations} \label{eq:Leff}
\begin{align}
\mathcal{L}_X&=\frac{x}{\sqrt{2}}X^{i\dagger}\left(\bar PV^i+P\bar V^i\right)+h.c. \\
\mathcal{L}_Y&=\frac{y}{\sqrt{2}}Y^{i\dagger}\left(\bar PV^i-P\bar V^i\right)+h.c. \\
\mathcal{L}_{Z_f}&=\frac{z_f}{\sqrt{2}}Z^{i\dagger}\left(\bar PV^i-P\bar V^i\right)+h.c. \\
\mathcal{L}_{Z_f^\prime}&=iz_f^\prime \epsilon^{ijk}\left(Z^{\prime}\right)^{i\dagger}\bar V^j V^k+h.c.
\end{align}
\end{subequations}
The fields $X^i$, $Y^i$ and $Z_f^{(\prime)i}$ annihilate the exotic mesons states while $P$ $(\bar P)$ and $V^i$ $(\bar V^i)$ annihilate a (anti-)pseudoscalar and a (anti-)vector state according to $P|P(k)\rangle = \sqrt{m_P}|0\rangle$ and $V^i|V(k,\epsilon)\rangle=\epsilon^i\sqrt{m_V}|0\rangle$. Also $i$, $j$ and $k$ are spatial indices and $x$, $y$ and $z_f^{(\prime)}$ are some unkown effective couplings. Lastly, $f=c,b$ is a flavor index.

\item The previous Lagrangians are dictated by symmetry considerations only -- \ie by the quantum numbers of the particles involved -- and hence they describe the interaction of exotic mesons regardless of their internal structure. The essential information on the hypotetical molecular nature of these states comes from the requirement that the $X$, $Y$ and $Z$ states \emph{only couple to their constituents}.
This automatically implies that every hadronic transition must occur via heavy meson loops like the ones shown in \figurename{~\ref{fig:loops}}.

\begin{figure}[t]
\centering
\includegraphics[scale=0.43]{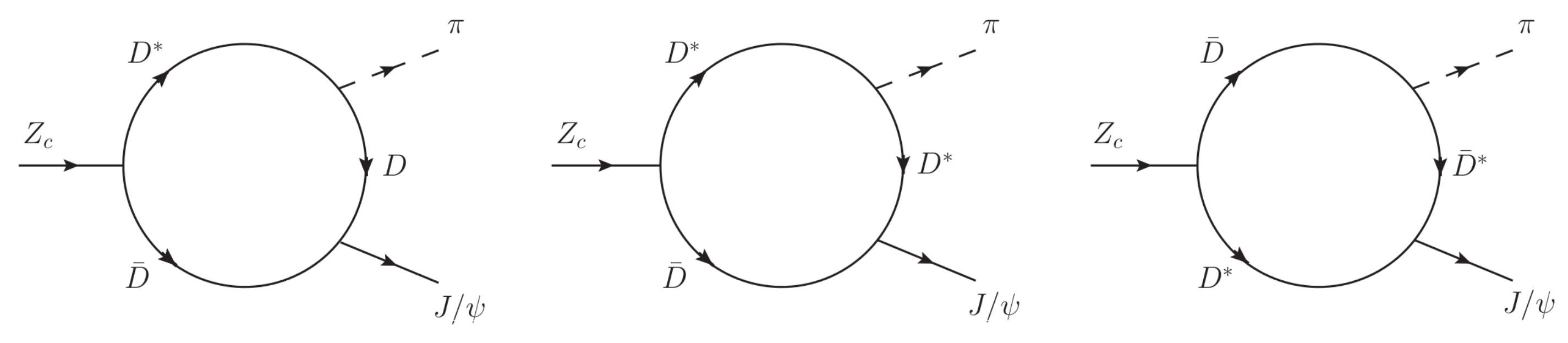}
\caption{Example of heavy meson loops used the NREFT formalism. In the picture the decay $Z_c(3900)\to \jpsi\pi$ can only happen via an intermediate $DD^*$ pair because of the molecular nature of the $Z_c$ itself.} \label{fig:loops}
\end{figure}

Since the problem is non-relativistic the propagators appearing in such loops must be the non-relativstic ones, namely:
\begin{align}
\frac{i}{p^2-m^2+i\epsilon} \longrightarrow \frac{1}{2m}\frac{i}{p^0-\frac{\vett p^2}{2m}-m+i\epsilon}
\end{align}

\item The typical velocities involved in the decay/creation of a certain particle with mass $M$ are given, in this context, by $v\simeq \sqrt{|M-2m|/m}$, where $m$ is the mass of the open flavor mesons appearing in the loop. Since our states are close to the threshold such velocities turn out to be small, thus allowing a the use of a non-relativistic approach and of a power counting procedure to estimate the relevance of a certain Feynman diagram\cite{Cleven:2013sq}. In particular, every meson loop counts as $v^5/(4\pi)^2$ while the heavy meson propagators scale as $1/v^2$. Moreover, depending on the possible presence of derivatives in the interaction vertices, the diagram might also scale as a power of one of the external momenta, $q$, or as an additional power of $v$. 

Since the interaction Lagrangians are non-perturbative some diagrams might be too challenging to be calculated and therefore, using this power counting technique, one can estimate the relevance of that particular process and hence determine an uncertainty related to its omission.
\end{itemize}

This formalism has been quite powerful in computing the decay width of many hadronic\cite{Cleven:2011gp,Cleven:2013sq,Wang:2013cya,Esposito:2014hsa} and radiative\cite{Cleven:2013sq,Guo:2013zbw} processes involving exotic mesons, assuming their internal structure to be a bound state of open flavor mesons. In particular, some attempts have been made to estimate the effective couplings appearing in Eqs.~\eqref{eq:Leff}. The $x$ and $y$ constants have been extrapolated from the experimental value of the binding energies\cite{Guo:2013zbw}:
\begin{align}
\left|x\right|=\left(0.97^{+0.40}_{-0.97}\pm0.14\right) \text{ GeV}^{-1/2}; \hspace{2em} \left|y\right|=\left(3.28^{+0.25}_{-0.28}\pm1.39\right) \text{ GeV}^{-1/2},
\end{align}
while $z_f$ and $z_f^\prime$ have been computed from experimental widths both in the charm and bottom sectors\cite{Esposito:2014hsa,Cleven:2013sq}:
\begin{subequations}
\begin{align}
\left|z_c\right|=\left(1.28\pm0.13\right) \text{ GeV}^{-1/2}; &\hspace{2em} \left|z_c^\prime\right|=\left(0.67\pm0.21\right) \text{ GeV}^{-1/2}; \\
\left|z_b\right|=\left(0.79\pm0.05\right) \text{ GeV}^{-1/2}; &\hspace{2em} \left|z_b^\prime\right|=\left(0.62\pm0.07\right) \text{ GeV}^{-1/2}.
\end{align}
\end{subequations}
It is interesting to note that $\left|z_c/z_c^\prime\right|=1.91\pm0.60$ and $\left|z_b/z_b^\prime\right|=1.27\pm0.16$ which indicates a large degree of spin symmetry violation. This is expected for very-near-threshold states, since small mass variations can lead to large changes in binding energies and hence in the couplings.

For the predictions made by this model about branching fractions and decay widths see Sec.~\ref{sec:comparison}.

\vspace{1em}

Lastly, it should be mentioned that another, slightly different, NREFT has been developed in some papers\cite{Fleming:2007rp,Braaten:2003he}. The main difference between such approach and the one explained above lies in how the molecular hypotesis is implemented. In particular, instead of requiring the presence of interemediate meson loops, the $X(3872)$ interpolating operator has been chosen to be explicitely $X^i\sim D\bar D^{*i}+\bar D D^{*i}$. We could refer to this model as a Non-Relativistic Effective Field Theory Type II (NREFT-II).

\subsection{Hadro-quarkonium} \label{sec:hadroq}
Another interpretation has been proposed\cite{Dubynskiy:2008mq,Dubynskiy:2008di} for the $J^{PC}=1^{--}$ resonances (namely $Y(4260)$, $Y(4360)$ and $Y(4660)$) and for the manifestly exotic $Z(4430)$. These states have always been observed in final states with a specific excitation of the charmonium spectrum, either $\jpsi$ or $\psiprime$. In particular, for the $Y(4260)$ all the observed decays contain a $\jpsi$, while for the other exotic particles their decay products only contain a $\psiprime$.
This feature motivated a model that describes these systems as composed of a heavy charmonium ``core'' sorrounded by a ``cloud'' of light hadronic matter. Such a configuration is known as \emph{hadro-charmonium} and it is an extension of a model for the binding of a $\jpsi$ or $\psiprime$ around a nucleus\cite{Brodsky:1989jd}. 
Note that, the distinction between a molecular states and a compact tetraquark is determined by the clustering of the constituents. For the case of the hadro-quarkonium, instead, the distinction between the heavy and light degrees of freedom is due to their size (instead of their superposition region), the light excitation being more extended that the quarkonium core.
 
The interaction between the central heavy quarks and the sorrounding excitation is a QCD analogous of the van der Waals force and is supposed to be strong enough to allow a bound state but also weak enough to mostly maintain the nature of the charmonium, thus explaining the absence of other excitations in the final states.
Since the $c\bar c$ state is color neutral, such an interaction can be treated using a multipole expansion, in close analogy with the well-known electromagnetic case. The heavy quark pair, that from now on we will call generically as $\psi$, has a chromo-electric dipole moment proportional to the chromo-electric gluon field generated by the sorrounding light excitation and this dipole will interact with the field itself, thus producing an effective Hamiltonian
\begin{align} \label{eq:Hdipole}
H_\text{eff}=-\frac{1}{2}\alpha^{(\psi)}E_i^aE_i^a,
\end{align}
where $E_i^a$ is the chromo-electric field generated by the sorrounding light matter and $\alpha^{(\psi)}$ is the chromo-electric polarizability. Here and in the following we indicate with $\alpha^{(\psi)}$ a generic element of the polarizability; in general we will have different components, $\alpha^{(\psi_1\psi_2)}$, both diagonal and off-diagonal. Such a polarizability is still unknown from first principles. We can only extimate its off-diagonal values for the charmonium and bottomonium case from the $\psiprime\to \jpsi \pi\pi$ and $\Upsilon(2S)\to\Upsilon\pi\pi$ transitions\cite{Voloshin:2004un}, where one finds $\alpha^{(\jpsi\psi')}\simeq 2$ GeV$^{-3}$ and $\alpha^{(\Upsilon\Upsilon')}\simeq0.6$ GeV$^{-3}$. The diagonal terms are usually expected to be larger that the off-diagonal ones.

Using the well-known expession for the conformal QCD anomaly in terms of the chromo-electric and chromo-magnetic fields, $E_i^a$ and $B_i^a$
\begin{align}
\theta_{\mu}^\mu=-\frac{9}{32\pi^2}F_{\mu\nu}^aF^{a\mu\nu}=\frac{9}{16\pi^2}\left(E_i^aE_i^a-B_i^aB_i^a\right),
\end{align}
one can compute a lower bound\cite{Dubynskiy:2008mq} for the expectation value of the previous Hamiltonian~\eqref{eq:Hdipole} over a generic hadron $X$:
\begin{align}
\left\langle X\right|\frac{1}{2}E_i^aE_i^a\left|X\right\rangle \geq \frac{8\pi^2}{9}M_X.
\end{align}
In particular, we used the fact that $\left\langle X\right|\theta_\mu^\mu(\vett{q}=0)\left|X\right\rangle=M_X$ and that the expectation value of $B_i^aB_i^a$ must be non-negative.
This can also be used to determine a condition for the presence of a bound state due to the van der Waals interection. One finds that it must be
\begin{align} \label{eq:condition}
\alpha^{(\psi)}\frac{M_X\bar M}{R}\geq C,
\end{align}
with $M_X$ the mass of the light hadronic excitation, $\bar M=M_XM_\psi/(M_X+M_\psi)$ the reduced mass of the charmonium-light hadron system and $C$ a (model dependent) constant of order 1. From Eq.~\eqref{eq:condition} one immediately notices that bound states are favoured for higher values of $M_X$, \ie for higher light hadronic excitations, but also for higher values of $\alpha^{(\psi)}$, which is in general considered to be larger for higher quarkonium levels. This last point would explain why three out of four of the previously mentioned exotic resonances decay into $\psiprime$.

Using a square well ansatz for the interaction potential and a reference value $\alpha^{(\psi)}=2$ GeV$^{-3}$ one finds\cite{Dubynskiy:2008mq} that bound states might appear for $M_X\gtrsim 2$ GeV or for lower $M_X$ but higher excitations of the central core. For the case of the bottomonium, since $\alpha^{(\Upsilon\Upsilon')}$ is much smaller, one needs much higher hadronic resonances in order to allow a bound state, making an experimental analysis quite challenging. However, it is still expected for lower values of $M_X$ but higher excitations of the $b\bar b$ pair (in particular with a $\Upsilon(3S)$ core).

It is worth noting that, so far we assumed that the nature of the heavy quarkonium does not change because of the gluonic field. However, it turns out that the interaction in Eq.~\eqref{eq:Hdipole} might cause a transition $\psiprime\to \jpsi$ via the off-diagonal polarizability $\alpha^{(\jpsi \psi')}$ with a width of a few MeV. Therefore, the present model also predicts the $Y(4360)$, $Y(4660)$ and the $Z(4430)$ to decay into $\jpsi$ but with a much lower (even though still detectable) branching ratios.

Lastly, using a holographic QCD approach\cite{Dubynskiy:2008di} one can show that the decays of hadroquarkonium states into open flavor mesons are suppressed by a factor $e^{-\sqrt{M_Q/\Lambda_{QCD}}}$ in the large $M_Q$ limit. This could explain why such final states are not observed experimentally. Recently, such a model has been applied to the $Y(4260)$ and $Y(4360)$ system by Voloshin and Li\cite{Li:2013ssa}.

\subsection{Hybrids} \label{sec:hybrids}
Quark model describes mesons as a quark and an antiquark which saturates color with each other. However, the QCD Lagrangian contains also the gluons, as dynamical degrees of freedom mediating strong interactions. From the point of view of the quark model, one might treat gluons as static degrees of freedom as well, belonging to the adjoint representation of the color group: since the tensor product of any number of adjoint fields always contains a singlet (${\bm 8}_c \otimes {\bm 8}_c \otimes \cdots = {\bm 1}_c \oplus \cdots$), we can form hadrons made up of just gluons, the so-called \emph{glueballs}. Moreover, we can add $q \bar q$ pairs in the color octet which saturates the gluon color, generating what it is usualy called a \emph{hybrid meson}. The addition of a gluon allows such mesons to have quantum numbers forbidden by ordinary quark model, \eg $0^{+-}$, $1^{-+}$ and so on. In the following we present a set of models developed during the years to describe these peculiar states.

The existence of hybrid mesons in the light sector was suggested in 1976 by Jaffe and Johnson\cite{Jaffe:1975fd} in the context of the MIT bag model. Some calculation\cite{Barnes:1982zs,Barnes:1982tx} predict the lightest hybrid multitplet to have a mass $\sim1.5\gev$ (it is worth noticing the observation of a exotic $\pi_1(1400)$ with the exotic $J^{PC}=1^{-+}$  exactly at $M=1354\mev$). The exotic $J^{PC}$ quantum numbers are due to the boundary conditions in the bag. 

For the heavy quarks a spherical bag would be quite unrealistic, and thus an adiabatic bag model was introduced
by Hasenfratz \etal\cite{Hasenfratz:1980jv}. In this model the bag was allowed to deform in the presence of a fixed $Q\bar Q$ source. 
The resulting potential is used in a Schr\"odinger equation to compute the mass of the states, as in usual quarkonium spectroscopy. 
The lightest hybrid was found at $\sim$3.9~GeV for $c\bar c$ and at $\sim$10.5~GeV for $b\bar b$. 
Some recent results on adiabatic potentials in QCD string models can be found in the literature\cite{Kalashnikova:2002tg}.

In the framework of constituent quark models, we can analogously consider constituent gluons.
These models were pioneered by Horn and Mandula\cite{Horn:1977rq} and later developedì\cite{Tanimoto:1982eh,Tanimoto:1982wy,Iddir:1988jd,Ishida:1991mx,Ishida:1989xh}.
The gluon has a fixed orbital angular momentum relatively to the $q\bar q$ pair, usually called $l_g$, 
and the $q\bar q$ is in a defined orbital configuration $l_{q\bar q}$ and spin configuration $s_{q\bar q}$. 
The quantum numbers of such bound states are
$P=(-1)^{l_g +l_{q\bar q}}$ and $C=(-1)^{l_{q\bar q}+s_{q\bar q}+1}$. 
The lightest hybrid state within this model has $l_{g}=0$ and thus non-exotic quantum numbers such as 
$1^{--}$ are obtained using {\it P}-wave $q\bar q$ states with $s_{q\bar q}=1$, while exotic $1^{-+}$ states have $s_{q\bar q}=0$. 

The most effective pictorial representation of hybrid mesons can be achieved via the flux-tube model. Lattice QCD simulations show that two static quarks at large distances are 
confined by approximately cylindrical regions of color fields. More specifically, if a gauge is fixed, the magnitude of chromoelectric field has cylindrical symmetry. The flux tube models this feature by approximating the confining region between quarks with an oscillating string. If one assumes Nambu-Goto action, \ie the action to be proportional to the area spanned by the string in coordinate space, one gets an exact potential for large values of the separation, $r$, between the sources:
\begin{equation} \label{eq:potential}
 V_\Lambda\left(r\right) = \sqrt{\sigma^2 r^2 - \frac{\pi\sigma\left(12 n - 1\right)}{6}},
\end{equation}
where $\sigma$ is the usual string tension, and $n$ parametrizes the quantized excitation of the string. For $n=0$ we get a linear rising potential, which corresponds to ordinary quarkonium spectrum. Higher excitations of the string would correspond to excitations of the color field, and so can be associated to hybrids. 

The previous potential is obtained as a function of the distance $r$ between the sources. In the first studies with this model an adiabatic separation of the quark and gluon degrees of freedom was carried on. Such approximation is allowed because of the large difference between the time scales of the fast dynamical 
response of the flux tube degrees of freedom and of the slow motion of the heavy quarks. 
This allows to fix the $Q\bar Q$ separation at some value $r$ (now considered as a parameter) and compute the eigenenergy  of the system in some fixed configuration of the flux tube: $E_{\Lambda}(r)$, $\Lambda$ being the quantum numbers of the flux tube. This eigenenergy is then treated as an effective potential $E_\Lambda(r)=V_\Lambda(r)$ acting on the heavy quark pair. The ground state $\Lambda=0$ 
gives the ordinary meson spectrum. Hybrids are obtained for $\Lambda>0$ and can be studied using the excited potential $V_{\Lambda}(r)$. This is nothing but the QCD analogous of the time-honored Born-Oppenheimer (BO) approximation for hydrogen molecules. This approximation has been successfully used since the first estimates of the charmonium spectrum on the lattice in the infinite mass limit (static potentials).
The lightest hybrid state is the one in which the string has a single orbital excitation about the $Q\bar Q$ axis. 
In initial models the adiabatic potentials were determined in the approximation of small fluctuations relatively to the $Q\bar Q$ axis. 
 This approximation was later removed by Barnes, Close and Swanson\cite{Barnes:1995hc}.

\vspace{1em}
Some insight on the spectrum of hybrids might be obtained from Lattice QCD simulations, which are supposed to give the most reliable predictions for absolute masses. In the heavy quark sector, when the $Q\bar Q$ pair is kept fixed while the gluonic degrees of freedom are allowed 
to be excited, the lightest charmonium hybrid was predicted\cite{Perantonis:1990dy} to have a mass of 4.2~GeV for $c\bar c$ and 10.81~GeV for $b\bar b$. In general, in the charmonium family hybrids are predicted in the mass region around 4.3~GeV, while the bottom sector they are predicted in the region 10.7-11.0~GeV.

Unfortunately many problems have to be faced when dealing with hybrids on the lattice since, from a field theory point of view, hybrids with ordinary quantum numbers suffer the same problem than tetraquarks: they are undistinguishable from mesons. 
One possible solution to this difficulty is to look at the overlap (the prefactors $\sqrt{Z_i^* Z_j}$ in Eq.~\eqref{2punti}) of those hybrid states with suitable operators.
This has been recently done\cite{Liu:2012ze} in lattice simulations (see \figurename{~\ref{ermafroditi}}) where a hybrid candidate with $J^{PC}=1^{--}$ 
is found close to the mass of the $Y(4260)$ resonance.
Although this evidence, it is not possible to conclude that the observed state is a hybrid meson instead of, for instance, a tetraquark.
\begin{figure}
\centering
\includegraphics[width=.80\textwidth]{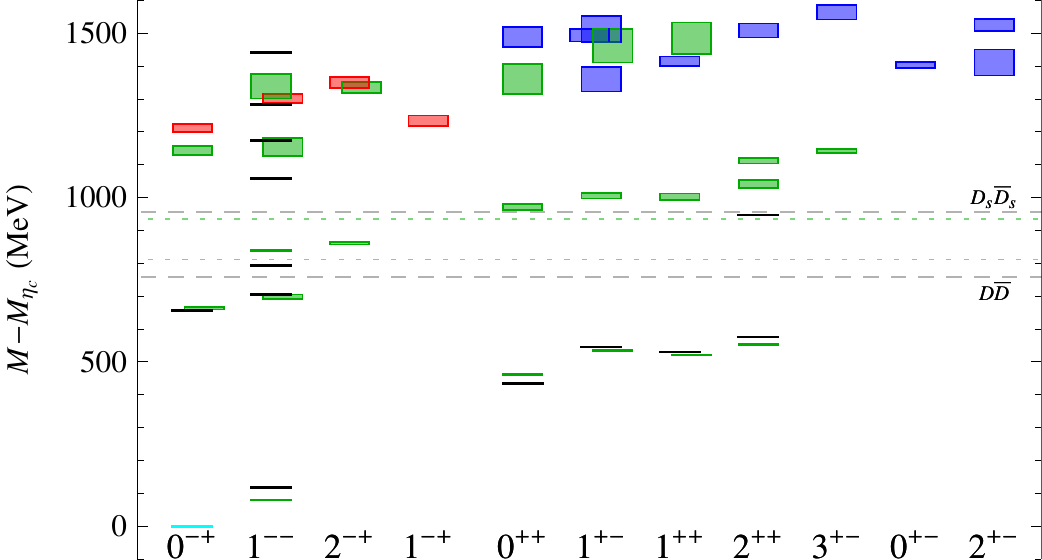}
\caption{Charmonium spectrum for masses around $4.5$ GeV. Red and blue boxes are identified as gluonic hybrids (ground and first excited states respectively). Green boxes are other charmonium states and black lines are experimentally observed levels. The $D\bar D$ and $D_s\bar D_s$ thresholds are also shown. From Liu \etal\cite{Liu:2012ze}} \label{ermafroditi}
\end{figure}
The observation of four hybrid candidates nearly degenerate with $J^{PC}=(0,1,2)^{-+}$ and $J^{PC}=1^{--}$ (see the red boxes in \figurename{~\ref{ermafroditi}}) is in agreement with the pattern predicted for the lightest states in the bag model\cite{Barnes:1982tx} and in the $P$-wave quasi-particle approach\cite{Guo:2008yz}. 
They appear at a mass scale $1.2-1.3$ GeV above the lightest conventional charmonia. 

\vspace{1em}

The picture of hybrids borrowed from Lattice QCD has been employed to try to explain some of the observed $XYZ$ resonances\cite{Braaten:2013boa,Braaten:2014qka}. In particular, it has been proposed that the $Y(4260)$ might indeed be an example of a hybrid composed of a $c\bar c$ pair with $J^{PC}=0^{-+}$ and a gluonic excitation with $J^{PC}=1^{+-}$. This interpretation would explain some of the striking properties of this resonance. In particular, the smallness of the $c\bar c$ wave function at the origin, $r=0$, would explain why the $Y(4260)$ is observed with a small production rate in $e^+e^-$ annihilation and why its decays into light hadrons are suppressed. Moreover, it is also known\cite{Kou:2005gt} that the decays of gluonic hybrids into a pair of $S$-wave mesons are suppressed and hence the dominant decay (if allowed) should be into an $S$-wave and a $P$-wave charmed mesons. However, for the $Y(4260)$ the decay into $D_1\bar D$ is phase space forbidden and the decay into $D^{*}\bar D$ is suppressed by a $D$-wave coupling. The only drawback of this interpretation was that the decays into charmonium plus light hadrons were also expected to be suppressed and this is in striking contrast with the observed large branching fraction for the $\jpsi\pi\pi$ channel. 

This problem found a solution with the discovery of the $Z_c(3900)$. It has, in fact, been hypotesized that this particle might be a different example of hybrid, a so-called \emph{tetraquark hybrid}. The main idea is that the excited gluon can be replaced with a $q\bar q$ pair of light quarks belonging to the adjoint, ${\bf 8}_c$, representation of the color group. In this context, the $Z_c$ would be made out of a $c\bar c$ pair with $J^{PC}=0^{-+}$ and a $q\bar q$ pair with $J^{PC}=1^{+-}$, this last assignement being motivated by the analogy with the gluonic hybrid, where the lowest energy excitation has these quantum numbers. If this idea were true, the $Y(4260)\to Z_c(3900)\pi$ decay would be explained as a transition of the gluon within the hybrid into a $q\bar q$ by pion emission, thus explaining the observed branching fraction.

A similar interpratation has also been given for the $Z_b$ and $Z_b^\prime$ states, even though in this case their closeness to the $\Bstarb B^{(*)}$ would also provide them with a strong molecular component.

As previously anticipated, the spectrum for gluonic and tetraquark hybrids can be computed under the BO approximation. To do that, one considers the $Q\bar Q$ pair to simply be a fixed source of color field, with a separation $r$ and solve for the eigenenergy of the gluonic (tetraquark) excitation. Once this is done this energy is taken as the effective potential suffered by the $Q\bar Q$ pair. Such a potential is given by Eq.~\eqref{eq:potential} for large value of $r$ and by a Coulomb-like expression for small $r$:
\begin{align}
V_\Lambda(r)=\frac{\alpha_s(1/r)}{6r}+E_\Lambda,
\end{align}
where $\alpha_s(1/r)$ is the strong coupling constant evaluated at a scale $\mu=1/r$ and $E_\Lambda$ is the so-called \emph{gluelump}, \ie an additive term that depends on the quantum numbers $(\Lambda)$ of the considered gluonic field (see for example Marsh and Lewis\cite{Marsh:2013xsa}). The parameters related to the previous potential can be fitted from lattice QCD results by Morningstar \etal\cite{Morningstar:1998xh}. Once this is done one can solve the Scr\"{o}dinger equation for the $Q\bar Q$ pair with this potential:
\begin{align}
\left[-\frac{1}{m_Q}\left(\frac{d}{dr}\right)^2+\frac{\left\langle \vett{L}^2_{Q\bar Q}\right\rangle_{\Lambda,r}}{m_Qr^2}+V_\Lambda(r)\right]rR(r) = E rR(r),
\end{align}
where $\left\langle \vett{L}^2_{Q\bar Q}\right\rangle_{\Lambda,r}$ is the orbital angular momentum of the $Q\bar Q$ pair computed for certain quantum numbers $\Lambda$ and for a separation $r$. $R(r)$ is the usual radial wave function. We will not go into the details of the this calculation since it is rather involved and does not add anything interesting to our discussion. In \figurename{~\ref{fig:ghybrid}} we report the spectrum for the excited gluonic hybrid obtained from this calculation in the charm and bottom sectors. For the charmonium case the lowest energy level is estimated to be $4246$ MeV, while for the bottomonium case it is $10559$ MeV.

In the tetraquark hybrid case we have no insight on the actual shape of the potential $V_\Lambda(r)$ generated by the two quark-antiquark in the adjoint representation. It has been proposed\cite{Braaten:2013boa,Braaten:2014qka} to assume a similar behavior as in the gluonic case. From this assumption and from a certain number of input values one can try again to derive a spectrum for this second kind of hybrids, and some generic selection rules. However, it is worth noticing that the hybrid potential computed on the lattice relies on quenched simulations (\ie without dynamical fermions), or on simulatios with unphysical light quarks masses (typically $m_\pi \sim 500$~\mev). The excited level corresponing to the hybrid state becomes more and more noisy, and the potential becomes more and more difficult to extract when approaching the physical point. In particular, at the physical pion mass the potential could be rather different from the present computations.

\begin{figure}[t]
\centering
\includegraphics[width=0.45\textwidth]{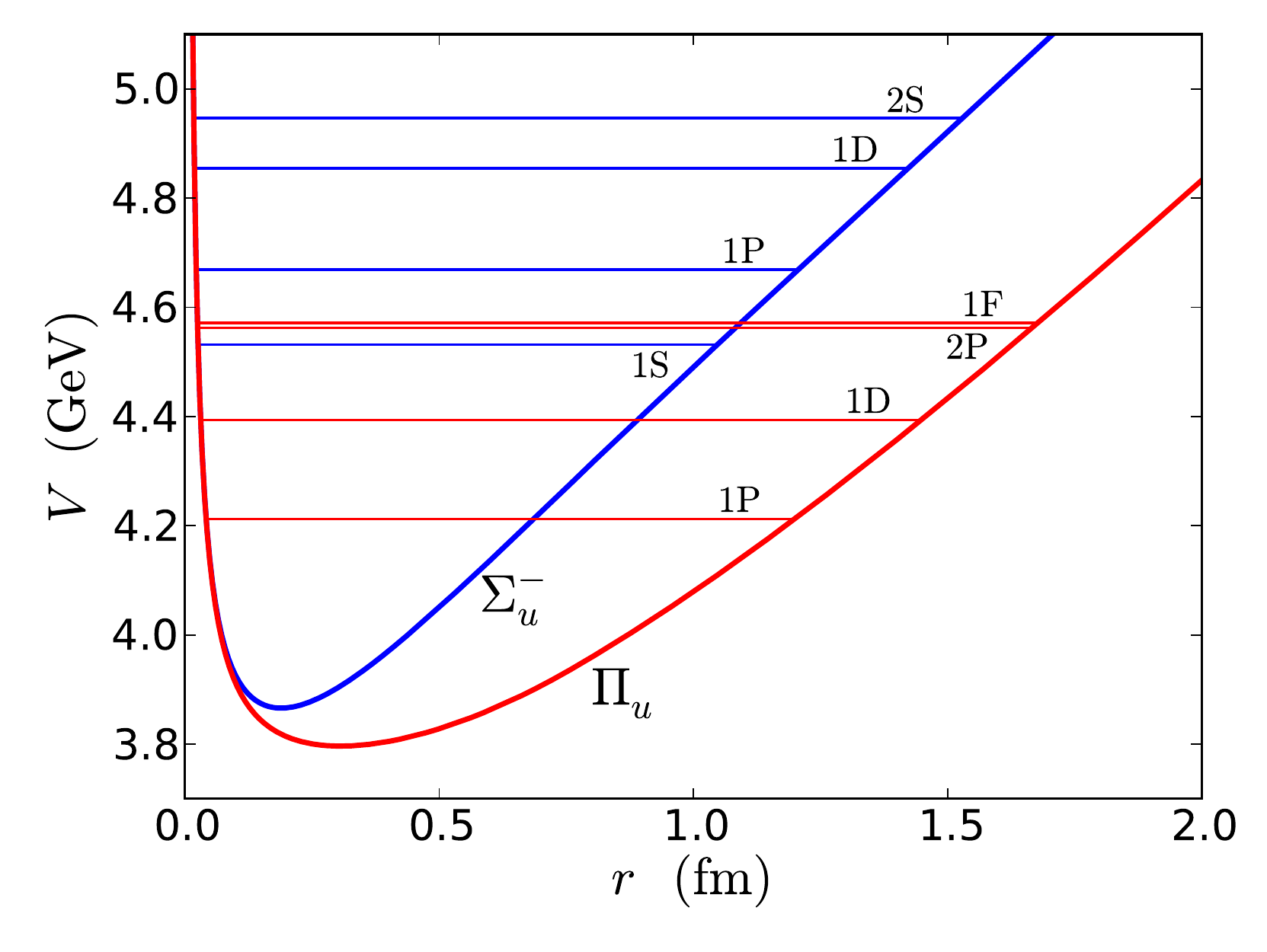}
\includegraphics[width=0.45\textwidth]{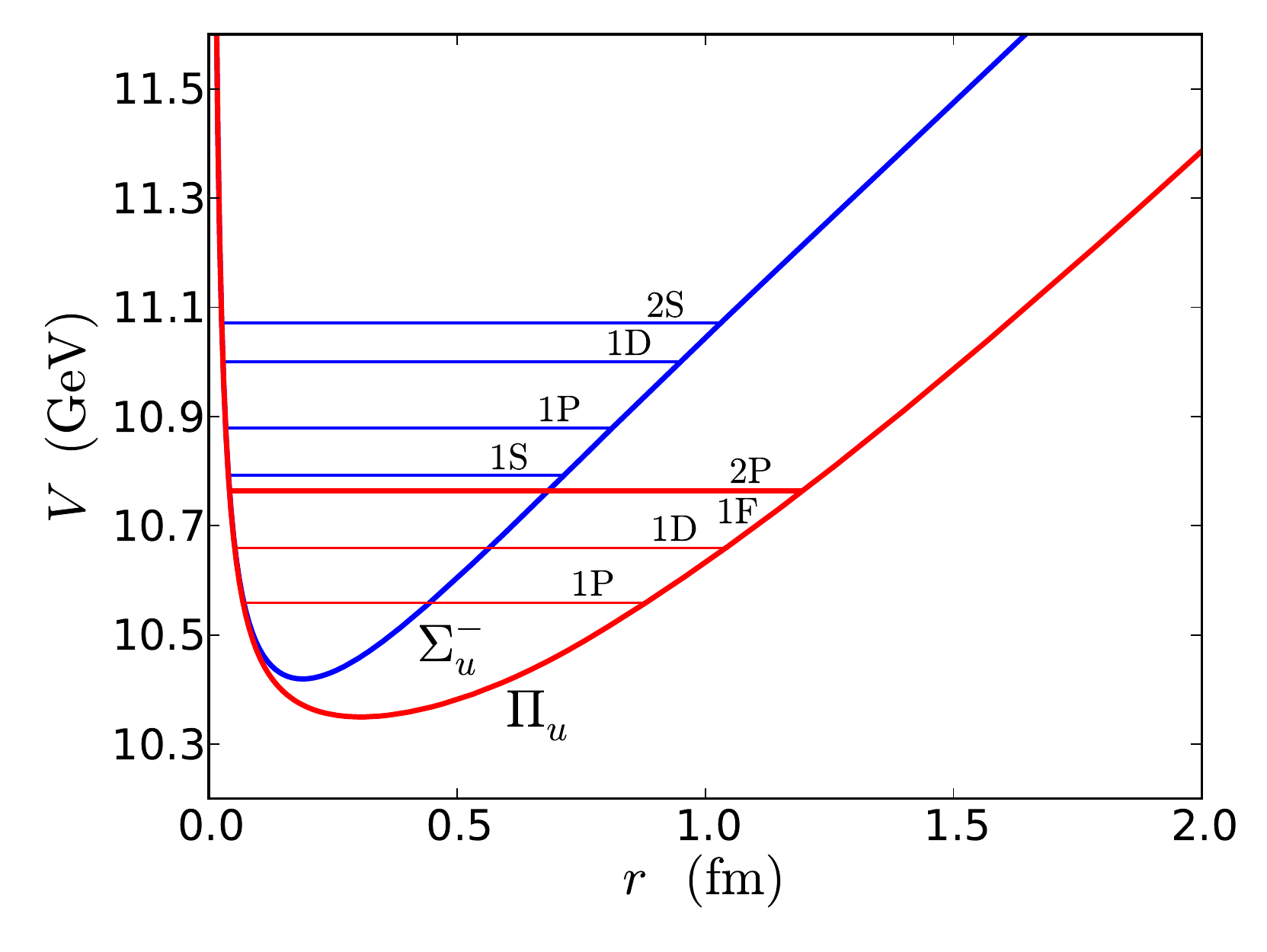}
\caption{Lowest energy levels for the gluonic hybrids in the charm (left panel) and bottom (right panel) sector. The notation for the quantum numbers of the gluonic degrees of freedom is borrowed from atomic physics. $\Pi_u$ has eigenvalue +1 for the operator $\left|\vett{r}\cdot\vett{J}_g\right|$, where $\vett{J}_g$ is the total angular momentum of the gluon excitation, and $(CP)_g=-1$ for the gluon with respect to the center of the $Q\bar Q$ system. $\Sigma_u^-$, instead, has eigenvalue zero for $\left|\vett{r}\cdot\vett{J}_g\right|$, $(CP)_g=-1$ and is also odd under reflection of the gluon field with respect to the plain containing the $Q\bar Q$ pair. The usual $nL$ notation for radial and angular quantum numbers has been used. From Braaten \etal\cite{Braaten:2014qka}} \label{fig:ghybrid}
\end{figure}

\vspace{1em}

While the masses of hybrid mesons are computable in all the models listed above, and in particular in Lattice QCD, the decay dynamics is more difficult to study. 

The only model which offers a description of the decay dynamics is again the flux-tube model. 
In fact, in this picture, the decay occurs when the flux-tube breaks at any point along its length, 
producing in the process a $q\bar q$ pair in a relative $J^{PC}=0^{++}$ state. Again this is just the well-know Lund model for ordinary mesons. The distance from the $Q\bar Q$ axis at which the light pair is created is controlled by the transverse distribution of the flux-tube. 
This distribution varies when going from the non-excited flux-tube to the first excited flux-tube configuration. 
Exploiting the empirical success of this model in describing the ordinary mesons decay dynamics, Close and Page\cite{Close:1994hc}
derived the decay pattern for hybrids. They found that in a two-meson decay the unit of orbital angular momentum of the incoming hybrid around the $Q\bar Q$ axis
is exactly absorbed by the component of the angular momentum of one of the two outgoing mesons along this axis. 
They treated explicitly the light flavor case\cite{Close:1994hc}, but a generalization to hybrid charmonia is straightforward. 
The final state should be in this case $D^{(*,**)}\bar{D}^{*,**}$, where $D^{**}$ indicates $D$-meson which are formed from {\it P}-wave $c\bar q$ ($q=u,d$) pairs. 
However, since the masses predicted in the flux-tube model are about $\sim 4.3 \gev$, \ie below the $DD^{**}$ threshold, it is possible that this decay is kinematically 
forbidden giving a rather narrow resonance decaying in charmonium and light hadrons. 
These modes offer a clear experimental signature and furthermore should have large branching fractions if the total width is sufficiently small. 

\subsection{Alternative explanations}
It should be mentioned that there are other interpretations about the nature of the $XYZ$ states. In particular, it is worth spending a few words about cusp effect. Some of these exotic states, in fact, lie slightly above their open flavor threshold. This suggested to some authors\cite{Bugg:2008wu,Swanson:2006st,Bugg:2004rk,Swanson:2014tra} that the experimental signals seen by the various collaborations might not be due to actual particles but to a dynamical effect. Cusps, in fact, can occur in amplitudes at threshold and these can manifest themselves as bumps in the cross sections right above the threshold. The proximity of many of these states to their open flavor threshold suggested to these authors that the cusp option might be taken seriously. Such possibility has been studied for the $X(3872)$\cite{Bugg:2008wu,Bugg:2004rk}, for the $Z(4430)$\cite{Bugg:2008wu,Pakhlov:2014qva} and most recently for the $Z_c^{(\prime)}$ and $Z_b^{(\prime)}$\cite{Swanson:2014tra}. This interpretation has been recently challenged by Hanhart \etal\cite{Guo:2014iya}.

Finally, some authors try to describe the exotic neutral candidates like $X(3872)$ as ordinary charmonia whose properties are deformed by the thresholds,  see for example the Unquenched Quark Model by Ferretti~\etal\cite{Ferretti:2013faa}.

\section{The prompt production of $X(3872)$} \label{sec:prompt}
In this section we report a brief summary of an old controversy about the molecular interpretation of the $X(3872)$. The main drawback of this picture, in fact, lies in the unexpectedly high production cross section measured at Tevatron and LHC that, for many years, has been seen as the definitive proof of the inconsistency of the molecular interpretation. 

\vspace{1em}

The description of the $X(3872)$ in terms of a very loosely bound meson molecule is often  compared to the well-known case of the deuteron, both being bound by strong interactions and having very small binding energies\cite{Tomaradze:2012iz,pdg}, $E_X=(-0.142\pm0.220)$ MeV and $E_d=(-2.2245\pm0.0002)$~MeV.

The deuteron can be described by means of the phenomenological \emph{coalescence model}\cite{Csernai:1986qf,Artoisenet:2010uu}, according to which a neutron and a proton will bind together if they are produced with a relative momentum smaller than a \emph{coalescence momentum}, $k_0\simeq 80$ MeV. Because of the close analogy between the deuteron and the $X(3872)$ we might wonder if a similar approach could be valid for the latter as well. In particular, in both cases, one might expect to have a very small yield of such loosely bound molecules in high energy hadron collisions, since their component will naturally tend to be produced with a very high relative momentum, thus preventing the system from binding.

The validity of this qualitative expectation and of the coalescence model for the case of the anti-deuteron has been successfully tested\cite{Guerrieri:2014gfa} using the well-known Monte~Carlo (MC) generator HERWIG\cite{Corcella:2000bw}. In particular this tool has been used to evaluate the number of $\bar p\bar n$ pairs produced with small relative momentum ($k_0<80$ MeV) in $pp$ collisions at $\sqrt{s}=7$ TeV in the interval $0.9$ GeV$<p_T<1.4$ GeV. The  results obtained have been compared to the preliminary data on anti-deuteron production from the ALICE experiment at LHC\cite{Sharma:2012zz}. The results of this analysis are reported in \figurename{~\ref{fig:deuteron}}. By anti-deuteron events we mean the number of $\bar p\bar n$ pairs produced with a momentum smaller than $80$ MeV. As one can see in \figurename{~\ref{fig:smallk0}}, the MC simulation describes the experimental data, thus providing a proof of the validity of the coalescence model. 

In \figurename{~\ref{fig:largek0}} the authors extrapolated the transverse momentum range up to $p_T=30$ GeV. In order to increase the statistics enough, they allowed the relative momentum between the $\bar p\bar n$ pair to be up to $(300-450)$ MeV since the simulation showed that the shape of the $p_T$ distribution was totally uncorrelated with $k_0^\text{max}$~\footnote{The ${(k_0^\text{max})}^3$ dependence coming from phase space is reabsorbed in the normalization factor used to tune the distributions on the experimental data.}. This rough estimate shows how the production cross section for anti-deuteron might turn out to be really small at high $p_T$, in agreement  to the intuitive picture mentioned before. There is, in fact, a close relation between the $p_T$ of the pair and their relative momentum, as it turns out that it can be written as $k_0\simeq p_T\tan(\phi/2)$, $\phi$ being the relative angle between $\bar p$ and $\bar n$ and with $p_T=|\vett{p}_n+\vett{p}_p|\sin\theta$, with $\theta$ being the angle from the beam axis. Hence, since the kinematical cuts used in the simulations give $\theta\simeq 45^\circ$, for a generic $\phi$, high $p_T$ is equivalent to high relative momentum.
Data from ALICE up to $p_T\approx 10$~GeV in anti-deuteron production might be available and would be highly discriminative for the molecular picture of the $X$. 

\begin{figure}[t] 
\centering
\subfigure[]{ \label{fig:smallk0}
\includegraphics[scale=.3]{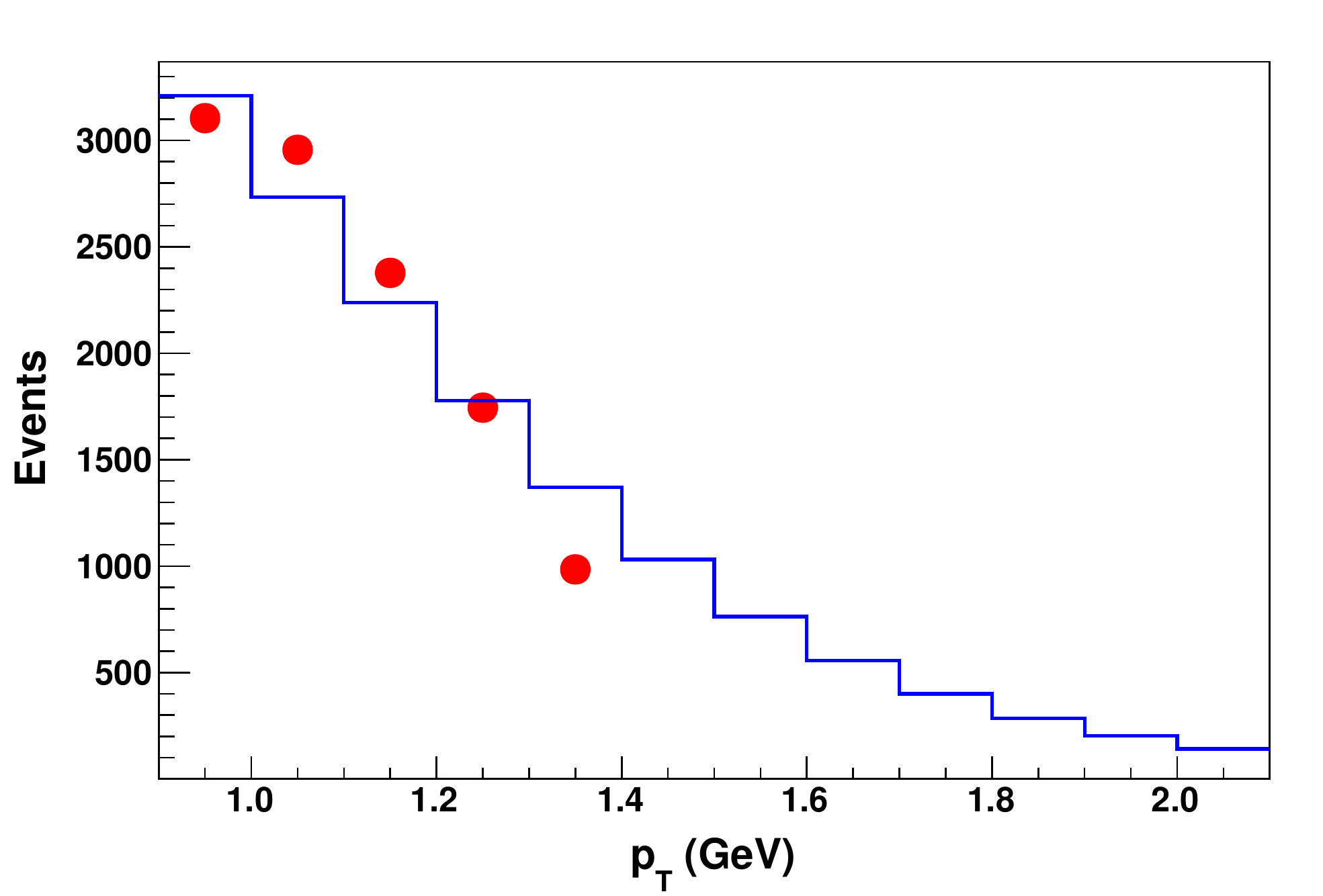}
}
\subfigure[]{ \label{fig:largek0}
\includegraphics[scale=.3]{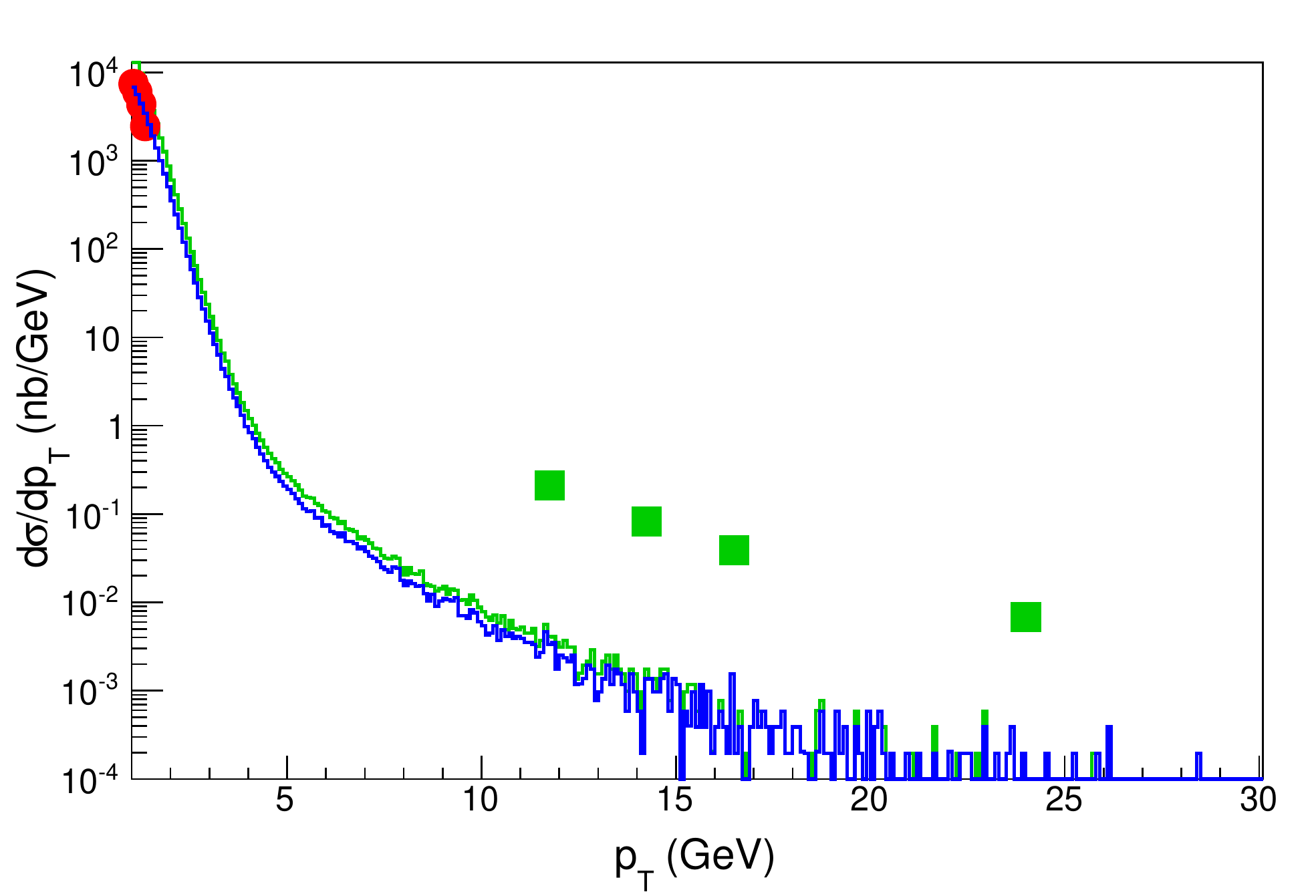}
}
\caption{Anti-deuteron events produced in $pp$ collisions according to $10^9$ HERWIG events with $|\eta|<0.9$ (blue solid line) and $|y|<1.2$ (green solid line) as a function of the transverse momentum of the molecule, from Guerrieri~\etal\cite{Guerrieri:2014gfa}. The MC simulation is compared to the ALICE deuteron production data\cite{Sharma:2012zz} (red circles) and with the \cms $X(3872)$ data\cite{Chatrchyan:2013cld} (green squares).} \label{fig:deuteron}
\end{figure}

It is exactly this qualitative expectation that casted many doubts on the interpretation of the $X(3872)$ in terms of loosely bound meson molecule. Such particle was, in fact, observed both by \cdf\cite{Abulencia:2006ma} and \cms\cite{Chatrchyan:2013cld} with a very large prompt -- \ie directly produced at the collision vertex -- production cross section, $\sigma \simeq 30$ nb. This experimental fact seems at odds with the previous conclusion about the deuteron and, more in general, about hadronic molecules with very small binding energy. 

In particular, it is possible to estimate an upper bound for the production cross section of the $X$ as follows\cite{Bignamini:2009sk}:
\begin{align} \label{eq:upperbound}
\sigma(p\bar p\to X(3872))\leq \sigma^\text{max}(p\bar p\to X(3872))\sim \int_\mathcal{R}\left|\langle D\bar D^*(\vett k)|p\bar p\rangle\right|^2,
\end{align}
where $\vett{k}$ is the relative momentum between the two $D$ mesons and $\mathcal{R}$ is the domain where the two-body wave function for the molecular $X(3872)$ is significantly different from zero. 

Such an upper bound can  be estimated by simply counting the number of $D^0\bar D^{*0}$~\footnote{Here and in what follows we will omit the charge conjugate system, $\bar D^0D^{*0}$, for simplicity.} produced with a relative momentum lower than a certain  $k_0$ value. This has been done\cite{Bignamini:2009sk}, again, using HERWIG and PYTHIA\cite{Sjostrand:2006za} (see \figurename{~\ref{fig:bigna}}), taking $\mathcal{R}$ to be a ball or radius $[0,35]$ MeV, on the basis of a na\"{i}ve gaussian shape for the two-body wave function of the $X$. The result of the MC simulation was a production cross section of $0.071$ nb for HERWIG and $0.11$ nb for PYTHIA, which are both smaller than the experimental value by more than two orders of magnitude. This is also confirmed by comparing the expected cross section at high $p_T$ for the anti-deuteron with the \cms data for the $X(3872)$ -- see \figurename{~\ref{fig:largek0}}. In fact, if we assume that the spin-interactions are second order effects, the yields for the anti-deuteron and for the $X(3872)$ should be similar. In \figurename{~\ref{fig:largek0}} the simulated cross section is again two orders of magnitude smaller that the experimental one. This seemed to be the definitive proof of the inconsistency of the molecular interpretation with the experimental data.

\begin{figure}
\centering
\subfigure[]{
\includegraphics[width=.45\textwidth]{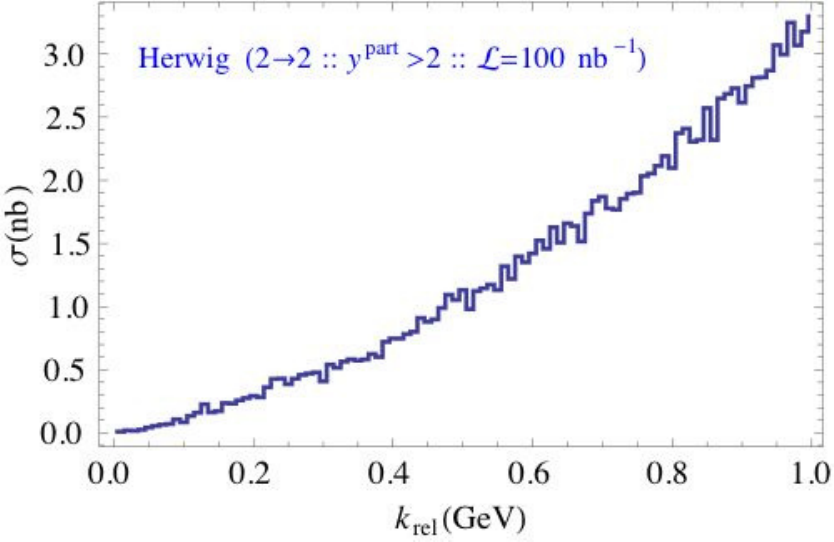}
}
\subfigure[]{
\includegraphics[width=.45\textwidth]{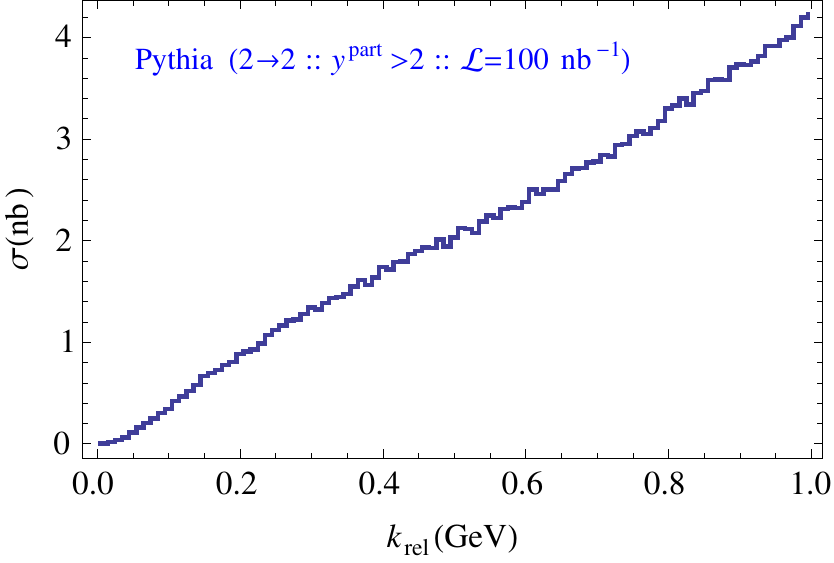}
}
\caption{Integrated cross section for the production of a $D^0\bar D^{*0}$ pair as a function of their relative momentum computed with HERWIG (a) and PYTHIA (b). These plots are obtained generating $55\times10^9$ events and applying final cuts on the $D$ mesons such that the produced molecule has $p_T>5$ GeV and $|y|<0.6$. From Bignamini \etal\cite{Bignamini:2009sk}} \label{fig:bigna}
\end{figure}

However, the previous approach was later criticised and it was shown\cite{Artoisenet:2009wk} that the theoretical and experimental cross sections might be matched resorting to Final State Interactions (FSI)\cite{book:17198}. The possible presence of FSI, in fact, casts doubts on the applicability of the simple coalescence picture to the case of the $X(3872)$, since the two components of the molecule could be bound by final state rescattering even when their relative momentum is large. In particular, the Migdal-Watson theory would change the previous results in two different ways:
\begin{enumerate}
\item The cross section for the production of the $X$ should be modified to
\begin{align} \label{eq:FSI}
\sigma(p\bar p\to X(3872))\simeq\left[\sigma(p\bar p\to X(3872))\right]_{k_0<k_0^\text{max}}\times\frac{6\pi\sqrt{-2\mu E_X}}{\Lambda},
\end{align}
where $\left[\sigma(p\bar p\to X(3872))\right]_{k_0<k_0^\text{max}}$ is the upper bound evaluated in~\eqref{eq:upperbound} and $\Lambda~\sim~m_\pi$ is the typical range of the interaction between the components;
\item Instead of being the inverse of the spread of the spatial wave function, the maximum value for the relative momentum should be given by the inverse of the range of the interaction, $k_0^\text{max}\simeq c\Lambda$, with $c=\mathcal{O}(1)$.
\end{enumerate}
By setting $k_0=2.7\Lambda\simeq 360$ MeV one can increase the theoretical cross section up to $32$ nb, which is in agreement  with the experimental value.

However, this approach has some flaws\cite{Bignamini:2009fn}: it can be shown that the use of Eq.~\eqref{eq:FSI} should enhance the occurrence of a new hypothetical molecule, the $D_s\bar D_s^*$, which otherwise would be suppressed, as one could infer by looking at data on $D_s$ production at Tevatron\cite{Acosta:2003ax}. In fact, the theoretical production cross section for this $X_s$ would be $\sigma\simeq 1\div 3$ nb and  should be detected by the \cdf experiment. No hint for such a particle has been found. Furthermore, the applicability of the Migdal-Watson theorem requires that, \emph{(i)} the two final particles should be in an $S$-wave state and \emph{(ii)} they should be free to interact with each other up to relative distances comparable to the interaction range.
First of all, the inclusion of relative momenta up to $k_0^\text{max}\simeq 360$ MeV means to include relative orbital angular momenta up to $\ell\sim k_0^\text{max}/m_\pi\simeq 2\div 3$, thus violating the hypothesis \emph{(i)}. Moreover, using again the MC softwares HERWIG and PYTHIA, one can show\cite{Bignamini:2009fn} that in high energy collisions, such as those occurring at Tevatron and LHC, there are on average $2\div 3$ more hadrons having a relative momentum with respect to one of the two components smaller that $100$ MeV, thus violating the hypothesis \emph{(ii)} -- see \figurename{~\ref{fig:altriadroni}}. Lastly, it seems really odd that the presence of FSI plays such an important role in the case of the $X(3872)$ but not in the case of the deuteron which, as we showed, is already very well described by a simple coalescence picture.

\begin{figure}
\centering
\includegraphics{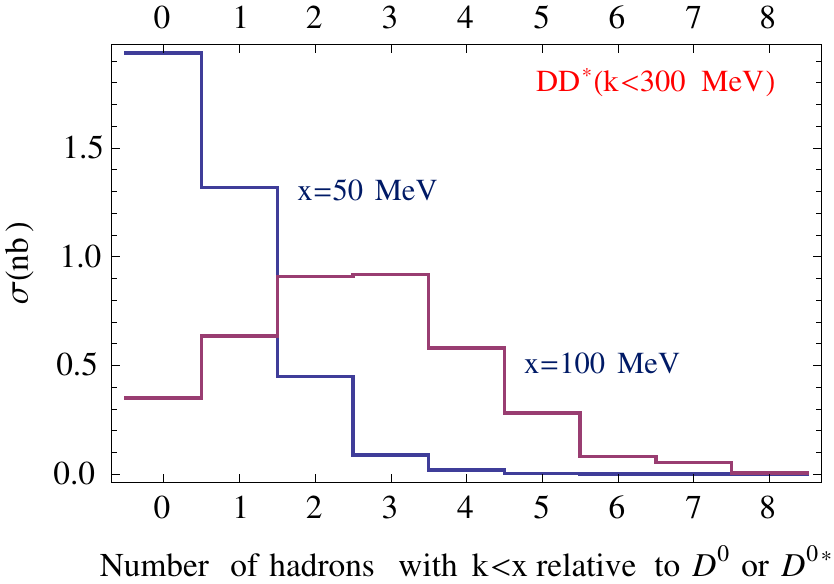}
\caption{Integrated cross section for the production of hadrons with relative momentum $k<x$ with respect to either the $D^0$ or the $\bar D^{*0}$ composing the molecular $X(3872)$. From Bignamini~\etal\cite{Bignamini:2009fn}} \label{fig:altriadroni}
\end{figure}

Even though the presence of other hadrons (mainly pions) sorrounding the system does not allow the use of FSI, it might still play an important role in explaining the unnaturally high prompt production of the $X(3872)$. 

It has been proposed\cite{Esposito:2013ada} that the possible elastic scattering of these ``comoving'' pions with one of the components of the molecule might decrease their relative momentum, hence increasing the number of would-be molecules. This possibility can be understood intuitively  referring to the distribution of $D$ meson pairs as a function of $k_0$ as reported in \figurename{~\ref{fig:bigna}}. The idea is that the interaction might push the pair both to higher and to lower values of $k_0$.  However, since the majority of would-be molecules are produced with high relative momenta, even if a small fraction of them would be pushed to smaller momenta, that could cause a feed-down of pairs towards the lower bins of the distribution, where the $X(3872)$ candidates live. For a pictorial representation of the considered rescattering mechanism see \figurename{~\ref{fig:rescattering}}.

\begin{figure}[t]
\centering
\includegraphics[scale=0.38]{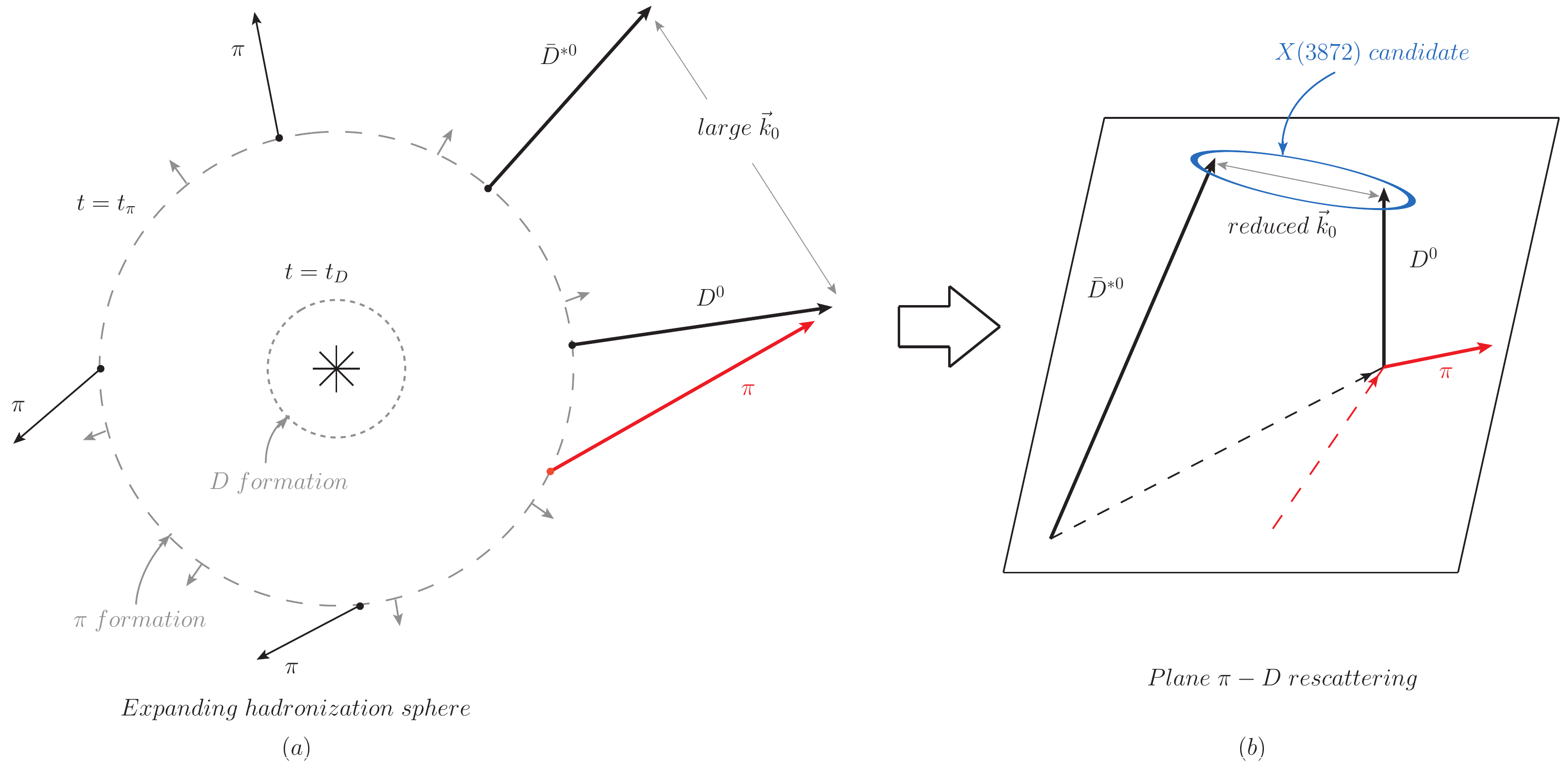}
\caption{Pictorical representation of the rescattering mechanism. After the main high-energy interaction has taken place, the final state particles can be thought of  as belonging to an expanding sphere. The hadronization time of a certain particle goes as $t_\text{hadr}\propto1/m$. Therefore the $D$ mesons hadronize at an earlier time $t_D$ whereas pions hadronize at a later time $t_\pi$ (dotted and dashed spheres respectively). In figure (a) the $D^0\bar D^{*0}$ pair starts with a large relative momentum $\vett{k}_0$. However, the $D^0$ might interact with one of the comoving pions (red arrow). The $\pi-D$ rescattering (figure (b)) can deviate the $D^0$ and reduce the relative momentum $\vett{k}_0$ thus producing a possible $X(3872)$ candidate.} \label{fig:rescattering}
\end{figure}

It is worth noting that, if we assume  the initial total energy $\mathcal{E}$ of the pair to be positive, the decrease in $k_0$ due to the elastic scattering may bring it to negative values, hence assuring the binding of the molecule. Therefore, in this model the $X(3872)$ would be a genuine, negative energy $D^0\bar D^{*0}$ bound state, whose lifetime would be entirely regulated by the lifetime of its shorter lived component, the $\bar D^{*0}$. Hence, this mechanism also predicts a narrow width, $\Gamma_X\sim\Gamma_{D^*}\simeq 65$ keV, in accordance with the experiments.

Once again, this idea has been tested\cite{Guerrieri:2014gfa,Esposito:2013ada} with the already mentioned MC algorithms. In particular, the recipe used to implement the interaction with the pions is as follows: first of all the 10 most coplanar pions to the $D^0\bar D^{*0}$ plane are selected, then the pion which will interact with (say the $D^0$) is randomly chosen and lasty the most parallel pion to the non-interacting meson (say the $\bar D^{*0}$) is selected. One expects this configuration to be the most effective in physical events. Moreover, in order to prevent that $D$ mesons belonging to different jets (separeted in coordinate space) would get closer by the scattering with a hard pion, one also requires $\Delta R_{DD^*}\equiv \sqrt{{(\Delta y_{DD^*})}^2+{(\Delta \phi_{DD^*})}^2}<0.7$. The $\pi D$ interaction in the centre of mass is given by
\begin{subequations}
\begin{align}
\langle \pi(p)D(q)|D^*(P,\eta)\rangle =& g_{\pi DD^*}\eta\cdot p, \\
\langle \pi(p)D^*(q,\lambda)|D^*(P,\eta)=&\frac{g_{\pi D^*D^*}}{M_{D^*}}\epsilon_{\alpha\beta\gamma\delta}\lambda^\alpha \eta^\beta p^\gamma q^\delta,
\end{align}
\end{subequations}
with $g_{\pi DD^*}\simeq 11$ and $g_{\pi D^*D^*}\simeq 17$\cite{Casalbuoni:1996pg}.

First of all, it has been checked\cite{Guerrieri:2014gfa} that this new mechanism does not spoil the high energy behavior of the relevant $D$ meson distribution, as shown in \figurename{~\ref{fig:spoil}}. It was actually showed that the inclusion of one elastic scattering improves the agreement of the simulation with the experimental data from \cdf -- see \tablename{~\ref{tab:chi2}}. This is a strong hint of the fact that this mechanism actually takes place in real physical events and should hence be considered when studying final hadronic distributions.

\begin{table}[b]
\centering
\tbl{Fit values referring to the distributions in \figurename{~\ref{fig:spoil}}.}
{\begin{tabular}{c|cc}\hline\hline
& $K$-factor & $\chi^2$/DOF \\ \hline
$0\pi$ (blue) & 1.35 & 45/11 \\
$1\pi$ (red) & 3.46 & 24/11 \\
\hline\hline
\end{tabular} \label{tab:chi2}}
\end{table}

\begin{figure}
\centering
\includegraphics[scale=.3]{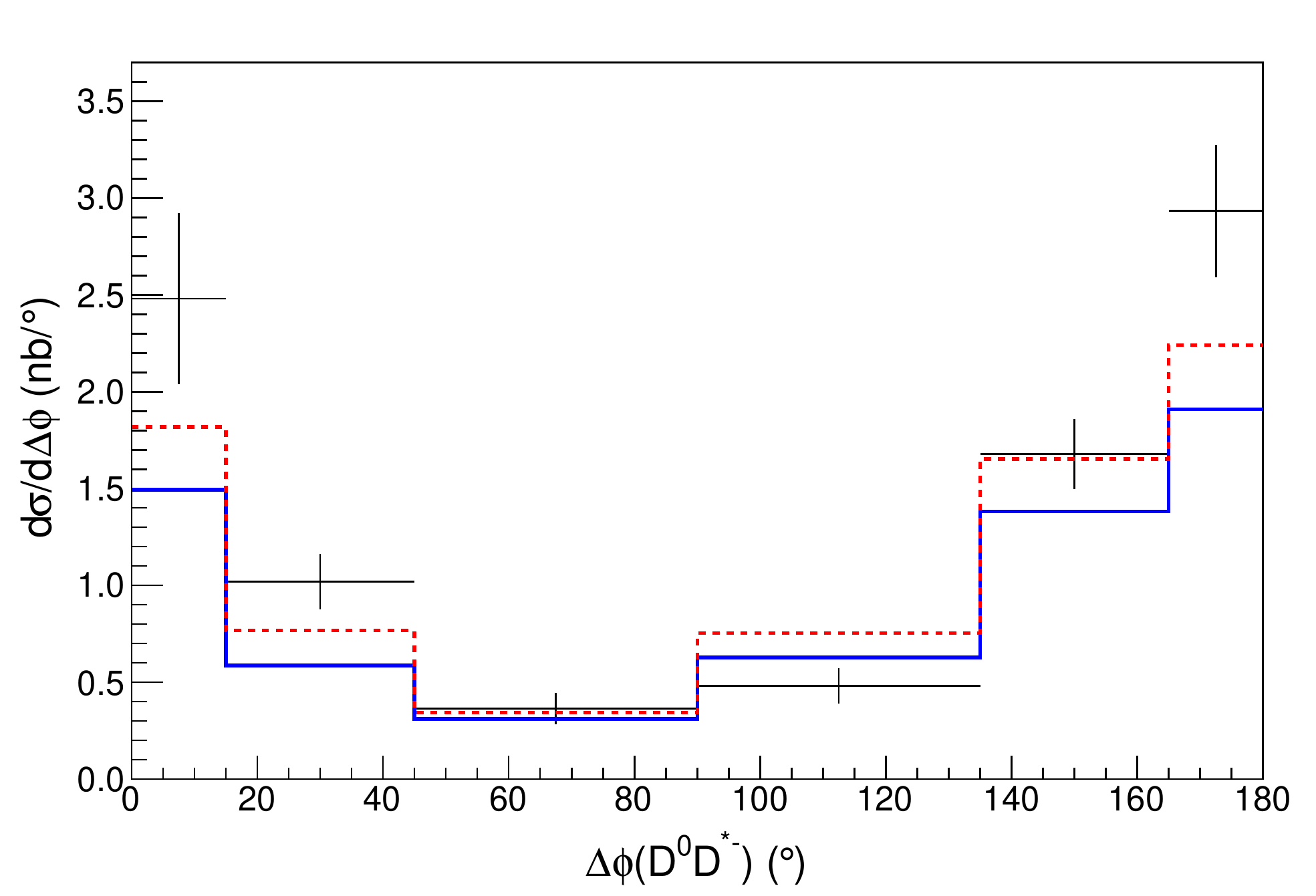}
\includegraphics[scale=.3]{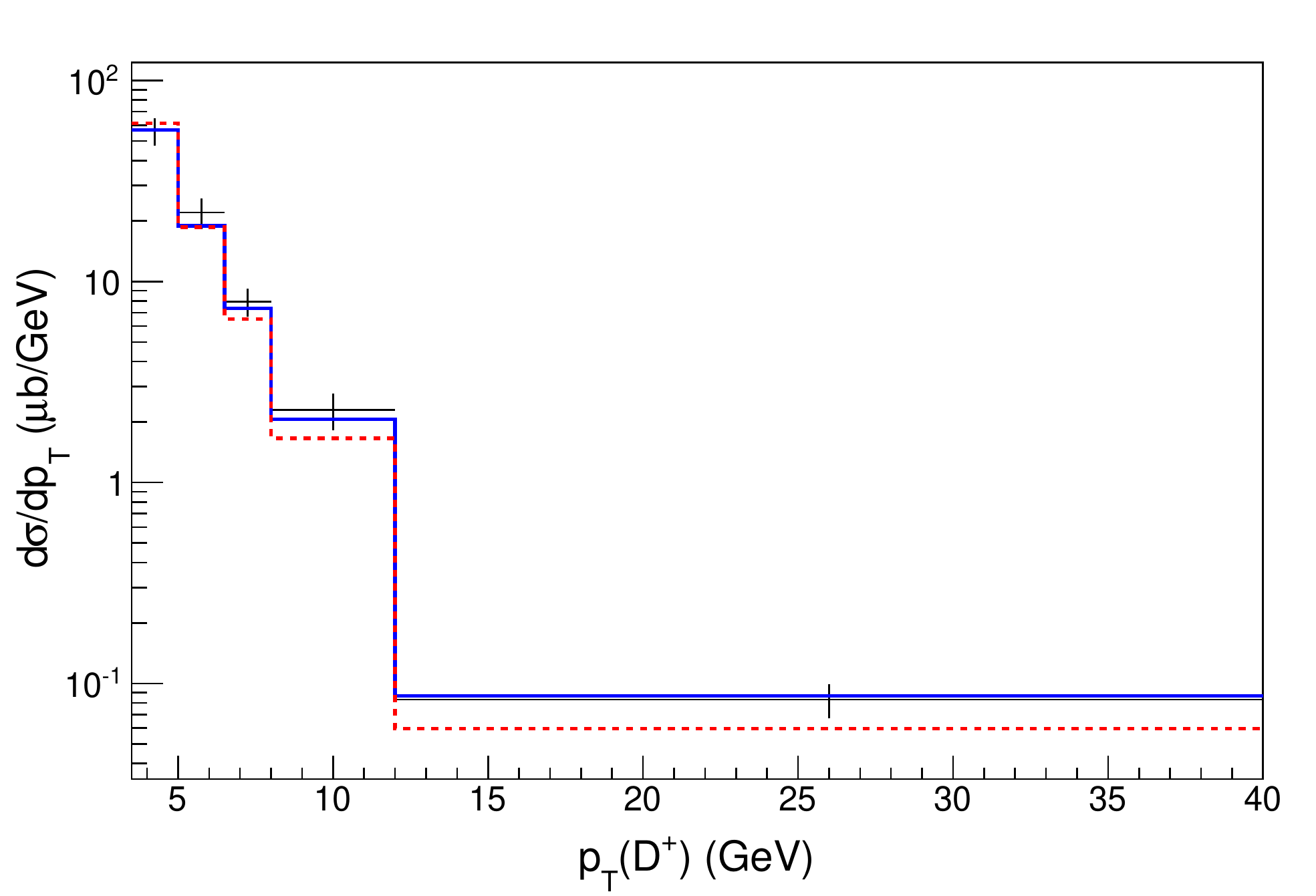}
\caption{Differential cross sections of $D^0$ and $D^0D^{*-}$ pairs at \cdf obtained not including (blue, solid) and including (red, dashed) the interaction with one pion per $D^0/D^0D^{*-}$ event, from Guerrieri~\etal\cite{Guerrieri:2014gfa}. Both distribution have been rescaled by the same normalization factor, $K$, obtained by minimizing the combined $\chi^2$. It is clear that the inclusion of one elastic scattering does not weaken the agreement of the MC simulations to the experimental data.} \label{fig:spoil}
\end{figure}

As one can see from \figurename{~\ref{fig:markov}} the proposed mechanism is actually effective in feeding down the lower $k_0<50$ MeV bin, the one containing the would-be molecules. It is also possible to estimate how many of these interactions may take place. In particular, considering a model where all the produced hadrons are flying away from each other on the surface of a sphere -- see again Fig.~\ref{fig:produzione} -- and taking into account the range of the interaction, one finds\cite{Esposito:2013ada} that the simulations suggest an average of 3 scatterings per event. These consecutive interactions can be reproduced by implementing a Markov chain\cite{Esposito:2013ada}. In \tablename{~\ref{tab:markov}} we report the values of the integrated cross section for the production of the $X(3872)$ varying both the number of interacting pions and the maximum $k_0$ allowed for the pair.

\begin{figure}
\centering
\includegraphics[scale=.3]{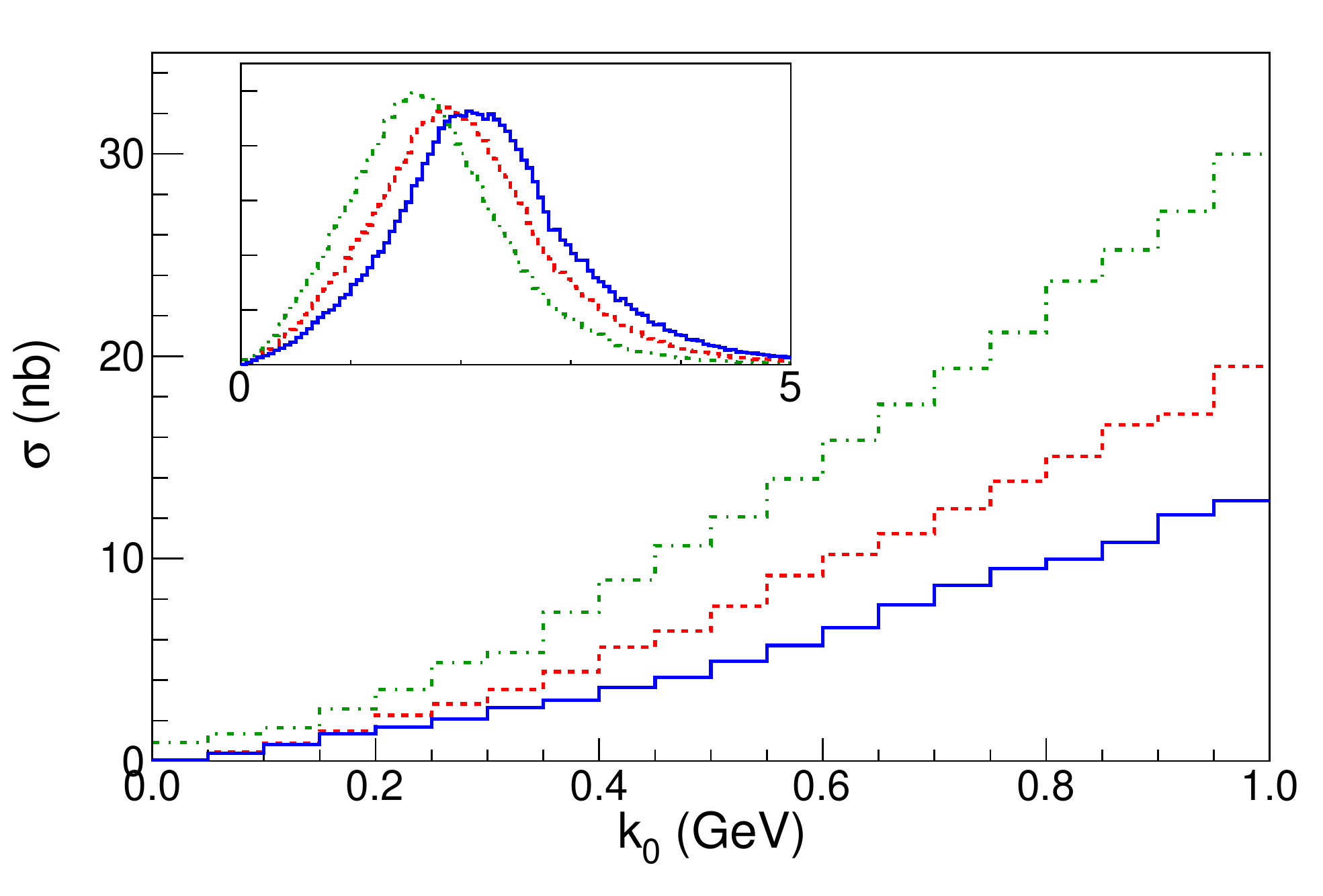}
\caption{Integrated corss section of $D^0\bar D^{*0}+h.c.$ pairs at \cdf obtained with HERWIG, without (blue, solid), with one (red, dashed) and with three (green, dot-dashed) interactions with pions, from Guerrieri~\etal\cite{Guerrieri:2014gfa}. In the inset the same plot on a wider range of $k_0$ values.} \label{fig:markov}
\end{figure}

\begin{table}[b]
\centering
\tbl{Effect of multiple scatterings in $X(3872)$ production cross section. $k_0^\text{max}$ indicates the integration region $k_0\in[0,k_0^\text{max}]$.}
{\begin{tabular}{c|ccc}
\hline\hline
$k_0^\text{max}$ & 50 MeV & 300 MeV & 450 MeV \\
\hline
$\sigma(0\pi)$ & 0.06 nb & 6 nb & 16 nb 	\\
$\sigma(1\pi)$ & 0.06 nb & 8 nb & 22 nb \\
$\sigma(3\pi)$ & 0.9 nb & 15 nb & 37 nb \\
\hline\hline
\end{tabular} \label{tab:markov}}
\end{table}

As one can see, if one trusts the coalescence model for the $X(3872)$ -- as the data on deuteron strongly suggest -- and hence consider $k_0^\text{max}\simeq 50$ MeV, not even the elastic scattering with three consecutive pions is able to raise the production cross section up to the experimental one ($\sigma\simeq 30$ nb). Moreover, if one considers the (questionable) use of FSI\cite{Artoisenet:2010uu,Artoisenet:2009wk} as explained previously, then it should be $k_0^\text{max}\simeq 360$ MeV. With this integration region, the simulations produce a cross section after the interaction with one pion -- and after a rescaling needed to take into account the different normalization factors between the two works\cite{Artoisenet:2010uu,Guerrieri:2014gfa}-- that is equal to $\sigma(1\pi)\simeq 52$ nb, larger than the experimental one.

\vspace{1em}

To summarise, the experimental value of the prompt production cross section of the $X(3872)$ casted serious doubts on its possible interpretation in terms of a $D^0\bar D^{*0}$ molecule. According to the expectations following from the phenomenological coalescence model -- that correctly describes the well-known deuteron -- the production of such a weakly bound state should be strongly suppressed in high energy collisions. Even though many ideas and models have been proposed during the years none of them has successfully reconciled the theoretical expectations with the experimental results.

It should also be enphasized that the inclusion of possible interactions between comoving pions and final state mesons\cite{Esposito:2013ada,Guerrieri:2014gfa} turned out to improve the accordance between the simulated MC distributions and the experimental ones and hence should be taken into account in future works.

\section{Tetraquarks} \label{sec:tetraquarks}
The previously described problem seemed a compelling evidence of the necessity for a new kind of interpretation for these exotic mesons. It has been proposed\cite{Maiani:2004vq} that these states might actually have a compact (point-like) four-quark structure -- the so-called \emph{tetraquark}. In this section we introduce this model starting from its consequences regarding the production of $XYZ$ mesons in high energy collisions, so that a more direct comparison could be done with the previous section. One of the simplest and more economic ways (in terms of new states predicted) of forming tetraquarks is the diquark-antidiquark realisation based on the binding of a diquark ${[q_1q_2]}_{\bar{\bm 3}_c}$ and an antidiquark ${[\bar q_3\bar q_4]}_{\bm 3_c}$. This was originally proposed by Jaffe and Wilczek to explain pentaquark baryons\cite{Jaffe:2003sg} (diquark-antidiquark-quark). In general tetraquark bound states have  been proposed a long time ago\cite{Jaffe:1976ih,Jaffe:1978bu,Alford:2000mm} to understand the nature of the light scalar mesons $a_0(980)$ and $f_0(980)$. An interpretation of light scalar mesons in terms of diquark-antidiquark states was instead proposed for the first time by Maiani \etal\cite{Maiani:scalars,thscal} following the revitalization of interest on the $\sigma$ meson (reappeared in heavy-light meson decays\cite{conal}) and contextually with an interesting reanalysis of $\pi\pi$ scattering\cite{colcapr}. 

Some aspect of the tetraquark models have been inspired by dibaryons\cite{Rossi:1977cy,Montanet:1980te}, in particular the feasibility of isospin violation\cite{Rossi:1977dp}. Different extensions of the constituent quark models for the tetraquark spectroscopy have been explored by Valcarce {\etal}\cite{Barnea:2006sd,Vijande:2007fc,Vijande:2007rf,FernandezCarames:2009zz,Vijande:2013qr}.

As we will show in the next sections, assuming a constituent quark model with color-spin interaction it is possible to study the spectroscopy of these states, in which the $J^{PC}=1^{+\pm}$ and $J^{PC}=1^{--}$ states discovered so far can be nicely accommodated. The main problem with this picture is that it also predicts other states which did not (yet?) show up in experimental researches. A possible solution to this problem is also presented.

\subsection{Tetraquark production} \label{subsec:prod_tetra}
From the studies on loosely bound molecule production at hadron colliders we are led to consider that multiquark hadrons should rather be initiated by the formation of compact quark clusters. The seed of a heavy-light tetraquark state state could be described by 
\begin{equation}
|\psi\rangle =\alpha |[Qq]_{\bar{\bm 3}_c}[\bar Q\bar q]_{\bm 3_c}\rangle+\beta |(Q\bar Q)_{\bm 1_c}(q \bar q)_{\bm 1_c}\rangle + \gamma |(Q\bar q)_{\bm 1_c}(\bar Q q)_{\bm 1_c}\rangle, 
\label{superp0}
\end{equation}
where by $Q$ and $q$ we represent the heavy and light quarks respectively.
In our scheme, the two-meson states will tend to fly apart, as strong Van der Waals-like forces between their meson components are not sufficient to produce bound states like $\jpsi \,\rho$ or $D\bar D^*$ -- depending on the spin and orbital quantum numbers of the original four-quark system.  In this sense such  states are in a ``open channel'' continuum. 

Most authors are convinced instead that some hadron molecule shallow potentials could allow for at least a single discrete level with almost zero energy $\sim -\epsilon$. Small binding energies in quantum field theory are possible and, if $g$ is the {\it strong} coupling, say in the $DD^*$ interaction,  one can connect $\epsilon$ to $g$ by
\be
\epsilon=\frac{g^4}{512\pi^2}\frac{\mu^5}{M_D^4M_{D^*}^4},
\label{weinlan}
\ee 
where $\mu$ is the reduced mass of the $DD^*$ system (in this formula we are treating $D$ and $D^*$  as if they were spinless particles. Accounting for the spin is simple and does not change our qualitative conclusions). This formula is obtained by resonance scattering theory at low energies supposing that there is a pole term associated with the virtual production of $X$ particle in the $f(DD^*\to DD^*)$ scattering amplitude
\be
\frac{1}{(p_D+p_{D^*})^2-M_X^2}
\ee
and supposing that ${\cal E}$, the barycentric energy after subtraction of rest energy, is, like $\epsilon$, a {\it small} quantity. We replace $(p_D+p_{D^*})^2\to (M_D+M_{D^*}+{\cal E})^2$.

Thus, if  ${\cal E}$ were indeed small enough, formula~\eqref{weinlan} would hold and, including spin and using the coupling $g$ as deduced in\cite{Brazzi:2011fq}, one would get
\be
|\epsilon|_{\rm exp}=0.1~{\rm MeV}\quad\text{vs.}\quad|\epsilon|=0.4~{\rm MeV}
\label{agr}
\ee 
from~\eqref{weinlan}.

However in the prompt production of $X$ at large hadron colliders, ${\cal E}$ is very far from being small.  The fact that the observed prompt production cross section at LHC is so much larger than expectations, as commented at length in previous sections, suggests then that the $X$ is not a loosely bound state of $DD^*$, and the rough agreement~\eqref{agr} simply  occurs outside of the kinematic conditions of $X$ production.

These kind of problems do not exist in the diquark-antidiquark picture for this particle would be  kept together by color interactions, the unknown being the effectiveness of the color force at producing diquarks. The diquark-antidiquark belongs to a ``closed'' channel.  The relative size of $\alpha,\beta,\gamma$ coefficients in~\eqref{superp0} is unknown. 

We might formulate different
hypotheses: $i)$ $\alpha,\beta,\gamma$ are all of the same order. In this case we should be observing the entire spectrum of diquark-antidiquark states which can be predicted using the color-spin Hamiltonian (see below). In a previous work\cite{Maiani:2004vq}, the Hamiltonian of the diquark-antidiquark model was supposed to contain both spin-spin interaction between quarks within each diquark and quarks in the two different diquarks. The resulting spectrum predicts a rich structure of states with some evident mismatches with the experimental findings. 

A `type-II' version of the diquark-antidiquark model, with  spin couplings suppressed between different diquarks, allows  a remarkable description of the $J^{PG}=1^{++}$ sector of charged tetraquarks as the $Z(4430)$, $ Z_c(4020)$, $Z_c(3900)$ and a good picture of the entire $J^{PC}=1^{--}$ sector (see next few sections). Some typical problems of the diquark-antidiquark model persist in the type-II  model. For example the $X(3872)$ should have charged partners and an hyperfine splitting between two neutral levels, to account for isospin violations\cite{Maiani:2004vq}. 

To solve this kind of difficulties we might formulate a different hypothesis for the hierarchy among $\alpha,\beta,\gamma$ coefficients. 
We might indeed assume that,  $ii)$
\begin{equation}
|\beta|^2,|\gamma|^2 \gg |\alpha|^2
\end{equation}
Such an assumption means that, in general, diquark-antidiquark states are  less likely to be formed in hadronization but a resonance could emerge as a result of the coupling between open and closed channels. This hypothesis introduces a selection rule in the diquark-antidiquark spectrum: 
especially those levels which are close enough to open channel thresholds (resonance conditions) are observed as physical resonances.
 
{\it More specifically, the diquark-antidiquark closed channel provides an effective attraction in the open channel which might lead to produce a resonance}. This phenomenon is effective if the energy level $E_n$, corresponding to the closed channel state $|[Qq]_{\bar{\bm 3}_c}[\bar Q\bar q]_{\bm 3_c},n\rangle_{{\mathcal C}}$, happens to be very close to one, or both, as in the $X$-particle case, of the open channel thresholds  (located, in the case of the $X$, at $E_{\mathcal O}=m_{\jpsi}+m_\rho$ or $E_{\mathcal O}=m_{D^0}+m_{\bar D^{*0}}$).  

Strong interactions provide the discrete spectrum for diquark-antidiquark states, however 
those levels correspond most likely  to physical states once the closed channel is hybridized with the open one, {\it i.e.} the difference in energy, or detuning parameter $\nu$, is small enough.  When this energy matching condition between the total energy in the open channel and the energy level in the closed channel takes place, the two hadrons in one open channel can undergo an elastic scattering, altered by the presence of the near closed channel level. The two hadrons in an open channel can scatter to the intermediate state in the closed channel, which subsequently decays into two particles in one of the open channels. 
 
 The contribution to the scattering length due to this phenomenon is of the form 
 \begin{equation}
 a\sim |C|\sum_n \frac{\langle [Qq]_{\bar{\bm 3}_c}[\bar Q\bar q]_{\bm 3_c}, n | 
 H_{{\cal C}{\cal O}}|(Q\bar q)_{\bm 1_c}(\bar Q q)_{\bm 1_c}\rangle }{E_{{\cal O}}-E_{n}}
 \label{fesh}
 \end{equation}
where $H_{\cal{CO}}$ couples the open and closed channels; the discrete levels of the closed channel are labeled by $n$. This sum is dominated by the term which minimizes the denominator $E_{{\cal O}}-E_{n} \equiv -\nu$, {\it i.e.}  the one with the smallest detuning. The width of the resulting resonance is naturally proportional to the detuning $\Gamma\sim \sqrt{\nu}$ for phase space arguments.

Since the $X(3872)$ is the narrowest among all $XYZ$ mesons, it must have  $\nu\simeq 0$, which means the highest possible hybridization between channels given the (unknown) inter-channel interaction Hamiltonian $H_{{\cal C}{\cal O}}$. 
 
The $D^+D^{*-}$ open channel level is found  to be at a mass {\it above} the $X$ diquark-antidiquark level, by about 8~MeV. Coupling between channels gives rise to a repulsive interaction if the energy of the scattering particles is larger than that of the bound state (and an attractive one if it is smaller).  For this reason we might conclude that the neutral particle has no $d\bar d$ content in its wavefunction explaining the well-known isospin breaking pattern in  $X$ decays.  

The diquark-antidiquark $X^{+}$ levels (the charged partner of the $X(3872)$), might also fall below $D^+\bar D^{*0}$ and $\bar D^0 D^{*+}$ thresholds by about $3\div 5$~MeV, which could be enough for inhibiting the resonance phenomenon described. This might be the reason why the $X^{+}$ particles, although present in the diquark-antidiquark spectrum, are more difficult to be produced.

The $\jpsi\,\rho^0$ open channel level is also perfectly matching the closed channel one for the $X(3872)$. However because of the large $\rho$ width, the modification in the scattering length~\eqref{fesh} is much less effective if compared to the open charm threshold: the sum in~\eqref{fesh} has to be smeared with an integral convoluting the $\rho$ Breit-Wigner.  Therefore we would expect that the $X^{+}$ particles are less likely to be formed or they could simply be too broad to be observed. Some examples are shown in \tablename{~\ref{tab:feshbach}}.

The mechanism here described is known in nuclear and atomic physics as the Feshbach resonance formation\cite{book:17652}. 

\begin{table}[b]
 \tbl{Exotic states in term of Feshbach resonances. The width is related to the detuning by $\Gamma = A \sqrt{\nu}$. An exception is given by the $Z(4430)$, whose width is forced to be larger than its constituent width, \ie $\Gamma_\rho \sim 150$~\mev.}{
 \begin{tabular}{lcccc}\hline\hline
 State & Open channel & $\nu$ (\mev) & $\Gamma_\text{th}$ (\mev) & $\Gamma_\text{exp}$ (\mev) \\ \hline
$X(3872)$  & $\Dz \Dstarzb$ & $-0.16 \pm 0.31$ & $0$ & $< 1.2$ \\ 
$Z_c(3900)$ & $\Dp \Dstarzb$ & $12.1 \pm 3.4$ & $30$ & $35 \pm 7$ \\ 
$Z_c^\prime(4020)$ & $\Dstarp \Dstarzb$ & $6.7 \pm 2.4$ & $22$ & $10 \pm 6$ \\ 
$Z(4430)$ & $\eta_c(2S) \rho$ & $64 \pm 17$ & $\gtrsim 150$ & $180 \pm 30$\\
$Z_b(10610)$ & $\Bp\Bstarzb$ & $2.7 \pm 2.0$ & $14$ & $18.4 \pm 2.4$ \\ 
$Z_b^\prime(10650)$ & $\Bstarp\Bstarzb$ & $1.8 \pm 1.5$ & $12$ & $11.5 \pm 2.2$ \\ 
 \hline\hline
 \end{tabular}
 \label{tab:feshbach} }
\end{table}
Recently two {\it charged} resonances have been confirmed to a high level of precision. The $Z(4430)$ and the $Z_c(3900)$. These are genuine tetraquark states. We need to recall that the prediction of charged states of this kind was exclusively formulated in the context of compact tetraquark states\cite{Maiani:2004vq}.  

In particular, when the first hint of a $Z(4430)$ charged tetraquark was provided by the \belle collaboration in the $\psiprime\,\pi^+$ channel, back in 2007, it was observed that another state  at 3880 MeV ({\it i.e.} lighter by the $\psiprime-\psi(1S)$ mass difference) was expected in the tetraquark model\cite{tetraz} with the same quantum numbers (the former being the radial excitation of the latter). The lower state was expected to decay into 
$\jpsi \,\pi^{+}$ or $\eta_c \, \rho^{+}$. A charged $Z_c(3900)$ with $J^{PG}=1^{++}$ decaying into $\jpsi\,\pi^+$ was discovered by \bes and \belle in 2013\cite{Ablikim:2013mio}. It was also shown that \bes and \belle data might be compatible with the presence of another peak about 100~MeV below that of the $Z_c(3900)$\cite{Faccini:2013lda}. That was also predicted by the tetraquark model.

The tetraquark model in its first diquark-antidiquark version\cite{Maiani:2004vq,Faccini:2013lda} predicts one more $J^{PG}=1^{++}$ level, at a mass of $3755$~MeV (these mass values are locked to the input mass value of the $X(3872)$).  We might predict that no resonance will be found at this level because there are no open channels nearby to make the Feshbach mechanism possible. The $Z(4430)$ is instead made possible by the presence of the $\eta_c(2S)\rho$ open channel. The expected width, driven by the $\rho$, is expected to be as large as $150$~MeV, to be compared with the $\sim 170$~MeV observed. 

The tetraquark model in its `type-II' version has no $3755$~MeV, but a level perfectly compatible with the observed $Z_c^\prime (4020)$ by the \bes Collaboration\cite{Ablikim:2013wzq}, which is also compatible with a Feshbach generated state. 
A $Z(4430)^0$ isosinglet resonance could be due to the vicinity of the $\eta_c(2S)\;\omega$ open channel, with a narrower width of about $70$~MeV. For a tetraquark interpretation of the states in the bottomonium sector, see Ali \etal\cite{Ali:2011ug,Ali:2009pi,Ali:2009es,Ali:2014dva}.

These considerations about the interplay between a closed (diquark-antidiquark) and open channel (molecular thresholds) are to be considered in a preliminary stage and possibly object of future developments. In the following we will focus instead on the description of the diquark-antidiquark closed channel listing its states by quantum numbers and showing how to estimate expected masses (the position of levels). 

We believe that recent experimental findings are clearly spelling in favor of tetraquark particles and the diquark-antidiquark model apparently has many features matching very well the present phenomenology. The Feshbach mechanism here sketched might be a viable  way for  implementing those selection rules still missing in the tetraquark Hamiltonian approach to be described below. 

\subsection{Diquarks}
One-gluon interaction in the $t$-channel between two quarks in the $SU(3)_c$ representation $R=\bm 3$ (antiquarks $R=\bar{\bm 3}$) involves a (tensor) product of color charges $T_{R_1}\otimes T_{R_2}$ which can be expressed as the direct sum of  diagonal blocks with the dimensions of the irreducible representations $S_i$ in $T_{R_1}\otimes T_{R_2}=S_1\oplus S_2$. According to the general rule
\be
T_{R_1}\otimes T_{R_2}=\bigoplus_i \frac{1}{2}(C_{S_i}-C_{R_1}-C_{R_2})\mathds{1}_{S_i}
\ee
 we only need the Casimir values $C_{\bm 3}=C_{\bar{\bm 3}}=4/3$ and $C_{\bm 1}=0$ to appreciate that one-gluon exchange generates a quark-quark {\it attraction} 
 in the $\bar{\bm 3}$ channel ($-2/3$)  which is just half of  that in the quark-antiquark singlet channel ($-4/3$). 
Even if one-gluon-exchange interaction is a primitive model of low energy strong interactions, correlating it with indications from lattice computations\cite{reticolo} on diquarks gives reasonable  support to the possibility of  diquark-antidiquark hadrons. 

Diquarks carry the same color charge of  antiquarks. The opposite for antiquark-antiquark pairs.
We represent a spin zero diquark with the bispinor notation
\be
[cq]_i=\epsilon_{ijk}(c^j)^T\sigma^2 q^k,
\ee
where $i$, $j$ and $k$ are color indices. In the quadrispinor notation we would have written $\epsilon_{ijk}\bar c^{j}_{\bm c}\gamma_5 q^{k}$, where $\bm c$ 
indicates the charge conjugated spinor.   In the next sections color index will be  left implicit.
Relaying on spin-flavor symmetry of heavy-light mesons, a spin-1 heavy-light diquark could equally be formed
\be
[cq]_i=\epsilon_{ijk}(c^j)^T\sigma^2\sigma^\lambda q^k.
\ee

The color-spin Hamiltonian is, {\it e.g.} Eq. (4.3) in Jaffe's report\cite{Jaffe:2004ph}
\be
H=-2\sum_{i\neq j,a} \kappa_{ij} \; \bm S_i\cdot \bm S_j\;\frac{\lambda^a_i}{2}\cdot \frac{\lambda^a_j}{2}.
\ee
Here we will discuss the color interaction only, leaving the spin  to 
the next sections.
We introduce the (normalized) color singlet/octet states using the following notation which turns out to be rather practical for calculations:
\begin{subequations}
\bea
\label{singlet}
&&|\bar c c_{\bm 1},\bar q q_{\bm 1}\rangle:=\frac{1}{3}\;\mathbbm{1}_{\bar c c}\otimes\mathbbm{1}_{\bar q q}; \\
&&|\bar c c_{\bm 8},\bar q q_{\bm 8}\rangle:= \frac{1}{4\sqrt{2}}\;\lambda^a_{\bar c c}\otimes\lambda^a_{\bar q q },
\label{octet}
\eea
\end{subequations}
where by $\lambda^a_{\bar c c}$, for example, we mean $\bar c_i \;(\lambda^{a})^i_{j} \;c^j$ using latin letters for color indices.

With the notation $|cq_{\bm{\bar{3}}},\bar c\bar q_{\bm 3}\rangle$ we mean {\it an overall color singlet} state of a diquark-antidiquark pair:
\be
[cq]_i[\bar c\bar q]^i=c_j\bar c^jq_k\bar q^k- c_j\bar q^j q_k\bar c^k.
\ee
Using the relation 
\be
\label{fierz}
(\lambda^a)^i_j(\lambda^a)^k_l=2(\delta^i_l\delta^k_j-1/3\;\delta^i_j\delta^k_l)
\ee
one  obtains 
\be
|cq_{\bm{\bar{3}}},\bar c\bar q_{\bm 3}\rangle=\frac{2}{3}\;\mathbbm{1}_{\bar c c}\otimes\mathbbm{1}_{\bar q q}-\frac{1}{2}\;\lambda^a_{\bar c c }\otimes\lambda^a_{\bar q q}=2|\bar c c_{\bm 1},\bar q q_{\bm 1}\rangle-2\sqrt{2} |\bar cc_{\bm 8},\bar q q_{\bm 8}\rangle,
\label{2octet}
\ee
{\it i.e.} the octet-octet component  has {\it  twice the probability}   of the singlet-singlet one. The previous state can itself  be normalized in the following way (multiply by $1/\sqrt{12}$)
\be
|cq_{\bm{\bar{3}}},\bar c\bar q_{\bm 3}\rangle
=\frac{1}{\sqrt{3}}\left(\frac{1}{3}\;\mathbbm{1}_{\bar c c}\otimes\mathbbm{1}_{\bar q q}-T^a_{\bar c c}\otimes T^a_{\bar q q}\right)
\label{atriplet}
\ee
and use  the $T=\lambda/2$ matrices. 

Let us represent  states of the fundamental representation with the symbol $|_i\rangle$ whereas those of the anti-fundamental are $|^j\rangle$. Then we have
\begin{subequations}
\bea
&&\langle_j|T^a|_i\rangle=(T^a)^j_i;\\
&& \langle^j|T^a|^i\rangle=-(T^a)^i_j,
\eea
\end{subequations}
{\it i.e.} one is the opposite of the transpose (complex-conjugate) of the other.

From the latter equation we get
\be
T^a|^i\rangle=-|^j\rangle(T^a)^i_j.
\ee
Consider a generic state $|v\rangle$
\be
|v\rangle=|^i\rangle v_i,
\ee
then
\be
|T^a v\rangle=T^a|v\rangle=T^a|^i\rangle v_i=-|^j\rangle(T^a)^i_j v_i.
\ee
We thus conclude that (multiply the latter by $\langle_k|$ and then rename $k\to i$)
\be
\label{anti}
T^a v_i=-(T^a)^j_i v_j,
\ee
whereas 
\be
\label{tri}
T^a v^i=(T^a)^i_j v^j.
\ee
If, for example, we consider the action of the Hamiltonian term  $H_{\bar c c}\,\propto\, T^a_{\bar c}T^a_{c}$, according to~(\ref{anti},\ref{tri}) we have to replace
\be
\label{op1}
\mathbbm{1}_{\bar cc} 
\longrightarrow  -(T^a)^j_i \bar c_j(T^a)^i_k c^k=-\bar c_j(T^aT^a)^j_k c^k.
\ee
Similarly if we start with some  $\bar c_i{\mathcal O}^i_j c^j$, where ${\cal O}$ is some combination of $T$'s,  we have to replace 
\be
\label{op2}
\bar c_i{\mathcal O}^i_j c^j 
\longrightarrow -\bar c_j(T^a{\cal O}T^a)^j_k c^k.
\ee
With this rules we can compute the action of $H_{c q}$ on a diquark state defined in~\eqref{atriplet}
\be
H_{ c q} |cq_{\bm{\bar{3}}},\bar c\bar q_{\bm 3}\rangle\; \propto \; \frac{1}{\sqrt{3}}\left(\frac{1}{3}\; T^b\otimes T^b-T^a T^b\otimes T^aT^b\right),
\ee
and thus 
\be
\langle cq_{\bm{\bar{3}}},\bar c\bar q_{\bm 3}|H_{c q} |cq_{\bm{\bar{3}}},\bar c\bar q_{\bm 3}\rangle\;\propto\; -\frac{1}{3}\left(-2\;\frac{2}{3}-\frac{2}{3}\right)=\frac{2}{3},
\ee
where we have used 
\be
\mathrm{Tr}(T^aT^bT^c)=\frac{1}{4}(d^{abc}+if^{abc})
\ee
and
\begin{subequations}
\bea
&& f^{abc}f^{abd}=3\delta^{ab}; \\
&&d^{abc}d^{abd}=\frac{5}{3}\delta^{ab}.
\eea
\end{subequations}
As for the color, taking matrix elements on diquark-antidiquark color-neutral states  amounts to redefine the chromomagnetic couplings by some numerical factor:  $2/3$ when the $H_{cq}$ and $H_{\bar c\bar q}$ terms are considered. Actually we assume that the dominant couplings in the Hamiltonian are $\kappa_{cq}$ and $\kappa_{\bar c\bar q}$, {\it i.e.},  intra-diquark interactions.  We take them to be equal 
$\kappa=\kappa_{cq}=\kappa_{\bar c\bar q}$~\footnote{
If extra-diquark couplings were considered we could determine them, {\it e.g.} $\kappa_{c\bar c}$, from the masses of standard $L=0$ mesons observing that
\be
\langle cq_{\bm{\bar{3}}},\bar c\bar q_{\bm 3}|H_{\bar c c} |cq_{\bm{\bar{3}}},\bar c\bar q_{\bm 3}\rangle = \frac{1}{4} \langle \bar c c_{\bm 1},\bar q q_{\bm 1}|H_{\bar c c} |\bar c c_{\bm 1},\bar q q_{\bm 1}\rangle
\ee.
}.

\subsection{Diquark-antidiquark States with $L=0$}
The following discussion is mostly based on a recent paper\cite{type2} where, as anticipated, a `type-II' tetraquark model is introduced.

\subsubsection{The $X$ tetraquark}
Consider a tetraquark made up of two $c$ quarks and two light quarks, with the same flavor: a neutral component.
Using explicit spin indices $s,b,r,d$ we write them in the order:
\be
c_s \quad q_b\quad \bar q_r\quad \bar c_d.
\label{quarkstring}
\ee
Assume that the $cq$ pair has spin 1 whereas the antidiquark $\bar q\bar c$  has spin 0. Then the spin indices are saturated by the operators
\be
\sigma^2_{sa}\sigma^i_{ab} 
\ee
for the $cq$ pair and
\be
\sigma^2_{rd}
\ee
for $\bar q\bar c$, where repeated indices are summed. We might write the operators as
\be
\sigma^2_{sa}\sigma^i_{aq}\delta_{qb}\quad \sigma^2_{rt}\delta_{td}
\label{ops}
\ee
and recall that
\be
\delta_{qb}\delta_{td}=\frac{1}{2}\delta_{qd}\delta_{tb}+\frac{1}{2}\sigma^{\ell}_{qd}\sigma^{\ell}_{tb}.
\label{delta}
\ee
Let us consider the first term in~\eqref{delta} and plug it into~\eqref{ops}
\be
\frac{1}{2}\sigma^2_{sa}\sigma^i_{ad}\quad \sigma^2_{rb},
\label{xprimores}
\ee
therefore forcing the $c\bar c$ pair to be spin 1 and $q\bar q$ to be spin 0. Strong interactions are not supposed to change the heavy spin, thus we may assume that  the color octet components of the $c\bar c$ will maintain spin 1 configuration whereas light quarks can rearrange their spins, when in the octet configuration (twice as probable as the singlet one; see Eq.~\eqref{2octet}), in such a way to fulfill decay quantum number conservation laws.

Consider now the second term in~\eqref{delta} and plug it into~\eqref{ops} to obtain
\be
\frac{1}{2}\sigma^2_{sa}\sigma^i_{aq}\sigma^\ell_{qd}\quad \sigma^2_{rt}\sigma^\ell_{tb}.
\label{ssig0}
\ee
Here we use that
\be
(\sigma^i\sigma^\ell)_{ad}=\delta^{i\ell}\delta_{ad}+i\epsilon^{i\ell m}\sigma^m_{ad}.
\label{ssig}
\ee
Consider the first term in~\eqref{ssig} and plug it back into~\eqref{ssig0} to obtain
\be
\frac{1}{2}\sigma^2_{sd}\quad\sigma^{2}_{rt}\sigma^i_{tb}.
\label{xsecondores}
\ee
This term forces the $c\bar c$ pair to be spin 0 and the $q\bar q$ pair to be spin 1.  Inserting the second term on the right-hand-side of~\eqref{ssig}
into~\eqref{ssig0} we have instead
\be
-\frac{i}{2}\epsilon^{i m\ell }\sigma^2_{sa}\sigma^m_{ad}\quad \sigma^2_{rt}\sigma^\ell_{tb},
\label{xterzores}
\ee
forcing both pairs to be spin 1 and the tetraquark to be  spin 1.   

Here we may introduce the definitions:
\begin{subequations}
\bea
\label{def1}
|1_{\bm q},0_{\bar{\bm q}}\rangle&=&\frac{1}{2}\sigma^2\sigma^i\otimes\sigma^2; \\ 
\label{def2}
|0_{\bm q},1_{\bar{\bm q}}\rangle&=&\frac{1}{2}\sigma^2\otimes\sigma^2\sigma^i; \\
\label{def3}
|1_{\bm q},1_{\bar{\bm q}}\rangle_{J=1}&=&\frac{i}{2\sqrt{2}}\epsilon^{ijk}\sigma^2\sigma^j\otimes\sigma^2\sigma^k.
\eea
\end{subequations}
With the symbol $\bm q$ we either mean a diquark in the order $cq$ or $\bar c \bar q$ or a quark-antiquark pair in the order 
$c\bar c$ or $q\bar q$. The ordering is relevant. 
The normalizations in~(\ref{def1},\ref{def2},\ref{def3}) are obtained using  that $\langle Q_a|Q_b\rangle=\delta_{ab} ...$, where $Q=c,q$. Therefore, taking for example~\eqref{def3}, we have (summing over repeated indices)
\bea
&&(\delta^{j\lambda}\delta^{k\rho}-\delta^{j\rho}\delta^{k\lambda})\left[(\sigma^2\sigma^j)_{rs}(\sigma^2\sigma^\lambda)_{rs}(\sigma^2\sigma^k)_{tu}(\sigma^2\sigma^\rho)_{tu}\right]=\notag\\
&&=\tr((\sigma^j)^T\sigma^j)\tr((\sigma^k)^T\sigma^k)-\tr((\sigma^j)^T\sigma^k)\tr((\sigma^k)^T\sigma^j)=\notag\\
&&=(2-2+2)(2-2+2)-(2\cdot2 +2\cdot2+2\cdot2)=-8,
\label{ord0}
\eea
therefore we choose the normalization $i/(2\sqrt{2})$.

According to the quark ordering in~\eqref{quarkstring}, the spin 0 component will be
\be
\bar q\,\sigma^2\,\bar c
\label{ord}
\ee
whereas in the definition of a heavy light diquark state, see~\eqref{def1}, might also be $\bar c\sigma^2 \bar q$ (heavy quark on the left). On the other hand:
\be
\bar q\,\sigma^2\,\bar c=-\bar c\,(\sigma^2)^T \bar q=\bar c\,\sigma^2\, \bar q.
\ee
This is not the case if we consider $\bar q\, \sigma^2\sigma^i\,\bar c$: 
\be
\bar q\, \sigma^2\sigma^i\,\bar c=-\bar c\,(\sigma^2\sigma^i)^T\,\bar q=\bar c\, (\sigma^i)^T\sigma^2\,\bar q=-\bar c\,\sigma^2\sigma^i\,\bar q. 
\label{segno}
\ee

Therefore, putting together~(\ref{xprimores},\ref{xsecondores},\ref{xterzores}), and keeping in mind that the following states are defined up to an overall minus sign (depending on the initial definition of diquark), we obtain
\be
2|1_{cq},0_{\bar c \bar q}\rangle=|1_{c\bar c},0_{q\bar q}\rangle-|0_{c\bar c},1_{q\bar q}\rangle+\sqrt{2}|1_{c\bar c},1_{q\bar q}\rangle_{J=1}.
\label{prima}
\ee
On the other hand,  if we restart from~\eqref{quarkstring} but with $cq$ taken with spin 0 and $\bar q\bar c$ with spin 1 we get
\be
-2|0_{cq},1_{\bar c \bar q}\rangle=|1_{c\bar c},0_{q\bar q}\rangle-|0_{c\bar c},1_{q\bar q}\rangle-\sqrt{2}|1_{c\bar c},1_{q\bar q}\rangle_{J=1}.
\label{seconda}
\ee
Subtracting~\eqref{seconda} from~\eqref{prima} we therefore obtain the result
\be
\frac{|1_{cq},0_{\bar c \bar q}\rangle+|0_{cq},1_{\bar c \bar q}\rangle}{\sqrt{2}}=|1_{c\bar c},1_{q\bar q}\rangle_{J=1}\equiv X.
\label{ics}
\ee
Since diquarks are defined to be positive parity states, overall we have  $J^{P}=1^+$ and   $C=+1$. This diquark-antidiquark arrangement is a natural candidate to 
describe the $X(3872)$, which is a $1^{++}$ resonance decaying into $\jpsi+\rho/\omega$, compatibly  with the $|1_{c\bar c},1_{q\bar q}\rangle_{J=1}$
assignment -- especially for what concerns the heavy spin. 

If we had started in~\eqref{quarkstring} with the ordering
\be
c_s \quad q_b\quad \bar c_r\quad \bar q_d 
\label{secondpar}
\ee
then in place of~\eqref{prima} and~\eqref{seconda} (exchange labels $\bar c\leftrightarrow \bar q$) we would have obtained:
\begin{subequations}
\bea
\label{xsec}
2|1_{cq},0_{\bar c \bar q}\rangle&=&|1_{c\bar q},0_{q\bar c}\rangle-|0_{c\bar q},1_{q\bar c}\rangle+\sqrt{2}|1_{c\bar q},1_{q\bar c}\rangle_{J=1}; \\
2|0_{cq},1_{\bar c \bar q}\rangle&=&|1_{c\bar q},0_{q\bar c}\rangle-|0_{c\bar q},1_{q\bar c}\rangle-\sqrt{2}|1_{c\bar q},1_{q\bar c}\rangle_{J=1}.
\label{xsec2}
\eea
\end{subequations}
or
\be
\frac{|1_{cq},0_{\bar c \bar q}\rangle+|0_{cq},1_{\bar c \bar q}\rangle}{\sqrt{2}}=\frac{|1_{c\bar q},0_{q\bar c}\rangle-|0_{c\bar q},1_{q\bar c}\rangle}{\sqrt{2}},
\ee
 which is compatible with the $DD^*$ decay mode of the $X(3872)$. Anyway, light quark spins in $Q\bar q$ or $\bar Q q$ configurations 
 might rearrange also to allow $DD$ or $D^*D^*$ decays although the latter is phase space forbidden and the former is simply forbidden by quantum numbers.  

\subsubsection{The $Z$ tetraquark}
The orthogonal combination to the lhs of~\eqref{ics}  might be formed, namely:
\be
Z=\frac{|1_{cq},0_{\bar c \bar q}\rangle-|0_{cq},1_{\bar c \bar q}\rangle}{\sqrt{2}}.
\ee
This has $J^P=1^+$ and $C=-1$ for the neutral component (if an isospin triplet is to  be considered, the $G-$parity has to be $G=+1$). Using~\eqref{prima}
and~\eqref{seconda} we obtain 
\be
Z=\frac{|1_{cq},0_{\bar c \bar q}\rangle-|0_{cq},1_{\bar c \bar q}\rangle}{\sqrt{2}}=\frac{|1_{c\bar c},0_{q\bar q}\rangle-|0_{c\bar c},1_{q\bar q}\rangle}{\sqrt{2}}.
\ee
The state  in the quark-antiquark basis has $C=-1$ since $C=(-1)^{L+s_{q\bar q}+s_{c\bar c}}$. In the quark-antiquark basis there is another state with $C=-1$, orthogonal to $Z$
\be
Z^\prime=\frac{|1_{c\bar c},0_{q\bar q}\rangle + |0_{c\bar c},1_{q\bar q}\rangle}{\sqrt{2}}.
\label{zprime0}
\ee
From what just found (reversing the reasoning leading to Eq.~\eqref{ics}) this state, in the diquark-antidiquark basis, corresponds to
\be
Z^\prime=|1_{cq},1_{\bar c \bar q}\rangle_{J=1},
\label{zprime}
\ee
which is indeed a $1^{+-}$ state. Exchanging the coordinates, spins and charges of two fermions/bosons having each spin $s$, the total wavefunction has to be completely 
antisymmetric/symmetric under this exchange:
\be
(-1)^L(-1)^{2s+S}C=\mp 1,
\ee
which in the case of~\eqref{zprime} is
\be
(-1)^0(-1)^{2+1}C=+1,
\ee
giving $C=-1$. The case of $X=|1_{c\bar c},1_{q\bar q}\rangle_{J=1}$ is different as the charge conjugation operator concerns the distinct $c\bar c$ and $q\bar q$ pairs. 

Linear combinations of $Z$ and $Z^\prime$ which diagonalize the spin-spin Hamiltonian can be  identified with $Z(3900)$ and $Z(4020)$.

 If on the other hand we had started with~\eqref{secondpar}, using~(\ref{xsec},\ref{xsec2}), we would have found 
 \be
 Z=\frac{|1_{cq},0_{\bar c \bar q}\rangle-|0_{cq},1_{\bar c \bar q}\rangle}{\sqrt{2}}=|1_{c\bar q},1_{q\bar c}\rangle_{J=1},
 \label{zprec}
 \ee
suggesting a $D^*D^*$ decay for the color singlet component, which is pahse-space suppressed for the $Z(3900)$.
Again, light quarks might rearrange their spins and decay into $DD^*$, so that nothing prevents us to assign $Z=Z(3900)$.

Similarly we obtain that (exchange $q\leftrightarrow \bar q$ in~(\ref{xsec},\ref{xsec2}) or simply in~\eqref{zprec})
 \be
 Z^\prime=|1_{cq},1_{\bar c \bar q}\rangle_{J=1}=\frac{|1_{c\bar q},0_{q\bar c}\rangle + |0_{c\bar q},1_{q\bar c}\rangle}{\sqrt{2}},
 \ee
apart from an overall minus sign (from~\eqref{segno}); we will anyway assign $Z^\prime=Z(4020)$  which might 
preferably decay into $D^*D^*$ rearranging light quark spins. 
  
\subsubsection{Scalar and Tensor states}
The diquark-antidiquark model also allows for $J^P=0^+,2^+$ states with $C=+1$. We have two $J^P=0^+$ states and a tensor one:
\begin{subequations}
\bea
X_0&=&|0_{cq},0_{\bar c\bar q}\rangle;\\
X_0^\prime &=&|1_{cq},1_{\bar c\bar q}\rangle_{J=0};\\
X_2 &=&|1_{cq},1_{\bar c\bar q}\rangle_{J=2},
\eea 
\end{subequations}
which are all charge-conjugation even. We use the definitions
\begin{subequations}
\bea
\label{zeroz}
|0_{\bm q},0_{\bm q}\rangle&=&\frac{1}{2}\sigma^2\otimes \sigma^2; \\
\label{zero}
|1_{\bm q},1_{\bm q}\rangle_{J=0}&=&\frac{1}{2\sqrt{3}}\sigma^2\sigma^i\otimes \sigma^2\sigma^i; \\
\label{duez}
|1_{\bm q},1_{\bm q}\rangle_{J=2}&=&
\frac{1}{2}\left(\sigma^2\sigma^{(i}\otimes \sigma^2\sigma^{j)}-\frac{1}{3}\delta^{ij}\sigma^2\sigma^{\ell}\otimes \sigma^2\sigma^{\ell}\right),
\eea
\end{subequations}
where  $ij$ indices are symmetrized; in the latter equation  (a factor of $1/2$ has to be included in the symmetrization) and the trace is subtracted. 

The normalization $1/\sqrt{N}$ in~\eqref{zero} is chosen in such a way that the square of
\be
\frac{1}{\sqrt{N}}(\sigma^2\sigma^i)_{rs}(\sigma^2\sigma^i)_{rs}=\frac{1}{\sqrt{N}}{\rm Tr}\left[ (\sigma^2\sigma^i)^T(\sigma^2\sigma^i)\right]
\ee
is equal to 1:
\be
\frac{1}{N}\left({\rm Tr}\left[ (\sigma^2\sigma^i)^T(\sigma^2\sigma^i)\right]\right)^2\equiv\frac{1}{N}\sum_i\left({\rm Tr}\left[ (\sigma^i)^T \sigma^i\right]\right)^2=1,
\ee
thus $N=12$. In the $J=2$ case we have for the first term in parentheses~\eqref{duez}:
\be
\frac{2}{4}{\rm Tr}\left[ (\sigma^i)^T \sigma^i\right] {\rm Tr}\left[ (\sigma^j)^T \sigma^j\right]+\frac{2}{4}  {\rm Tr}\left[ (\sigma^i)^T \sigma^j\right] {\rm Tr}\left[ (\sigma^j)^T \sigma^i\right] = 8,
\ee
whereas the second term squared gives $1/9 \times 3\times  12=4$. The crossed term is $-2\times 1/3\times 12=-8$ so that $N=8+4-8=4$. 

Now we  recall that
\be
\frac{1}{2}\bm \sigma_{ad}\cdot \bm \sigma_{cb}+\frac{1}{2}\delta_{ad}\delta_{cb}=\delta_{ab}\delta_{cd}
\label{complet}
\ee
and observe that~\eqref{zeroz} may be written as
\be
\frac{1}{2}(c_s\,\sigma^2_{sa}\delta_{ab}\,q_b)\otimes (\bar q_r\, \sigma^2_{rc}\delta_{cd}\bar c_d),
\ee
which contains $\delta_{ab}\delta_{cd}$. Substituting the completeness relation~\eqref{complet} in place of $\delta_{ab}\delta_{cd}$
in the latter expression we get
\be
X_0=\frac{1}{2}|0_{c\bar c},0_{q\bar q}\rangle-\frac{\sqrt{3}}{2} |1_{c\bar c},1_{q\bar q}\rangle_{J=0},
\ee
where the minus sign arises to preserve the $q\bar q$ ordering (instead of $\bar q q$ -- see discussion before Eq.~\eqref{ord}) whereas the factor of $\sqrt{3}$ 
is introduced in agreement with~\eqref{zero}.

Consider now~\eqref{zero} which might be written as (because of~\eqref{segno})
\be
-\frac{1}{2\sqrt{3}}(c_s\, \sigma^2_{sa}\sigma^i_{ab}\, q_b)\otimes(\bar q_r\sigma^2_{rc}\sigma^i_{cd}\, \bar c_d)
\ee
and make use of the relation 
\be
\frac{3}{2}\delta_{ad}\delta_{cb}-\frac{1}{2}\bm \sigma_{ad}\cdot\bm \sigma_{cb}=\bm \sigma_{ab}\cdot\bm \sigma_{cd},
\ee
which immediately leads to
\be
X_0^\prime=\frac{\sqrt{3}}{2}|0_{c\bar c},0_{q\bar q}\rangle+\frac{1}{2} |1_{c\bar c},1_{q\bar q}\rangle_{J=0},
\ee
up to an inessential overall $-1$ sign.  Considering the conservation of the heavy quark spin, we see that both scalar states found might
decay into a spin 0 or spin 1 charmonium. 

Finally  consider~\eqref{duez}:
\be
\left(c\sigma^2\sigma^iq\right)\otimes\left(\bar c\sigma^2\sigma^j\qbar\right)
\ee
With the usual Fierz transformation, we have
\be
\frac{1}{2}\left(c\sigma^2\sigma^i\cbar\right)\otimes\left(\qbar\sigma^2\sigma^jq\right)+\frac{1}{2}\left(c\sigma^2\sigma^i\sigma^l\cbar\right)\otimes\left(\qbar\sigma^2\sigma^j\sigma^lq\right),
\ee
and symmetrizing and using the Pauli matrices properties, we get
\be
-\left(c\sigma^2\sigma^{(i}\cbar\right)\otimes\left( q\sigma^2\sigma^{j)}\qbar\right)+\frac{1}{2}\delta^{ij}\left(c\sigma^2\cbar\otimes q\sigma^2\qbar\right)+\frac{1}{2}\delta^{ij}\left(c\sigma^2\sigma^m\cbar\right)\otimes \left(q\sigma^2\sigma^m\qbar\right).
\ee
The term proportional to $\delta^{ij}$ in Eq. (148c) after a Fierz transformation cancels the singlet terms in Eq. (160). We conclude that 
\be
|1_{cq},1_{\bar c\bar q}\rangle_{J=2}=|1_{c\bar c},1_{q\bar q}\rangle_{J=2}.
\ee

We summarize these results in \tablename{~\ref{primatab}}.
\begin{table}[h]
\tbl{
We list the states obtained together with possible assignments and decay modes. We refer here to the neutral components. A $G=+1$ parity may be assigned to the $Z,Z^\prime$ particles. Searches by \babar and \belle still exclude a $I=1$ assignment for the $X(3872)$, however a mixed $I=1$ and $I=0$ seems possible as well as very broad charged partners of $X(3872)$.}{
\begin{tabular}{lllll}
    \hline\hline
$J^{PC}$ & $cq\;\bar c\bar q$&$c\bar c\;q\bar q$&Resonance Assig. &Decays\\ \hline
$0^{++}$ & $|0,0\rangle$ & $1/2|0,0\rangle+ \sqrt{3}/2|1,1\rangle_0$ & $X_0 (\sim 3770~\mathrm{MeV})$ & $\eta_c, \jpsi$ + light mesons\\
$0^{++}$ & $|1,1\rangle_0$ &  $\sqrt{3}/2|0,0\rangle-1/2|1,1\rangle_0$ & $X_0^\prime (\sim 4000~\mathrm{MeV})$ & $\eta_c,\jpsi$ + light mesons\\
$1^{++}$ & $1/\sqrt{2}(|1,0\rangle+|0,1\rangle)$ & $|1,1\rangle_1$ & $X_1=X(3872)$& $\jpsi+\rho/\omega$, $DD^*$\\
$1^{+-}$ & $1/\sqrt{2}(|1,0\rangle-|0,1\rangle) $ &  $1/\sqrt{2}(|1,0\rangle-|0,1\rangle) $ & $Z=Z(3900)$ & $\jpsi +\pi,\,h_c/\eta_c+\pi/\rho$\\
$1^{+-}$ & $|1,1\rangle_1$ & $1/\sqrt{2}(|1,0\rangle+|0,1\rangle) $ & $Z^\prime=Z(4020)$ & $\jpsi +\pi,\,h_c/\eta_c+\pi/\rho$\\
$2^{++}$ & $|1,1\rangle_2$ & $|1,1\rangle_2$ & $X_2(\sim4000~\mathrm{MeV})$ & $\jpsi$ + light mesons\\\hline
\hline
    \end{tabular}
\label{primatab}
}
\end{table}

\subsection{Spectrum of $L=0$ states}
We assume that the spin-spin interactions within the diquark shells are dominant with respect to quark-antiquark interactions.
Then the Hamiltonian would be 
\be
H\approx 2 \kappa (\bm S_q\cdot \bm S_c+\bm S_{\bar q}\cdot \bm S_{\bar c}).
\label{ham1}
\ee 
Consider for example
\be
4\bm S_q\cdot \bm S_c |1_{cq},0_{\bar c \bar q}\rangle=\bm \sigma_{(q)}\cdot \bm \sigma_{(c)} |1_{cq},0_{\bar c \bar q}\rangle := \frac{1}{2}(\sigma^j)^T\sigma^2 \sigma^i\sigma^j\otimes\sigma^2,
\ee
where summation over $j$ is understood. The matrix $(\sigma^j)^T$ works on $c$ whereas $\sigma^j$ on $q$. Considering that 
\be
\frac{1}{2}(\sigma^j)^T\sigma^2 \sigma^i\sigma^j\otimes\sigma^2=-\frac{1}{2}(\sigma^2 \sigma^j \sigma^i\sigma^j)\otimes\sigma^2\equiv \frac{1}{2} \sigma^2\sigma^i\otimes \sigma^2=|1_{cq},0_{\bar c \bar q}\rangle,
\ee
where we have used $i\epsilon^{ijk}\sigma^j\sigma^k=i\epsilon^{ijk}i\epsilon^{jk\ell}\sigma^\ell=-2\sigma^i$.
Considering also the antidiquark contribution one readly finds
\be
4\bm S_{\bar q}\cdot \bm S_{\bar c} |1_{cq},0_{\bar c \bar q}\rangle=-3|1_{cq},0_{\bar c \bar q}\rangle,
\ee
thus 
\be
4(\bm S_q\cdot \bm S_c+\bm S_{\bar q}\cdot \bm S_{\bar c}) |1_{cq},0_{\bar c \bar q}\rangle=-2|1_{cq},0_{\bar c \bar q}\rangle
\ee
and
\be
\label{h11}
H|1_{cq},0_{\bar c \bar q}\rangle=-\kappa |1_{cq},0_{\bar c \bar q}\rangle.
\ee
Similarly:
\begin{subequations}
\begin{align}
\label{h12}
H|0_{cq},1_{\bar c \bar q}\rangle=&-\kappa |0_{cq},1_{\bar c \bar q}\rangle; \\
H|1_{cq},1_{\bar c \bar q}\rangle_{J=1}=&\kappa |1_{cq},1_{\bar c \bar q}\rangle_{J=1}.
\end{align}
\end{subequations}
We can also determine
\begin{subequations}
\bea
\label{scal0}
H|0_{cq},0_{\bar c\bar q}\rangle=&-3\kappa|0_{cq},0_{\bar c\bar q}\rangle;\\
H|1_{cq},1_{\bar c\bar q}\rangle_{J=0}=&\kappa |1_{cq},1_{\bar c\bar q}\rangle_{J=0}.
\label{scal0p}
\eea
\end{subequations}

The Hamiltonian~\eqref{ham1} is diagonal in the diquark-antidiquark  basis formed by the $1^{+-}$ states  of \tablename{~\ref{primatab}}:
\be
(H)_{1^{+-}}=\bpm -\kappa & 0\\ 0 & \kappa\epm,
\ee
$|1\rangle=1/\sqrt{2}(|1,0\rangle-|0,1\rangle)$, $|2\rangle=|1,1\rangle_{J=1}$.  This requires $|1\rangle$ 
to be lighter than $|2\rangle$.
Similarly:
\begin{subequations}
\bea
\label{unozz}
(H)_{1^{++}}&=-\kappa;\\
(H)_{2^{++}}&=\kappa,
\label{duezz}
\eea 
\end{subequations}
so that we conclude that $X(3872)$ and $Z(3900)$ are degenerate in first approximation their masses being
twice the diquark mass plus the same spin-spin interaction correction
\be
M(X,Z)=2m_{[cq]}-\kappa.
\ee
The $Z^\prime$ (and the hypothetical tensor state) is instead heavier by a gap of $2\kappa$:
\be
M(Z^\prime)=2m_{[cq]}+\kappa.
\ee

As for the scalar case, from~(\ref{scal0},\ref{scal0p})  we have
\begin{subequations}
\bea
(H)_{0^{++}}=&-3\kappa;\\
(H)_{0^{++\prime}}=&\kappa,
\eea 
\end{subequations}
showing what was anticipated in \tablename{~\ref{primatab}}, \ie that $M(X_0^\prime)\sim M(X_2)$. 

Considering an average mass value between $X(3872)$ and $Z(3900)$ we can solve finding $m_{[cq]}\simeq 1976$~MeV and $\kappa\simeq 67$~MeV, indicating that $M(0^{++})\simeq 3750$~MeV and $M(X_2)\sim M(X_0^\prime)\simeq 4000$~MeV -- for a pictorial representation see \figurename{~\ref{fig:brodsky2}}.

In this scheme we propose that the newly discovered $Z(4430)$ is the first radial excitation of the $Z(3900)$ as $M(Z(4430))-M(Z(3900))\simeq M(\psiprime)-M(\jpsi)$. 

\begin{figure}[htb!]
\centering
\includegraphics[scale=0.8]{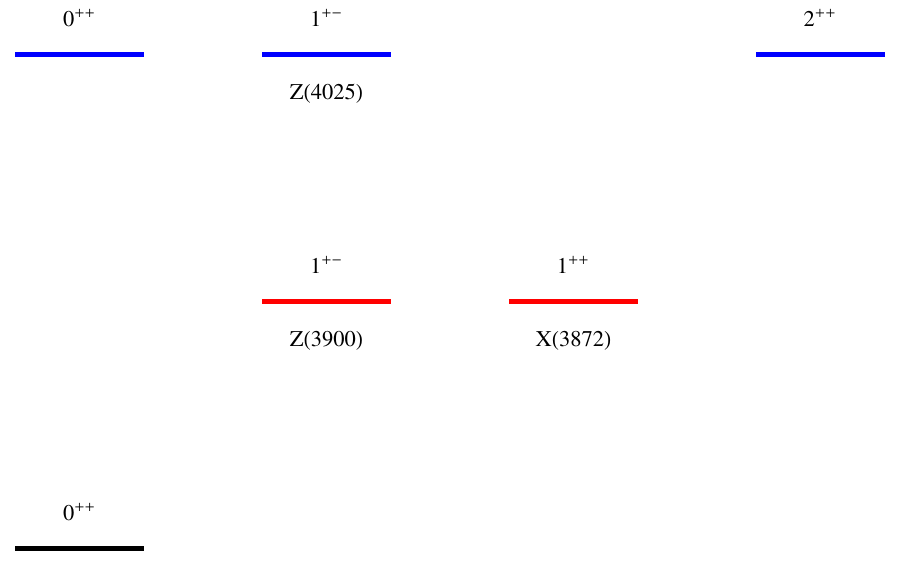}
\caption{The mass pattern dictated by the color-spin Hamiltonian and the construction of states is shown and level degeneracies are highlighted. } \label{fig:brodsky2}
\end{figure}

\subsection{Diquark-antidiquark States with $L=1$}
Tetraquarks with $J^{PC}=1^{--}$ can be obtained with odd values of the angular momentum; here we set $L=1$ and select charge-conjugation odd states.

In the diquark-antidiquark basis of $cq\;\bar c \bar q$ we have:
\begin{subequations}
\bea
\label{y1}
&&Y_1=|0,0\rangle;\quad\quad\quad\quad\;\; C=(-1)^{L=1}\\
\label{y2}
&& Y_2=\frac{|1,0\rangle+|0,1\rangle}{\sqrt{2}};\quad  C=(-1)^{L=1}\\
&& Y_3=|1,1\rangle_{S=0};\quad\quad\quad  (-1)^L (-1)^{2s+S}C=+1\Rightarrow C=(-1)^1(-1)^{2\cdot 1+0}\\
&& Y_4=|1,1\rangle_{S=2};\quad\quad\quad  C=(-1)^1(-1)^{2\cdot 1+2}
\eea
\end{subequations}
Aside from orbital angular momentum considerations we can still make use of \tablename{~\ref{primatab}} to read the $c\bar c$ (conserved) spin; see \tablename{~\ref{secondatab}}.
Observe that the spin structure of $Y_2$ and $X$ in~\eqref{ics} is exactly the same. The mass difference between $Y_2$ and $X$ might entirely be attributed to the orbital excitation of $Y_2$.  The fact that $Y_2$ and $X$ have the same spin structure also suggests that radiative transitions with $\Delta L=1$ and $\Delta S_{c\bar c}=0$ might occur:
\be
Y_2\to \gamma X,
\ee  
as confirmed by the conspicuous radiative decay mode\cite{Ablikim:2013dyn} 
\be
Y(4260)\to X(3872)+\gamma.
\ee
Other transitions are reported in \tablename{~\ref{secondatab}}.

\begin{table}
\tbl{The relative probability of having spin 1 versus spin 0 in the $c\bar c$ pair as read by \tablename{~\ref{primatab}}. Observe that $Y_3$ 
is predicted to decay preferably in $h_c(1P)$ where $S_{c\bar c}=0$.  The states $Y(4290)$ and $Y(4220)$ correspond either to the broad structure in the $h_c$ channel as described by Yuan\cite{yuan1,yuan2} or the narrow one. In this respects the mass ordering can be reversed $Y_3$ becoming lighter than $Y_2$ if the $Y(4220)$ assignement is taken. Radiative decays are suggested by conservation of the heavy quark spin $\Delta S_{c\bar c}=0$ -- see \tablename{~\ref{primatab}}.}
{    \begin{tabular}{cccc}
    \hline\hline
State &$P(S_{c\bar c}=1):P(S_{c\bar c}=0)$& Assignment & Radiative Decay\\ \hline
$Y_1$& 3:1 & $Y(4008)$ & $\gamma +X_0$ \\
$Y_2$&1:0 & $Y(4260)$ & $\gamma+X$\\
$Y_3$&1:3 & $Y(4290)/Y(4220)$ & $\gamma+ X_0^\prime$ \\
$Y_4$&1:0 & $Y(4630)$ & $\gamma+X_2$\\
\hline\hline
    \end{tabular} \label{secondatab}}
\end{table}

The experimentally well established $Y(4360)$ and $Y(4660)$ are interpreted as radial excitations of $Y_1=Y(4008)$ (see \tablename{~\ref{secondatab}}) and $Y_2=Y(4260)$.
We may note correspondences as $M(\chi_{bJ}(2P))-M(\chi_{bJ}(1P))\simeq M(Y(4360))-M(Y((4008)))$ and $M(\chi_{cJ}(2P))-M(\chi_{cJ}(1P))\simeq M(Y(4660))-M(Y((4260)))$. For the identification of the $Y_3$ state as the structures seen in $e^+e^- \to h_c \pi \pi, \chi_{c0} \omega$, see Faccini {\it et al.}\cite{Faccini:2014pma}.

As for the $Y(4630)$, decaying predominantly into $\Lambda\bar \Lambda$, we recall that there is also the possibility of its assignment to a tetraquark 
degenerate with $Y(4660)$\cite{Cotugno:2009ys}.

\subsection{Spectrum of $L=1$ states}
We use the same Hamiltonian form~\eqref{ham1} with the addition of a spin-orbit and a purely orbital term -- here  the chromomagnetic coupling $\kappa^\prime$ is taken to be different from $\kappa$ used in~\eqref{ham1}. We have then: 
\be
H\approx 2 \kappa^\prime (\bm S_q\cdot \bm S_c+\bm S_{\bar q}\cdot \bm S_{\bar c})-2 A\, \bm S\cdot \bm L+\frac{1}{2} B\,\bm L^2,
\label{ham2}
\ee 
in such a way that energy increases for increasing $\bm L^2$ and $\bm S^2$, provided $\kappa^\prime, A, B$ are positive; indeed $2\bm L\cdot \bm S=2-L(L+1)-S(S+1)$ and the masses of $Y$ states will be given by
\be
M=M_0^\prime+\kappa^\prime (s(s+1)+\bar s(\bar s+1)-3)+A(L(L+1)+S(S+1)-2)+B\frac{L(L+1)}{2},
\ee
where $S,\bar S$ are the total spins of diquark and antidiquark. The latter equation can be simplified to
\be
M=M_0+(A+B/2)L(L+1)+AS(S+1)+\kappa^\prime(s(s+1)+\bar s(\bar s+1))
\label{massumm}
\ee
with 
\be
M_0=M_0^\prime-2A-3\kappa^\prime.
\ee
From Eq.~\eqref{massumm}, the mass of the state $Y_1$ in~\eqref{y1} is given by
\be
M_1=M_0+2(A+B/2)
\ee
for $s=\bar s=0$, therefore implying $S=0$, and $L=1$.  The $Y_2$ state in~\eqref{y2} has $s=1$ or $\bar s=1$, thus $S=1$ -- considering that $M_0$ contains $-3\kappa^\prime$ we can determine the mass gap between $Y_2$ and $Y_1$:
\be
\label{dmu}
M_2-M_1=2\kappa^\prime+2A,
\ee
requiring $M_2>M_1$.
The $Y_3$ state has both spins $s=\bar s=1$ but $S=0$ so that 
\be
\label{tmd}
M_3-M_2=2\kappa^\prime-2A,
\ee
which can take either sign depending on $\kappa^\prime-A$ difference; $\kappa^\prime$ and $A$ have in principle similar size.
Finally $Y_4$ has both spins $s=\bar s=1$ and $S=2$ so that 
\be
\label{qmt}
M_4-M_3=6A
\ee
requiring $M_4>M_3$. So the mass ordering is $M_1,M_2,M_3,M_4$ or $M_1,M_3,M_2,M_4$ from lighter to heavier. Using the assignments in \tablename{~\ref{secondatab}} (chosing $Y_3=Y(4290)$), from~(\ref{dmu},\ref{tmd}) we obtain:
\begin{subequations}
\bea
&&4008+2\kappa^\prime+2A=4260;\\
&&4260+2\kappa^\prime-2A=4290,
\eea
\end{subequations}
in units of MeV, giving 
\be
\kappa^\prime=71;\quad\quad\quad A=56.
\ee
If we had chosen $Y_3=Y(4220)$ we would have obtained
\be
\kappa^\prime=53;\quad\quad\quad A=73.
\ee
The values found for  $\kappa^\prime$ have to be compared with the value of $\kappa=67$~MeV  obtained studying the spectrum of $L=0$ states. Both choices are reasonably consistent with it also in consideration of the simplicity of the model described. With respect to the results found in the original paper\cite{Maiani:2004vq}, we can conclude that diquarks in tetraquarks are expected to behave in a different way from diquarks in baryons: in the latter case the coupling $\kappa$ is found to be rather smaller $\kappa\simeq 22$~MeV. 

As a crosscheck, observe that from~\eqref{qmt} we get a reasonable agreement with the assigned mass of $Y_4$:
\be
M(Y_4=Y(4630))=4290(4220)+6\times 56(73)=4626(4658).
\ee

In formula~\eqref{massumm}, the orbital contribution is $2A+B$. Considering that $X$ in~\eqref{ics} and $Y_2$ have the same spin structure, we can conclude that the difference in mass $Y(4260)-X(3872)=2A+B$, giving a value of $B$ in good agreement with what discussed in\cite{ynoi}.

\subsection{Amplitudes in the compact tetraquark model} \label{sec:amptetra}
Although one can use the phenomenological constituent quark model to predict the mass spectrum of the tetraquark states, our lack of knowledge on the actual internal structure of such particles is still almost total. The exact solution of this problem would require to solve a four-body problem, having at least a hint on the nature of the strong potential binding the four constituents. This implies that, until now, we have no methods to compute scattering amplitudes for these states from first principles. 

The typical approach is to gain as much information as possible from available experimental data. In particular, one usually takes into account the kinematics of a decay parameterizing the matrix elements in terms of an \emph{unknown} effective strong coupling times the most general Lorentz-invariant combination of polarization and momenta with the right behaviour under parity and charge conjugation. The effective strong coupling is typically fitted from experimental data when available or, otherwise, it can be estimated by dimensional analysis and under the assumption for it to be of ``natural size''. In the latter case one can clearly just give an order of magnitude estimate of the amplitude considered. 

To be more definite let us make a classic example: the decay $X(3872)\to D^0\bar D^{*0}$. This is a $1^{+}\to 0^-1^-$ strong decay and its matrix element can be parametrized in the following way:
\begin{align}
\langle D^0(q)\bar D^{*0}(k,\lambda)|X(P,\epsilon)\rangle=g_{XDD^*} \epsilon\cdot\lambda,
\end{align}
where $g_{XDD^*}$ is the effective coupling and $\epsilon$ and $\lambda$ are the polarization vectors of the $X(3872)$ and of the $\bar D^{*0}$ respectively. This decay already conserves total angular momentum and parity when happening in $S$-wave. The next parity-conserving contribution to the matrix element would be the $D$-wave one, which however must be proportional to a momentum squared and hence is suppressed by the small $Q$-value for the reaction. Therefore, we want a Lorentz-invariant combination of the available quantities with no momentum dependence. Such combination is clearly just the product of the polarization. The effective coupling can be fitted from the known experimental width for the process considered, obtaining\cite{Faccini:2013lda} $g_{XDD^*}\simeq 2.5$ GeV.

Very recently, an interesting paper by Brodsky \etal\cite{Brodsky:2014xia} proposed a model for the internal dynamics of a tetraquark to compute the effective coupling of the exotic states to quarkonia ($Q\bar Q$). The idea is that after the diquark-antidiquark pair is created, it tends to convert all its kinetic energy into potential energy of the color flux tube until it comes to rest at a relative distance $\bar r$. Such distance must satisfy $V(\bar r)=M-2m_{Qq}$, where $M$ is the mass of the exotic particle, $m_{Qq}$ is the constituent diquark mass and $V(r)$ is the spinless Cornell potential. This essentially means that the mass difference between the exotic meson and its diquark-antidiquark constituents is given by the potential energy at $r=\bar r$. Once $\bar r$ is computed, one can evaluate the quarkonium component of the diquark-antidiquark wave function, \ie the overlap $\langle \psi h|\delta\bar\delta\rangle$, $\psi$ being a generic quarkonium, $h$ a light hadron and $\delta\bar\delta$ the diquark-antidiquark pair. The larger is this component at $\bar r$ the more probable the decay of the exotic state into that particular quarkonium -- see \figurename{~\ref{fig:brodsky}}. In other words, the effective coupling can be taken to be proportional to $\left|\psi_{Q\bar Q}(\bar r)\right|^2$.

\begin{figure}[t]
\centering
\includegraphics[scale=0.6]{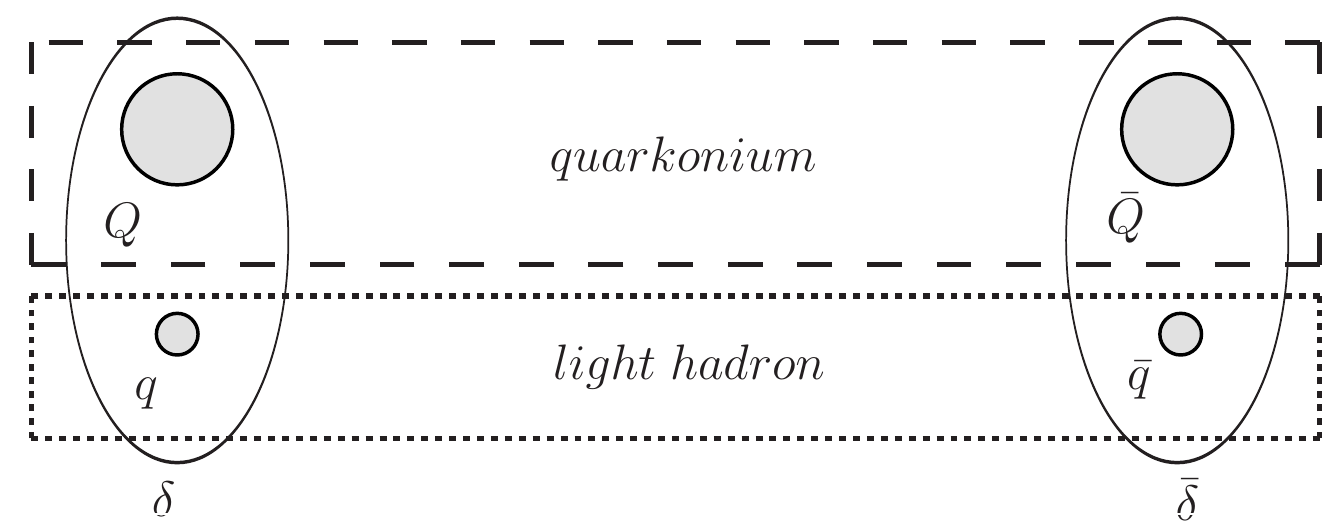}
\caption{Pictorical representation of the overlapping region between the diquark-antidiquark pair and the quarkonium wave function. The larger this overlapping, the more probable the decay.} \label{fig:brodsky}
\end{figure}

\subsubsection{Brief review on QCD sum rules} \label{sec:QCDSR}
Another technique used to compute the mass, widths and coupling constants for these exotic states is to employ the well-known QCD Sum Rules (QCDSR)\cite{Nielsen:2009uh}. As we will see shortly, their use is not limited to tetraquarks only but they can be used also assuming different internal structures or even a mixture of them. 

The method of QCDSR was introduced for the first time by Shifman, Vainshtein and Zakharov\cite{Shifman:1978bx} and used to study the properties of mesons. The main concept\footnote{Here we just review very quickly the main concepts about QCDSR. For a deeper understanding one should refer to a review on the topic\cite{Nielsen:2009uh}.} is based on the evaluation of a two-point correlation function given by
\begin{align}
\Pi(q)\equiv i\int d^4x e^{iq\cdot x}\langle0|T\left(j(x)j^\dagger(0)\right)|0\rangle,
\end{align}
where $j(x)$ is a current with the quantum numbers of the hadron we want to study. The important assumption is that this correlator can be evaluated both at the quark level (the so-called \emph{OPE side}) and at the hadron level (the so-called \emph{phenomenological side}). On the OPE side, as the name suggests, one expands the function as a series of local operators:
\begin{align}
\Pi(q^2)=\sum_n C_n(Q^2) \hat{\mathcal{O}}_n,
\end{align}
with $Q^2=-q^2$ and where the set $\{\hat{\mathcal{O}}_n\}$ includes all the local, gauge-invariant operators that can be written in terms of the gluon and quark fields. As usual, the information about the short-range (perturbative) part of the correlator is contained in the $C_n(Q^2)$. The matrix elements for the operators $\hat{\mathcal{O}}_n$ are non-pertubative and must be evaluated through Lattice QCD or using some phenomenological model.
On the phenomenological side, instead, one writes the two-point function in terms of a spectral density $\rho(s)$:
\begin{align}
\Pi(q^2)=-\int ds\,\frac{\rho(s)}{(q^2-s+i\epsilon)}+\cdots,
\end{align}
with the dots representing subtraction terms. One usually assumes that, over a set of hadrons with certain quantum numbers, the spectral density has a pole correspondent to the mass of the ground-state hadron, while higher mass states are contained in a smooth, continuous part:
\begin{align}
\rho(s)=\lambda^2\delta(s-m^2)+\rho_\text{cont}(s),
\end{align}
$\lambda$ being the coupling of the current to the lowest mass hadron, $H$, $\langle0|j|H\rangle=\lambda$. The main assumption is that in a certain range of $Q^2$ (to be determined) the OPE and phenomenological sides can be matched to extrapolate the values of the mass and width of the hadrons of interest.

The choice of the current $j(x)$ is only dictated by the $(I^G)J^{PC}$ quantum numbers of the hadron and by the assumptions on its internal structure. For example, the currents for a pure $J^{PC}=1^{++}$ tetraquark and molecule can be written as
\begin{subequations}
\begin{gather}
j_\mu^\text{(4q)}=\epsilon_{abc}\epsilon_{dec}\frac{i}{\sqrt{2}}\left(\left(q_a^TC\gamma_5c_b\right)\left(\bar q_d\gamma_\mu C\bar c_e^T\right) + \left(q_a^TC\gamma_\mu c_b\right)\left(\bar q_d\gamma_5C\bar c_e^T\right) \right); \\
j_\mu^\text{(mol.)}=\frac{1}{2}\left(\left(\bar q\gamma_5 c\right)\left(\bar c\gamma_\mu q\right)-\left(\bar q\gamma_\mu c\right)\left(\bar c\gamma_5 q\right)\right),
\end{gather}
\end{subequations}
where $C$ is the charge conjugation matrix and lower case latin letters are color indices. As we mentioned before, one can also build a current for a pure $c\bar c$ state or even take a current for a mixture of these states through a certain mixing angle\cite{Matheus:2009vq}. 

Lastly, one can estimated dacay widths, \ie coupling contants, throught the study of an analogous three-point function. To be definite, let us consider tha decay of the $X(3872)$ into $\jpsi$ plus a vector mesons $V$ (say, $\rho$ or $\omega$). One can compute the coupling constant, $g_{X\psi V}$, for this process using the following correlator:
\begin{align}
\Pi_{\mu\nu\alpha}(p,q)\equiv\int d^4xd^4y e^{ip\cdot x}e^{iq\cdot y}\Pi_{\mu\nu\alpha}(x,y),
\end{align}
with
\begin{align}
\Pi_{\mu\nu\alpha}(x,y)\equiv \langle0|T\left(j_\mu^\psi(x)j_\nu^V(y)j_\alpha^{X\dagger}(0)\right)|0\rangle,
\end{align}
and $j_\mu^\psi$, $j_\nu^V$ and $j_\alpha^X$ are the interpolating currents for the $\jpsi$, the vector meson and the $X(3872)$ respectively.

In Sec.~\ref{sec:comparison} we report some of the prediction made using QCDSR for different exotic states and assuming different mixtures of molecule, tetraquark and charmonium.

\section{Production of exotic states at hadron colliders}
\label{sec:exotic}
Now that we have introduced a large spectrum of possible interpretations for the exotic $XYZ$ states we can describe a couple of circumstances that might give some hint on the real nature of such particles. In particular, we will focus on their production mechanisms, showing how they can give some criteria to distinguish between a compact tetraquark and a loosely bound molecule. 

In Sec.~\ref{subsec:cc} we focus on the potential appearence of exotic states carrying a double flavor charge (\eg $cc$ or $bb$). We show how the production branching fractions and decay widths of these particles are of the right order of magnitude to allow them to be detected by the current experimental facilities. In particular, the spectrum of such states contains doubly charged particle. If they were to be observed that would be almost a full-proof of the existence of compact tetrquarks since, because of the strong Coulomb repulsion, hadronic molecules would be forbidden.

In Sec.~\ref{subsec:HI}, instead, we describe a few models used to predict the production rate of exotic particles, in particular the $X(3872)$, in relativistic heavy ion collisions as those performed at RHIC and LHC. As it will be clear soon, the production cross sections for a molecular states and for a compact tetraquark are expected to be largely different, thus providing a good way to discriminate between the two.

\subsection{Possible production of doubly charmed states} \label{subsec:cc}

The key for the discrimination between the molecular and the compact tetraquark model might be the search for particles with even more exotic properties. It has been pointed out\cite{Moinester:1995tt,DelFabbro:2004ta,Valcarce:2010zs,Esposito:2013fma} that such exotic particles might appear in doubly charmed/bottomed configurations. Because of their peculiar flavor quantum numbers such particles would be clarly distinguished from ordinary mesons and would have a very neat experimental signature. Moreover, as we will show briefly, their spectrum allows the precence of doubly charged states that can only be interpreted in terms of a compact four-quark particle, since the Coulomb repulsion between the two (like-charged) mesons would prevent any possible molecular binding.

In the following section we will focus on doubly charmed states\cite{Esposito:2013fma}. Their existence is predicted within the constituent diquark-antidiquark model\cite{Jaffe:2004ph}:
\begin{align}
\mathcal{T}\equiv [cc][\bar q_1\bar q_2],\hspace{1em} \text{ with } q_1,q_2=u,d,s.
\end{align}
The one-gluon-exchange model suggests that the two quarks (antiquarks) combine in the attractive $\bar{\mathbf{3}}_c$ ($\mathbf{3}_c$) color representation. The total wave function for the diquark (antidiquark) must be completely anti-symmetric because of Fermi statistics. For the $[cc]$ diquark we only have one possibility since the flavor wave function can only be symmetric:
\begin{align}
[cc]=\left|\bar{\mathbf{3}}_c(A), J^P=1^+(S)\right\rangle,
\end{align}
where with $(S)$ and  $(A)$ we indicate the symmetry and anti-symmetry of a configuration. For the light antidiquark, instead, we can have
\begin{subequations}
\begin{align}
\left[\bar q_1\bar q_2\right]_G&=\left|\mathbf{3}_c(A),\mathbf{3}_f(A),J^P=0^+(A)\right\rangle; \\
\left[\bar q_1\bar q_2\right]_B&=\left|\mathbf{3}_c(A),\mathbf{6}_f(S), J^P=1^+(S)\right\rangle,
\end{align}
\end{subequations}
where with the subscript $f$ we indicate the flavour $SU(3)$ group. According to the phenomenological color-spin Hamiltonian, the ``good'' ($G$) scalar state is expected to be lighter than the ``bad'' ($B$) vectorial state, and hence should be more likely produced.

Combining the $1^+$ diquark with both good and bad antidiquarks one obtains the configurations reported in \tablename{~\ref{tab:Tconf}}. As previously pointed out among those states we can find the very peculiar doubly charged ones. Moreover, while the good states can only be produced with $J^P=1^+$, the bad ones can be found with $J=0,1,2$, although one expects the scalar configuration to be the lighter and, hence, more probable one.

\begin{table}[h]
\centering
\tbl{Expected $\mathcal{T}$ states. $A$ and $S$ stand for the anti-symmetric and symmetric flavor configurations. Quantum numbers in {\color{red} \textbf{red}} are the most likely produced.} 
{\begin{tabular}{c|c}
\hline\hline
\multicolumn{2}{c}{$\mathcal{T}$ states} \\
\hline
``Good'', {\color{red} $\mathbf{1}^+$} & ``Bad'', {\color{red} $\mathbf{0}^+$}, $1^+$, $2^+$ \\
\hline
$\mathcal{T}^+$ $\left([cc][\bar u\bar d]_A\right)$ & $\mathcal{T}^0$ $\left([cc][\bar u\bar u]\right)$ \\
$\mathcal{T}^+_s$ $\left([cc][\bar u\bar s]_A\right)$ & $\mathcal{T}^{++}$ $\left([cc][\bar d\bar d]\right)$ \\
$\mathcal{T}^{++}_s$ $\left([cc][\bar d\bar s]_A\right)$ & $\mathcal{T}^{++}_{ss}$ $\left([cc][\bar s\bar s]\right)$ \\
& $\mathcal{T}^+$ $\left([cc][\bar u\bar d]_S\right)$ \\
& $\mathcal{T}^+_s$ $\left([cc][\bar u\bar s]_S\right)$ \\
& $\mathcal{T}^{++}_s$ $\left([cc][\bar d\bar s]_S\right)$ \\
\hline\hline
\end{tabular} \label{tab:Tconf}}
\end{table}

\vspace{1em}

The allowed decay channels for doubly charmed tetraquarks depend crucially on whether or not their masses lie above the open charm threshold. Many analyses\cite{Moinester:1995fk,DelFabbro:2004ta,Valcarce:2010zs,Hyodo:2012pm,Bicudo:2012qt} have been studying the case in which $\mathcal{T}$ particles are below threshold, one of the reasons being that, under this assumption, they would be stable against the (flavor conserving) strong and electromagnetic interactions, hence favoring lattice studies -- see also the Large-$N$ discussion in \sectionname{~\ref{sec:narrow}}. However, in this case, the weak decay channels would present a too complicated pattern, making an experimental analysis very challenging. Since we are interested in the possible detection of these multi-quark states in hadronic colliders we will assume that they lie above the open-charm threshold\cite{Esposito:2013fma,Guerrieri:2014nxa}. Also, for a matter of simplicity, we will focus our study on the $\mathcal{T}^{++}_s$, the extension to the other states being straightforward. In \tablename{~\ref{tab:Tdecay}} we report the $S$-wave decay channels into the lightest $0^+$ and $1^+$ open charm mesons ($P$-wave decays are forbidden by partity conservation).

\begin{table}[h]
\centering
\tbl{Possible $\mathcal{T}_s^{++}$ decay channels. The configurations in {\color{red}\textbf{red}} are the most likely ones. The $1^+$ bad configuration cannot decay into a vector-vector state because of heavy quark spin conservation.}
{\begin{tabular}{cccc}
\hline\hline
\multicolumn{4}{c}{$\mathcal{T}_s^{++}$ decays} \\
\hline
{\color{red} $\mathbf{0}^+$ \textbf{bad}} & {\color{red} $\mathbf{1}^+$ \textbf{good}} & $1^+$ bad & $2^+$ bad \\
\hline
$D_s^+D^+$ & $D_s^{*+}D^+$ & $D_s^{*+}D^+$ & $D_s^{*+}D^{*+}$ \\
$D_s^{*+}D^{*+}$ & $D_sD^{*+} $ & $D_sD^{*+} $ & \\
& $D_s^{*+}D^{*+}$ & & \\
\hline\hline
\end{tabular} \label{tab:Tdecay}}
\end{table}

As explained in Sec.~\ref{sec:amptetra}, the decay amplitudes can be parametrized in terms of a color Fierz coefficient, a kinematical term and an unkonwn strong effective coupling, $g_\mathcal{T}$. The lack of theoretical understanding on the internal structure of tetraquarks makes impossible to exactly compute the value of this coupling. However, one can obtain an order of magnitude estimate by setting $g_\mathcal{T}\simeq M_\mathcal{T}$ by dimensional analysis and under the assumption for the coupling to be of ``natural'' size. Another possible choice could be to set $g_\mathcal{T}$ to be the same as in the $X(3872)\to D^0D^{*0}$ case -- which can be estimated from experimental data\cite{Faccini:2013lda} -- that is $g_\mathcal{T}\simeq g_{XDD^*}=2.5$ GeV. In \figurename{~\ref{fig:Tdecay}} we report the computed decays widths as a function of the $\mathcal{T}^{++}_s$ mass for both good and bad states and for both choices of the coupling. It is worth noting that the value of $g_\mathcal{T}$ does not change the order of magnitude of such widths. Moreover, they are narrow enough to be experimentally measured in the present hadronic experimental facilities. 

\begin{figure}[h]
\centering
\includegraphics[width=.45\textwidth]{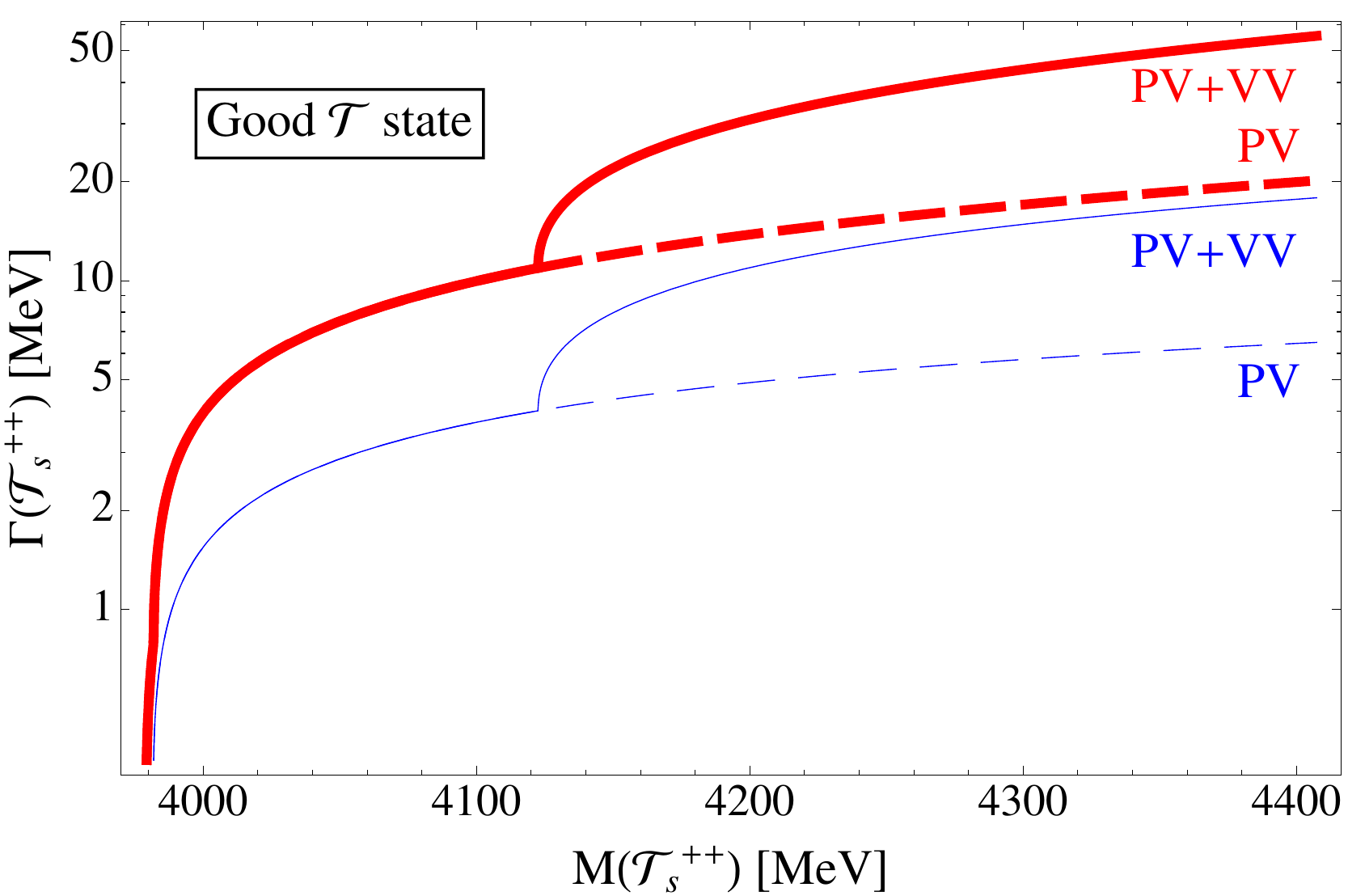}
\includegraphics[width=.45\textwidth]{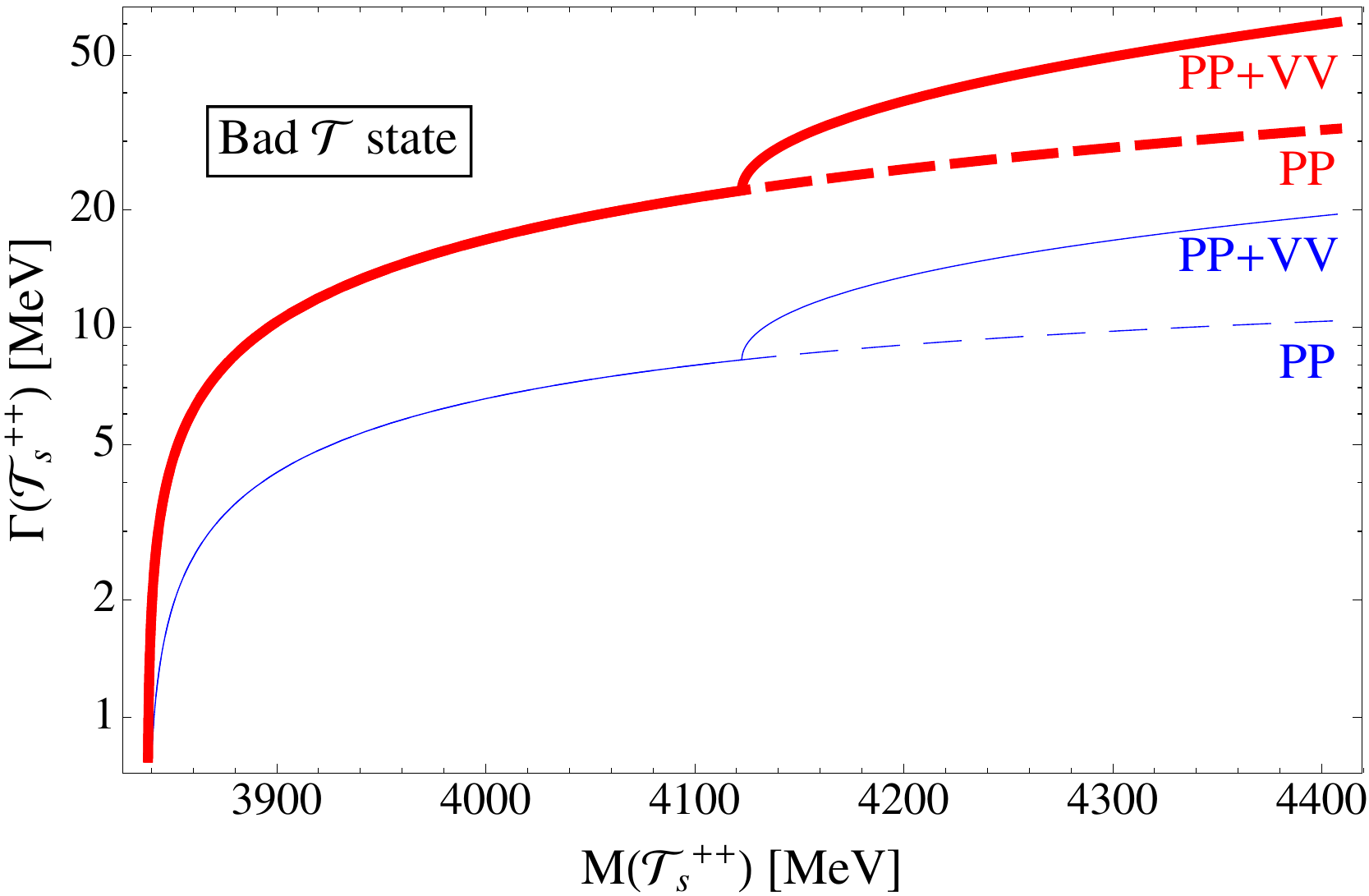}
\caption{Width of good $1^+$ (left panel) and bad $0^+$ (right panel) $\mathcal{T}_s^{++}$ as a function of the mass for both $g_\mathcal{T}= M_\mathcal{T}$ (red thick) and $g_\mathcal{T}=2.5$ GeV (blue thin), from Esposito~\etal\cite{Esposito:2013fma}. With $P$ and $V$ we indicate the $D_{(s)}$ and $D_{(s)}^*$ final states respectively.} \label{fig:Tdecay}
\end{figure}

We can now turn on the study of the production of these particles. They could be created both promptly from the main partonic interaction or from the decay of some other particle. The prompt production has been studied as a three-step process\cite{DelFabbro:2004ta}:
\begin{enumerate}
\item \emph{Creation of a $cc$ pair from the main interaction.} The two analysed possibilities are the single parton interaction, dominated by gluon-gluon fusion $gg\to c\bar cc\bar c$, and double parton interaction, dominated by $(gg)+(gg)\to (c\bar c)+(c\bar c)$, where the two dinstinct interactions occur in the same hadronic event. These proceses are dominant in the small transverse momenta region and the presented results are computed in that range. In particular, the cross section for the production of a $cc$ pair has been calculated\cite{DelFabbro:2004ta} for quarks with relative momentum $|p_{1i}-p_{2i}|<\Delta$, with $i=x,y,z$. The result as a function of $\Delta$ is shown in \figurename{~\ref{fig:ccprod}}. The chosen kinematical cuts for the different experiments are: ($\sqrt{s}=14$ TeV, $1.8<\eta<4.9$) for LHCb, ($\sqrt{s}=14$ TeV, $|\eta|<0.9$) for ALICE, ($\sqrt{s}=1.8$ TeV, $|y|<1$) for Tevatron and ($\sqrt{s}=200$ GeV, $|\eta|<1.6$) for RHIC.

\begin{figure}[t]
\centering
\includegraphics[width=.5\textwidth]{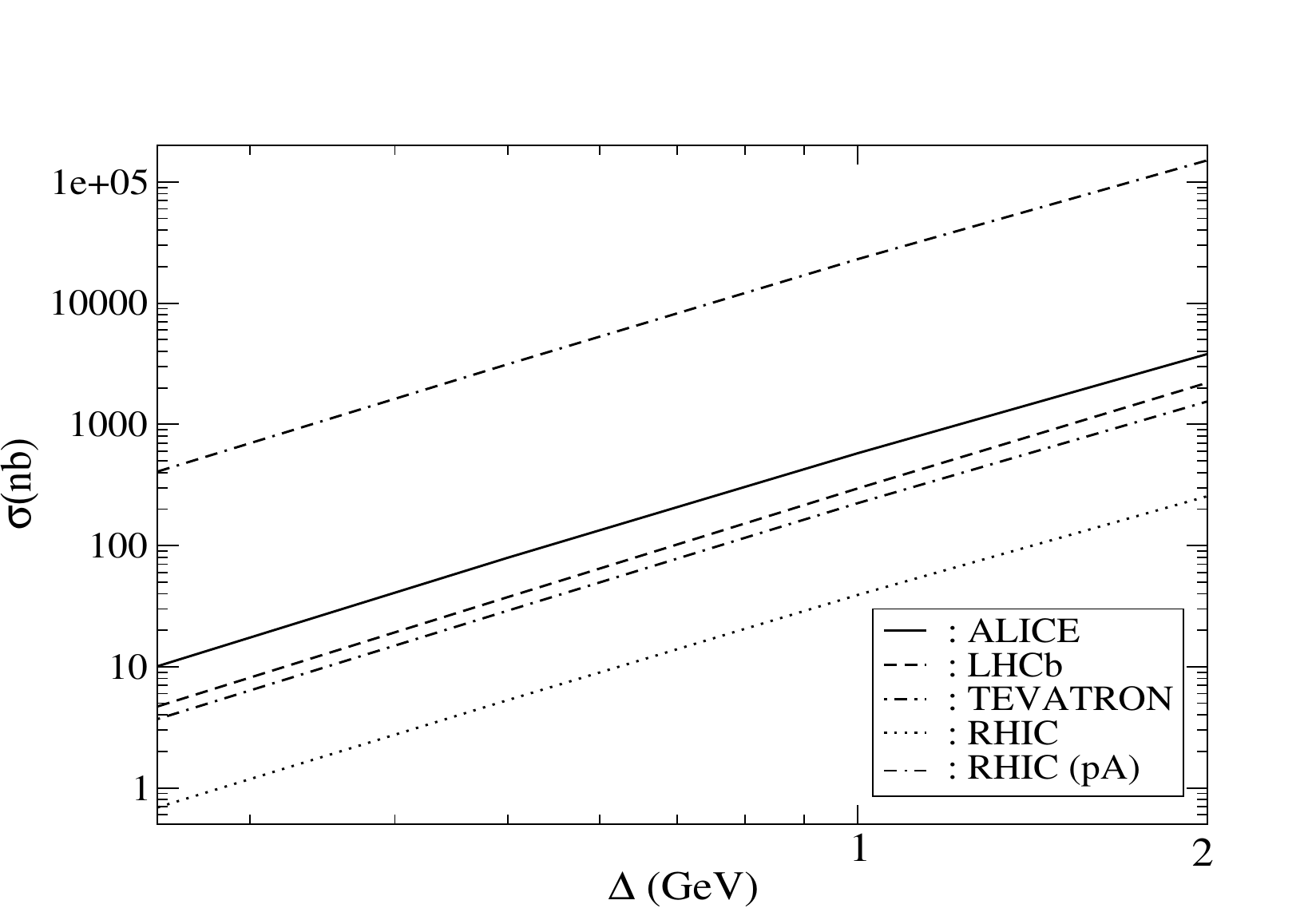}
\caption{Estimated production cross section of two $c$ quarks in momentum space $\Delta$ for different experimental facilities at LHC, Tevatron and RHIC. From Del Fabbro~\etal\cite{DelFabbro:2004ta}.} \label{fig:ccprod}
\end{figure}

\item \emph{Binding of the two charm quarks into a diquark.} To compute that, one can consider the overlap between the two quarks wave function with the diquark one. In particular, the wave function for the two quarks can be taken to be gaussian and expressing it as a function of the relative, $\vett{r}$, and of the center-of-mass, $\vett{R}$, coordinates one gets, aside from a normalization factor:
\begin{align}
\psi_{cc}(\vett{r},\vett{R})\propto e^{-\vett{R}^2/2(B/\sqrt{2})^2+i\vett{P}\cdot\vett{R}}e^{-{(\vett{r}-\vett{r}_a)}^2/2(B\sqrt{2})^2+i\vett{p}\cdot\vett{r}},
\end{align}
with the ``oscillator parameter'' being $B=0.69$ fm\cite{DelFabbro:2004ta}. Moreover, $\vett{r}_a=1$ or 0 fm depending if we are dealing with a proton-nucleus or a proton-proton collision.

Approximating the diquark wave function with a gaussian with an oscillator parameter $\beta=0.41$ fm, one gets an amplitude for the conversion of the $cc$ pair into a diquark equal to
\begin{align}
\mathcal{M}(p)\propto \int d^3r\, e^{-{(\vett{r}-\vett{r}_a)}^2/2{(B\sqrt{2})}^2-i\vett{p}\cdot\vett{r}} e^{-\vett{r}^2/2\beta^2},
\end{align}
while the cross section is given by
\begin{align}
\sigma(cc\to[cc])\simeq \frac{1}{4}\frac{d\sigma_{cc}}{d^3p}\left(\frac{2\sqrt{\pi}}{\sqrt{2B^2+\beta^2}}\right)^3e^{-r_a^2/2B^2},
\end{align}
where $d\sigma_{cc}/d^3p$ is the (approximately constant) cross section in \figurename{~\ref{fig:ccprod}}.
\item \emph{Dressing of the heavy diquark with two light antiquarks.} Neglecting the possible dissociation of the diquark into a $DD$ pair -- and hence providing an upper estimate for the production of $\mathcal{T}$ particles -- one can assume the probability for ``dressing'' the diquark with a light antidiquark to be 0.1. Such probability has been estimated in analogy with the single heavy quark fragmentation. In particular, it has been assumed to be the same as in the $b\to\Lambda_b$ case at Tevatron\cite{Affolder:1999iq}.
\end{enumerate}

Putting thsese three steps together, the expected yield for the $\mathcal{T}$ particles are 20900, 9700, 600 and 1 events/hour for LHC at luminosity $10^{33}$ cm$^{-2}$s$^{-1}$, Tevatron at luminosity $8\times 10^{31}$ cm$^{-2}$s$^{-1}$ and RHIC at d-Au luminosity $0.2\times10^{28}$ cm$^{-2}$s$^{-1}$, respectively.

\vspace{1em}

However, as we previously mentioned, $\mathcal{T}$ particles might also be produced from the decay of other particles. In particular, it seems reasonable to expect this production to be more likely if from particles that already contain a charm quark. In what follows we will consider the possible production from $B_c$ decays\cite{Esposito:2013fma}. In \figurename{~\ref{fig:prodBc}} we report the Feynman diagram for these decays.

\begin{figure}[t]
\centering
\includegraphics[width=.3\textwidth]{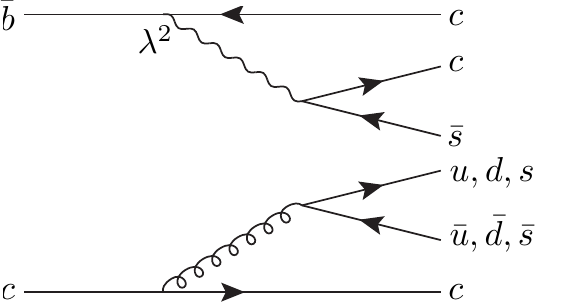}
\includegraphics[width=.3\textwidth]{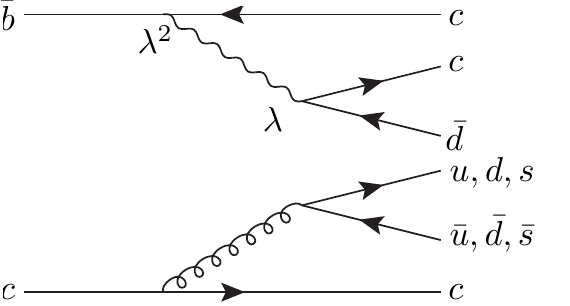}
\includegraphics[width=.3\textwidth]{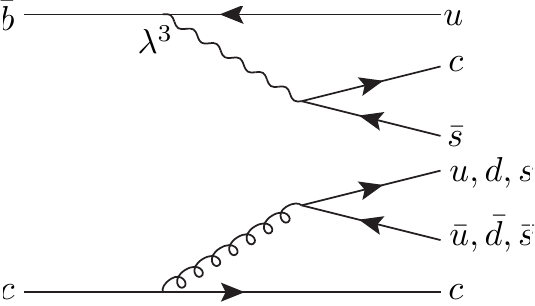}
\caption{Feynman diagrams for the production of $\mathcal{T}$ particles from $B_c^+$. $\lambda=\sin\theta_C$ is the sine of the Cabibbo angle associated to each vertex. From Esposito \etal\cite{Esposito:2013fma}} \label{fig:prodBc}
\end{figure}

We will focus on the $B_c^+\to\mathcal{T}^{++}_sD^{(*)-}$ decay, avoiding the use of specific models. Heavy meson decays into two baryons are particularly suitable to extract the effective strong coupling which we expect to determine also the process we are interested in -- both indeed contain six quarks confined in a two-hadron final state. In particular, one can consider\cite{Aubert:2008ax,Gabyshev:2002dt,Abe:2004sr}
\begin{subequations} \label{eq:BR}
\begin{align} 
\mathcal{BR}(B^0\to\bar\Lambda_c^-p)=(2.0\pm0.4)\times 10^{-5};\\
\mathcal{BR}(B^+\to\bar\Sigma_c^0p)=(3.7\pm1.5)\times 10^{-5}.
\end{align}
\end{subequations}

These interactions can be described by mean of the following heavy meson chiral Lagrangian\cite{Casalbuoni:1996pg}:
\begin{align}
\mathcal{L}_{eff}=\frac{g_B}{2M_B^2}\partial_\mu B\bar p\gamma^\mu\left(1-\frac{g_A}{g_V}\gamma_5\right)\Lambda,
\end{align}
where $g_B$ is a strong effective coupling and we take $g_A/g_V\simeq1.27$ as for the $\beta$-decay. $\Lambda$ represents both the $\bar\Lambda_c^-$ and the $\bar\Sigma_c^0$, the dynamics of the two processes being the same. Fitting from the experimental data in Eqs.~\eqref{eq:BR} one finds
\begin{subequations}
\begin{align}
g_{B^0}&=(4\pm1)\times 10^{-3} \text{ MeV}; \\
g_{B^+}&=(5\pm3)\times 10^{-3} \text{ MeV},
\end{align}
\end{subequations}
which are compatible within the errors, thus suggesting that the internal dynamics might indeed be similar. 

Extending this assuption to the $B_c^+\to\mathcal{T}^{++}_sD^{(*)-}$ decay we can take the effective coupling for this case, $g_{B_c^+}$, to be the average of the previous two. The decay amplitudes can again be parametrized in terms of color structure, kinematics and the effective coupling. In \figurename{~\ref{fig:produzione}} we report the obtained results for both good and bad states and for a production associated with both a $D^-$ and a $D^{*-}$. One can notice that, if the $\mathcal{T}$ is near threshold, than the branching ratio for an $S$-wave production is just one order of magnitude smaller of the observed $B_c^+\to \jpsi D_s^{(*)+}$ decays\cite{Aaij:2013gia}. 

\vspace{1em}

Summarizing, the theoretical study and the experimental search for possible exotic mesons with double flavor quantun numbers might be an interesting idea to further understand the nature of these particles. In particular, such particles might appear with double electric charge, in which case the only possible interpretation would be that of a compact tetraquark. 

We also showed, that their widths and production branching fractions are large enought to be accessible at the present hadron facilities such as LCHb, ALICE, Tevatron and RHIC.

\begin{figure}
\centering
\includegraphics[width=.45\textwidth]{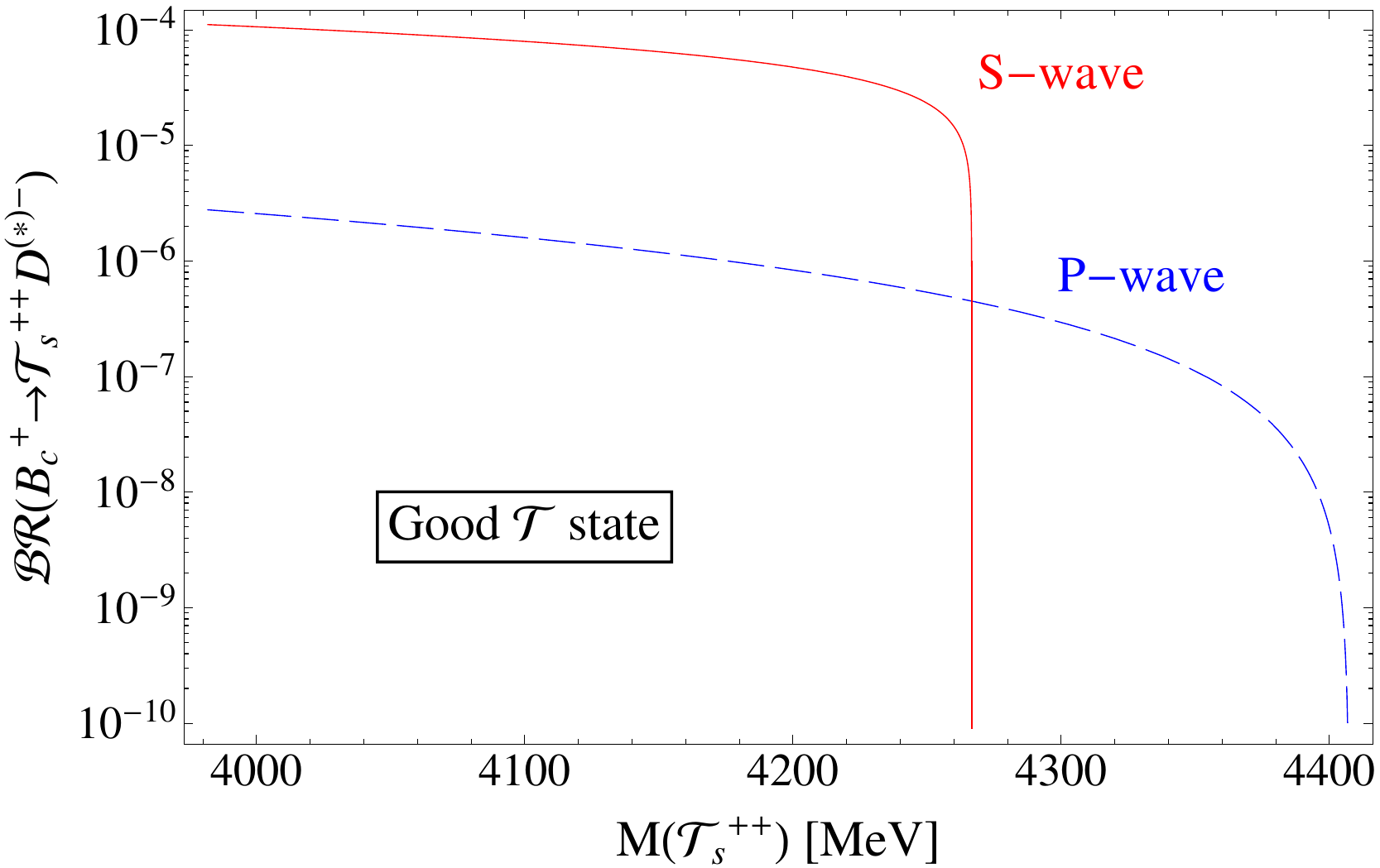}
\includegraphics[width=.45\textwidth]{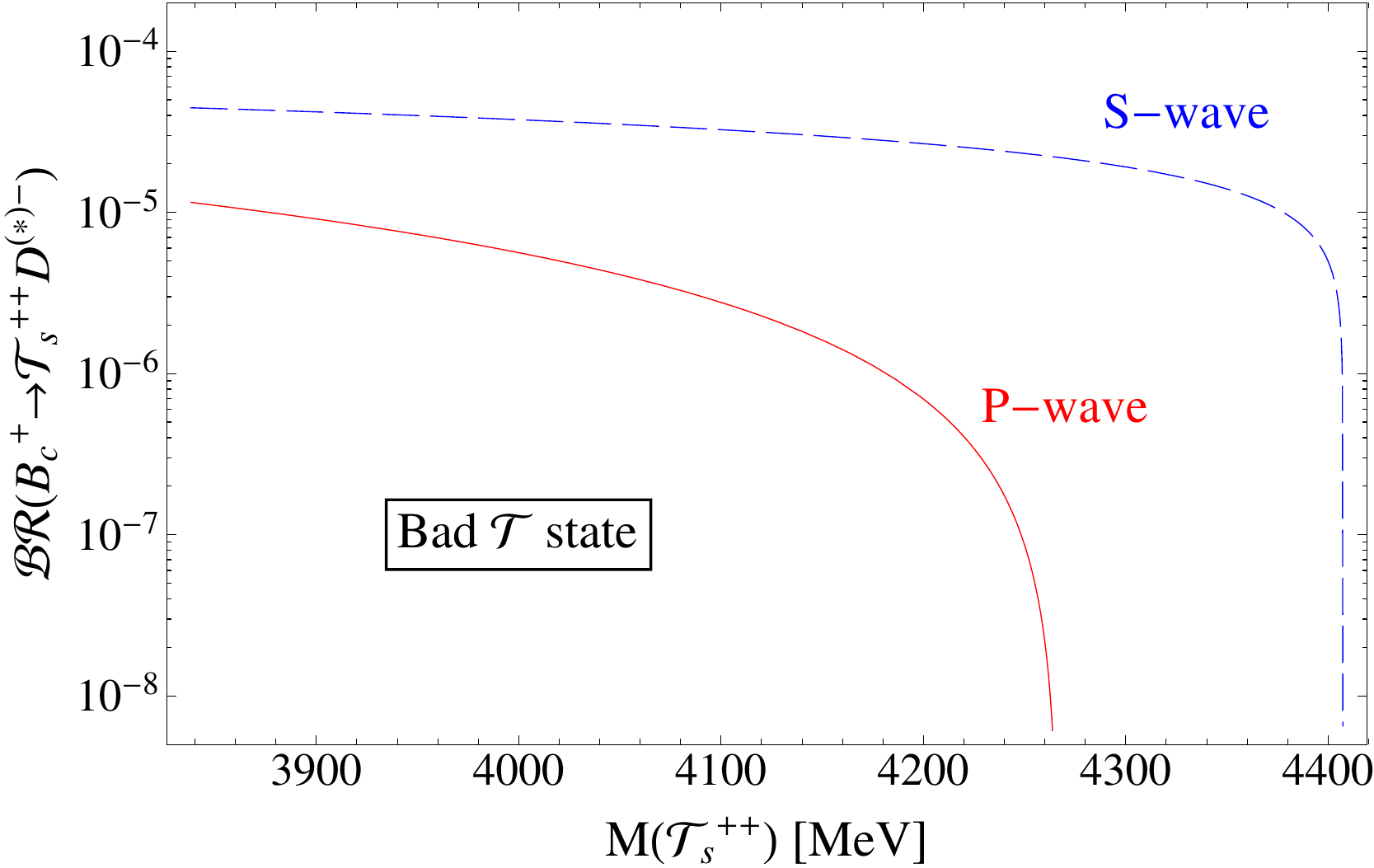}
\caption{Branching ratios for the production of $B_c^+\to \mathcal{T}^{++}_sD^-$ (dashed curve) and $B_c^+\to\mathcal{T}^{++}_sD^{*-}$ (solid curve) for the good $1^+$ state (left panel) and for the bad $0^+$ state (right panel) as a function of the mass of $\mathcal{T}^{++}_s$, in the above-threshold region. From Esposito~\etal\cite{Esposito:2013fma}.} \label{fig:produzione}
\end{figure}

\subsection{Compact tetraquarks and meson molecules in heavy ion collisions} \label{subsec:HI}

Another possible tool to gain some insight about the nature of candidate tetraquarks is to study their behavior in the extreme conditions of relativistic heavy ion collisions at RHIC and LHC.

When two nuclei (Au+Au and Pb+Pb for RHIC and LHC respectively) collide at relativistic speed, the resulting system reaches extremely high temperatures. In particular, if those temperatures are higher than a critical value\cite{Bazavov:2014pvz}, $T_C=(154\pm9)$ MeV, quark and gluons are liberated from hadrons and a new state of matter is created, the so-called Quark-Gluon-Plasma (QGP). It is a QCD plasma of deconfined quarks and gluons which seems to behave as a nearly perfect fluid\cite{Iancu:2012xa,CasalderreySolana:2011us}. After a certain amount of time this ``fire-ball'' expands and cools down to temperatures below $T_C$ and hence the partons confine again. In this phase the system looks like an expanding gas of interacting hadrons, the so-called Hadron Resonance Gas (HRG). When the temperature drops below the so-called freeze-out temperature, $T_F\simeq 120$ MeV\cite{Cho:2010db}, these hadrons simply fly apart without interacting anymore.
In the following we will indicate with the subscripts $C$, $H$ and $F$ quantities at the critical, hadronization and freeze-out temperatures respectively (see for example \tablename{~\ref{tab:param}}).

\vspace{1em}

It has been proposed\cite{Lee:2007tn,Cho:2010db,Cho:2013rpa,Cho:2011ew} that the study of the produced number of exotic mesons, and in particular the time-honored $X(3872)$,  in heavy ion collisions might help to distinguish between the compact tetraquark picture and the molecular one.
In particular, the two main techniques to estimate the yield of a particle in hot QGP are:
\begin{itemize}
\item \emph{The statistical model\cite{Andronic:2005yp}}: it assumes that the matter produced in heavy ion collisions is in thermodynamical equilibrium and it is know to describe the relative yields of ordinary hadrons very well. In this model the number of hadrons of a given type, $h$, produced is given by
\begin{align} \label{eq:Nstat}
N_h^\text{stat}=V_H\frac{g_h}{2\pi^2}\int_0^\infty \frac{p^2dp}{\gamma_h^{-1}e^{E_h/T_H}\pm 1},
\end{align}
with $g_h$ being the degeneracy of $h$ and $V_H$ $(T_H)$ the volume (temperature) of the source when the statistical production of the hadron occurs. $\gamma_h=\gamma_c^{n_c+n_{\bar c}}e^{(\mu_BB+\mu_S S)/T_H}$ is the fugacity, with $n_c$ and $n_{\bar c}$ the number of charm and anti-charm in the hadron, $B$ and $S$ the baryon and strangeness numbers of the hadron and $\mu_B$ and $\mu_S$ the corresponding chemical potentials.

This model does not contain any information about the actual internal structure of $h$ and, in the following, it will be used as a normalization factor.
\item \emph{The coalescence model\cite{Sato:1981ez}}: it is based on the sudden approximation by calculating the overlap of the density matrix for the constituents of the hadron $h$ with the Wigner function for the produced particle. It is built to take into account the inner structure of $h$, such as angular momentum, multiplicity of quarks, etc. This picture has successfully explained many different experimental data (\eg enhancement of baryon production in the intermediate $p_T$ region\cite{Adcox:2001mf}, quark number scaling of the elliptic flow\cite{Adler:2003kt}). In this context, the number of hadrons produced is given by
\begin{align} \label{eq:Ncoal}
N_h^\text{coal}\simeq g_h\prod_{j=1}^n\frac{N_j}{g_j}\prod_{i=1}^{n-1}\frac{{(4\pi\sigma_i^2)}^{3/2}}{V(1+2\mu_i T\sigma_i^2)}\left[\frac{4\mu_i T\sigma_i^2}{3(1+2\mu_iT\sigma_i^2)}\right]^{l_i},
\end{align}
if one uses the non-relativistic approximation, neglect the transverse flow and considers only the unit rapidity. Moreover, one assumes an harmonic oscillator ansatz for the hadron internal structure. Here $g_j$ and $N_j$ are the degeneracy and number of the $j$-th constituent and $\sigma_i=1/\sqrt{\mu_i\omega}$, with $\omega$ the oscillator frequency and $\mu_i$ the reduced mass given by $\mu_i^{-1}=m_{i+1}^{-1}+\left(\sum_{j=1}^i m_j\right)^{-1}$. Lastly $l_i=0,1$ for a $S$-wave and a $P$-wave constituents respectively.

Note that from Eq.~\eqref{eq:Ncoal} follows that hadrons with more constituents are, in general, more suppressed and that S-wave coalescence if favored with respect to the P-wave one.
\end{itemize}

A large part of the information about the nature of the considered hadron is hence somehow embedded in the frequency $\omega$ of the harmonic oscillator. In the case of a compact multiquark state one can fit $\omega$ by requiring the coalescence model to reproduce the reference normal hadron yields in the statistical model. For the case of interest, the $X(3872)$ with light and charm quarks, one finds $\omega_c=385$ MeV by requiring the matching with the yield of $\Lambda_c(2286)$\cite{Cho:2010db}. The final result is a yield $N_X^\text{4q}=4.0\times 10^{-5}$.

For the case of a meson molecule, instead, one can fix $\omega$ by using $\omega=3/(2\mu_1\langle r^2\rangle)$ for a two-body $S$-wave state, together with the equation that relates the binding energy of a loosely bound molecule with its scattering length, $a$, $E\simeq1/(2\mu_1a^2)$ and $\langle r^2\rangle\simeq a^2/2$. For the case of the $X$ one gets $\omega= 3.6$ MeV\cite{Cho:2010db}. It is worth noting that $\omega\propto E$ and hence, according to Eq.~\eqref{eq:Ncoal}, the smaller the binding energy of the molecule, the smaller $\omega$ and thus the larger is the $N_h^\text{coal}$. In this case it turns out to be $N_X^\text{mol}=7.8\times 10^{-4}$.

\begin{figure}
\centering
\includegraphics[scale=0.6]{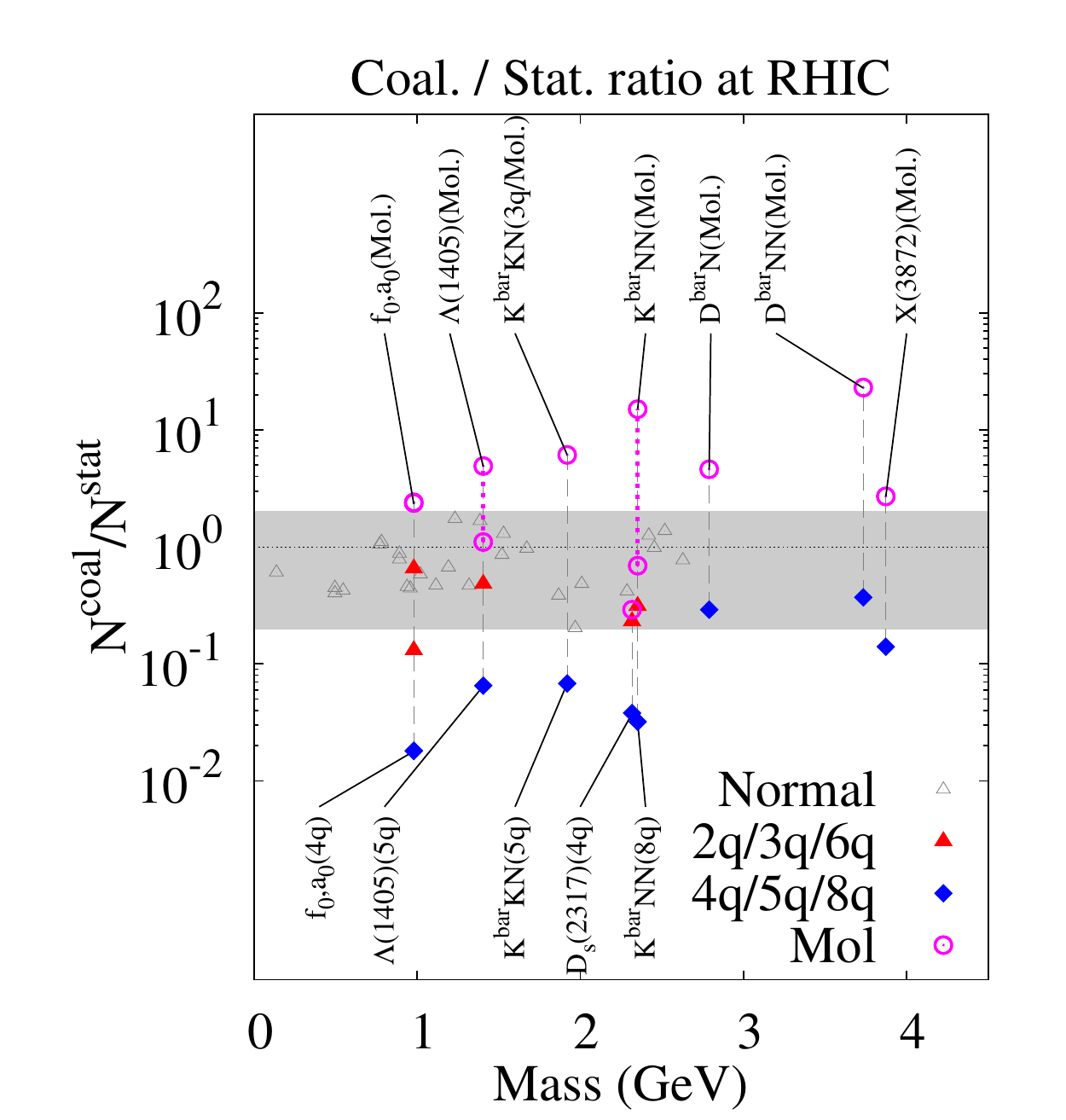}
\caption{Hadron yields in the coalescence model normalized with respect to the statistical one at RHIC, from ExHIC Collaboration\cite{Cho:2010db}. Note the sharp difference between the predictions for a compact four-quark structure and for a molecular structure of the $X(3872)$. The grey band represents the range of yields for ordinary hadrons.} \label{fig:Xcoal}
\end{figure}

In \figurename{~\ref{fig:Xcoal}} the predicted yield for different hadrons in the coalescence model are shown\cite{Cho:2010db}. As one immediately notices, the predictions for the compact tetraquark and for the molecule are completely different. In particular, a molecular structure for the $X(3872)$ implies a yield which is higher than ordinary hadrons, while a compact structure implies a lower yield. This difference is essentially due to the small binding energy of the molecular state and to the high number of constituents of the compact states. There is also another extremely striking feature, \ie the predicted behaviors for the molecule and the compact tetraquark in relativistic heavy ion collisions are opposite to those predicted for $pp$ collisions, as discussed in \sectionname{~\ref{sec:prompt}}. Also note that both yields are close enough to the ordinary ones to be experimentally measured at RHIC and LHC.

\vspace{1em}

The previous description of exotic mesons in heavy ion collision can be further improved\cite{Cho:2013rpa}. So far we only studied the production during the QGP phase. However, the number of exotic mesons can also vary during the HRG phase, when disintegration/creation processes due to the interaction with other particles can occur. In particular, the most effective processes are -- see \figurename{~\ref{fig:Xscatt}}:
\begin{subequations} \label{eq:Xdestr}
\begin{gather}
X\pi\to \bar D^*D^*;\hspace{1em}X\pi\to\bar D D;\hspace{1em} X\rho\to\bar D^* D; \\
X\rho\to D^*\bar D;\hspace{1em} X\rho\to D\bar D;\hspace{1em}X\rho\to D^*\bar D^*,
\end{gather}
\end{subequations}
and the inverse ones for the creation of a $X(3872)$. The vertices for such reaction can be obtained from an effective Lagrangian approach, with a combination of Heavy Quark Effective Theory (HQET) and chiral theory\cite{Cho:2013rpa,Brazzi:2011fq}. In \figurename{~\ref{fig:Xdestr}} we report the cross sections for the processes in Eq.~\eqref{eq:Xdestr}.

\begin{figure}
\centering
\includegraphics[scale=0.45]{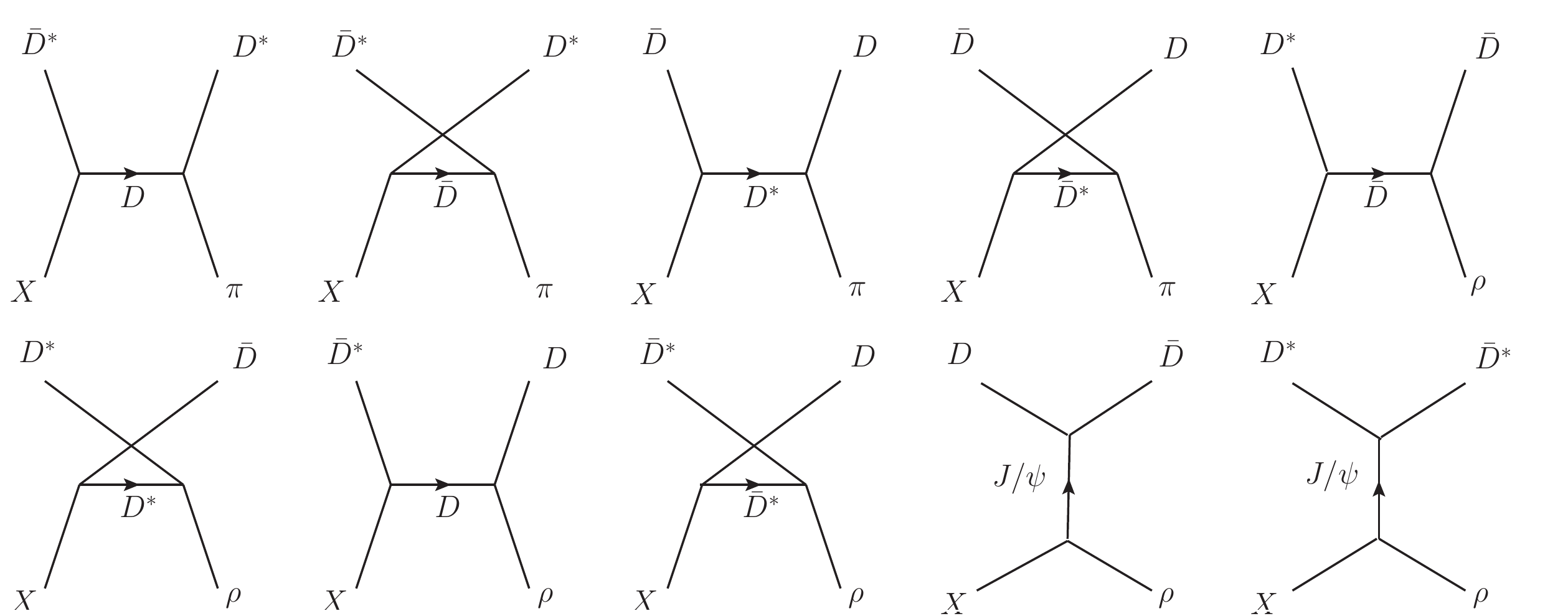}
\caption{Possible disintegration processes of the $X(3872)$ in the hadron resonance gas phase.} \label{fig:Xscatt}
\end{figure}

\begin{figure}
\centering
\includegraphics[width=.5\textwidth]{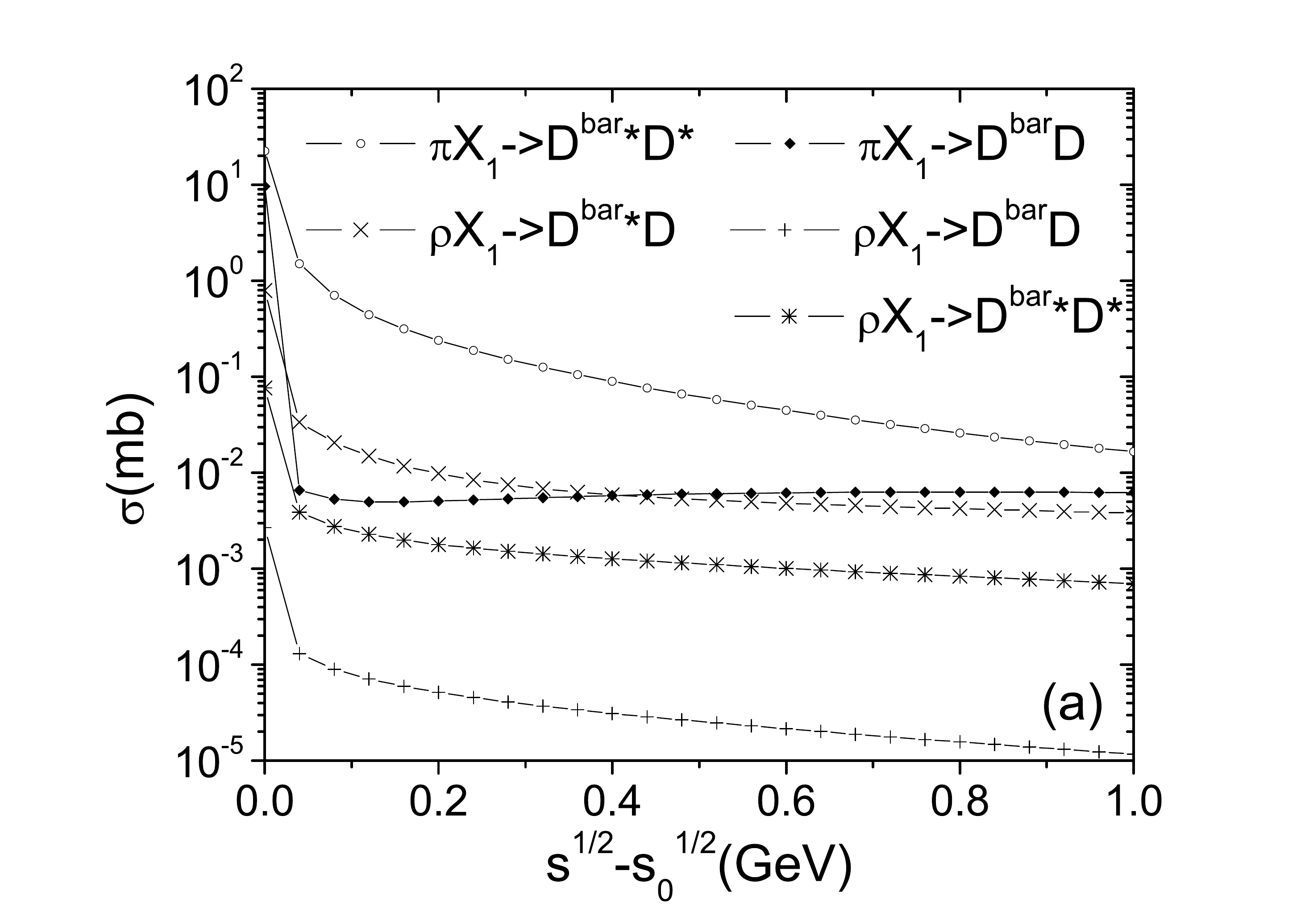}
\caption{Cross sections for the absorption of a $X(3872)$ with $J^{PC}=1^{++}$ as a function of the difference between the total center-of-mass energy $\sqrt{s}$ and the energy threshold for the process $\sqrt{s_0}$. From Cho and Lee\cite{Cho:2013rpa}.} \label{fig:Xdestr}
\end{figure}

The computed cross sections can be used to estimate the change of the number of $X(3872)$ in the HRG as a function of the propert time, $\tau$, by mean of kinetic theory\cite{Cho:2013rpa}:
\begin{align} \label{eq:dNX}
\frac{dN_X}{d\tau}=R_{QGP}(\tau)+\sum_{\ell, c, c^\prime}\left(\left\langle \sigma_{cc^\prime\to \ell X}v_{cc^\prime}\right\rangle n_c(\tau)N_{c^\prime}(\tau)-\left\langle \sigma_{\ell X\to cc^\prime}v_{\ell X}\right\rangle n_\ell(\tau) N_X(\tau)\right),
\end{align}
where the subscripts $\ell$, $c$ and $c^\prime$ stand for a light meson and the two charmed mesons respectively. $n_a(\tau)$ and $N_a(\tau)$ are the density and abundancy of the particle $a$ at proper time $\tau$ calculated using the statistical model Eq.~\eqref{eq:Nstat} with $\tau$-dependent volume and temperature\cite{Cho:2013rpa}:
\begin{align}
V(\tau)&=\pi\left[R_C+v_C\left(\tau-\tau_C\right)+a_C/2\left(\tau-\tau_C\right)^2\right]^2\tau_C, \nonumber\\
T(\tau)&=T_C-\left(T_H-T_F\right)\left(\frac{\tau-\tau_H}{\tau_F-\tau_H}\right)^{4/5}. \label{eq:Bjork}
\end{align}
These equations are obtained following the boost invariant Bjorken picture with an accelerated transverse expansion\cite{Chen:2007zp}. In particular, $R_C$ is the radius of the system at $T_C$, and $v_C$ and $a_C$ are its expansion velocity and acceleration. In \tablename{~\ref{tab:param}} we report the values used for the present analysis.

\begin{table}[h]
\centering
\tbl{Values for the volume and temperature profiles in the schematic model of Eq.~\eqref{eq:Bjork}.} 
{\begin{tabular}{ccc}
\hline\hline
& Temp. (MeV) & Time (fm/c) \\
\hline
$R_C=8.0$ fm & $T_C=175$ & $\tau_C=5.0$ \\
$v_C=0.4c$ & $T_H=175$ & $\tau_H=7.5$ \\
$a_C=0.02c^2/$fm & $T_F=125$ & $\tau_F=17.3$ \\
\hline\hline
\end{tabular}\label{tab:param}}
\end{table}

The averages in Eq.~\eqref{eq:dNX} can be evaluated by mean of the kinetic theory:
\begin{align}
\langle \sigma_{ab\to cd} v_{ab}\rangle=\frac{\int d^3p_ad^3p_b f_a(\vett{p}_a)f_b(\vett{p}_b)\sigma_{ab\to cd}v_{ab}}{\int d^3p_ad^3p_b f_a(\vett{p}_a)f_b(\vett{p}_b)},
\end{align}
with $f_a(\vett{p}_a)$ being the single particle density in momentum space. Lastly, the term $R_{QGP}(\tau)$ is included to take into account the effect of the production of the $X(3872)$ through hadronization from the quark-gluon-plasma and is given by

\begin{align}
R_{QGP}(\tau)=\begin{cases}
N_X^0/(\tau_H-\tau_C),&\tau_C<\tau<\tau_H \\
0,&\text{otherwise}
\end{cases};
\end{align}

$N_X^0$ is the number of $X(3872)$ produced by the quark-gluon-plasma as explained in Eqs.~\eqref{eq:Nstat} and \eqref{eq:Ncoal}. Once all the ingredients are set one can compute the number of $X$ as a function of proper time, \ie of the evolution of the hot expanding system. In \figurename{~\ref{fig:Xevtetra}} we report the results for central Au-Au collisions at $\sqrt{s_{NN}}=200$ GeV.

\begin{figure}
\centering
\includegraphics[width=.75\textwidth]{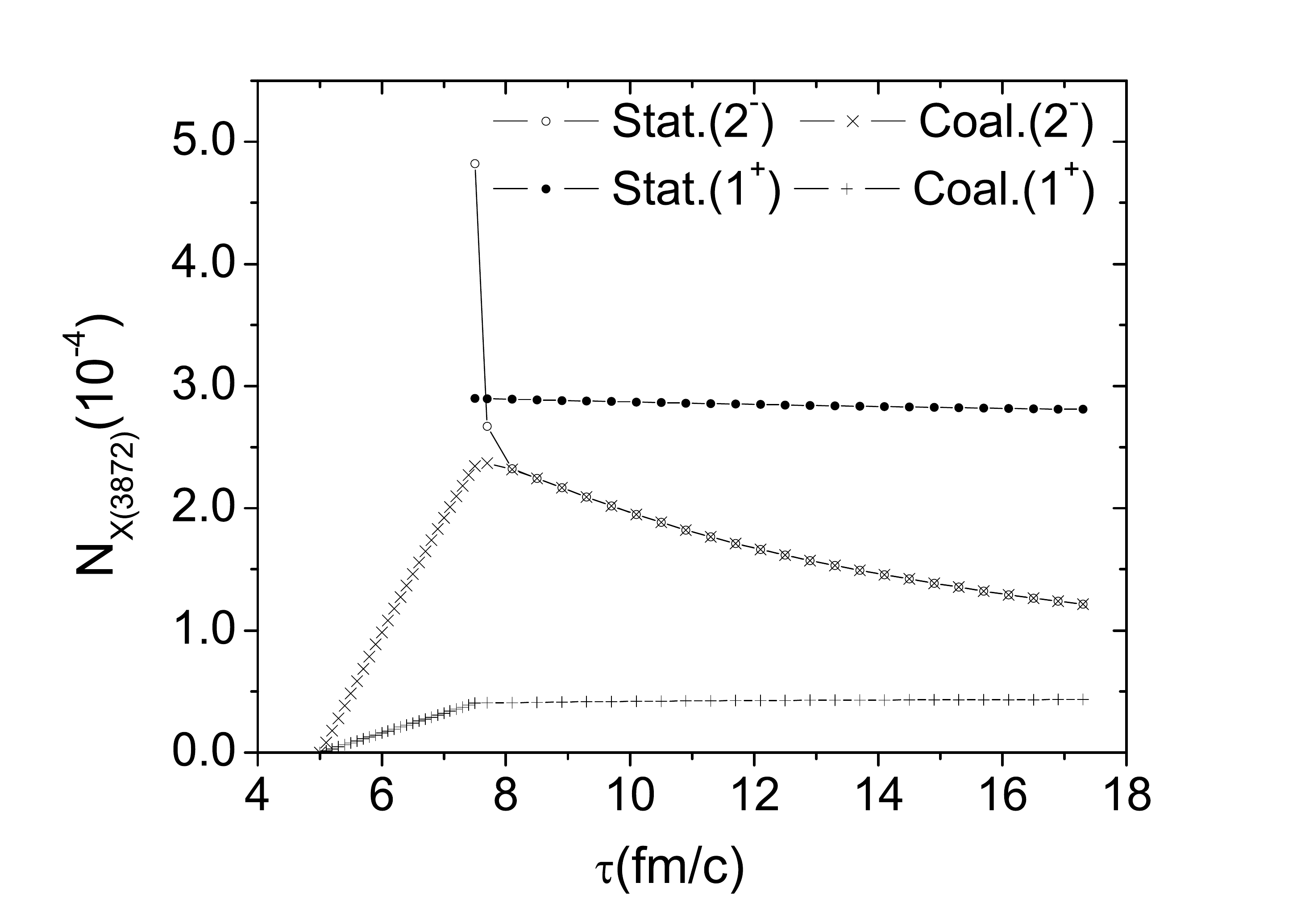}
\caption{Number of $X(3872)$ as a function of proper time under the assuption of a compact tetraquark and of a molecular nature, from Cho and Lee\cite{Cho:2013rpa}. For comparison, the prediction obtained with a pure statistical approach (which is blind to the interal structure) is drawn as well.} \label{fig:Xevtetra}
\end{figure}

The number of $X(3872)$ in the assumption of a $D^0\bar D^{*0}$ molecular nature can be computed by solving the evolution equation~\eqref{eq:dNX} backward in time, starting from the yield found before using the coalescence model, $N_X^\text{mol}=7.8\times10^{-4}$. The result is again shown in \figurename{~\ref{fig:Xevtetra}}.

As one can see, the inclusion of this further possible mechanism of creation/destruction of the $X(3872)$, \ie the interaction of this meson in the hadron resonance gas, leads to a yield for an eventual molecular state which is a factor of $\sim 18$ larger than that for a compact four-quark structure.

\vspace{1em}

It should be noted that the previous discussion completely neglects transverse flow effects. However, this phenomena turned out to be the key ingredient to explain some puzzling experimental results like, for example, the observation made at RHIC that medium-induced suppression for values of the transverse momentum $p_T\simeq2$ GeV is not as effective on protons as it is for pions\cite{Adler:2003ii,Adler:2003cb}. Particularly surprising was the fact that the ratio $p/\pi^+$ of protons over charged pions for transverse momenta above $2$ GeV reaches or even exceeds unity. 
The explanation to this phenomenon was the following. The expectation from the use of coalescence model is that it is less likely to bind states made by a larger number of components simply because the convolution of their wave functions is smaller. In other words, it is hard to find the components at small value of the relative momentum \emph{and} at small relative positions. However, if one takes into account flow effects it turns out that, exactly because of collectivity, if the previously mentioned components are found close in momentum space, they are also likely to be close in coordinate space, thus increasing the yield of states with higher number of constituents. As we said, this effect turned out to be relevant at explaining the observed $p/\pi^+$ ratio for $p_T\simeq2$ GeV.

In our case, this phenomenon might be quite important at increasing the number of tetraquarks produced with respect to the number of molecules~\cite{concarlos,concarlos2}.  Therefore, in our view, the conclusions drawn in Fig.~\ref{fig:Xcoal} are merely  partial.

\vspace{1em}

In conclusion, we showed how the study of the yield of $X(3872)$ -- and possibly of other exotic mesons -- in heavy ion collisions might have an impact on the determination of its nature. In particular, it turns out that, according to the coalescence model\cite{Sato:1981ez}, the number of $X$ produced should be much larger if it is a loosely bound molecule than if it is compact tetraquark, in strking contrast to what expected for $pp$ collisions. Moreover, the predicted yields\cite{Cho:2010db,Cho:2013rpa} are large enough to be measurable by the current experimental facilities, RHIC and LHC. Future studies should try to include collective flow effects as well.

\section{Comparisons} \label{sec:comparison}

In this section we would like to recollect the various predictions made by different models in a way as complete as possible to the best of our knowledge. Such predictions will include both production and decay of the exotic mesons considered in this review. Moreover, we would like to stress how, in our opinion, because of the unavoidable amount of assumptions/approximations made by every model, the following results should mainly be considered as order of magnitude estimates, even when statistical uncertainties are quoted. In \tablename{s~\ref{tab:riassuntone}} and \ref{tab:riassuntone2} we report the predictions for many quantities and different models, together with the references where they can be found.

\begin{table}[th]
\centering
\tbl{Recollection of predictions for different models and different channels involving both the production and decay of exotic $XYZ$ states. We remind to the reader that NREFT stands for Non-Relativistic Effective Field Theory and that NREFT-II stands for Non-Relativistic Effective Field Theory Type II -- see Sec.~\ref{sec:NREFT}.}
{\footnotesize \begin{tabular}{cccc}
\hline\hline
Quantity & Model & Ref. & Prediction  \\
\hline\hline
$\Gamma(\psi(4040)\to\gamma X)$ & Mol. (NREFT) (Sec.~\ref{sec:NREFT}) &\cite{Guo:2013zbw} & $<0.25$~keV \\
$\Gamma(\psi(4415)\to\gamma X)$ & Mol. (NREFT) (Sec.~\ref{sec:NREFT})&\cite{Guo:2013zbw} & $<0.63$ keV \\
$\Gamma(\psi(4160)\to\gamma X)$ & Mol. (NREFT) (Sec.~\ref{sec:NREFT})&\cite{Guo:2013zbw} & $<3$ keV \\
$\Gamma(Y(4260)\to\gamma X)$ & Mol. (NREFT) (Sec.~\ref{sec:NREFT})&\cite{Guo:2013zbw} & $\simeq20$ keV \\
\hline
$\Gamma(Z_b^\prime\to\Upsilon(1S)\pi)/\Gamma(Z_b\to\Upsilon(1S)\pi)$ & Mol. (NREFT) (Sec.~\ref{sec:NREFT})&\cite{Cleven:2013sq} & $\simeq0.7$ \\
$\Gamma(Z_b^\prime\to\Upsilon(2S)\pi)/\Gamma(Z_b\to\Upsilon(2S)\pi)$ & Mol. (NREFT) (Sec.~\ref{sec:NREFT})&\cite{Cleven:2013sq} & $\simeq0.9$ \\
$\Gamma(Z_b^\prime\to\Upsilon(3S)\pi)/\Gamma(Z_b\to\Upsilon(3S)\pi)$ & Mol. (NREFT) (Sec.~\ref{sec:NREFT})&\cite{Cleven:2013sq} & $2\pm2$ \\
$\Gamma(Z_b^\prime\to h_b(1P)\pi)/\Gamma(Z_b\to h_b(1P)\pi)$ & Mol. (NREFT) (Sec.~\ref{sec:NREFT})&\cite{Cleven:2013sq} & $\simeq0.8$ \\
$\Gamma(Z_b^\prime\to h_b(2P)\pi)/\Gamma(Z_b\to h_b(2P)\pi)$ & Mol. (NREFT) (Sec.~\ref{sec:NREFT})&\cite{Cleven:2013sq} & $1.0\pm0.4$ \\
$\Gamma(Z_b\to\chi_{b0}(1P)\gamma)/\Gamma(Z_b\to h_b(1P)\pi)$ & Mol. (NREFT) (Sec.~\ref{sec:NREFT})&\cite{Cleven:2013sq} & $\simeq5\times10^{-3}$ \\
$\Gamma(Z_b\to\chi_{b1}(1P)\gamma)/\Gamma(Z_b\to h_b(1P)\pi)$ & Mol. (NREFT) (Sec.~\ref{sec:NREFT})&\cite{Cleven:2013sq} & $\simeq1\times10^{-2}$ \\
$\Gamma(Z_b\to\chi_{b2}(1P)\gamma)/\Gamma(Z_b\to h_b(1P)\pi)$ & Mol. (NREFT) (Sec.~\ref{sec:NREFT})&\cite{Cleven:2013sq} & $\simeq2\times10^{-2}$ \\
$\Gamma(Z_b\to\chi_{b0}(2P)\gamma)/\Gamma(Z_b\to h_b(2P)\pi)$ & Mol. (NREFT) (Sec.~\ref{sec:NREFT})&\cite{Cleven:2013sq} & $(6.3\pm1.8)\times10^{-3}$ \\
$\Gamma(Z_b\to\chi_{b1}(2P)\gamma)/\Gamma(Z_b\to h_b(2P)\pi)$ & Mol. (NREFT) (Sec.~\ref{sec:NREFT})&\cite{Cleven:2013sq} & $(1.3\pm0.4)\times10^{-2}$ \\
$\Gamma(Z_b\to\chi_{b2}(2P)\gamma)/\Gamma(Z_b\to h_b(2P)\pi)$ & Mol. (NREFT) (Sec.~\ref{sec:NREFT})&\cite{Cleven:2013sq} & $(1.9\pm0.5)\times10^{-2}$ \\
$\Gamma(Z_b^\prime\to\chi_{b0}(1P)\gamma)/\Gamma(Z_b^\prime\to h_b(1P)\pi)$ & Mol. (NREFT) (Sec.~\ref{sec:NREFT})&\cite{Cleven:2013sq} & $\simeq4\times10^{-3}$ \\
$\Gamma(Z_b^\prime\to\chi_{b1}(1P)\gamma)/\Gamma(Z_b^\prime\to h_b(1P)\pi)$ & Mol. (NREFT) (Sec.~\ref{sec:NREFT})&\cite{Cleven:2013sq} & $\simeq1\times10^{-2}$ \\
$\Gamma(Z_b^\prime\to\chi_{b2}(1P)\gamma)/\Gamma(Z_b^\prime\to h_b(1P)\pi)$ & Mol. (NREFT) (Sec.~\ref{sec:NREFT})&\cite{Cleven:2013sq} & $\simeq2\times10^{-2}$ \\
$\Gamma(Z_b^\prime\to\chi_{b0}(2P)\gamma)/\Gamma(Z_b^\prime\to h_b(2P)\pi)$ & Mol. (NREFT) (Sec.~\ref{sec:NREFT})&\cite{Cleven:2013sq} & $(4.2\pm1.2)\times10^{-3}$ \\
$\Gamma(Z_b^\prime\to\chi_{b1}(2P)\gamma)/\Gamma(Z_b^\prime\to h_b(2P)\pi)$ & Mol. (NREFT) (Sec.~\ref{sec:NREFT})&\cite{Cleven:2013sq} & $(1.3\pm0.4)\times10^{-2}$ \\
$\Gamma(Z_b^\prime\to\chi_{b2}(2P)\gamma)/\Gamma(Z_b^\prime\to h_b(2P)\pi)$ & Mol. (NREFT) (Sec.~\ref{sec:NREFT})&\cite{Cleven:2013sq} & $(1.8\pm0.5)\times10^{-2}$ \\
\hline
$\Gamma(X\to D^0\bar D^0\pi^0)+\Gamma(X\to D^0\bar D^0\gamma)$ & Mol. (Low-energy univ.) (Sec.~\ref{sec:lowenergy}) &\cite{Braaten:2003he} & $\simeq74\pm14$ keV \\
$\Gamma(X\to D^+D^-\gamma)$ & Mol. (Low-energy univ.) (Sec.~\ref{sec:lowenergy}) &\cite{Braaten:2003he} & suppressed as $1/a$ \\
$\Gamma(X\to \psiprime\gamma)$ & Mol. (Low-energy univ.) (Sec.~\ref{sec:lowenergy}) &\cite{Braaten:2003he} & suppressed as $1/a$ \\
$\Gamma(X\to\eta_c(2S)\gamma)$ & Mol. (Low-energy univ.) (Sec.~\ref{sec:lowenergy}) &\cite{Braaten:2003he} & suppressed as $1/a$ \\
$\Gamma(X\to p\bar p)$ & Mol. (Low-energy univ.) (Sec.~\ref{sec:lowenergy}) &\cite{Braaten:2003he} & suppressed as $1/a$ \\
\hline
$\Gamma(X\to D^0\bar D^0\pi^0)$ & Mol. (NREFT-II) (Sec.~\ref{sec:NREFT}) &\cite{Fleming:2007rp} & $\simeq45$ keV \\
\hline
$N_{X(3872)}^\text{coal.}/N_{X(3872)}^\text{stat.}$ & Mol. in Heavy-Ion coll. (Sec.~\ref{subsec:HI}) &\cite{Cho:2013rpa} & $ \simeq 2$ \\
$N_{X(3872)}^\text{coal.}/N_{X(3872)}^\text{stat.}$ & Tetraq. in Heavy-Ion coll. (Sec.~\ref{subsec:HI}) &\cite{Cho:2013rpa} & $\simeq 10^{-1}$ \\
\hline\hline
\end{tabular} } \label{tab:riassuntone}
\end{table}

\begin{table}[th]
\centering
\tbl{{\it (Continued).} We remind to the reader that QCDSR stands for QCD Sum Rules. Moreover, with ``no dynamics'' and ``dynamics incl.'' we refer to the exclusion or inclusion of the recent dynamical model developed by Brodsky \etal\cite{Brodsky:2014xia} to compute the tetraquark effective couplings to quarkonia.}
{\footnotesize \begin{tabular}{cccc}
\hline\hline
Quantity & Model & Ref. & Prediction  \\
\hline\hline
$\Gamma(Z_c\to\eta_c\rho)/\Gamma(Z_c\to \jpsi\pi)$ & Mol. (NREFT) (Sec.~\ref{sec:NREFT}) &\cite{Esposito:2014hsa} & $0.053\pm0.011$ \\
$\Gamma(Z_c^\prime\to\eta_c\rho)/\Gamma(Z_c^\prime\to h_c\pi)$ & Mol. (NREFT) (Sec.~\ref{sec:NREFT}) &\cite{Esposito:2014hsa} & $0.012\pm0.002$ \\
$\Gamma(Z_c\to h_c\pi)/\Gamma(Z_c^\prime\to h_c\pi)$ & Mol. (NREFT) (Sec.~\ref{sec:NREFT}) &\cite{Esposito:2014hsa} & $1.46\pm0.30$ \\
$\Gamma(Z_c\to \jpsi\pi)/\Gamma(Z_c^\prime\to \jpsi\pi)$ & Mol. (NREFT) (Sec.~\ref{sec:NREFT}) &\cite{Esposito:2014hsa} & $1.71\pm0.68$ \\
$\Gamma(Z_c\to \eta_c\rho)/\Gamma(Z_c\to \jpsi\pi)$ & Tetraq. (Type-I no dynam.) (Sec.~\ref{sec:amptetra}) &\cite{Esposito:2014hsa} & $\simeq560$ \\
$\Gamma(Z_c\to \eta_c\rho)/\Gamma(Z_c\to \jpsi\pi)$ & Tetraq. (Type-II no dynam.) (Sec.~\ref{sec:amptetra}) &\cite{Esposito:2014hsa} & $\simeq 0.68$ \\
$\Gamma(Z_c\to \eta_c\rho)/\Gamma(Z_c\to \jpsi\pi)$ & Tetraq. (Type-I dynam. incl.) (Sec.~\ref{sec:amptetra}) &\cite{Esposito:2014hsa} & $\simeq 390$ \\
$\Gamma(Z_c\to \eta_c\rho)/\Gamma(Z_c\to \jpsi\pi)$ & Tetraq. (Type-II dynam. incl.) (Sec.~\ref{sec:amptetra}) &\cite{Esposito:2014hsa} & $\simeq 0.48$ \\
$\Gamma(Z_c^\prime\to\eta_c\rho)/\Gamma(Z_c^\prime \to h_c\pi)$ & Tetraq. (Type-I no dynam.) (Sec.~\ref{sec:amptetra}) &\cite{Esposito:2014hsa} & $\simeq200$ \\
$\Gamma(Z_c^\prime\to\eta_c\rho)/\Gamma(Z_c^\prime \to h_c\pi)$ & Tetraq. (Type-II no dynam.) (Sec.~\ref{sec:amptetra}) &\cite{Esposito:2014hsa} & $\simeq200$ \\
$\Gamma(Z_c^\prime\to\eta_c\rho)/\Gamma(Z_c^\prime \to h_c\pi)$ & Tetraq. (Type-I dynam. incl.) (Sec.~\ref{sec:amptetra}) &\cite{Esposito:2014hsa} & $\simeq36$ \\
$\Gamma(Z_c^\prime\to\eta_c\rho)/\Gamma(Z_c^\prime \to h_c\pi)$ & Tetraq. (Type-II dynam. incl.) (Sec.~\ref{sec:amptetra}) &\cite{Esposito:2014hsa} & $\simeq36$ \\
\hline
$\mathcal{BR}(B_c^+\to\mathcal{T}_s^{++}(\text{good})D^-)$ & Tetraq. (no dynam.) (Sec.~\ref{subsec:cc}) &\cite{Esposito:2013fma} & $\simeq10^{-6}$ (near thresh.) \\
$\mathcal{BR}(B_c^+\to\mathcal{T}_s^{++}(\text{good})D^{*-})$ & Tetraq. (no dynam.) (Sec.~\ref{subsec:cc}) &\cite{Esposito:2013fma} & $\simeq10^{-4}$ (near thresh.) \\
$\mathcal{BR}(B_c^+\to\mathcal{T}_s^{++}(\text{bad})D^-)$ & Tetraq. (no dynam.) (Sec.~\ref{subsec:cc}) &\cite{Esposito:2013fma} & $\simeq5\times10^{-5}$ (near thresh.) \\
$\mathcal{BR}(B_c^+\to\mathcal{T}_s^{++}(\text{bad})D^{*-})$ & Tetraq. (no dynam.) (Sec.~\ref{subsec:cc}) &\cite{Esposito:2013fma} & $\simeq5\times10^{-6}$ (near thresh.) \\
\hline
$\Gamma(Y(4260)\to \jpsi\pi\pi)$ & Hadro-charm. (Sec.~\ref{sec:hadroq}) &\cite{Dubynskiy:2008mq} & $\simeq$ few MeV \\
$\Gamma(Y(4360)\to \psiprime\pi\pi)$ & Hadro-charm. (Sec.~\ref{sec:hadroq}) &\cite{Dubynskiy:2008mq} & $\simeq$ few MeV \\
$\Gamma(Y(4660)\to \psiprime\pi\pi)$ & Hadro-charm. (Sec.~\ref{sec:hadroq}) &\cite{Dubynskiy:2008mq} & $\simeq$ few MeV \\
$\Gamma(Z(4430)\to \psiprime\pi\pi)$ & Hadro-charm. (Sec.~\ref{sec:hadroq}) &\cite{Dubynskiy:2008mq} & $\simeq$ few MeV \\
\hline
$\Gamma(X\to \jpsi\pi\pi\pi)/\Gamma(X\to \jpsi \pi\pi)$ & Pure mol. or tetraq. (QCDSR) (Sec.~\ref{sec:QCDSR}) &\cite{Matheus:2009vq} & $\simeq0.15$ \\
$\Gamma(X\to \jpsi\pi\pi\pi)$ & Pure tetraq. (QCDSR) (Sec.~\ref{sec:QCDSR}) &\cite{Matheus:2009vq} & $(50\pm15)$ MeV \\
$\Gamma(X\to \jpsi\pi\pi\pi)$ & Mix. of mol. and $c\bar c$ (QCDSR) (Sec.~\ref{sec:QCDSR}) &\cite{Matheus:2009vq} & $(9.3\pm6.9)$ MeV \\
\hline
$\Gamma(X\to \jpsi\gamma)/\Gamma(X\to \jpsi\pi\pi)$ & Mix. of mol. and $c\bar c$ (QCDSR) (Sec.~\ref{sec:QCDSR}) &\cite{Nielsen:2010ij} & $0.19\pm0.13$ \\
\hline
$\Gamma(Z_c^+\to \jpsi\pi^+)$ & Pure tetraq. (QCDSR) (Sec.~\ref{sec:QCDSR}) &\cite{Navarra:2014aba} & $(29.1\pm8.2)$ MeV \\
$\Gamma(Z_c^+\to \eta_c\rho^+)$ & Pure tetraq. (QCDSR) (Sec.~\ref{sec:QCDSR}) &\cite{Navarra:2014aba} & $(27.5\pm8.5)$ MeV \\
$\Gamma(Z_c^+\to D^+\bar D^{*0})$ & Pure tetraq. (QCDSR) (Sec.~\ref{sec:QCDSR}) &\cite{Navarra:2014aba} & $(3.2\pm0.7)$ MeV \\
$\Gamma(Z_c^+\to \bar D^0 D^{*+})$ & Pure tetraq. (QCDSR) (Sec.~\ref{sec:QCDSR}) &\cite{Navarra:2014aba} & $(3.2\pm0.7)$ MeV \\
\hline
$\Gamma(Z_b^+\to\Upsilon\pi^+)$ & Pure tetraq. (QCDSR) (Sec.~\ref{sec:QCDSR}) &\cite{Wang:2013zra} & $4.77^{+3.27}_{-2.46}$ MeV \\
$\Gamma(Z_b^{\prime+}\to\eta_b\rho^+)$ & Pure tetraq. (QCDSR) (Sec.~\ref{sec:QCDSR}) &\cite{Wang:2013zra} & $13.52^{+8.89}_{-6.93}$ MeV \\
\hline\hline
\end{tabular} } \label{tab:riassuntone2}
\end{table}

\section{Conclusions}
The field of $XYZ$ phenomenology has  impressively been growing on the experimental side. On the other hand it seems that
the theoretical models to explain the rich amount of information nowadays available on these states are  lagging behind. One of the reasons for this is
maybe due to the narrow number of active researcher in the field, especially those culturally close to high-energy physics. $XYZ$ particles are observed in high-energy reactions, at high transverse momenta, as we know from the recent findings by LHCb, CMS and, even more recently, by  ATLAS. For these reasons  they are unlikely to be the manifestations of some  nuclear force type dynamics. We are also very skeptical about  interpreting  most of them in terms of kinematical effects, as claimed by some authors. 

In this review we have tried to highlight the reasons for the most fundamental quark picture, suggesting these states to be new kinds of hadrons with respect to standard mesons and baryons, namely new bodyplans of quarks arranged into tetraquarks. The possibility of having long-lived tetraquarks is not excluded by the large number of colors limit of QCD and, in addition, some of the observed charged resonances  appear as striking evidence that compact tetraquarks have  already   been {\it observed}. The simplified diquark-antidiquark model reviewed in this paper is not the definitive explanation of the $XYZ$ resonances, but we believe  it must be at the core of the picture.
One still has  to explain what prevents some of the states predicted by that model to be formed/observed in experiment. We started by looking for the origin of such selection rules in the {\it accidental} matchings of diquark-antidiquark levels with open charm (beauty) meson thresholds. We do not think that there is anything profound 
in these matchings, given the huge number of thresholds which can be formed with the known pairs of charmed or beauty mesons. However tetraquark discrete levels might correspond to narrow hadron resonances whenever anyone of these matchings happens to be realized.
This is probably the passage to be done to fill the gap between the tetraquark interpretation and the actual phenomenology of these resonances.    

In this respects it may be true that we are looking at different aspects of the same problem using different descriptions, as considered by someone. It may also be  true that the definitive answer could be in the non-perturbative intricacies of strong interactions which might eventually be captured only using  lattice QCD methods. Yet, in the meanwhile, we think that we made some interesting and solid progress, and especially  we believe that there is indeed a problem in $XYZ$ resonance physics: whatever side you are looking at it, it is not a ``mirage''  made of effects, cusps, particular cases. It looks like there is a pattern behind.

\section*{Acknowledgements}
\addcontentsline{toc}{section}{Acknowledgements}
We wish to thank R.~Faccini, G.~Filaci, B.~Grinstein, F.~Renga, M.~Papinutto, C.~Sabelli, N.~Tantalo, M.~Testa and especially L.~Maiani and V.~Riquer-Ramirez for many stimulating discussions and a very fruitful collaboration. 

\clearpage
\addcontentsline{toc}{section}{References}
\bibliographystyle{ws-ijmpa}
\bibliography{tetra}
\end{document}